\documentclass[%
 rsi,
 amsmath,amssymb,
 reprint,%
]{revtex4-1}
\usepackage{rotating}
\usepackage{printlen}
\usepackage{braket}
\usepackage{graphicx}
\usepackage{dcolumn}
\usepackage{bm}
\usepackage[dvipsnames]{xcolor}
\usepackage{color, colortbl}
\usepackage[utf8]{inputenc}
\usepackage[T1]{fontenc}
\usepackage{mathptmx}
\usepackage{array,multirow,graphicx}
\usepackage{longtable}
\usepackage{dcolumn}
\usepackage{soul}

\newcommand\clearrow{\global\let\rowmac\relax}
\clearrow

\begin{document}

\preprint{AIP/123-QED}

\title{Entangled Photon-Pair Sources based on three-wave mixing in bulk crystals}

\author{Ali Anwar$^1$}
\author{Chithrabhanu Perumangatt$^1$}
\author{Fabian Steinlechner$^{3,4}$}
\author{Thomas Jennewein$^5$}
\author{Alexander Ling$^{1,2}$}
 \email{alianwarma@gmail.com}
\affiliation{ 
$^1$Centre for Quantum Technologies, National University of Singapore, 3 Science Drive 2, S117543,
Singapore
}%
\affiliation{$^2$Physics Department, National University of Singapore, 2 Science Drive 3, S117542, Singapore}

\affiliation{%
$^3$ Fraunhofer Institute for Applied Optics and Precision Engineering IOF, Albert-Einstein-Straße 7, 07745 Jena, Germany 
}%
\affiliation{%
$^4$ Friedrich Schiller University Jena, Abbe Center of Photonics, Albert-Einstein-Str. 6, 07745 Jena, Germany
}

\affiliation{$^5$Institute of Quantum Computing and Department of Physics and Astronomy, University of Waterloo, 200 University Ave W, Waterloo, ON N2L 3G1, Canada}%

\date{\today}
\begin{abstract}
Entangled photon-pairs are a critical resource in quantum communication protocols ranging from quantum key distribution to teleportation. The current workhorse technique for producing photon-pairs is via spontaneous parametric down conversion (SPDC) in bulk nonlinear crystals. The increased prominence of  quantum networks has led to growing interest in deployable high performance entangled photon-pair sources. This manuscript provides a review of the state-of-the-art for bulk-optics-based SPDC sources with continuous wave pump, and discusses some of the main considerations when building for deployment.
\end{abstract}

 \maketitle

\section{\label{intro}Introduction}
The nonlocal correlations of entangled systems play a central role  in quantum technologies across domains as varied as communication, sensing and computation. Generating entangled states in a reliable and reproducible manner while pushing the limits of repetition rates and quality of entanglement can thus be considered a key prerequisite towards practical applications.

From an applications perspective, even the most basic form of entanglement - bi-partite entanglement in two-level quantum systems - promises to play a major role in quantum information science and technology. In the context of quantum communications for example (but also for some scenarios in sensing and information processing), entangled photon-pairs which can readily be distributed to remote parties are a basic building block for many envisioned applications.

The last two decades have seen significant efforts dedicated to the proof-of-concept demonstration and then refinement of methods for photon-pair generation on a variety of  technological platforms. Technologies ranging from atom- and ion-based systems to engineered defects in solid-state devices and 2-D materials are steadily improving in performance such that the ultimate goal of scalable and on-demand emission of single- and multi-photon states is well within reach. At present, however, the most widely employed approach remain probabilistic photon-pair sources based on \textit{spontaneous} wave-mixing processes in nonlinear optical materials. Decades of research in  nonlinear optics have provided a vast set of tools for engineering wave-mixing processes,  resulting in highly efficient and versatile photon-pair generation schemes for entanglement in a variety of photonic degrees of freedom.  

The most common wave-mixing techniques employed in photon-pair sources are based on accessing either the second order ($\chi^{(2)}$) or third order ($\chi^{(3)}$) optical nonlinearities, which lead to the three-wave or four-wave mixing regimes, respectively. Higher-order wave mixing for direct generation of photon multiplets has been studied and proposed, for instance in \cite{Corona:11}, yet is not commonly employed in the optical regime due to the low conversion rates but has recently been achieved in microwave systems~\cite{PhysRevApplied.10.044019}. Note however, that equivalent higher-order nonlinear interactions have already been realized via cascading of $\chi^{(2)}$ nonlinearities \cite{Huebel2010}.

The major differences between three- and four-wave mixing is in the magnitude of the optical nonlinearity governing the process as well as the number of fields involved in the process. The $\chi^{(3)}$ nonlinearity on which the spontaneous four-wave mixing (SFWM) process occurs is very weak and efficient SFWM sources are thus most commonly implemented in guided wave systems such as integrated photonics ring resonators,  pulsed pump in single mode fibers, or highly nonlinear photonic crystal fibers \cite{fulconis2005high}. Even higher nonlinear interactions may be achieved via mode-confinement in waveguides, which significantly improves the rate of photon-pair production, but also limits the degrees-of-freedom accessible. Spontaneous three-wave mixing processes (STWM), on the other hand,  have much stronger optical nonlinearity and are thus routinely achieved  within both bulk and waveguide optical systems.

An important feature of three-wave mixing in bulk material is the access to various degrees-of-freedom and this has been utilised to demonstrate entanglement in polarization \cite{kwiattype2,kwiattype1dual}, spatial mode \cite{krenn2017orbital,forbes2019quantum}, time \cite{brendel1999pulsed} and frequency \cite{avenhaus2009experimental}. Moreover, these sources have been shown to provide  entanglement  across degrees-of-freedom simultaneously in a so-called hyper-entangled state \cite{fujiwara:261103,Ma:2009}. As a result of this flexibility, the spontaneous three-wave mixing process, commonly referred to as Spontaneous Parametric Down Conversion (SPDC), is a widely used tool in quantum information science implementations. 

The supporting technology needed for implementing SPDC are readily obtainable and setups work reliably at room temperature. The components can be easily modified to explore multiple applications. For these reasons, bulk-based SPDC systems have a higher technological maturity when compared to waveguide based systems \cite{meyer2018high,orieux2017semiconductor,huber2018semiconductor}.

In this review article, we focus mainly on the development of entangled photon sources based on the process of SPDC in bulk nonlinear materials, and make a brief comparison to waveguided SPDC.  For a more detailed discussion of nonlinear quantum optics on integrated platforms, we refer the interested reader to the comprehensive review articles, see e.g. \cite{vergyris2017fully}. Furthermore, we will primarily discuss continuous wave pumped SPDC, where  the pump is generally considered a monochromatic wave.  In particular, SPDC pumped with ultra-fast laser pulses is widely used in studies of multi-partite photonic entanglement  where the sub-coherence pumping \cite{Zukowski93a} allows the coherent interaction between multiple photon-pairs. This technique has been  used for instance in the implementation of four-photon entanglement \cite{Pan01b}, or quantum state teleportation \cite{bouwmeester1997experimental}, and we refer the reader to the following review article \cite{pan2012multiphoton}.
We begin with a description on the principles governing SPDC production, followed by identifying the relevant performance parameters for an entangled photon source (section \ref{theory}). We provide a detailed description of polarization entangled photon sources in different phase-matching geometries in section \ref{polarizationentlpairsources}. In section \ref{entlotherdof}, we discuss the progress of entangled photon sources in other degrees of freedom. Finally, our review will conclude with a discussion of the main application areas for entangled photon sources and an outlook (section \ref{applications}). 

\section{\label{theory}Spontaneous Parametric Down Conversion}

\begin{figure}[h]
	\centering
	\includegraphics[width=0.48\textwidth]{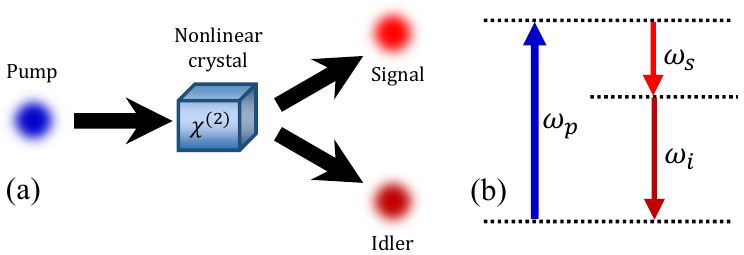}
	\caption{(a) Generation of photon-pairs in the SPDC process. A \textit{pump} photon of frequency $\omega_p$ decays to two photons of frequencies $\omega_s$ and $\omega_i$, known as \textit{signal} and \textit{idler} respectively. (b) Illustration of energy conservation in SPDC process.}
	\label{SPDC-process}
\end{figure}

The response of a dielectric medium to an electromagnetic field is given in terms of polarization of the material written as a sum of electric fields $E_j\ (j=1,2,3,..)$
\begin{equation}
    P=\epsilon_0(\chi^{(1)}E_1+\chi^{(2)}E_1E_2+\chi^{(3)}E_1E_2E_3+...)
    \label{NLpolarization}
\end{equation}
where $\epsilon_0$ is the vacuum electric permittivity, $\chi^{(1)}$ is the linear susceptibility and $\chi^{(2)},\chi^{(3)},...$ are the nonlinear susceptibilities of the medium. The second order ($\chi^{(2)}$) process enables three electromagnetic fields to interact in a non-centrosymmetric medium leading to a transfer of energies between the fields. This is known as three-wave mixing. The specific case where the two lower energy fields are initially vacuum modes is known as Spontaneous Parametric Down Conversion and is illustrated in Fig. \ref{SPDC-process}.
The process is \textit{spontaneous} as there are no input signal and idler photons to stimulate the process, but only the inherent vacuum fluctuations of the modes. SPDC is therefore a direct manifestation of the quantized description of the electro-magnetic fields. The term parametric refers to the fact that the interaction medium does not add or subtract energy or momentum in the process. 

\subsection{\label{hamiltonian]}The Interaction Hamiltonian}

 The Hamiltonian governing SPDC inside an nonlinear interaction volume can be written as \cite{mandel1995optical,glauber2007quantum,ou2007multi,torres2011engineering}:
\begin{equation}
    \mathcal{H}_I=\epsilon_0\int d^3\mathbf{r}\ \chi^{(2)}\ \mathbf{E}_p^{(+)}(\mathbf{r},t)\mathbf{\hat{E}}_s^{(-)}(\mathbf{r},t)\mathbf{\hat{E}}_i^{(-)}(\mathbf{r},t)+\text{H.c.}
    \label{interaction-hamiltonian}
\end{equation}

The interaction of the pump and target fields is commonly expressed using the non-depleted pump approximation. This is appropriate as the $\chi^{(2)}$ susceptibility has a very small magnitude, and most of the pump energy is not down converted.
The non-depleted pump field with wave-vector $\mathbf{k}_p$ and frequency $\omega_p$ may be expressed classically as
\begin{equation}
    \mathbf{E}_p(\mathbf{r},t)=\mathbf{e}_p\int d\omega_p\ d\mathbf{k}_p^\perp\ N_p\ A_p(\mathbf{k}_p^\perp)S(\omega_p)\ e^{i(\mathbf{k}_p\cdot \mathbf{r}-\omega_pt)}+\text{c.c}
    \label{Ep}
\end{equation}
where $N_p$=$\sqrt{\hbar\omega_p/(2\epsilon_0cn_p(\omega_p))}$ is the normalization factor, $A_p(\mathbf{k}_p^\perp)$ is the spatial mode of the pump in transverse momentum coordinates $\mathbf{k}_p^\perp\equiv(k_{px},k_{py})$, $S(\omega_p)$ is the spectral function and $\mathbf{e}_p$ is the polarization vector of the pump. This expression is  generally known as the squeeze-operator \cite{Loudon83a}, and SPDC is considered as the result of the squeeze-operator acting on vacuum modes, and thus the output states of SPDC are also referred to as {\it squeezed vacuum}. The electric field operators for the signal and idler photons have the form
\begin{equation}
    \mathbf{\hat{E}}_{s,i}(\mathbf{r},t)=\frac{\mathbf{e}_{s,i}}{(2\pi)^{3/2}}\int d\omega_{s,i}\ d\mathbf{k}_{s,i}^\perp\ N_{s,i} e^{i(\mathbf{k}_{s,i}\cdot \mathbf{r}-\omega_{s,i}t)}\hat{a}_{k_{s,i}}+\text{c.c}
    \label{Esi}
\end{equation}
where $N_{s,i}$=$\sqrt{\hbar\omega_{s,i}/(2\epsilon_0cn_{s,i}(\omega_{s,i}))}$ is the normalization factor and $n_{s,i}(\omega_{s,i})$, $\mathbf{e}_{s,i}$ are the respective refractive indices and polarization vectors. The quantum state of SPDC output at time $t$ is then given by
\begin{equation}
    \ket{\Psi(t)}=\exp\left[-\frac{i}{\hbar}\int_0^t\mathcal{H}_I(t')\ dt'\right]\ket{0}_s\ket{0}_i
    \label{SPDC-qtm-state}
\end{equation}
where $\ket{0}_s\ket{0}_i$ is the initial vacuum state. The Taylor series expansion of the above equation suggests that multi-pair events in SPDC are possible \cite{takesue2010effects,takeoka2015full,tsujimoto2018optimal}, with a strong dependence on pump power. The probability of single pair emission is obtained by assuming the weak-pump regime, which allows to truncate the series expansion to first order, 
\begin{equation}
    \ket{\Psi(t)}\propto\int_0^t\mathcal{H}_I(t')\ dt'\ \ket{0}_s\ket{0}_i.
    \label{SPDC-biphoton-qtm-state}
\end{equation}
Keep in mind that the higher-order emissions (multi-pair emissions) of SPDC must  be considered in some scenarios as they could reduce the entanglement quality at sufficiently large pump intensities. The general description of the bi-photon  state emitted by SPDC (or  squeezed vacuum) is \cite{Loudon83a, RevModPhys.77.513}
\begin{eqnarray}
    \ket{\Psi} &=& \sqrt{1 - \lambda} \biggl(\ket{0}_s\ket{0}_i + \lambda^{1/2}\ket{1}_s\ket{1}_i + \nonumber\\ 
    & & + \lambda^{2/2}\ket{2}_s\ket{2}_i +  
     \lambda^{3/2}\ket{3}_s\ket{3}_i + ... \biggr),
\end{eqnarray}
where $\ket{n}$ are the photon number states in the signal and idler modes, $\lambda = \tanh^2 r$, and $r$ is the interaction parameter in the Hamiltonian. In many applications of SPDC pair sources the truncation to single-pairs is a valid approach because the interaction strength is chosen relatively low, e.g. $\lambda<0.01$, less than 1\% of all emission events contain more than one photon-pair.

The integral in \eqref{SPDC-biphoton-qtm-state} can be evaluated by decomposing the respective wave-vectors into transverse and longitudinal components. The integral is explicitly written as
\begin{align}
    \ket{\Psi(t)}\propto& \frac{\epsilon_0\mathbf{e}_p\chi^{(2)}\mathbf{e}_s\mathbf{e}_i}{(2\pi)^3}\int d\omega_s\ d\omega_i\ d\mathbf{k}_s^\perp d\mathbf{k}_i^\perp \nonumber \\
    &\times\biggl[A_p(\mathbf{k}_p^\perp)S(\omega_p)N_pN_sN_i\int_0^tdt\ e^{i(\omega_p-\omega_s-\omega_i)t} \nonumber \\
    &\quad\times\int_{-\infty}^\infty dx\ e^{i(k_{px}-k_{sx}-k_{ix})x}\int_{-\infty}^\infty dy\ e^{i(k_{py}-k_{sy}-k_{iy})y} \nonumber\\
    &\qquad\times\int_{-L/2}^{L/2}dz\ e^{i(k_{pz}-k_{sz}-k_{iz})z}\biggr]\ket{\mathbf{k}_s^\perp,\omega_s}\ket{\mathbf{k}_i^\perp,\omega_i}
    \label{full-SPDC-integral-expression}
\end{align}

One simplifying condition is that the pump and SPDC transverse modes are much smaller than the cross-section of the non-linear material, which is usually a crystal. The crystal structure further reduces the susceptibility tensor $\chi^{(2)}$ to a single coefficient $d$ where $2d=\mathbf{e}_p\chi^{(2)}:\mathbf{e}_s\mathbf{e}_i$. 
Over a suitable interaction time, the integral along the crystal volume can be expressed as
\begin{align}
    \ket{\Psi(t)}\propto\int d\omega_s\ d\omega_i\ d\mathbf{k}_s^\perp d\mathbf{k}_i^\perp\ &\Phi(\mathbf{k}_s^\perp,\mathbf{k}_i^\perp,\omega_s,\omega_i)\nonumber\\
    &\times\ket{\mathbf{k}_s^\perp,\omega_s}_s\ket{\mathbf{k}_i^\perp,\omega_i}_i,
    \label{SPDC-biphoton-qtm-state-final}
\end{align}
where $\Phi(\mathbf{k}_s^\perp,\mathbf{k}_i^\perp,\omega_s,\omega_i)$ is the SPDC mode function given by
\begin{equation}
    \Phi(\mathbf{k}_s^\perp,\mathbf{k}_i^\perp,\omega_s,\omega_i)=L\sigma A_p(\mathbf{k}_s^\perp+\mathbf{k}_i^\perp)S(\omega_s+\omega_i)\text{sinc}\left(\frac{\Delta k_zL}{2}\right).
    \label{SPDC-biphoton-modefn}
\end{equation}

The effective nonlinear coefficient is now
\begin{equation}
    \sigma=d\sqrt{\frac{\hbar^3\omega_p\omega_s\omega_i}{32\pi^4\epsilon_0c^3n_p(\omega_p)n_s(\omega_s)n_i(\omega_i)}}
\end{equation}
and $\Delta k_z$=$k_{pz}-k_{sz}-k_{iz}$ is the longitudinal wave-vector mismatch. The mode function in Eqn.\eqref{SPDC-biphoton-modefn}
obeys
energy and momentum conservation :
\begin{align}
    \omega_p&=\omega_s+\omega_i \qquad\ \ \text{Energy} \\
    \label{energy-conserv}
    \mathbf{k}_p^\perp&=\mathbf{k}_s^\perp+\mathbf{k}_i^\perp \qquad \text{Transverse momentum} \\
    k_{pz}&=k_{sz}+k_{iz} \qquad\ \ \text{Longitudinal momentum}
    \label{SPDC-conservation}
\end{align}

The transverse function of the pump and targeted SPDC modes can be modeled in order to optimize brightness and heralding efficiency in practical sources \cite{ling2008absolute,bennink2010optimal}.
Note that for a continuous-wave pump (monochromatic), $S(\omega_s+\omega_i)=\delta(\omega_s+\omega_i)$, which will further simplify the bi-photon mode generated by SPDC, expression~(\ref{SPDC-biphoton-modefn}). If the SPDC pump is a wide-band and/or pulsed laser, the bi-photon mode will have an internal structure and potentially have mixed spectra, which has been extensively studied, see for instance \cite{grice1997spectral,pan2012multiphoton}.



\begin{figure}[h]
	\centering
	\includegraphics[width=0.48\textwidth]{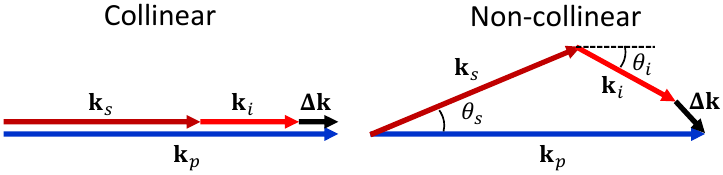}
	\caption{Vector diagrams of collinear and non-collinear phase matching configurations in SPDC sources.}
	\label{colli-noncolli-types}
\end{figure}

Based on phase-matching geometry, SPDC sources have collinear or non-collinear emission of photon-pairs. Figure \ref{colli-noncolli-types} shows the wave-vector diagrams for collinear and non-collinear cases of phase-matching. Here, $\theta_s$ and $\theta_i$ are the angles of signal and idler wave-vector with reference to the pump, which are different for non-degenerate photon-pairs. $\Delta\mathbf{k}$ is the mismatch among the pump, signal and idler wave-vectors, mathematically expressed as $\Delta\mathbf{k}=\mathbf{k}_p-\mathbf{k}_s-\mathbf{k}_i$. This wave-vector mismatch has great significance in designing entangled photon sources.

\subsection{\label{phasematching} Phase matching techniques}

When using bulk crystal SPDC there are two main ways to achieve phase-matching: (a) by exploiting birefringence to compensate for phase mismatch, or (b) using periodic poling of the material which leads to quasi-phase matching (QPM). Typically this involves at least one of the fields propagating at an angle with respect to the optical axis (critical phase matching configuration), and the resulting transverse walk-off limiting the maximum interaction length for the process. The more desirable non-critical phase matching condition, where propagation of all fields is along an axis perpendicular to the optical axis is usually only fulfilled for very specific materials and wavelength combinations. Alternatively, non-critical collinear interaction can also be achieved quasi phase-matching in periodically-poled nonlinear materials. All of these methods try to achieve the momentum conservation condition:
\begin{equation}
\frac{n_p(\lambda_p)}{\lambda_p}=\frac{n_s(\lambda_s)}{\lambda_s}+\frac{n_i(\lambda_i)}{\lambda_i}
\label{crystal_phase_matching_eqn}.
\end{equation}

This condition cannot be achieved with normal dispersion but requires the above mentioned approaches, i.e. either materials exhibiting birefringence, or the use of quasi-phase matching. Note that SPDC in waveguides can also make use of the  propagation parameters of guided modes to achieve modal phase-matching, as will be discussed later. 

In a birefringent optical material, the wave polarized orthogonal to the principal plane is denoted as ordinary ($o$) beam and the wave polarized along the plane is denoted as extra-ordinary ($e$). For commonly used uniaxial crystals the refractive index of the ordinary wave ($n_o$) is independent of the orientation of its wave-vector inside the crystal, whereas the extra-ordinary refractive index depends on the angle $\theta$ between the wave-vector and the optic axis ($n_{\text{eff}}(\theta)$). In sources based on biaxial crystals the orientation of all wave-vectors to the optic axis must be considered \cite{boeuf2000calculating}. This effective extra-ordinary refractive index is expressed as:
\begin{equation}
    \frac{1}{n_{\text{eff}}^2(\theta)}=\frac{\cos^2\theta}{n_o^2}+\frac{\sin^2\theta}{n_e^2}.
    \label{index-ellipsoid-eqn}
\end{equation}

Critical phase matching is then achieved by tuning the angle of the pump wave vector with respect to the optic axis.
There are two possible configurations known as type-I or type-II.
In type-I phase-matching the generated photon-pairs are co-polarized with their polarization orthogonal to that of the pump.
In type-II, the photon-pairs are orthogonally polarized to each other.

The other phase-matching technique is quasi phase-matching (QPM) where there is a spatial modulation of the linear and nonlinear properties in the medium along the propagation direction \cite{fejer1992quasi}. A common QPM method uses periodic flipping of the crystal lattice orientation to achieve an effective nonlinearity. This method makes use of a multi-domain material, unlike in critical phase-matching where the entire material is a single domain with the optical axis fixed in one direction. Moreover, the periodic domains compensate the spatial walk off between the three waves which effectively increases the interaction length \cite{shichijyo2004total}.
\begin{figure}[h]
	\centering
	\includegraphics[width=0.48\textwidth]{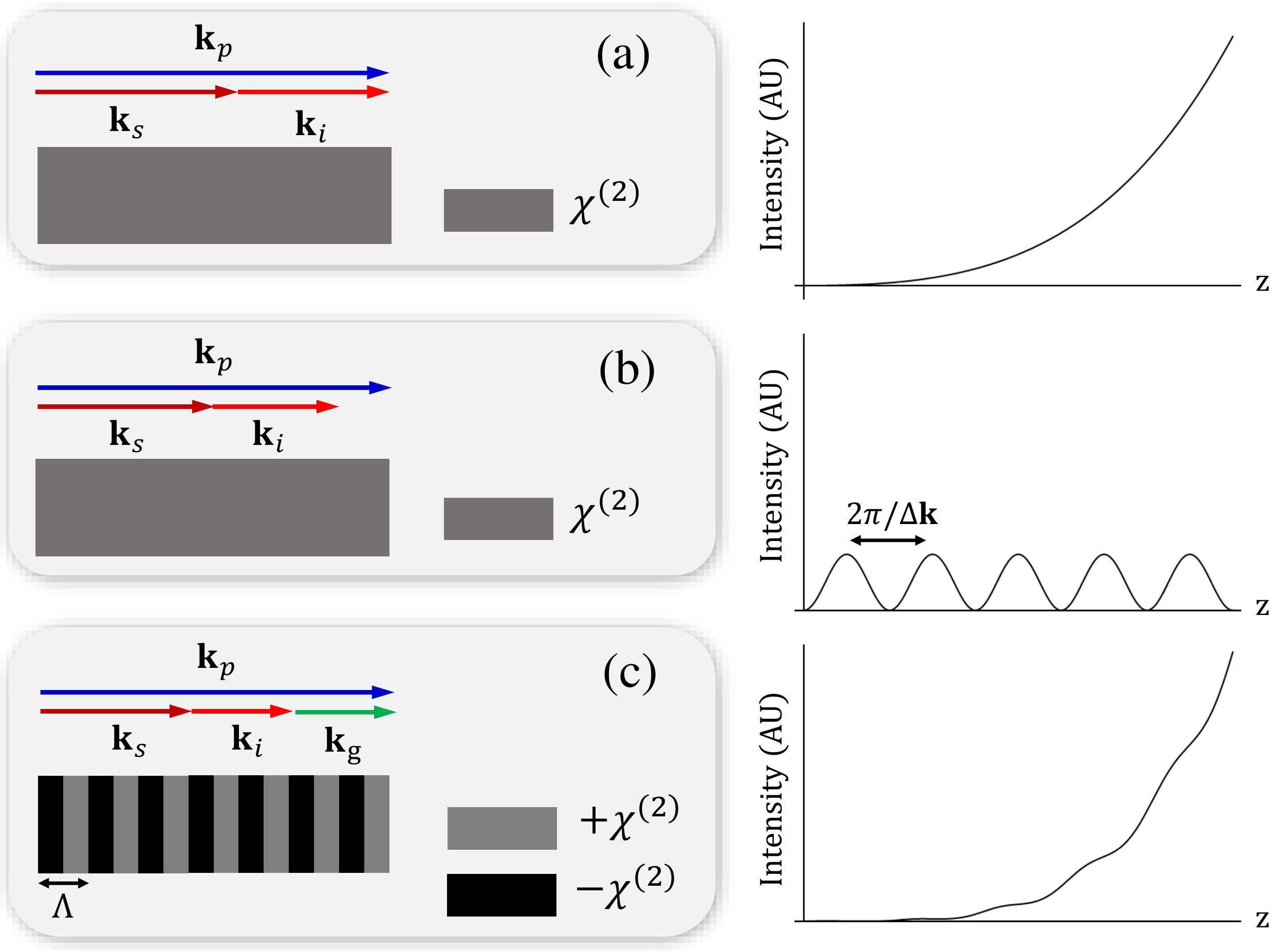}
	\caption{Vector representation (left) and variation of intensity along the nonlinear medium (right) for (a) phase matched, (b) non-phase matched and (c) quasi-phase matched cases, in the weakly-pumped regime.}
	\label{phase-matching-curves}
\end{figure}

Due to this periodic domain structure the intensity of generated photons grows in a step-wise manner along the interaction medium (Fig. \ref{phase-matching-curves}). The phase mismatch in QPM is mathematically expressed as
\begin{equation}
    \Delta\mathbf{k}_{\text{QPM}}=\Delta\mathbf{k}_{\text{bulk}}-\mathbf{k}_\text{g}
    \label{momconserv_qpm}
\end{equation}
Here, $\Delta\mathbf{k}_{\text{bulk}}=\mathbf{k}_p-\mathbf{k}_s-\mathbf{k}_i$, is the phase mismatch of the bulk material. The grating wave-vector $\mathbf{k}_\text{g}$ has a magnitude of $2\pi/\Lambda$, with $\Lambda$ being the grating or poling period. This grating wave-vector allow us to engineer the spectral and spatial properties of the down-converted photon-pairs. Moreover, QPM can allow type-0 phase matching where pump, signal and idler photons are co-polarized.  


\subsection{\label{performparameter}Basic performance parameters for photon-pair sources}

This section discusses the performance parameters commonly used to characterise photon-pair sources.
\vspace{10pt}\\
1. \textit{Brightness}: The brightness of a source is defined as the number of pairs produced in one second for unit pump power typically written as the number of pairs/s/mW. The intrinsic brightness of a source can be  traced back to the nonlinear coefficient of the material which determines the probability that a pump photon will be converted to a photon-pair \cite{ling2008absolute}. It is important to make a distinction between \textit{observed} and \textit{generated} brightness. The observed brightness is highly dependent on heralding efficiency (see below) and other system parameters such as detector efficiency. The generated brightness can often be inferred once the system losses are taken into account. In terms of applications, the observed brightness is often the more relevant parameter. 

The SPDC process is typically a broadband process, and for specific application it is important to understand the \textit{spectral brightness} where the typical unit is pairs/s/nm/mW. This is the brightness divided by the spectral bandwidth of the SPDC output. It should be noted that brightness values reported for SPDC sources often ignores the contribution of multiple pairs, and so is valid only in the weakly pumped regime.
\vspace{10pt}\\
2. \textit{Heralding efficiency}: Here, we define the heralding efficiency as the ratio of observed coincidence counts to the signal/idler counts \footnote{In the context of heralded single photon generation, the heralding efficiency is typically normalized with respect to detection efficiency.}. This parameter, also known as the coincidence-to-singles ratio or Klyshko efficiency, is a very practical and readily accessible performance estimator that depends upon both the efficiency of collection and detection. Note that, while detectors can be changed in principle, the collection efficiency is closely related to the spatial properties of the SPDC emission, specifically the spatial mode distribution of the signal (idler) conditioned on the partner photon coupling into a specific spatial mode \cite{bennink2010optimal}, \footnote{In the context of characterizing the spatial mode distribution in the SPDC process, the concept of conditional spatial mode overlap, may also be considered equivalent to the heralding efficiency.}. The number of photons registered by signal ($s$) and idler ($i$) detectors for a given resolution time are given by \cite{takesue2010effects} 
\begin{align}
    N_s&=\eta_sN \\
    N_i&=\eta_iN,
    \label{singles}
\end{align}
and the rate of coincidence events between the two detectors is
\begin{equation}
    N_c=\eta_s\eta_iN
    \label{coincidences}
\end{equation}
where $N$ is the number of photon-pairs produced in the given resolution time and $\eta_s$, $\eta_i$ are the heralding efficiencies of signal and idler respectively. The heralding efficiencies are expressed as $\eta_s=\xi_s\mu_s$ and $\eta_s=\xi_s\mu_s$, where $\xi_{s,i}$ and $\mu_{s,i}$ are the detection and collection efficiencies.

\subsection{Photonic Entanglement and its experimental measures}
In this article, we mainly focus on two qubit entanglement in different degrees of freedom such as polarization, time bin etc. The target state of such entangled photon-pair sources fall to any of the maximally entangled bell states, given as 
\begin{equation}
\vert \Psi^{\pm}\rangle = \frac{1}{\sqrt{2}}\left(\vert 10\rangle\pm\vert 01\rangle\right); \ 
\vert \Phi^{\pm}\rangle = \frac{1}{\sqrt{2}}\left(\vert 00\rangle\pm\vert 11\rangle\right)
\end{equation}
The $\vert 0\rangle$ and $\vert 1\rangle$ are two orthogonal states. Below we discuss some of the experimentally observable measures of entanglement.

\paragraph{Polarization correlation visibility / contrast:} Entangled photon-pairs exhibits non-local correlation which can be observed from their measurement outcomes. In case of polarization entanglement, signal and idler photons are projected to different linear polarization states
\begin{align}
    \ket{\theta_1\theta_2}\equiv(\cos\theta_1&\ket{H}_1+\sin\theta_1\ket{V}_1)\nonumber\\
    \otimes&(\cos\theta_2\ket{H}_2+\sin\theta_2\ket{V}_2)
\end{align}
The probability of finding the entangled photon-pair in the projected state $\ket{\theta_1\theta_2}$ is given by
\begin{align}
    p(\theta_1,\theta_2)
    =
    \begin{cases}
     \frac{1}{2}\cos^2(\theta_1\mp\theta_2) & \text{for } \ket{\Phi^{\pm}} \\[8pt]
     \frac{1}{2}\sin^2(\theta_1\pm\theta_2) & \text{for } \ket{\Psi^{\pm}}
  \end{cases}
  \label{bellprob}
\end{align}
\begin{figure}[h]
	\centering
	\includegraphics[width=0.48\textwidth]{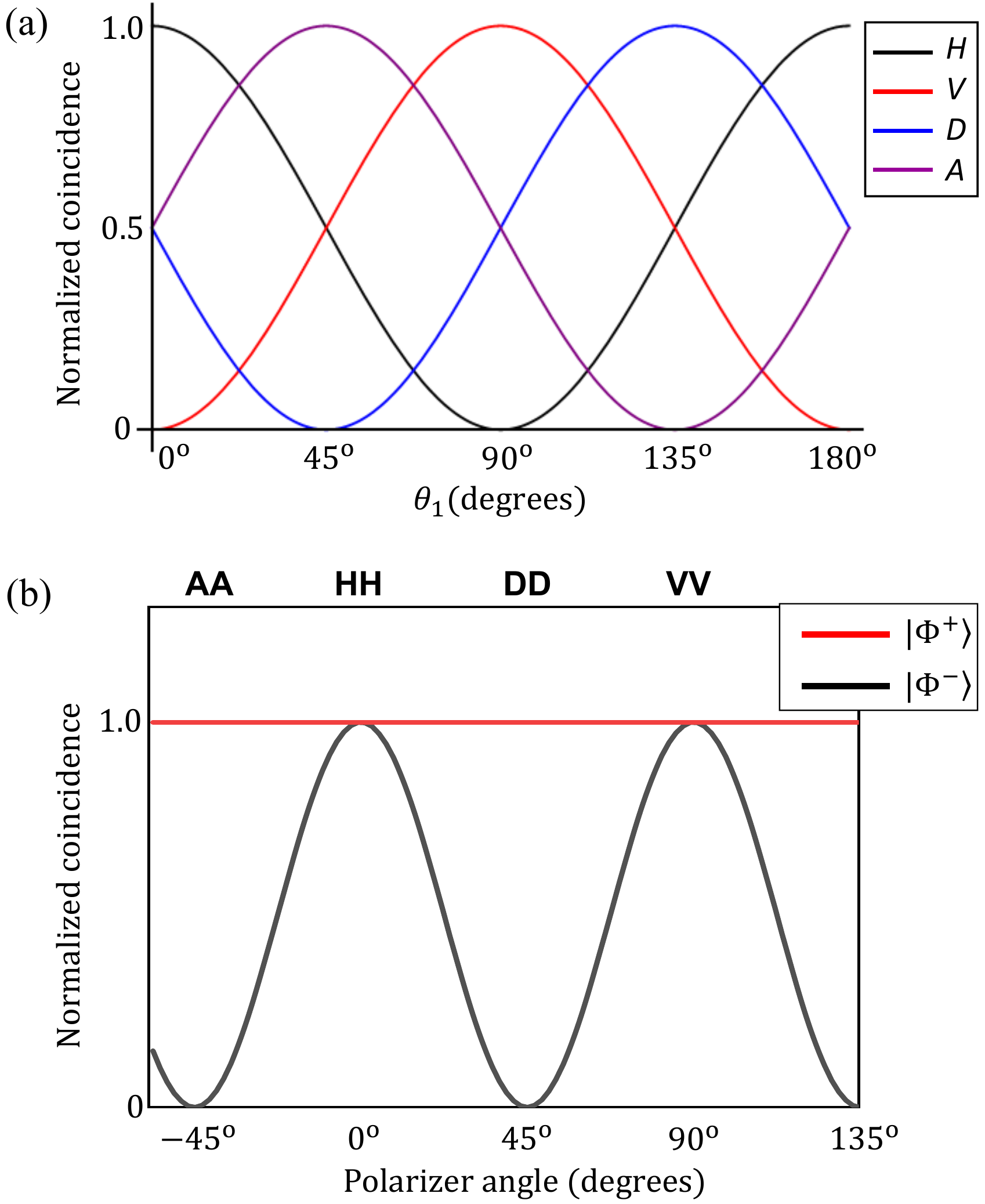}
	\caption{(a) Sample plot of variation of coincidence counts with signal polarizer angle $\theta_1$ for the Bell state $\ket{\Phi^+}$. Here the black, red, blue and green shades corresponds to idler photon projected to $\ket{H}$, $\ket{V}$, $\ket{D}$ and $\ket{A}$ polarizations, respectively. (b) Plots of single polarizer measurements for distinction between $\ket{\Phi^+}$ and $\ket{\Phi^-}$ states.}
	\label{bell_curve_plots}
\end{figure}
To examine polarization entanglement, polarization in one arm is projected to one among the four horizontal, vertical, diagonal or anti-dagonal polarizations represented by the states $\ket{H}$, $\ket{V}$, $\ket{D}$ or $\ket{A}$ respectively, and the coincidence counts are plotted for different polarizer angles of the second arm.  Figure \ref{bell_curve_plots}(a) represents the polarization curves corresponding to the Bell state $\ket{\Phi^+}$. The polarization visibility for each basis is determined by the expression
\begin{equation}
    V=\frac{N_{max}-N_{min}}{N_{max}+N_{min}}
    \label{visibility}
\end{equation}
where $N_{max}$ and $N_{min}$ are the maximum and minimum values of the curve in a given measurement basis. A major contribution to error when estimating visibility is the rate of accidental coincidences across the detectors, due to stray light, multi-pair emissions and internal detector noise. The rate of accidental coincidences is approximated as
\begin{equation}
    N_{acc}=N_sN_i\tau
    \label{accidentalcounts}
\end{equation}
where $\tau$ is the coincidence window used in coincidence counting. By subtracting $N_{acc}$, the rate of true coincidences can be determined. Note that the Eqn. \eqref{accidentalcounts} is an approximation which works for lower values of heralding efficiencies. In the regime of high photon pair rate and higher detection efficiencies, accidental coincidences due to multi-pair emission needs to be accounted separately for estimating the visibility  \cite{takesue2010effects}. Many single photon detectors exhibit saturation behaviour at higher detection counts and this needs to be addressed while estimating the accidental coincidences
\cite{grieve2016correcting}.

Note that it is also possible to obtain practical estimates on the quality of entanglement using only common polarization measurement acting on both signal and idler photons. This is particularly useful for quick characterization of entangled photon-pair sources in collinear design using a single polarizing element that is placed in the common path of the co-propagating signal and idler photons. For example, in  \cite{villar2018experimental}, it is shown that a joint linear polarization measurements allow to differentiate the state $\Ket{\Phi^+}$ from $\Ket{\Phi^-}$ or a mixed state by observing the polarization correlations.  The photons prepared in state $\ket{\Phi^-}$ are co-polarized in one basis (H/V) and anti-polarized in the other linear basis (D/A). So, in the case of $\ket{\Phi^-}$, the curve from single polarizer measurements show high contrast between minima and maxima, while $\ket{\Phi^+}$ show a constant pair rate for any polarizer setting (Fig. \ref{bell_curve_plots}(b)). Similarly, for maximally entangled states of the form $\ket{HH}+e^{i\phi}\ket{VV}$, recording the visibility of coincidence fringes over the relative phase $e^{i\phi}$ (e.g. by tilting a birefringent plate or a Soleil Babinet compensator) it is possible to certify entanglement with a common diagonally oriented polarizer in the signal and idler path \cite{steinlechner2017distribution}. 

\paragraph{Bell's inequality:} The violation of Bell's inequality is one of the most commonly used parameters to qualify or witness polarization entanglement. An experimentally very accessible form of Bell’s inequality is given by the inequality proposed by Clauser, Horne, Shimony, and Holt (CHSH) in 1969 \cite{clauser1969proposed}. The CHSH version of Bell’s inequality is expressed as
\begin{align}
    S=& \vert E(\theta_1,\theta_2)-E(\theta_1,\theta_2')+E(\theta_1',\theta_2)+E(\theta_1',\theta_2')\vert \nonumber \\
    \leq & \quad 2
    \label{bellviolation}
\end{align}
where  correlation values, $E(\theta_1,\theta_2)$,  can conveniently be normalized using the measured coincidences given by \cite{aspect1982aexperimental}
\begin{align}
    E(\theta_1,\theta_2)&= \nonumber \\
    &\frac{N_c(\theta_1,\theta_2)+N_c(\theta_1^\perp,\theta_2^\perp)-N_c(\theta_1^\perp,\theta_2)-N_c(\theta_1,\theta_2^\perp)}{N_c(\theta_1,\theta_2)+N_c(\theta_1^\perp,\theta_2^\perp)+N_c(\theta_1^\perp,\theta_2)+N_c(\theta_1,\theta_2^\perp)}
    \label{bellparameter}
\end{align}
Here $N_c(\theta_1,\theta_2)$ is the coincidence counts for each polarization angle and $\theta_i^\perp=\theta_i+90^\circ$. All Bell states clearly violates the inequality given in Eqn.\eqref{bellviolation}. For example, consider a source emitting polarization entangled photons in the state $\ket{\Phi^+} =(\Ket{HH}+\Ket{VV})/\sqrt{2}$. Substituting the probability corresponding to the state from Eqn.\eqref{bellprob} in Eqn.\eqref{bellparameter}, we obtain $S = 2\sqrt{2}>2$ for angles $\theta_1'-\theta_1$ = $45^\circ$, $\theta_2-\theta_1$ = $22.5^\circ$ and $\theta_2'-\theta_1$ = $67.5^\circ$. The CHSH values can be estimated from the visibility curves (Fig. \ref{bell_curve_plots}).

\paragraph{State fidelity:} The quantum state fidelity determines the closeness of the experimental quantum state expressed by the density matrix $\rho$ to a particular target state expressed by the density matrix $\rho_0$. It is mathematically expressed as
\begin{equation}
    F(\rho)=\left(\text{Tr}\sqrt{\sqrt{\rho_0}\rho\sqrt{\rho_0}}\right)^2
    \label{fidelity}
\end{equation}
$F$ will have values ranging from 0 to 1 depending on the overlap of the experimental with the desired target state. For an ideal state $\rho=\rho_0$, $F=1$. Typically, the experimenter's goal is that the target state be a pure maximally-entangled Bell state. In this case, one often refers to the Bell-state Fidelity or simply the entanglement fidelity of a state. The Bell-state fidelity of a classical mixed state to the maximally entangled state is 0.5. In order to determine how to calculate the Bell state Fidelity, let us first consider the density matrix of a perfect target state  for example, the density matrix of the Bell state $\ket{\Phi^+}$: 
\begin{equation}
    \rho=\ket{\Phi^+}\bra{\Phi^+} =\frac{1}{2}
    \begin{pmatrix}
    1 & 0 & 0 & 1 \\
    0 & 0 & 0 & 0 \\
    0 & 0 & 0 & 0 \\
    1 & 0 & 0 & 1
    \end{pmatrix}
\end{equation}

In order to evaluate the Bell state Fidelity of the two-photon SPDC state produced in the source according to the definition above, we must first reconstruct the experimental density matrix $\rho$. This can be accomplished by  performing a series of polarization correlation measurements on the two photon state. In general, at least 16 such polarization measurements are required to reconstruct the density matrix using a procedure called \textit{quantum state tomography}. However, errors and noise in the measurement data can lead to unphysical density matrices. A physically allowed density matrix can be obtained using \textit{maximum likelihood estimation} \cite{banaszek1999maximum,kwiat2001measure}. For more detailed description of the experimental estimation of photonic qubits, the reader may refer to Ref. \cite{altepeter2005photonic}. 

The aforementioned procedure is not always the most efficient way of characterizing entanglement generated in an SPDC source. Whenever the fidelity of an unknown experimental state with a particular Bell state is the main property of interest, then a complete reconstruction of the density matrix is not strictly required. From the properties of the density matrix, it can be shown that the following expression \cite{jennewein2009performing,guhne2009entanglement} gives the the Bell-state fidelity as:
\begin{equation}
F(\hat{\rho}) = \frac{1 + V_{H/V} + V_{D/A} + V_{L/R}}{4}
\label{eq:fidelitywitness-exp}
\end{equation}
\begin{figure*}
	\centering
	\includegraphics[width=1.0\textwidth]{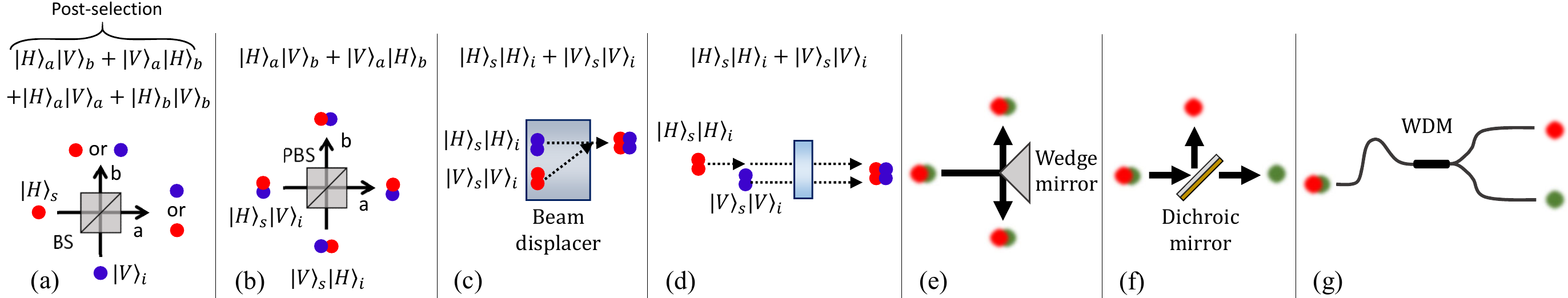}
	\caption{Some major techniques for generation and splitting of polarization entangled photon-pairs using SPDC sources: (a) Post-selection at a beam splitter after a signal photon from one port interfere with the corresponding idler photon at the other port. (b) Post-selection free generation by combining photon-pairs from each port of a polarizing beam splitter (PBS). (c) Orthogonal decay states can also be combined using a beam-displacer. (d) Coherent addition of photon-pairs in collinear fashion using compensation crystals. (e) Splitting of non-degenerate photon-pairs by momentum using a wedge mirror. Entangled photon-pairs in collinear fashion can by split by wavelength using either (f) a dichroic mirror or (g) a wavelengtth division multiplexer. The red and blue dots denote $H$ and $V$ polarizations respectively.}
	\label{big-figure1}
\end{figure*}
where $V\propto |E(\theta_1,\theta_2)|$ refers to the absolute values of the two-photon correlation functions in three mutually unbiased measurement bases (here linear H/V, D/A and circular L/R), each of which can derived from four protective measurements, thus requiring only a total of 12 settings. Note that when a lower bound on the Bell-state fidelity is sufficient the number of required measurements can be further reduced. It can be shown \cite{steinlechner2017distribution} that the Bell-state fidelity is lower bounded by experimental visibilities in two mutually unbiased measurement bases, e.g. in the linear horizontal vertical and diagonal/antidiagonal bases via the relation:
\begin{equation}
  F \geq \frac{ V_{D/A} + V_{H/V} }{2}
\end{equation}

Other frequently employed measures of entanglement are the tangle and the concurrence, and entanglement of formation  \cite{coffman2000distributed,wootters1998entanglement}. The concurrence can be calculated from the tomographically reconstructed density matrix, or lower bounded via:
\begin{equation}
  C(\rho) \geq V_{D/A} + V_{H/V} - 1 
\end{equation}

The entanglement of formation is particularly relevant to quantum information processing as it has clear operational meaning; the entanglement of formation represents the minimal number of maximally entangled bits required to produce produce the state via local operations and classical communication (LOCC) procedure. The entanglement of formation may be lower bounded by the concurrence according to:
\begin{equation}\label{Eofdef}
E_\text{oF}(\rho)\geq-\log\left(1-\frac{C(\rho)^2}{2}\right),
\end{equation}
The interested reader is referred to Refs. \cite{guhne2009entanglement,horodecki2009quantum,bavaresco2018measurements} for more detailed discussion on measures of entanglement, in particular extension to high-dimensional and multi-partite entanglement. 

\section{\label{polarizationentlpairsources}Polarization Entangled photon-pair sources - variations on a theme}

Photons generated via SPDC in a single decay channel are not inherently entangled in the polarization degree of freedom.  To see how we may obtain polarization entangled photon-pairs, let us consider two generic SPDC sources, each producing orthogonally polarized photon-pairs in a sate $\ket{H,1}_s\ket{V,1}_i$ and $\ket{V,2}_s\ket{H,2}_i$, where the labels (1,2) denote any spatial, spectral, or temporal features of the respective source. In many experimental settings these properties are not directly accessible, e.g. the characteristic features of the temporal wave function occur on the ps timescale, whereas the detectors' timing resolution is on the order of ns. In order to determine the experimentally accessible polarization correlations for the combined SPDC emission from the two sources
\begin{equation}
\ket{\Psi} \propto \ket{H,1}_s\ket{V,1}_i + \ket{V,2}_s\ket{H,2}_i,
\end{equation}
let us consider the result of a joint projective polarization measurement of the signal and idler photons in a local superposition basis  $\ket{\phi_{s,i}} = \frac{1}{\sqrt{2}}( \ket{H}_{s,i}+e^{i\phi_{s,i}}\ket{V}_{s,i})$. The probability for coincidence detection is given by
\begin{align}
 p(\phi_s,\phi_i) & = |\bra{\phi_i}\bra{\phi_s}\Psi\rangle|^2 \nonumber \\
                  & = \frac{1}{2}\left(1 + \text{Re}\left(e^{i (\phi_s-\phi_i)} \mathstrut_i\bra{1} \mathstrut_s\langle 1|2\rangle_s\ket{2}_i \right)\right).
\end{align}
Hence, if the photon-pairs emitted from the two sources are  distinguishable, i.e. the labels  correspond to orthogonal states ($\mathstrut_{s,i}\langle 1|2\rangle_{s,i}=0$),  the phase- dependent term vanishes, and we can not observe the characteristic non-local coherence of a polarization-entangled state. In order to directly obtain a maximally entangled Bell state $(\ket{H}_s\ket{V}_i+\ket{V}_s\ket{H}_i)/\sqrt{2}$, the labels indicating the source of origin must factor out of the picture. Only then can we consider the state as coherent superposition of two, otherwise identical, orthogonal polarization states $\ket{H}_s\ket{V}_i$ and $\ket{V}_s\ket{H}_i$.  

Engineering the pairs to exhibit polarization entanglement is thus generally associated with the concept of a 'quantum eraser' \cite{scully1982quantum,scully1991quantum,aharonov2005time} to remove any distinguishing features. These concepts have been applied to polarization entangled photon sources by using  one or two beam splitters (Fig. \ref{big-figure1}(a))  to erase the photon path information. The photons are detected after the beam splitter(s) and the non-local correlations can be observed by employing   post-selection and only considering a certain subset of all possible detection outcomes. This approach, while elegant, however reduces the efficiency of the source by at least 50\% and is therefor not suitable for loophole-free Bell-tests. First demonstrations of entanglement from single-path SPDC include the work by Ou and Mandel 1988 \cite{Ou88a} or Rarity and Tapster in 1990 \cite{Rarity90b}.

By using nonlinear interferometers \cite{chekhova2016nonlinear}, researchers have come up with a large variety of entangled photon-pair sources. What follows are the general steps to be taken when realizing a polarization entangled photon-pair source. 

\paragraph{Selection of two pump decay paths:} Two decay paths with orthogonal polarization states from Type-I/0 ($\ket{H}_s\ket{H}_i$ and $\ket{V}_s\ket{V}_i$) or Type-II ($\ket{H}_s\ket{V}_i$ and $\ket{V}_s\ket{H}_i$) phase matching are used. In any case, the two decay paths must be indistinguishable in spectral and spatial modes. It is important to note that the pump decay paths can occur within a single process (e.g.  from a single crystal \cite{kwiattype2}), or from different processes (in different crystals e.g. crossed crystal source \cite{kwiattype1dual}).

\paragraph{Superposition:} The two decay paths need to be put in a coherent superposition to obtain the entangled state. 
This can be achieved by using regular or polarizing beam splitters. In Fig. \ref{big-figure1}(a), the interference of orthogonally polarized photons from two input ports of the beam splitter generates entanglement after post-selection. Two orthogonally polarized photon-pairs each at the two input ports of a polarizing beam splitter can be superposed to generate entangled state, as shown in Fig. \ref{big-figure1}(b). Superposition can also be achieved by making use of spatially overlap from beam walk-off within birefringent crystals (Fig. \ref{big-figure1}(c)). Spatial filtering (using pinholes or single mode fibers) can be used to remove residual spatial distinguishability.

\paragraph{Temporal compensation:} SPDC is a broadband process which  allows phase-matching over a broad spectrum of wavelengths. In many source configurations,  the generated SPDC output travel through dispersive materials where the photon-pairs will pick up a wavelength dependent phase. This results in spectral distinguishability, and thereby reducing the quality of entanglement. This wavelength dependent phase can be removed by employing additional 'compensation' crystals (e.g. Yttrium orthovanadate, YVO$_4$,). Methods of temporal compensation using birefringent crystals is illustrated in Fig. \ref{big-figure1}(d). 

\paragraph{Splitting:} The signal and idler photons must be separated before they are useful for downstream applications.
These photons can be separated by emission angle \cite{kwiattype2,kwiattype1dual}, polarization property \cite{kim2006phase}, or wavelength \cite{hentschel2009three}.
Emission angle splitting is typically associated with non-collinear SPDC. However, this is also possible in collinear cases where the signal and idler can be split using a wedge mirror \cite{perumangatt2019experimental,lohrmann2019PAPA}, as shown in Fig. \ref{big-figure1}(e). Many modern sources prepare the signal and idler photons with different wavelengths (non-degenerate) for easy separation. For wavelength based separation, dichroic mirrors are a natural choice (Fig. \ref{big-figure1}(f)). However, many dichroic mirrors comes with group delay dispersion which introduce additional wavelength dependent phase between $\vert H\rangle$ and $\vert V\rangle$ polarization states degrading the quality of entanglement.
Wavelength division multiplexers (WDM) based on fused fibers employing evanescent coupling are an  alternative technology for splitting the two wavelengths without introducing the wavelength dependent phase \cite{trojek2008collinear,steinlechner2013phase} (Fig. \ref{big-figure1}(g)). 


In the following we detail concrete examples involving these basic steps in a variety of geometric arrangements and phase-matching configurations. 

\subsection*{\label{noncollpolentlpairsources}Non-collinear phase-matched sources}

Photon-pair sources with non-collinear emission can be grouped into three general types: 1. Single pass of the pump through a single crystal, 2. Single pass of the pump through two crystals, and 3. Two pump passes through a single crystal. In these sources, the signal and idler photons have degenerate wavelengths. The major characteristics for each category are discussed below.

\subsubsection{Single crystal, single pass configuration}

In this geometry, 
\begin{figure}[h]
	\centering
	\includegraphics[width=0.48\textwidth]{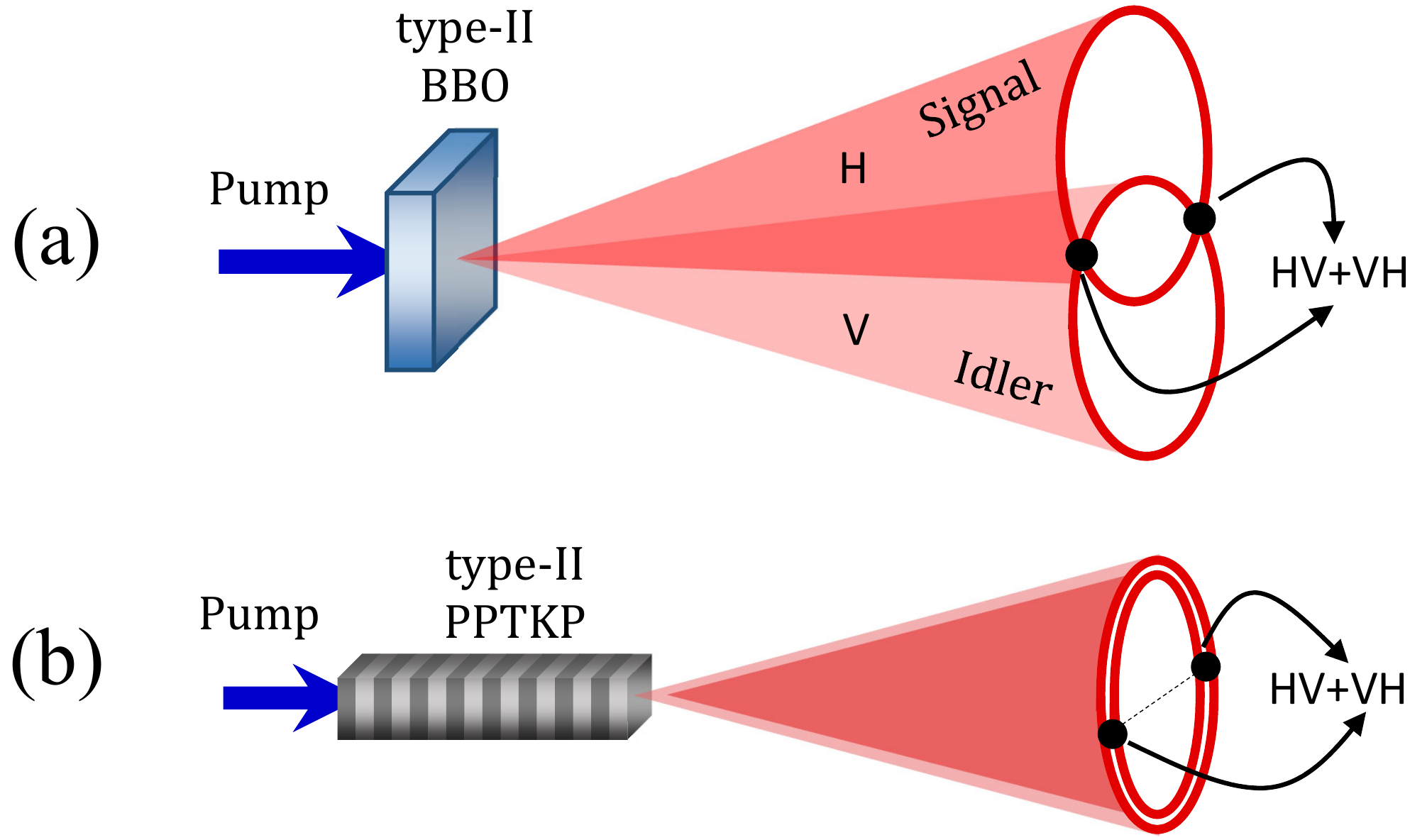}
	\caption{(a) Photon-pairs emitted from a single type-II BBO crystal \cite{kwiattype2,trojek2004compact} and (b) from a single type-II PPKTP crystal \cite{fiorentino2005source}, in non-collinear geometry.}
	\label{pol-entl-type2}
\end{figure}
\begin{figure}[h]
	\centering
	\includegraphics[width=0.48\textwidth]{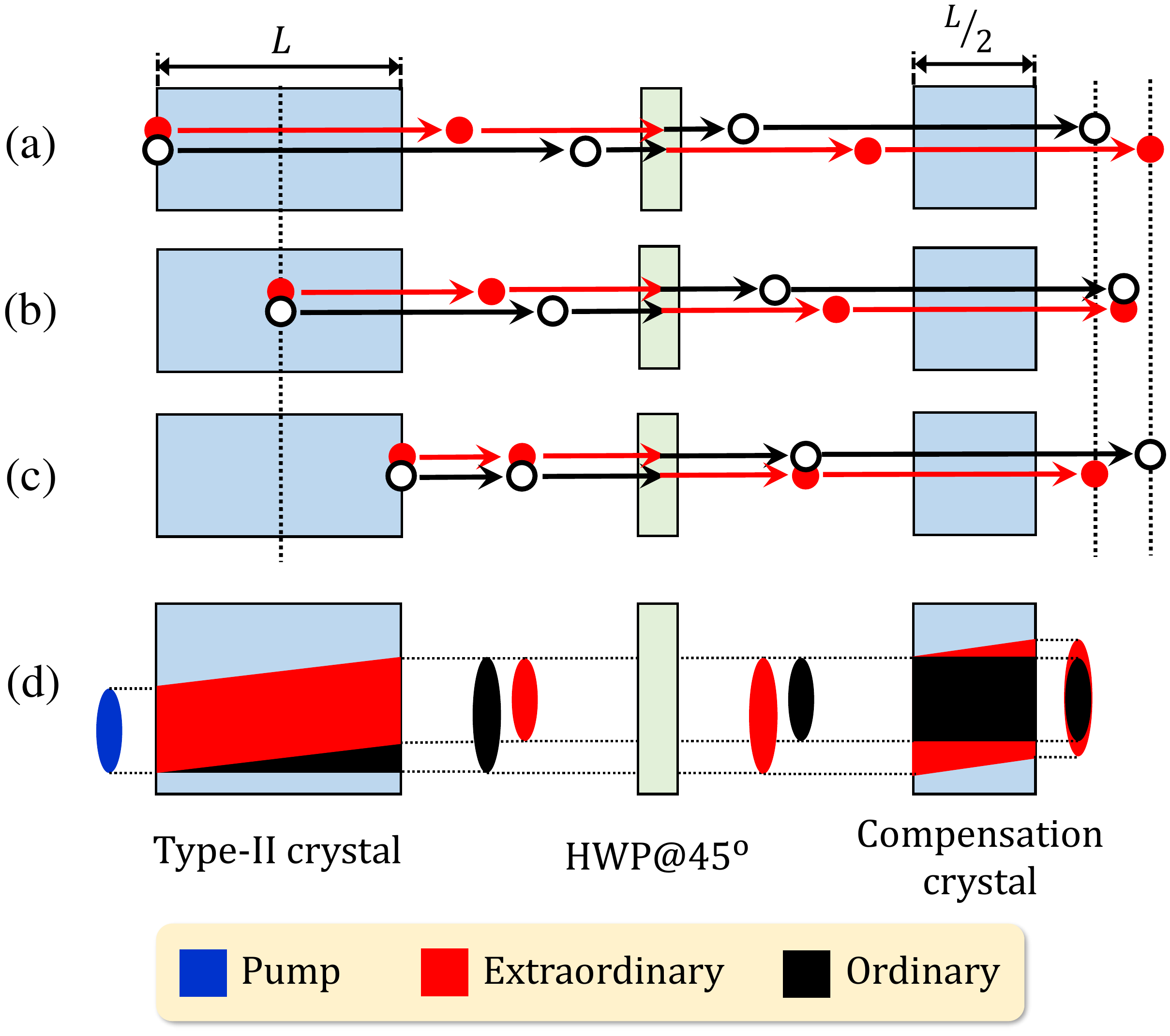}
	\caption{Temporal walk-off compensation between photons generated at the (a) entrance plane, (b) center and (c) exit plane of a SPDC crystal. (a) Photons generated at the entrance plane will experience maximum temporal walk-off. A half-wave plate (HWP) swaps their polarization so that the relative delay between the photons is reduced by half after passing through the compensation crystal. (c) Photons generated at the exit-plane will experience a similar amount of delay. This method makes the SPDC process between any two complementary planes indistinguishable in the temporal degree of freedom. Photons generated at the center are unaffected. (d) Spatial walk-off compensation between photons generated using a type-II crystal. The extraordinary polarized pump and SPDC photons walk-off from the ordinary polarized photons. As SPDC takes place throughout the crystal, the ordinary photons' output will have an elongated distribution compared to the pump and extraordinary photons. The half-wave plate swaps the polarization of the SPDC photons and after traversing the compensation crystal, there is better overlap of the output. Further compensation may be possible if birefringent lenses could be employed to further shape the output to get better overlap. However, there is no evidence in the literature of this having been studied.}
	\label{type2-compensation}
\end{figure}
type-II phase matching is preferred where a single crystal emitting photon-pairs polarized orthogonal to each other can be a source of polarization entanglement (Fig. \ref{pol-entl-type2}) \cite{kwiattype2,trojek2004compact}.
The signal and idler photons are emitted in separate cones, which can be made to intersect at two points where the polarization of the emitted photon is undefined (Fig. \ref{pol-entl-type2}(a)). The photon-pairs collected from these intersection points are entangled in polarization with the state
\begin{equation}
\ket{\Psi}=\frac{1}{\sqrt{2}}(\ket{H}_s\ket{V}_i+e^{i\phi}\ket{V}_s\ket{H}_i)
\label{psistate}
\end{equation}

By using a type-II quasi phase-matched crystal, the same state can be obtained, but where the signal and idler cones complete overlap (Fig.\ref{pol-entl-type2}(b)). This geometry has been very popular with a large number of ground-breaking experiments adopting the design. Its simple alignment requirements, and small part number also make it a useful teaching tool. The quasi phase-matching allows access to higher nonlinearity leading to a relatively brighter source \cite{fiorentino2005source,lee2016polarization}. Overlap of two concurrent type-I QPM structures in a single crystal can produce highly non-degenerate entangled photon-pairs \cite{de2006non}.

The temporal walkoff between ordinary and extra-ordinary photons travelling through the crystal degrades the entanglement. This is corrected by the use of a half-wave plate and a compensation crystal that is half the thickness of the SPDC crystal \cite{kwiattype2} (Fig. \ref{type2-compensation}(a)-(c)).
This compensation technique partially corrects for spatial walk-off (Fig. \ref{type2-compensation}(d)). The source can act as a 'universal Bell state synthesizer' if a half-wave plate is introduced in one arm and the two paths are combined at a polarizing beam splitter \cite{kim2003experimental}.

Focusing of the pump brings the SPDC conditions towards the thick-crystal regime, and the spatial-walkoff can be quite pronounced \cite{lee2005focused}. However, with proper spatial filtering, the entanglement visibility can be preserved at the cost of the brightness. 
In general, type-II non-collinear SPDC sources employ $\beta-$Barium Borate (BBO) as the nonlinear material. The brightness of the sources could be improved by using other materials with a larger nonlinear coefficient, e.g. Bismuth Borate (BIBO) \cite{halevy2011biaxial}. 

\subsubsection{Double crystal, single pass configuration}

\begin{figure}[h]
	\centering
	\includegraphics[width=0.48\textwidth]{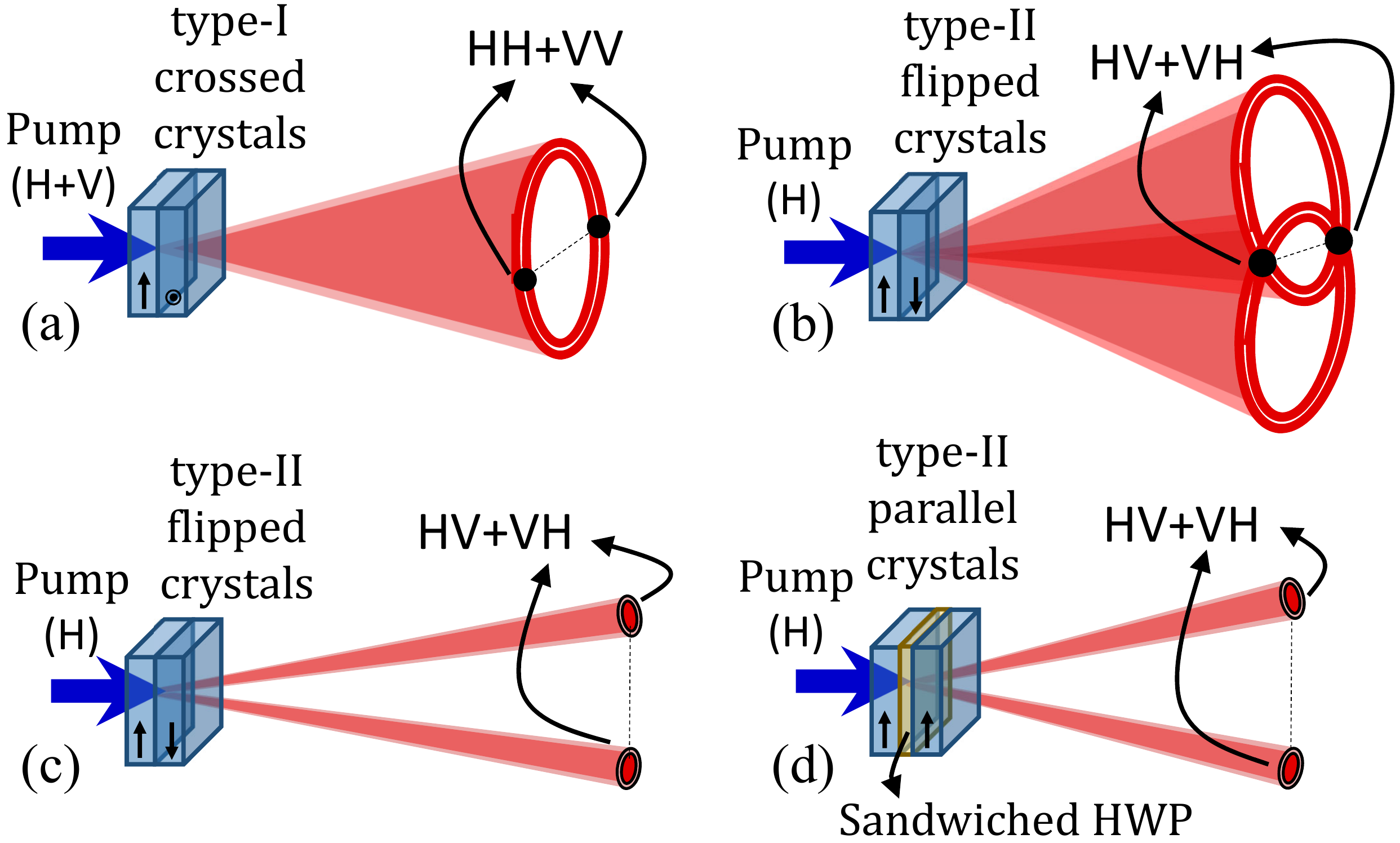}
	\caption{Non-collinear emission of photon-pairs two crystals: (a) crossed type-I crystals \cite{kwiattype1dual}, (b) flipped type-II crystals \cite{bitton2002cascaded}, (c) 'beam-like' output \cite{niu2008beamlike} and (d) parallel-crystal configuration with a half-wave plate (HWP) sandwiched between the crystals \cite{zhang2015experimental}.}
	\label{cascaded-crystals-noncolli}
\end{figure}

Polarization entanglement can be achieved by superposition of photons from two successive non-linear crystals. This geometry accepts both type-I and type-II phase matching. This technique was first reported using two thin Type-I BBO crystals \cite{kwiattype1dual} (Fig. \ref{cascaded-crystals-noncolli}(a)). A diagonally polarized pump photon incident on the crossed crystals have an equal chance to be down converted in either crystal. These two possible SPDC processes are coherent, and when their spatial output overlap, the photon-pairs can be in the maximally entangled state
\begin{equation}
\ket{\Phi}=\frac{1}{\sqrt{2}}(\ket{H}_s\ket{H}_i+e^{i\phi}\ket{V}_s\ket{V}_i)
\label{phistate}
\end{equation}
 
This geometry enables non-maximally entangled states \cite{white1999nonmaximally} by tuning the pump polarization.
Various optimization techniques can be found in the literature \cite{migdall1997polarization,rangarajan2011migdall,nambu2002generation,rangarajan2009optimizing,rangarajan2011engineering}.

Apart from the original scheme using two crossed crystals, there has been several other two crystal alignment techniques that have been studied.
In Ref.\cite{bitton2002cascaded}, two crystals are used with flipped optical axes (Fig. \ref{cascaded-crystals-noncolli}(b)). 
The emission cones with orthogonal polarization overlap completely removing the dependence of photons' polarization on their azimuthal positions within the cone. This ensures entanglement of photon-pairs at any point of the cone if proper temporal compensation is used. Since polarization of the pump is same for both crystals, this doubles the brightness of the SPDC output compared to \cite{kwiattype1dual}.
Instead of emission cones, overlapping of beam-like emission from two crystals can also be used \cite{niu2008beamlike} (Fig. \ref{cascaded-crystals-noncolli}(c)). 
Finally, a thin half wave plate sandwiched between two type-II parallel crystals can also give a beam-like emission of entangled photon-pairs \cite{zhang2015experimental} (Fig. \ref{cascaded-crystals-noncolli}(d)).
Notice that in all of these configurations, the pump field driving the second pair-emission process, co-propagates with the SPDC emission from the first source. In configurations with a common optical path for pump, singal and idler, phase shifts due to changes of the path length between the two SPDC emitters perfectly cancel, and do not contribute to the overall phase of the polarization-entangled state. This auto-compensating feature of 'common-path configurations' leads to long-term stability without the need for active interferometric stabilization.

\subsubsection{Single crystal, double pass configuration}

\begin{figure}[h]
	\centering
	\includegraphics[width=0.46\textwidth]{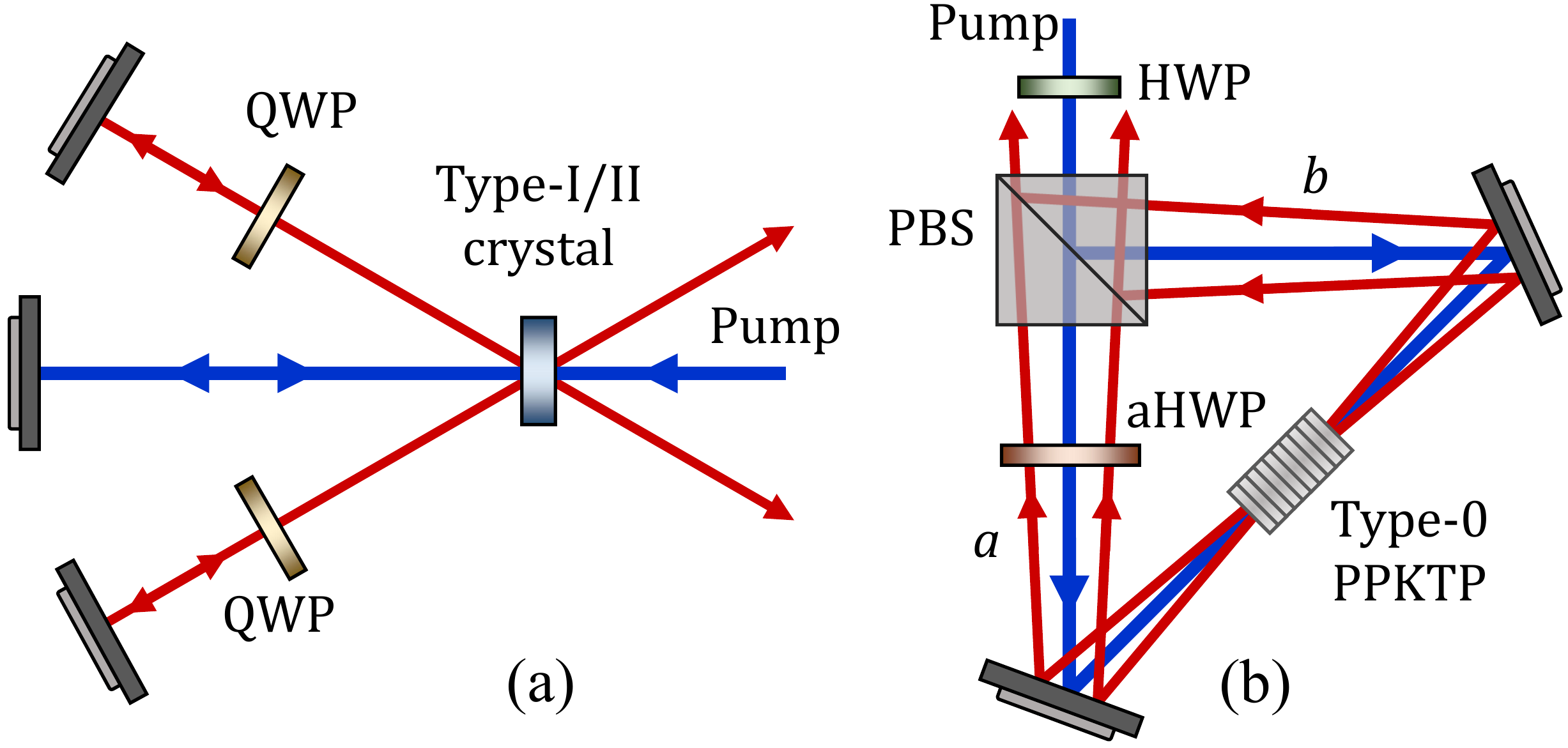}
	\caption{(a) The rail-cross scheme with a single crystal \cite{herzog1994frustrated,herzog1995complementarity,hodelin2006optimal,vallone2009polarization}. (b) An interferometer scheme with a single crystal in a polarizing Sagnac loop \cite{jabir2017robust,kim2019pulsed}. HWP - half wave plate; aHWP - achromatic half wave plate; PBS - polarizing beam splitter; QWP - quarter wave plate.}
	\label{SCDPfig}
\end{figure}

Another method for entanglement generation is to use two paths for the pump beam within the same crystal. There have been two schemes using this method. In the first scheme \cite{herzog1994frustrated,herzog1995complementarity,vallone2009polarization}, the pump, after passing through a type-I crystal, is reflected back (Fig. \ref{SCDPfig}(a)). Additional mirrors reflect the photon-pairs from the first pass to overlap with the output of the second pass producing the state $\vert HH\rangle_1+e^{i\Delta\phi}\vert VV\rangle_2$, where $\Delta\phi$ is the phase difference from the path length differences. Since any changes in the path length of the pump and the SPDC photons will also contribute to changes in this phase shift, interferometric stability of the system is required. This poses a critical challenge, which is not present in two-crystal configuration.
It has been shown that this scheme can be utilized for optimal generation of pulsed entangled photon-pairs that give better quantum interference visibility even without spectral filtering \cite{hodelin2006optimal}.

The second scheme (Fig. \ref{SCDPfig}(b)) employs a quasi phase-matched crystal inside a polarizing Sagnac interferometer \cite{slussarenko2010polarizing}. This can be realised using a Type-0 PPKTP crystal with either continuous-wave \cite{jabir2017robust} or pulsed \cite{kim2019pulsed} pump. Counter-propagating pump beams in the interferometer generate photon-pairs in the state $\ket{VV}$. In one arm, an achromatic half-wave plate (denoted as aHWP) rotates the polarization of all beams. The photon-pair passing through the aHWP will flip its polarization to the $\ket{HH}$ state. The output from both arms combine at the PBS to give the state $\vert HH\rangle_a+e^{i\phi}\vert VV\rangle_b$. The notations $a$ \& $b$ are used to distinguish the propagation direction with the loop. The inherent symmetry of the Sagnac loop configuration thus removes the need for additional compensation optics. Fine-tuning of the SPDC overlap can be done by adjusting the crystal position inside the loop.

\subsection*{\label{collpolentlpairsources}Collinear phase-matched sources}

Non-collinear emission of entangled photon-pairs enable straightforward separation of the photons, but are generally more complicated to align and also need a larger footprint. Collinear emission simplifies the collection leading to brighter and potentially more compact sources.
Collinear photon-pair sources found in the literature can be grouped into the same categories as the non-collinear sources apart from a recent design using a double pump pass through two crystals.

\subsubsection{\label{singlecrystalsinglepass}Single crystal, single pass configuration}

One of the earliest sources in this configuration made use of type-II SPDC with a temporal walk-off compensation crystal \cite{kuklewicz2004high,zhou2013ultra}. \begin{figure}[h]	
\centering	
\includegraphics[width=0.48\textwidth]{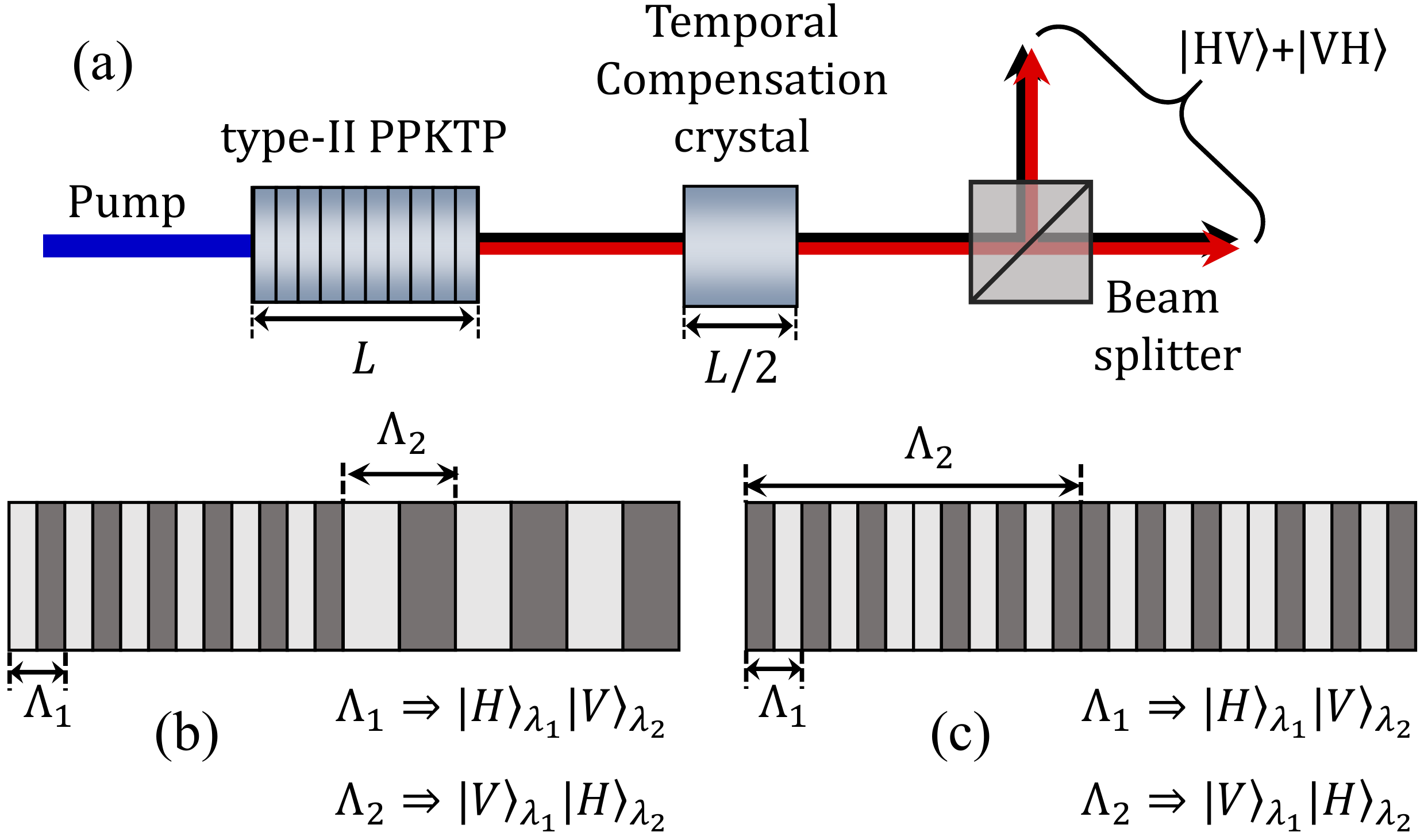}\caption{(a) Collinear source of polarization entangled photon-pairs using type-II PPKTP crystal \cite{kuklewicz2004high}. Photon-pairs are produced using (b) type-II poling along the two halves of the crystal \cite{ueno2012entangled}, and (c)  phase-modulated domain structure of a type-II QPM crystal with interlaced poling \cite{gong2011compact,laudenbach2017novel,kuo2020demonstration}. In both cases, different poling periods ($\Lambda_1$ \& $\Lambda_2$) results in the emission of photon-pairs at wavelengths $\lambda_1$ and $\lambda_2$ with their polarization swapped.}
\label{PPKTP-type2-and-domain-engg}
\end{figure}
In the original implementation, the entangled state $(\ket{H}\ket{V}+\ket{V}\ket{H})/\sqrt{2}$ was post-selected from the output of a 50:50 beamsplitter (Fig. \ref{PPKTP-type2-and-domain-engg}(a)). We note that post-selection can be avoided if nearly collinear signal and idler photons could also be separated using their momentum correlations \cite{prabhakar2020two} (see Fig. \ref{big-figure1}(e)).

A different approach uses two different poling periods in the same crystal. There are two reported schemes.
In the first scheme, two halves of the crystal have different poling periods (Fig. \ref{PPKTP-type2-and-domain-engg}(b)), while the second scheme employs a phase-modulated structure with two interlaced poling periods (Fig. \ref{PPKTP-type2-and-domain-engg}(c)). In both cases, the non-degenerate photon-pairs emitted due to first poling period ($\Lambda_1$) are orthogonal to the pairs from second poling period ($\Lambda_2$). The collinear propagation of both pairs gives the state
\begin{equation}
    \ket{\psi}=\frac{1}{\sqrt{2}}(\ket{H}_{\lambda_1}\ket{V}_{\lambda_2}+e^{i\phi}\ket{V}_{\lambda_1}\ket{H}_{\lambda_2})
    \label{pol-entl-state-two-domain}
\end{equation}

\begin{figure}[h]
	\centering
	\includegraphics[width=0.47\textwidth]{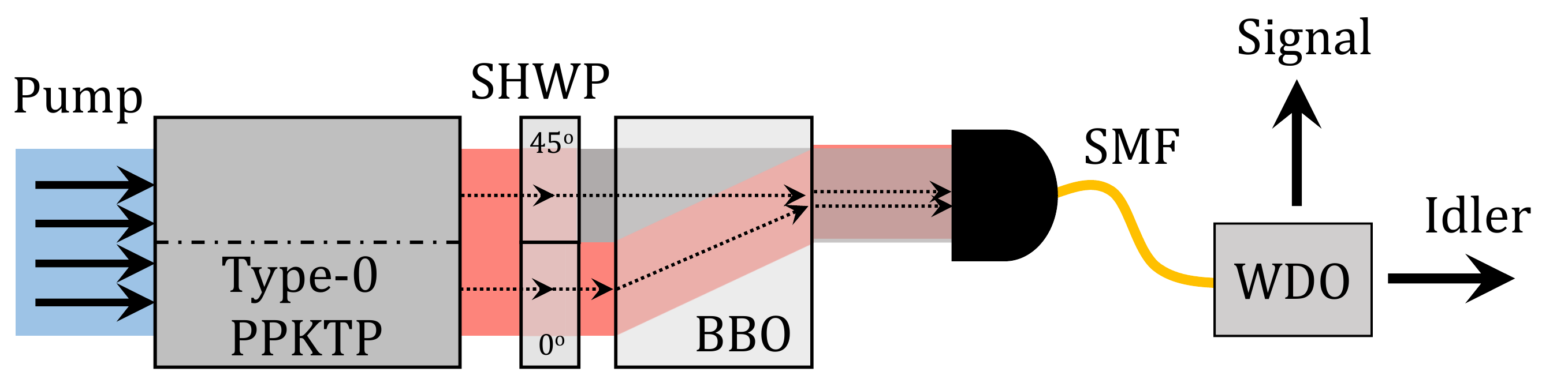}
	\caption{
Schematic for conversion of position correlation to polarization entanglement \cite{perumangatt2019experimental}. SHWP - segmented half-wave plate; SMF - Single mode fiber; WDO - wavelength division optics.}
	\label{pomo}
\end{figure}

A third approach is to convert the intrinsic position correlations in SPDC into polarization entanglement  \cite{shimizu2008generation,perumangatt2019experimental}.
For example, in a single type-0 PPKTP crystal (Fig. \ref{pomo}), photon-pairs may be generated in either the upper or lower half of the interaction volume. Using a segmented half-wave plate to swap the polarization state of photon-pairs from one half, followed by an $\alpha$-BBO crystal to overlap the output will generate an entangled state. 
This can be very compact as this geometry requires minimal components, and in principle can be extended to other physical systems (such as four-wave mixing in atomic vapor) with an extended interaction volume.

\subsubsection{\label{doublecrystalsinglepass}Double crystal, single pass configuration}

\begin{figure}[h]
	\centering
	\includegraphics[width=0.46\textwidth]{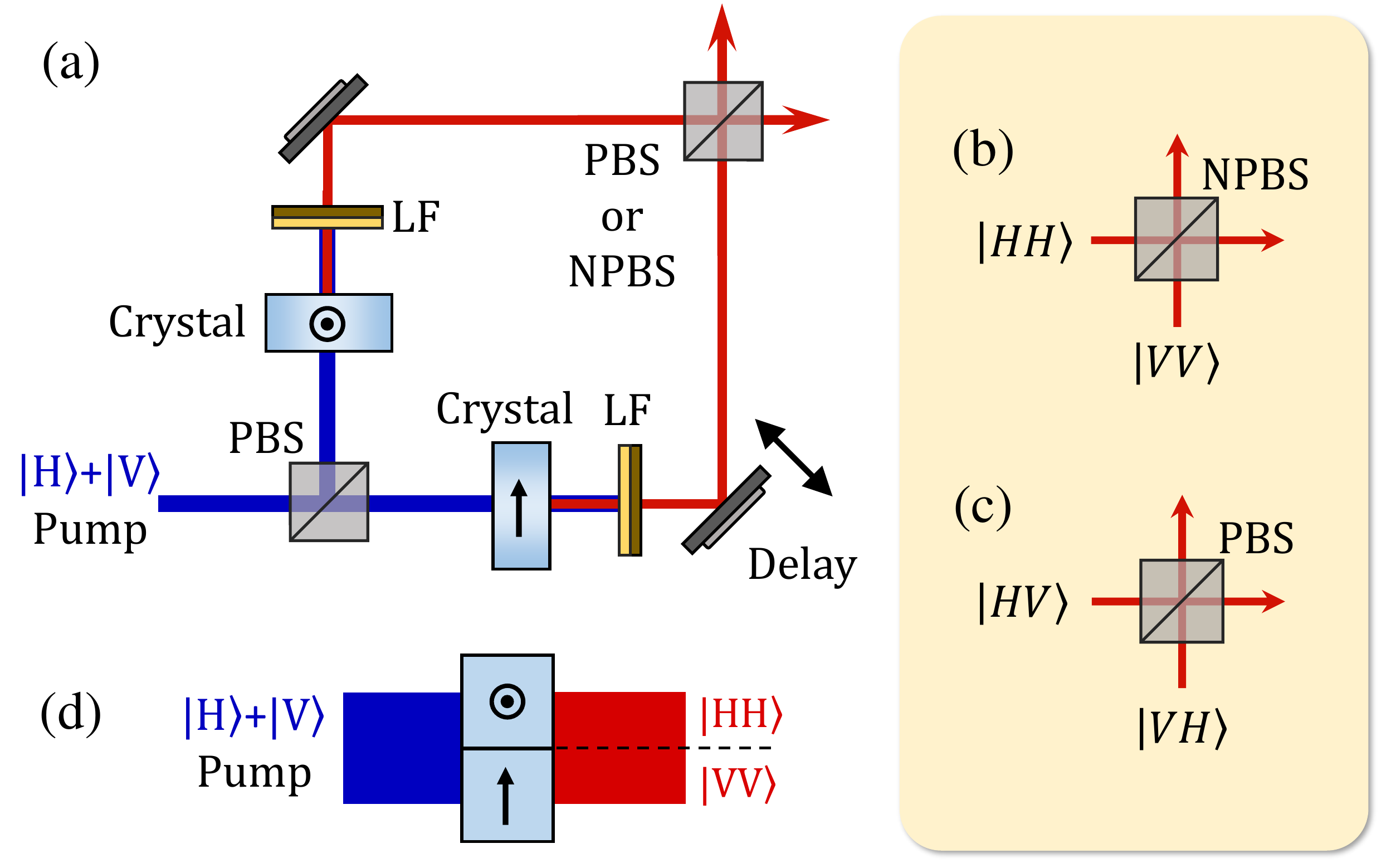}
	\caption{(a) Mach-Zehnder interferometer approach \cite{kim2001interferometric} for the generation of polarization entangled photons from two crystals with collinear phase-matching. (b) Non-polarizaing beamspliters (NPBS) are used in the case of type-I  phase matching. (c) Polarizing beamsplitters (PBS) are used for type-II phase matching. (d) Vertically stacked crossed-crystal configuration for polarization entangled photons \cite{shi2003generation}. A diagonally polarized pump passing through both the crystals equally produces $\Ket{HH}$ photon-pair from one crystal and $\Ket{VV}$ photon-pair from the other, which combine to give the entangled pair. LP - Longpass filter.}
	\label{twocrystal-machzender-setup}
\end{figure}

An early approach used a polarizing Mach-Zehnder interferometer (MZI) with a SPDC crystal in each arm (Fig. \ref{twocrystal-machzender-setup}(a)).
\cite{kim2001interferometric,kwiat1994proposal,shapiro2000ultrabright}. Entangled photons in all four Bell states can be generated with this scheme by selecting appropriate phase-matching combined with the correct type of (non-)polarizing beam splitter (Fig \ref{twocrystal-machzender-setup}(b) \& (c)).

\begin{figure}[h]
	\centering
	\includegraphics[width=0.48\textwidth]{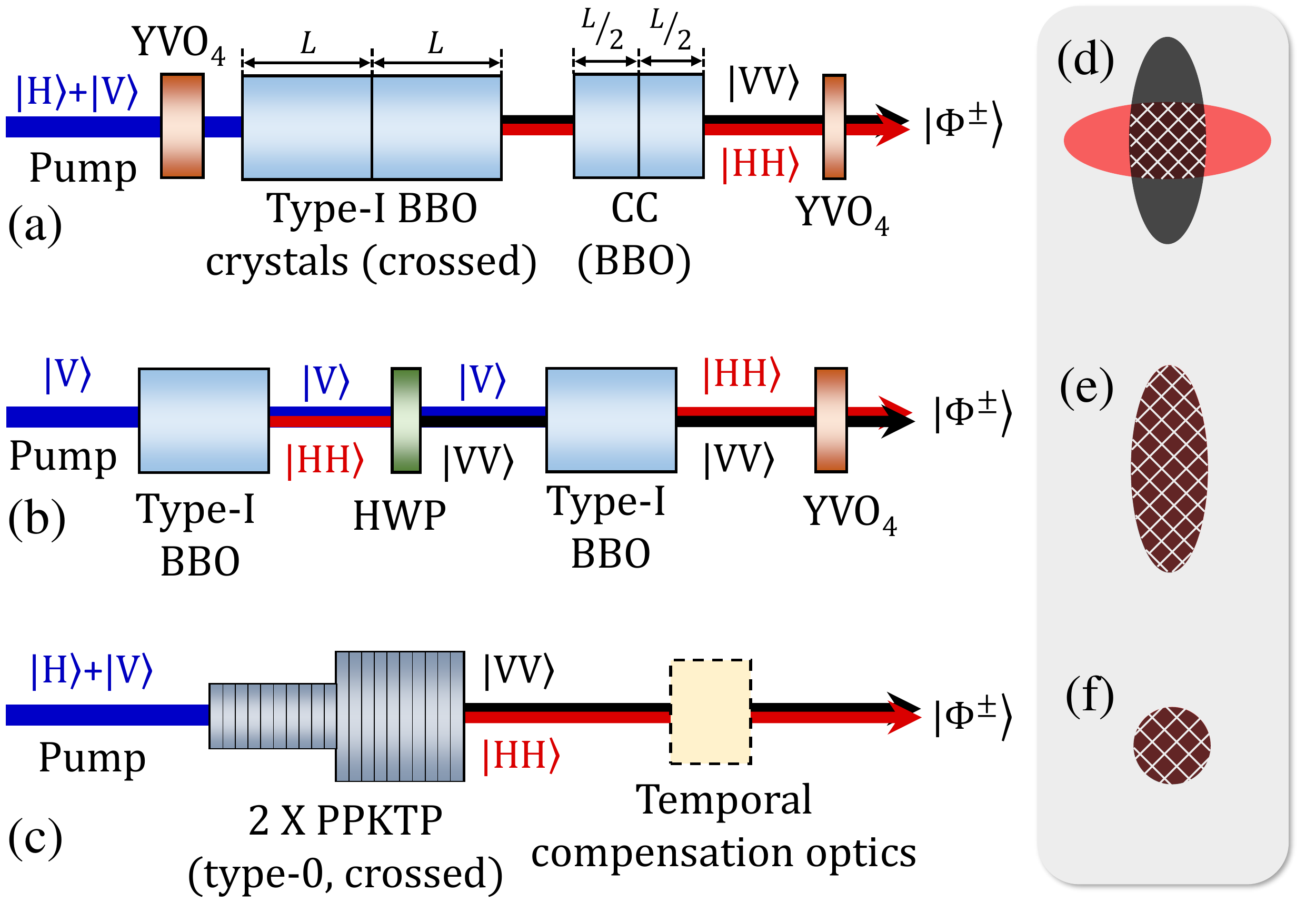}
	\caption{Photon-pair sources using two  crystals: (a) critically phase-matched crossed crystal \cite{trojek2008collinear} and (b) critically phase-matched parallel \cite{villar2018experimental} crystals using a half-wave plate to rotates only the polarization of SPDC photons from the first crystal (c) Crossed-crystal configuration with two quasi phase-matched crystals \cite{pelton2004bright,ljunggren2006theory,hubel2007high, steinlechner2012high, scheidl2014crossed}. (d)-(f) Spatial overlap of photon-pairs from (d) crossed crystals, (e) parallel crystals and (f) crossed quasi phase-matched crystals. The overlapping region is marked with cross-hatched pattern.}
	\label{crossedcrystal-collineartypes}
\end{figure}

Instead of building an interferometer, a single diagonally polarized pump can act on two crossed crystals to produce collinear photon-pairs in the state $\ket{\Phi^\pm}$. An early attempt for this configuration used two vertically stacked crystals with a diagonally polarized pump passing through both of them equally to generate entangled photon-pairs (Fig. \ref{twocrystal-machzender-setup}(d)) \cite{shi2003generation}. A similar method uses horizontally stacked crossed-crystals instead of vertical stacking to produce collinear photon-pairs (Fig. \ref{crossedcrystal-collineartypes}(a)).
This approach can use thick crystals to give high collection efficiency and brightness as the spatial walk-off can be readily compensated \cite{septriani2016thick}. The wavelength dependent phase associated with the pump and the SPDC photons are independently compensated using two YVO$_4$ crystals each in the pump and SPDC paths. This scheme enables the use of a broadband pump.

When performing critical phase-matching there are always spatial walk-off effects. This is more pronounced when two crossed crystals are used because the walk-offs are not in the same direction and produce an elongated SPDC output (Fig. \ref{crossedcrystal-collineartypes}(d)).
To get the centres of the output to overlap, additional crystals must be used.
Complete spatial mode overlap of the SPDC output is obtained in the ``parallel-crystal'' geometry (Fig. \ref{crossedcrystal-collineartypes}(b) \& (e)). In contrast to the crossed-crystal scheme where only half of the pump power is used in each crystal, the parallel-crystal configuration utilizes the full pump power. One drawback of using parallel crystals is that a broadband pump cannot be used.
The elongated SPDC output can be avoided using quasi phase-matching \cite{pelton2004bright,ljunggren2006theory,hubel2007high, steinlechner2012high, scheidl2014crossed} (Fig. \ref{crossedcrystal-collineartypes}(c) \& (f)). 

Using periodically poled crystals also enables type-0 and type-II phase-matching. In all cases, however, the wavelength of the SPDC output from both crystals must be matched carefully. For critical phase matching, the angle tuning and stability is important, while for periodically poled materials, it is the temperature control that must be maintained.

\subsubsection{\label{singlecrystaldoublepass}Single crystal, double pass configuration}

In the collinear emission type sources, another way to obtain two different decay paths is to propagate the pump beam twice through the same SPDC material.
This avoids having to compensate for material defects in different crystals (assuming the single crystal is uniform). This section reviews the set of schemes that use a double pass through the same crystal and they all employ interferometric techniques such as the Michelson, Mach-Zehnder and Sagnac designs.
In some of these cases, the full pump power can be used in each pass, leading to brighter sources.
\begin{figure}[h]
	\centering
	\includegraphics[width=0.48\textwidth]{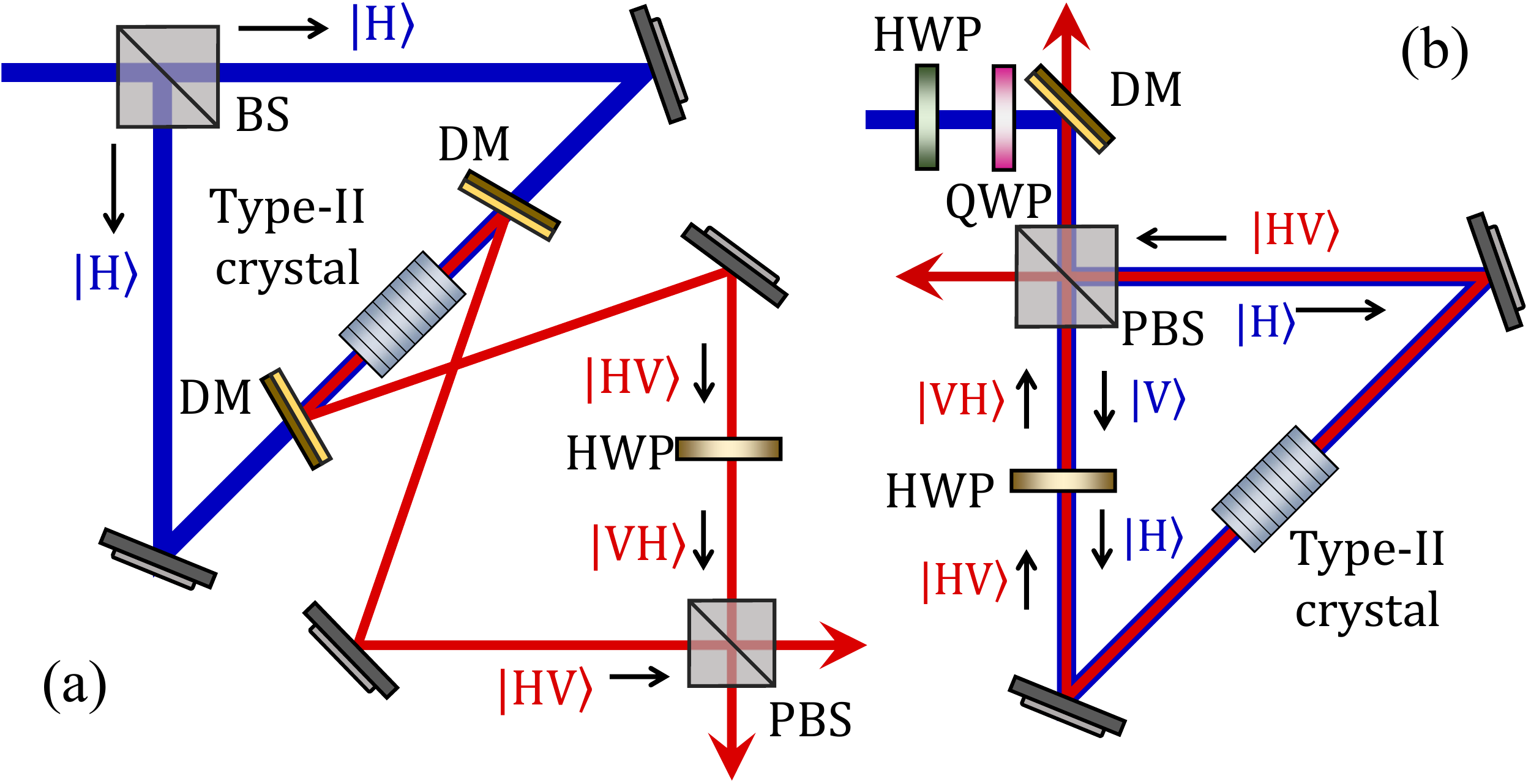}
	\caption{(a) Folded Mach-Zehnder \cite{fiorentino2004generation,konig2005efficient} and (b) Sagnac \cite{anderson1994sagnac,kim2006phase,kuzucu2008pulsed,fedrizzi2007wavelength,predojevic2012pulsed,steinlechner2014efficient} configuration using a single crystal. DM - Dichroic mirror}
	\label{foldedMZI-polSagnac}
\end{figure}

Figure \ref{foldedMZI-polSagnac}(a) shows the folded Mach-Zehnder scheme where entangled photons are generated by bidirectional pumping of a single crystal \cite{fiorentino2004generation}. The photons generated in each direction are filtered out using dichroic mirrors. 
An additional arm manipulates the photon polarization before recombination to produce the entangled state $\ket{\Psi^\pm}$. 
This scheme can also work with type-0/I crystals \cite{konig2005efficient}. 

The major disadvantage of the scheme in Fig. \ref{foldedMZI-polSagnac}(a) is that it requires two different interferometers to be stabilised. 
Common-path interferometers based on the Sagnac design \cite{anderson1994sagnac}) have improved stability.
The first entangled photon source using this scheme employed a type-I BBO crystal with post-selection at a beam-splitter \cite{shi2004generation}. Modern designs, however, do not need post-selection and instead use a polarization beam splitter as the first element (Fig. \ref{foldedMZI-polSagnac}(b)). This polarizing Sagnac configuration \cite{slussarenko2010polarizing} is currently one of the most common sources in modern studies as it supports the use of all types (0,I \& II) of phase-matching \cite{kim2006phase,fedrizzi2007wavelength,kuzucu2008pulsed,predojevic2012pulsed,hentschel2009three,steinlechner2014efficient}. This design also has progressed to field deployment \cite{beckert2019space}, and has been demonstrated on the Micius spacecraft \cite{yin2017satellite}.
\begin{figure}[h]
	\centering
	\includegraphics[width=0.48\textwidth]{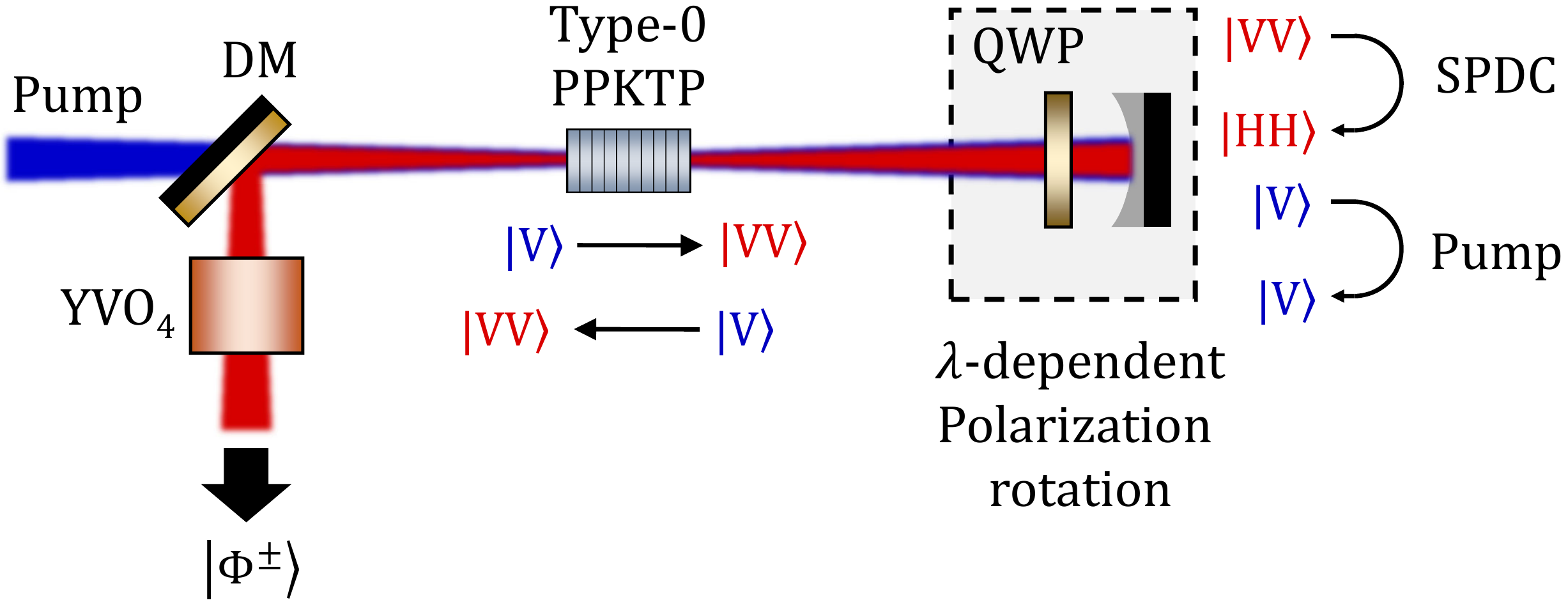}
	\caption{'Folded sandwich' configuration. This is similar to the scheme in Fig. \ref{crossedcrystal-collineartypes}(b) with a waveplate that is ``sandwiched'' between two SPDC decay-paths. The quarter waveplate in this diagram rotates only the polarization of the SPDC photons. This source holds the record for observed brightness in the literature. \cite{steinlechner2013phase,steinlechner2015sources}.}
	\label{folded-sandwich}
\end{figure}

The Sagnac scheme has advantages in stability, but has a larger physical footprint as there is always an enclosure that is unused. Furthermore, the separation of the pump beam into two parts means that the SPDC production in each arm is not utilising the full pump power. A compact and bright source fully utilising the pump power is the ``folded-sandwich'' design (Fig. \ref{folded-sandwich}) \cite{steinlechner2013phase,steinlechner2015sources}.
The folded-sandwich design uses less components and the full pump power is utilised in each pass. To date, this design holds the record in the literature for detected brightness. The design is not intrinsically stable and generally requires active stabilisation.

\begin{figure}[h]
	\centering
	\includegraphics[width=0.48\textwidth]{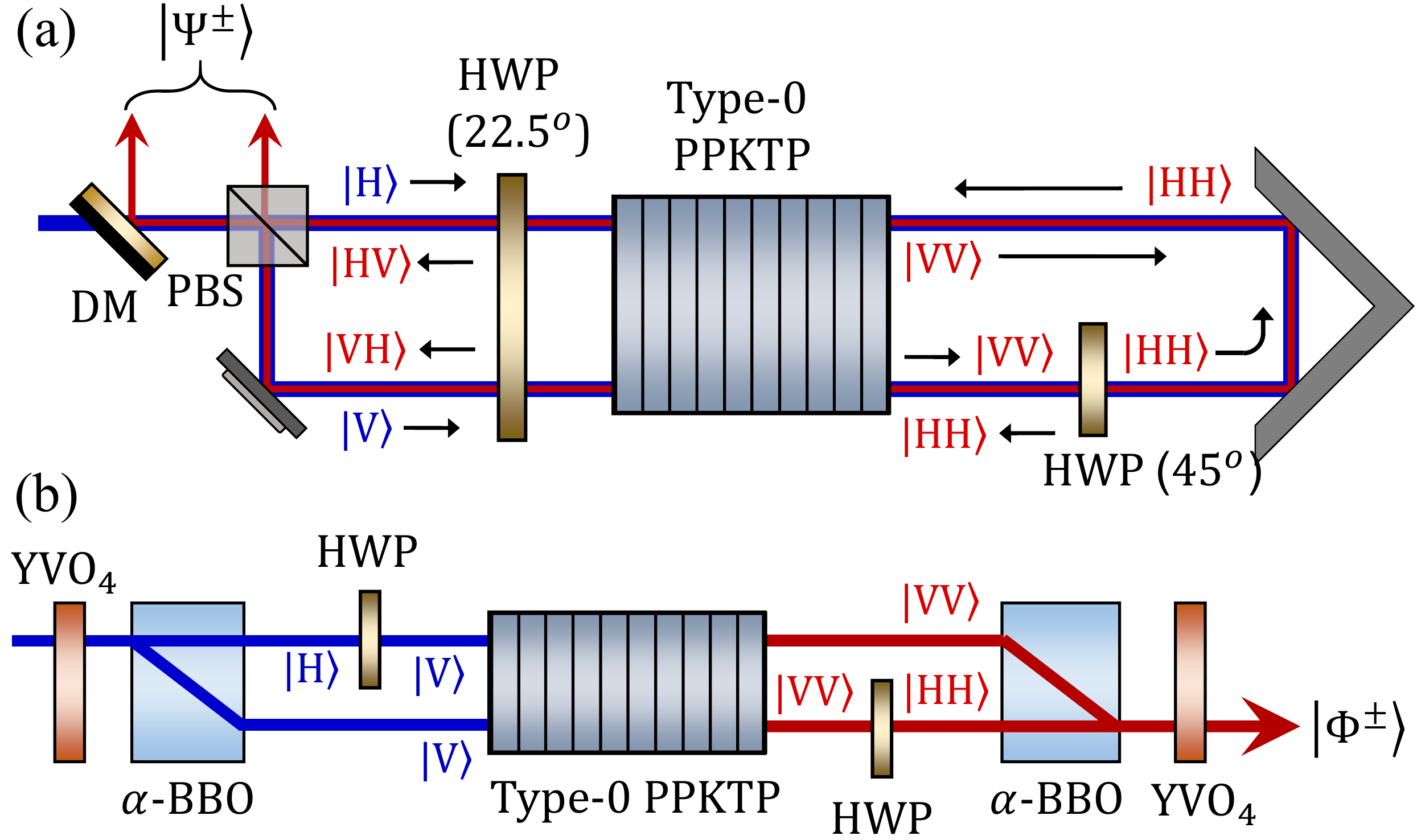}
	\caption{Schematic of (a) integrated double-pass polarization Sagnac interferometer \cite{terashima2018quantum} and (b) linear displacement interferometer  \cite{fiorentino2008compact,lohrmann2019PAPA}.}
	\label{type0-PPKTP-doublepass}
\end{figure}

To combine the advantages from the previous designs, i.e. better stability and reduced physical footprint, it is possible to use the concept of a linear displacement interferometer (Fig. \ref{type0-PPKTP-doublepass}).
In this scheme, the pump does not backtrack on itself, but takes two distinct paths through the material.

One approach uses a loop-back configuration (Fig. \ref{type0-PPKTP-doublepass}(a)) where the pump beams pass through two different regions of the same crystal.
The design uses achromatic waveplates to manipulate the polarization of both pump and SPDC photons.
This particular Sagnac design can be used for all phase-matching types. The other approach would separate the pump beams laterally to propagate in the same direction through a type-0 PPKTP crystal \cite{fiorentino2008compact, lohrmann2019PAPA} (Fig. \ref{type0-PPKTP-doublepass}(b)). This linear beam-displacement interferometer is intrinsically very stable. In a similar configuration, the two laterally displaced pump beams can also be passed through a type-II PPKTP crystal to produce entangled photons after combining signal photons and idler photons separately using two beam displacers \cite{evans2010bright}.

\subsubsection{\label{doublecrystaldoublepass}Double crystal, double pass configuration}

\begin{figure}[h]
	\centering
	\includegraphics[width=0.48\textwidth]{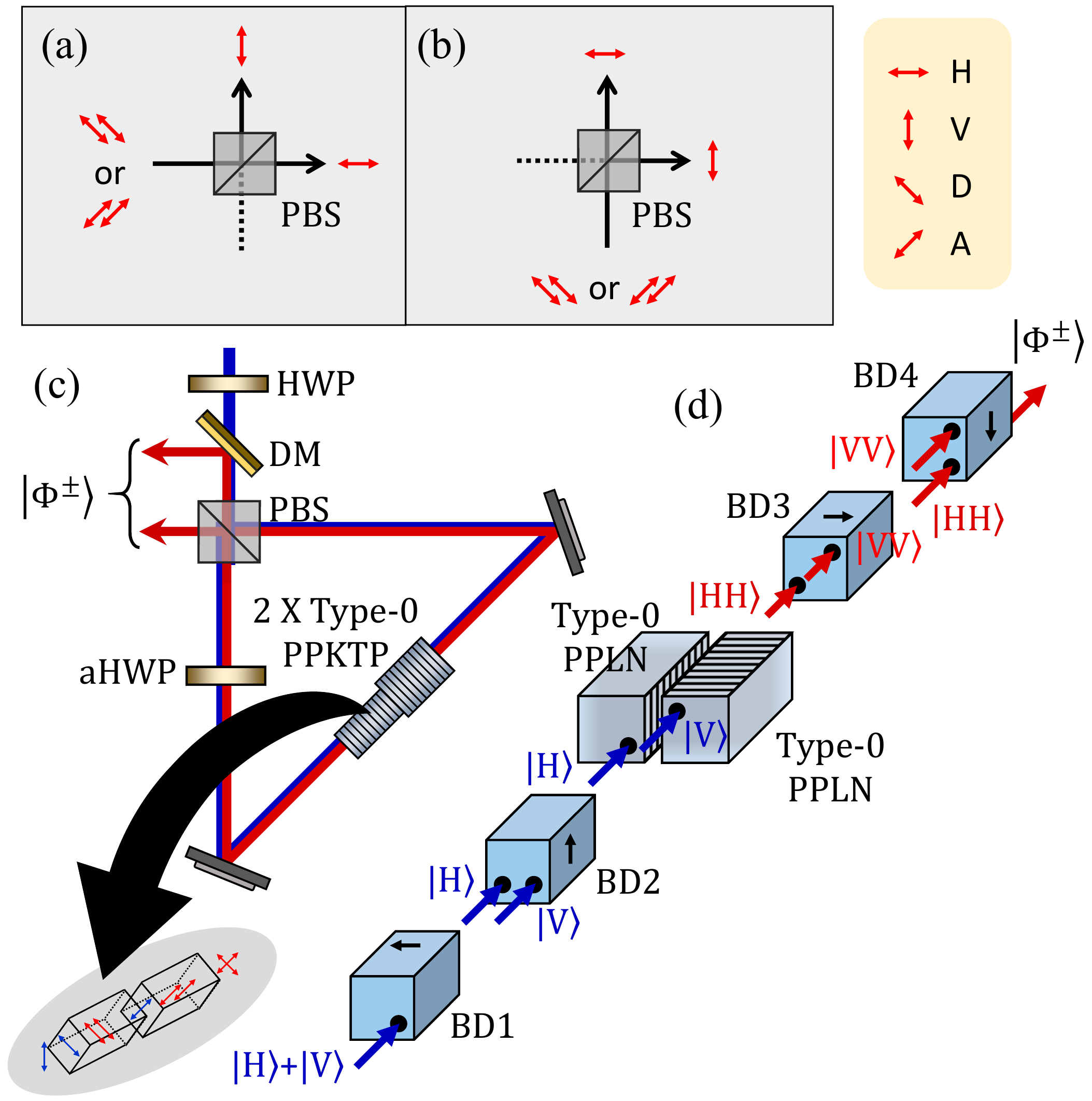}
	\caption{(a), (b) Two cases of time-reversed Hong-Ou-Mandel (HOM) interference \cite{chen2007deterministic}. (a) In the case for first input, two photons with diagonal or anti-diagonal polarization from one input port will split at a polarizing beam-splitter to $H$ and $V$ polarizations in first and second output ports. (b) In the case for second input, The photons will be separated with swapped polarizations at the output. (c) Two-crystal Sagnac scheme based on time-reversed HOM interference \cite{chen2018polarization}. (d) Source of entangled photon-pairs based on double-crystal interferometer \cite{horn2019auto}. BD - Beam displacer.}
	\label{TCDP}
\end{figure}

When two crossed crystals are oriented in a polarizing Sagnac loop, entangled photon-pairs can be obtained using time-reversed Hong-Ou-Mandel interference \cite{chen2007deterministic}  (Fig. \ref{TCDP}). 
This enables photon-pairs with the same color to be deterministically separated without using dichroic mirrors or momentum correlations. This was first reported for fiber-based systems, but has also been implemented for bulk crystals \cite{chen2018polarization}.

A modified linear interferometric source based on double displacement technique was recently introduced \cite{horn2019auto} (Fig. \ref{TCDP}). In this scheme, two beam displacing (BD) crystals are used in the input and output and this can also achieve spatial and temporal balance, leading to a high quality entangled photon-pair output.

A comparison of the performance for major polarization entangled photon-pair sources in near-infrared and telecom wavelengths with their performance parameters can be found in Tables \ref{table1} and \ref{table2}, respectively. 

\subsubsection{\label{C-SPDC} Polarization-entangled photon sources based on cavity-enhanced PDC}

Typically, single-pass SPDC  produces photon-pairs with a bandwidth that corresponds to the phase mismatch $\Delta \bf{k} < \pi$ which typically spans several  GHZ. The bandwidth of these sources can be reduced by passive filtering using narrow-band spectral filters or by placing the SPDC-crystal inside a resonator. The advantage of the passive filtering approach is its flexibility - almost any spectral profile can be 'cut out' from the emission spectrum. However, in order to counteract the filter losses that occur in this process, a correspondingly high number of emitted photons is required. 

Alternatively, it is also possible to imprint the desired spectral characteristics on the generation process itself, by embedding the nonlinear medium in a resonant cavity. In this case,  the cavity finesse and the ratio of its free-spectral range with respect to the natural phase-matching bandwidth - as well as which of the fields $(p,s,i)$ are on resonance - all act together in determining the photons' spectral properties.  Depending on how many of the interacting fields are resonantly enhanced, one distinguishes between singly-, doubly-, and triply resonant cavity PDC configurations. 

When the signal and/or idler photons are on resonance, the  SPDC emission is squeezed into frequency modes permitted by the optical cavity, resulting in a series of  emission peaks with reduced spectral bandwidth.  Cavity enhanced SPDC thus prolongs the coherence time of the created photon-pairs  \cite{ou1999cavity} and enhances the spectral brightness within the resonance bandwidth by a factor of up to $F^2$, where $F$ is the Finesse of the resonator. Another advantage of this approach is that locking the cavity to another material system,  allows to produce entangled photons that spectrally match another quantum system of interest such as atoms or ions which are relatively narrow band of just a few MHz.
 
Since the first theory papers on this topic, several groups have reported on narrowband photon-pair sources based on cavity-enhanced SPDC. A  detailed  account of the body of experimental research is beyond the scope of this article, andcan be found in Ref. \cite{slattery2019background}. Section \ref{WGSources} merely highlights some of the work on polarization-entangled photon sources with cavity-enhancement.

In 2000, Oberparleiter et al.  demonstrated a SPDC source for polarization entangled photons with a pump cavity \cite{oberparleiter2000cavity}, thereby enhancing the effective pump power for a 2 mm BBO crystal by a factor of 7, while observing a polarization correlation visibility of up to  96\%. The bandwidth of the 702 nm SPDC photons remained relatively broadband, about 5 nm FWHM.

Producing narrow-band SPDC, on the other hand, requires a cavity for the SPDC signals themselves, essentially representing a optical parametric oscillator (OPO) that is operated  below threshold \cite{Kuklewicz_2002}. An important application of narrow-band photons is compatibility with atomic resonances. One such design weas demonstrated by Kuklewiecz et al. \cite{Kuklewicz2006}, where a cavity around a PPKTP crystal with a free-spectral range of 1.21 GHZ generated 795nm SPDC signals with a bandwidth of about 22 MHz, and a spectral brightness of 0.7 pairs/s/mW/MHz. Polarization entanglement was not implemented in that design.

The generation of narrow-band polarization-entanglement in cavities is associated with the additional challenge that resonance conditions for orthogonal polarizations must be matched inside the cavity. Polarization-etnanglement can then be achieved in post-selection \cite{wolfgramm2010noon,zhang2011preparation} or by superimposing narrowband pair emission processes on a polarizing beam splitter. 

One example for cavity enhanced polarization entanglement was demonstrated by Bao et al. in 2008, who reported a narrow-band polarization entangled photon source \cite{bao2008generation} generating pairs at 790 nm with about 9.6 MHz bandwidth (or a 32 m coherence length), and a entanglement visibility of about 94\%. The spectral brightness was reported at about $6\times10^3$ pairs/s/MHz/mW. In this source, polarization entanglement was generated in post-selection on coincidence events after probabilistic separation on e.g. a beam splitter. 

The first report, to the best of our knowledge, of a narrow-band cavity-enhanced source without the requirement for post-selection was reported in Ref. \cite{arenskotter2017polarization}, where photons were deterministically separated by embedding a resonant down-conversion cavity inside a polarization Sagnac loop configuration. 

More recently, the other forms of entanglement have also been studied in cavities. In resonators with corresponding dispersion properties, complex emission profiles with multiple resonances can also be realized, a so-called bi-photon frequency comb \cite{lu2003mode,zavatta2008toward,xie2015harnessing}. Such an approach was recently used to demonstrate frequency-bin entanglement  \cite{rielander2017frequency}. Similar approaches have recently been demonstrated in micro-ring resonators in silica \cite{caspani2017integrated}, aided by the great flexibility in engineering dispersion on this platform.

\section{\label{WGSources} Comparison with waveguide sources of entangled photons} 

Sources of entangled photon-pairs based on non-resonant SPDC in periodically poled crystals and bulk-optical systems are highly flexible and still mark the state of the art with respect to several critical performance characteristics (see Appendix for a detailed comparison). They are, however, subject to practical limitations with respect to their scalability to higher mode and photon numbers, overall footprint, as well as to the complexity of the possible state transformations. To this end, integrated photonic technology offers a variety of promising platforms for ultra-compact turn-key photon sources that will be discussed briefly in the following. For the sake of consistency with the scope of the Section \ref{polarizationentlpairsources} on bulk SPDC sources,  this section will cover only source based on SPDC in second-order nonlinear materials, and refer to Ref. \cite{caspani2017integrated,wang2020integrated} for a  comprehensive review on the topic of integrated  sources based on SFWM in structured fibers, chip-integrated micro ring resonators and silicon waveguides.

Integrated waveguide structures may be realized in a number of  second-order nonlinear materials, such as poled silica fibers, planar waveguides in KTP and LN, and semiconductor Bragg reflection waveguides \cite{valles2013generation}. Due to stronger  mode-confinement in waveguides, the spectral efficiency of PDC scales beneficially with the length of the nonlinear interaction region ($\propto L^2$, as opposed to $\propto L$ in bulk crystals \cite{helt2012does}). The propagation parameters of the guided modes are critical in establishing modal phase-matching \cite{Christ:2009aa}, and they can actually be used to accomplish phase matching in non-birefringent materials such as  Gallium-Arsenide \cite{horn2013inherent}.

Additionally, guided wave technology promises additional functionalities such as the integration of narrow-bandwidth Bragg filters for pump suppression, fiber-based wavelength division multiplexers, tunable on-chip routing of photons via ultra-fast electro-optic modulators, and - most importantly - the possibility of dispersion engineering for biphoton states in order to tailor their spectral properties, e.g. spectrally pure biphoton states for multi-photon experiments at low pump powers. In the telecom wavelength range sources based on SPDC in periodically-poled lithium niobate waveguides have emerged as one of the key platforms for photonic integration, owing in particular to the maturity of fabrication processes adapted from classical telecommunication systems. It should, however, be mentioned that fabrication and poling technologies for other materials, such as Stoichiometric Lithium Tantalate (SLT) and Potassium Tantalate (KT) rapidly advancing, making these very promising platforms for future PDC sources.

Dating back as early as 2003, Yoshizawa et al. reported on a source consisting of two type-0 PPLN waveguides in a fiber-based Mach–Zehnder interferometer configuration \cite{yoshizawa2003generation}. Since then, several other groups have adapted the previously-mentioned bulk optics polarization-entangled photon  schemes to on-chip platforms. In-fiber Sagnac loop configurations being a particularly popular choice, owning to the intrinsic phase stability of this configuration, without the stringent alignment tolerances of the bulk optical counterpart \cite{vergyris2017fully}. A very different approach, was demonstrated by Hermann et al. \cite{herrmann2013post}. Using mature PPLN poling technology, an integrated waveguide chip with two interlaced poling periods was realized. The poling periods where chosen for type-II SPDC at slightly different wavelengths, so that a post-selection free, integrated optical source of non-degenerate, polarization-entangled photon-pairs was achieved. Refining this approach, Sun et al. \cite{sun2019compact} recently demonstrated polarization entanglement with a state fidelity of >95\% and an estimated pair generation rate inside the waveguide of $1.22\times10^7$ pairs/s/mW/nm.

Next to the adaptation of bulk optic source schemes discussed in Section \ref{polarizationentlpairsources}, readily available fiber-based wavelength-division multiplexers, with channel spacing far below the typical PDC bandwidth have enabled novel integrated sources that do not have a bulk-optics counterpart. Ref \cite{kaiser2012high} demonstrated polarization entangled photon source operating at 1540 nm employing type-II PPLN waveguide and a polarization-maintaining fiber for walk-off compensation, a cascade of two dense wavelength division multiplexers was used to determnistically separate signal and idler photons into distinct channels of the ITU wavelength grid.  Later, the same group reported a  source configuration based on  SPDC in a type-0 PPLN waveguide. Polarization entanglement was prepared by mapping temporal entanglement to the polarization domain using an unbalanced Mach-Zehnder interferometer with in-fiber polarization beam splitters and detection by post selection \cite{kaiser2014polarization}. 

Another promising platform for in-fiber photon-pair generation is based on inducing a second-order nonlinearity in glass by thermal poling or UV illumination in a strong (typically kV) field. Upon cooling of the sample, a residual electric field remains, thus breaking the materials centro-symmetry and inducing a potentially strong second-order nonlinearity. A significant advantage of this approach is that it allows for the generation of entangled photons directly within an optical single mode fiber. After the pioneering work  demonstrating parametric fluorescence in poled fibers in 1999 \cite{bonfrate1999parametric}, this approach has since led to several high-performance sources. For example, Ref. \cite{zhu2013poled} reported on the generation of polarization entangled photons with visibilities of more than 97\%  via  a type-II quasi phase-matching in a poled polarization-maintaining fiber.  Further optimization of this scheme  has led to sources with high-fidelity polarization-entanglement over spectral bandwidths approaching 100 nm, making these sources highly suitable for multiplexing in fiber networks \cite{chen2017compensation}.

In conclusion, integrated platforms such as on-chip waveguides, nonlinear fiber optics and resonant nonlinear microresonators offer not only practical advantages such as reduction of form factor and improvement of long-term stability, but also the possibility to significantly increase pair generation rates, mode numbers and coherence. These sources hold great promise for future applications - in particular, once the demand for efficient pair sources reaches levels that require scalable mass-producable devices. At present, however, with the exception of reported \emph{emitted pair rates},  integrated optics sources are yet to match the performance of their bulk-optical counterparts in  key parameters for quantum optics experimentation such as fiber coupling efficiency \cite{shalm2015strong,giustina2015significant} and entanglement fidelity \cite{poh2015approaching}. This is owing mostly to the larger total loss (as reflected in the lower heralding efficiency typically observed for on-chip sources, see comparison in Appendix) and will undoubtedly be tackled with increasing availability of ultra-low-loss on-chip components and maturity of fabrication processes. As an intermediate alternative solution, hybrid approaches   promise the best of the two worlds. In Ref. \cite{meyer2018high}, the authors reported on a hybrid bulk-waveguide source with performance close to the best bulk-optics sources. In this scheme  photon-pairs were produced in a type-II PPLN waveguide, which was bi-directionally pumped inside a bulk Sagnac loop configuration. A similar approach based on SPDC generated in a PPLN waveguide with the more effiecient type-0 QPM interaction was recently demonstrated to achieve even higher performance \cite{tsujimoto2018high}. 

The intention of this section was to highlight the vast amount of research dedicated to integrated entangled photon sources. For a more comprehensive review on the topic of integrated  source technology based on SPDC, the reader is referred to Refs. \cite{tanzilli2012genesis,alibart2016quantum}. 

\section{\label{entlotherdof}Photonic entanglement in other degrees of freedom}

The correlations between photon-pairs in SPDC extend beyond polarization to include position-momentum, energy-time and spatial modes. From these correlations it is possible to implement entanglement using different degrees of freedom. This can lead to \textit{hyper} and \textit{hybrid} entanglement. Hyper entanglement occurs when the photon-pair has entanglement in two or more degrees-of-freedom \cite{kwiat1997hyper}. Hybrid entanglement occurs when \textit{different} degrees of freedom are entangled (e.g. polarization and spatial mode \cite{nagali2010generation}). These non-polarization entanglement arise from the conservation of momentum and energy (Eqn. \eqref{energy-conserv}), and can be classified accordingly.
\subsection*{Entanglement from momentum conservation}

\subsubsection{Spatial entanglement}

\begin{figure}[h]	
\centering	
\includegraphics[width=0.48\textwidth]{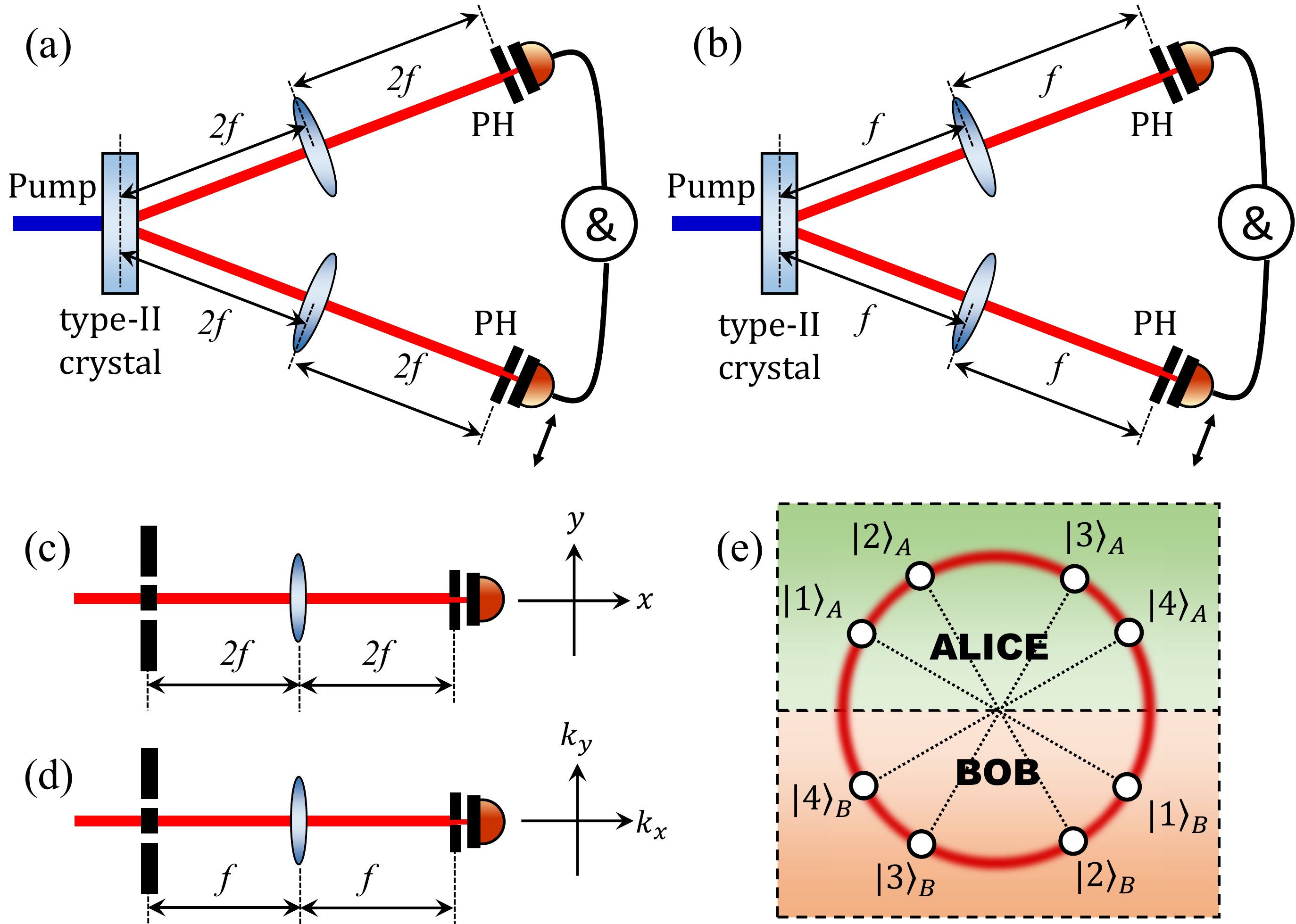}	\caption{Experimental schemes for measuring (a) position correlation of photon-pairs using $4f$-imaging and (b) momentum correlation of photon-pairs using $2f$-imaging. Spaitally entangled qubits can be prepared using imaging of double-slit \cite{neves2007characterizing} (c) in position space ($x,y$), or (d) in transverse momentum space ($k_x,k_y$). (e) Labeling of the correlated pairs of SPDC modes. Photon $A$ ($B$) can be collected with equal probability into one of the four modes $\ket{1}_j$, $\ket{2}_j$, $\ket{3}_j$ and $\ket{4}_j$ ($j=A,B$). The four photon-pairs lead to multi-path entanglement \cite{rossi2009multipath}. PH - Pinhole.}
\label{position-mom-entanglement}
\end{figure}

Photon-pairs produced by SPDC are correlated in position, as they are generated at the same position in the interaction volume, and are anti-correlated in momentum due to momentum conservation \cite{walborn2010spatial}. Position correlation is commonly measured using $4f$-imaging (Fig. \ref{position-mom-entanglement}(a)), whereas  momentum correlations are measured using $2f$-imaging (Fig. \ref{position-mom-entanglement}(b)). These correlations have been used to demonstrate the Einstein-Podolsky-Rosen paradox \cite{howell2004realization, chen2019realization}.

These correlations can give rise to spatial entanglement (sometimes also known as transverse entanglement, or pixel entanglement) when the position or momentum state is discretized \cite{law2004analysis,just2013transverse}. Experimentally, this is done using multiple slits \cite{neves2007characterizing,neves2005generation}, coupling into a fiber array \cite{o2005pixel}  or with programmable optical devices \cite{lima2009manipulating}. Imaging of transverse spatial entanglement uses either a raster scanning method \cite{ostermeyer2009two} or low-light imaging devices \cite{edgar2012imaging}. An alternative approach to spatial entanglement is by proper domain engineering and linear chirping of quasi phase-matched crystals \cite{yu2008transforming,svozilik2012high}. 

Photon-pairs from diametrically opposite points on a non-collinear SPDC output are used for path encoding for multi-path entanglement (Fig. \ref{position-mom-entanglement}(e))  \cite{rossi2009multipath}. Spatial entanglement has applications in non-local imaging \cite{gomes2009observation} and demonstration of higher dimensional quantum protocols \cite{tasca2011continuous,zhang2008secure}.

\subsubsection{Orbital angular momentum entanglement}

\begin{figure}[h]
	\centering
	\includegraphics[width=0.48\textwidth]{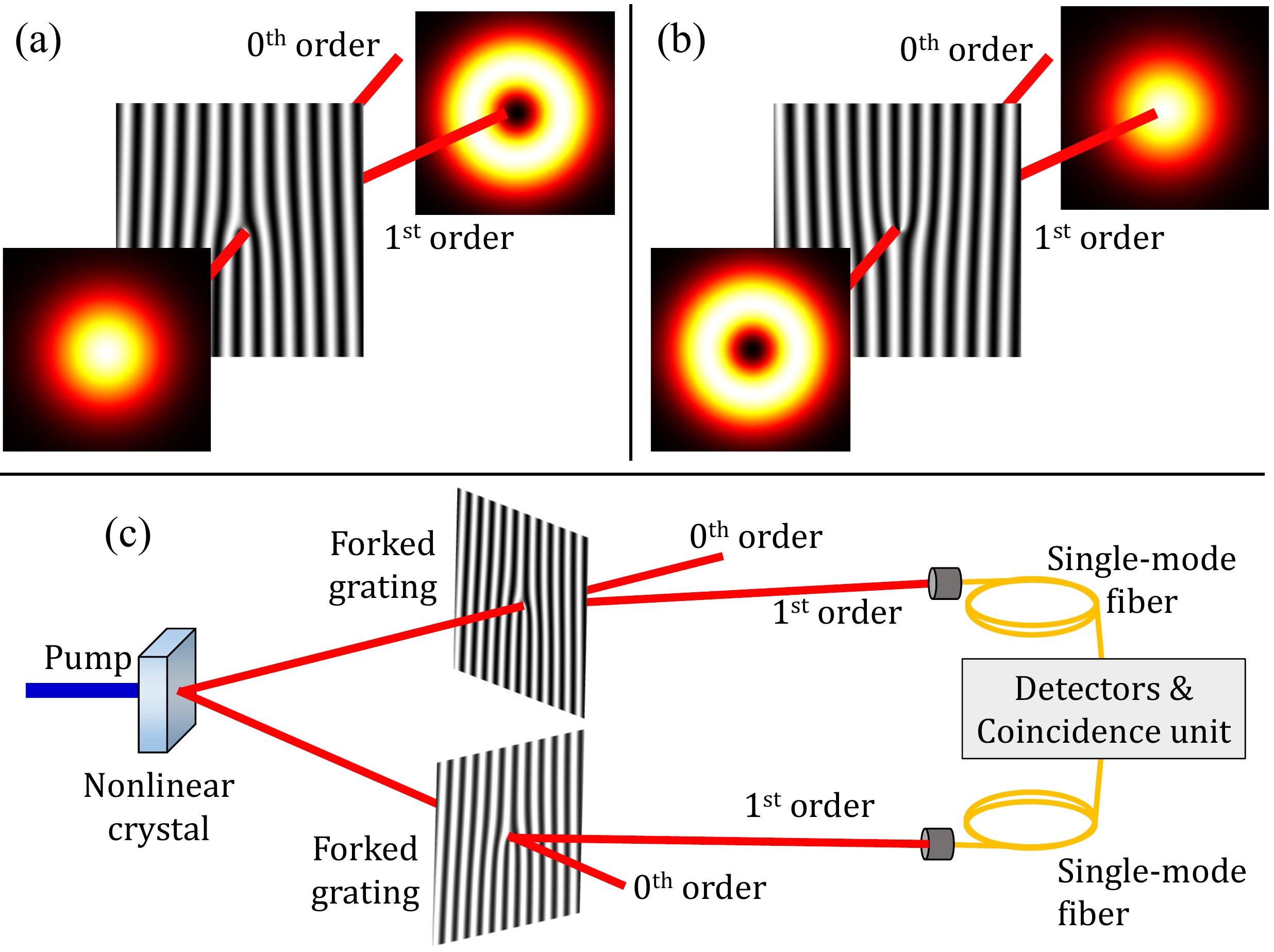}
	\caption{(a) Generation of OAM modes using grating holograms. (b) OAM modes can be detected using 'phase-flattening' technique where the incoming azimuthal phase of the mode ($e^{il\phi}$) is nullified after passing through a hologram containing azimuthal phase in opposite order ($e^{-il\phi}$). (c) Generic scheme for the detection and analysis of OAM entanglement in SPDC \cite{mair}.  Using phase-flattening technique, the output phase-nullified distribution in each arm is coupled to a single-mode fiber.}
	\label{SPDC-OAM-scheme}
\end{figure}

Light beams carrying orbital angular momentum (OAM) have received attention due to applications in optical tweezers and optical communication. \cite{padgett2017orbital}. The Laguerre-Gauss mode was the first to be identified to carry a well defined OAM \cite{allen1992orbital,beijersbergen1993astigmatic,padgett2017orbital}. The OAM modes have a doughnut-like intensity distribution and a helical phase structure where the phase changes azimuthally around the optic axis. Such beams are also known as optical vortices. These beams can be generated using holography where a Gaussian beam diffracted through a forked holographic grating  generate an OAM mode in the first order (Fig. \ref{SPDC-OAM-scheme}(a)).

The use of OAM modes enables photons to access a degree-of-freedom which spans an infinite dimensional Hilbert space. In contrast, polarization is restricted to two-level systems. In SPDC orbital angular momentum is also conserved, $l_p=l_s$+$l_i$, where $l_p$, $l_s$ are $l_i$ are OAMs of pump, signal and idler respectively \cite{walborn2004entanglement}. A pump beam with a Gaussian intensity distribution has $l_p=0$ and imposes a condition of $l_s =-l_i$. The OAM state of the SPDC output is then given by 
\begin{equation}
     \ket{\psi}=\sum_{l=-\infty}^\infty a_l\ket{l}_s\ket{-l}_i
     \label{SPDC-OAM-state}
\end{equation}
where $a_l$ is the probability amplitude for the state $\ket{l}_s\ket{-l}_i$. Realization of OAM entanglement is achieved by restricting the dimension via post selection. 

In the first experimental verification of OAM entanglement \cite{mair}, the Hilbert space is restricted to $\{\vert 1\rangle ,\vert -1\rangle \}$. To measure the correlations in OAM, the photon-pair is diffracted through a pair of holograms before collection into a single mode fiber (Fig. \ref{SPDC-OAM-scheme}(b)). An alternative method uses spiral phase plates \cite{oemrawsingh2005experimental} and sector plates \cite{oemrawsingh2006high,pors2008shannon,pors2011high}. 
The use of dynamic spatial light modulators (SLM) instead of static mode projectors gave a significant boost in the OAM entanglement experiments. SLMs allow the use of any desired projected spatial mode which can be changed dynamically without any realignment in the experiment \cite{franke2002two,leach2009violation,dada2011experimental}. This leads to the development of various methods for generation and analysis of muti-dimensional quantum states with a single pair of photons \cite{vaziri2002experimental,torres2003preparation,vaziri2003concentration} as well as multi-pair systems \cite{malik2016multi}. 
Alternatively even and odd states of OAM could be used as measurement basis in quantum information tasks involving OAM entanglement \cite{perumangatt2017quantum}. 

In the above examples, there is only one sub-space for the OAM states, corresponding to $l_p=l_s$+$l_i$.
Increasing the number of $l_p$ states in the pump, would increase the number of sub-spaces
 \cite{romero2012orbital,kovlakov2018quantum,liu2018coherent,anwar2020selective}. The bandwidth of the OAM spectrum can be controlled by adjusting the pump waist \cite{torres2003quantum}.
The two photon OAM spectrum of photon-pairs is not uniform as the contribution of higher OAM in Laguerre-Gaussian basis is less due to the dependence of OAM on the radial part of the mode \cite{qassim2014limitations}. Also, there is significant effect of atmospheric turbulence on the OAM content in the spectrum \cite{tyler2009influence,ibrahim2013orbital}, which causes decay of high dimensional entanglement \cite{zhang2016experimentally}. To overcome some of these effects, entanglement of OAM were investigated with other elegant spatial modes \cite{kovlakov2017spatial,mclaren2012entangled,mclaren2013two,mclaren2014self,krenn2013entangled}.

\subsection*{Entanglement from energy conservation}

The conservation of energy in SPDC ensures frequency correlation and simultaneity in the generation of the photons in each pair. The time-bandwidth product of a photon obeys the uncertainty principle, $\Delta E\Delta t\ge \hbar/2$, where $\Delta E$ and $\Delta t$ are uncertainties in energy and time of the photon, respectively. However, for a photon-pair in SPDC, the joint uncertainties $\Delta E_{si}=\Delta(E_s+E_i)$ and $\Delta t_{si}=\Delta(t_s-t_i)$ of signal ($s$) and idler ($i$) violate inequality for the classical separability bound \cite{howell2004realization}, leading to different forms of entanglement in the temporal degree-of-freedom. This can allow high-dimensional entangled states which can be distributed over fibre, along with polarization encoding, to increase the capacity of quantum communications.

\subsubsection{Energy-time entanglement}

Energy-time entangled photonic states have been realized through SPDC  \cite{rarity1990two,kwiat1990correlated,ou1990observation,brendel1992experimental,chiao1995quantum}, following the proposal by Franson \cite{franson1989bell}. Entanglement in such systems is characterized by analysis of Bell states \cite{weinfurter1994experimental} from correlated interference effects.
\begin{figure}[h]
	\centering
	\includegraphics[width=0.48\textwidth]{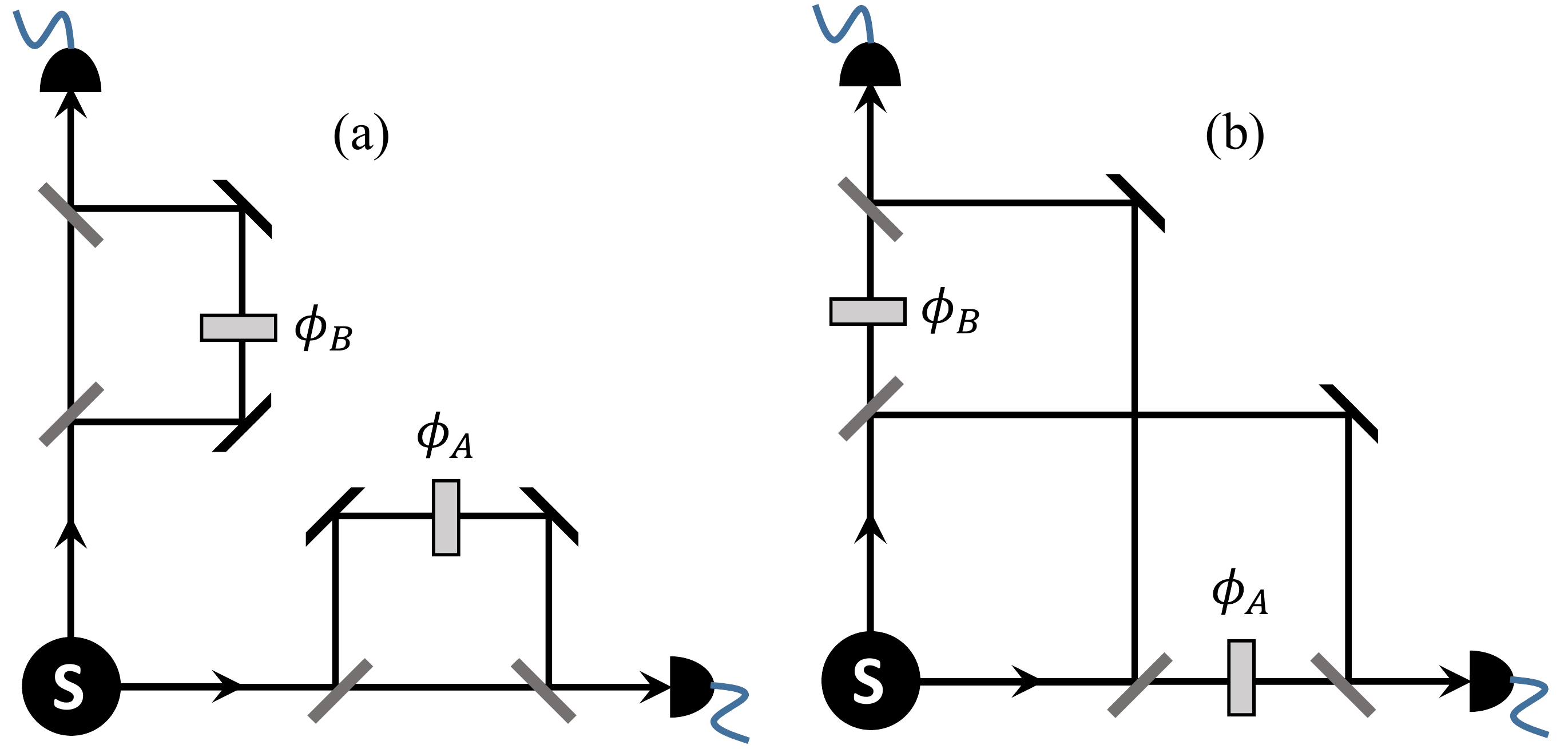}
	\caption{(a) Experimental schematic for generation of time-energy entangled photons using SPDC \cite{franson1989bell,rarity1990two,kwiat1990correlated,ou1990observation,brendel1992experimental,chiao1995quantum}. (b) The 'hug' type scheme of time-energy entangled photon source \cite{cabello2009proposed}.}
	\label{time-entanglement-schemes}
\end{figure}

The basic experimental scheme for energy-time entanglement using SPDC is given in Fig. \ref{time-entanglement-schemes}(a). An unbalanced interferometer is introduced in each SPDC arm such that the time delays in both interferometers are much larger than the coherence times of individual photons, but much smaller than the coherence times of photon-pairs. 
The interference effects show up from the joint measurement of photon-pairs. Coincidence events are detected when a photon-pair travels through either 'longer' arms ($l$) or 'shorter' arms ($s$) of the interferometers. 
So, the entangled state created by SPDC pairs is
\begin{equation}
    \ket{\psi}=\frac{1}{\sqrt{2}}(\ket{s}\ket{s}+e^{i\phi}\ket{l}\ket{l})
    \label{time-energy-entl-state}
\end{equation}

In the schemes discussed above, the fringe visibility is limited by slow detectors and lower counts due to restriction of coincidence window. The interference visibility was significantly improved by means of post-selection using fast photodetectors \cite{kwiat1993high}. The reduction in coincidence detection due to post-selection is eliminated by the use polarization entangled photon source with type-II crystal \cite{kwiattype2} to demonstrate energy-time entanglement \cite{strekalov1996postselection}. Energy-time entangled states in high dimensions were also demonstrated with continuous-wave SPDC sources \cite{khan2006experimental,ali2007large,shalm2013three,schwarz2015experimental} and discretization of frequency spectrum using spatial light modulators for labelling qudits \cite{bessire2014versatile}. It was proposed that the use of multiple Franson interferometers in Alice and Bob increases the sensitivity against certain photon localization attacks in high dimensional energy-time entanglement based quantum key distribution systems \cite{brougham2013security}. 

Some major schemes for energy-time entanglement discussed above suffer from an intrinsic post-selection loophole (PSL) \cite{aerts1999two} as well as a geometrical loophole (GL) that could affect the security of communication \cite{larson2002practical}. The polarization and energy-time hyper-entanglement realized in SPDC \cite{barreiro2005generation} is exploited in overcoming PSL \cite{strekalov1996postselection} and utilized in distributing entanglement through intra-city free-space link \cite{steinlechner2017distribution}. To address PSL and GL, a 'hug' type configuration for genuine time-energy entanglement based on interferometers wrapped around the photon-pair source, was introduced \cite{cabello2009proposed} (Fig. \ref{time-entanglement-schemes}(b)). Sources based on this have a mutual interlocking of interferometers and are convenient in table-top experiments \cite{lima2010experimental,vallone2011testing}. Although such scheme is practical in distribution of entanglement over larger distances \cite{cuevas2013long}, the source requires stabilization of longer interferometers.

As a concluding remark, we note that - much like spatial mode entanglement, the continuous energy-time entanglement may also be decomposed into field orthogonal temporal modes. To make the number of temporal modes experimentally tractable, discrete time-energy entanglement, so-called temporal-mode entanglement, typically involves SPDC with ps pump pulses. Detection of field orthogonal temporal modes can be accomplished either using SFG techniques or ultra-fast EOM. For a review refer to Ref. \cite{brecht2015photon}.

\subsubsection{Time-bin entanglement}
\begin{figure}[h]
	\centering
	\includegraphics[width=0.48\textwidth]{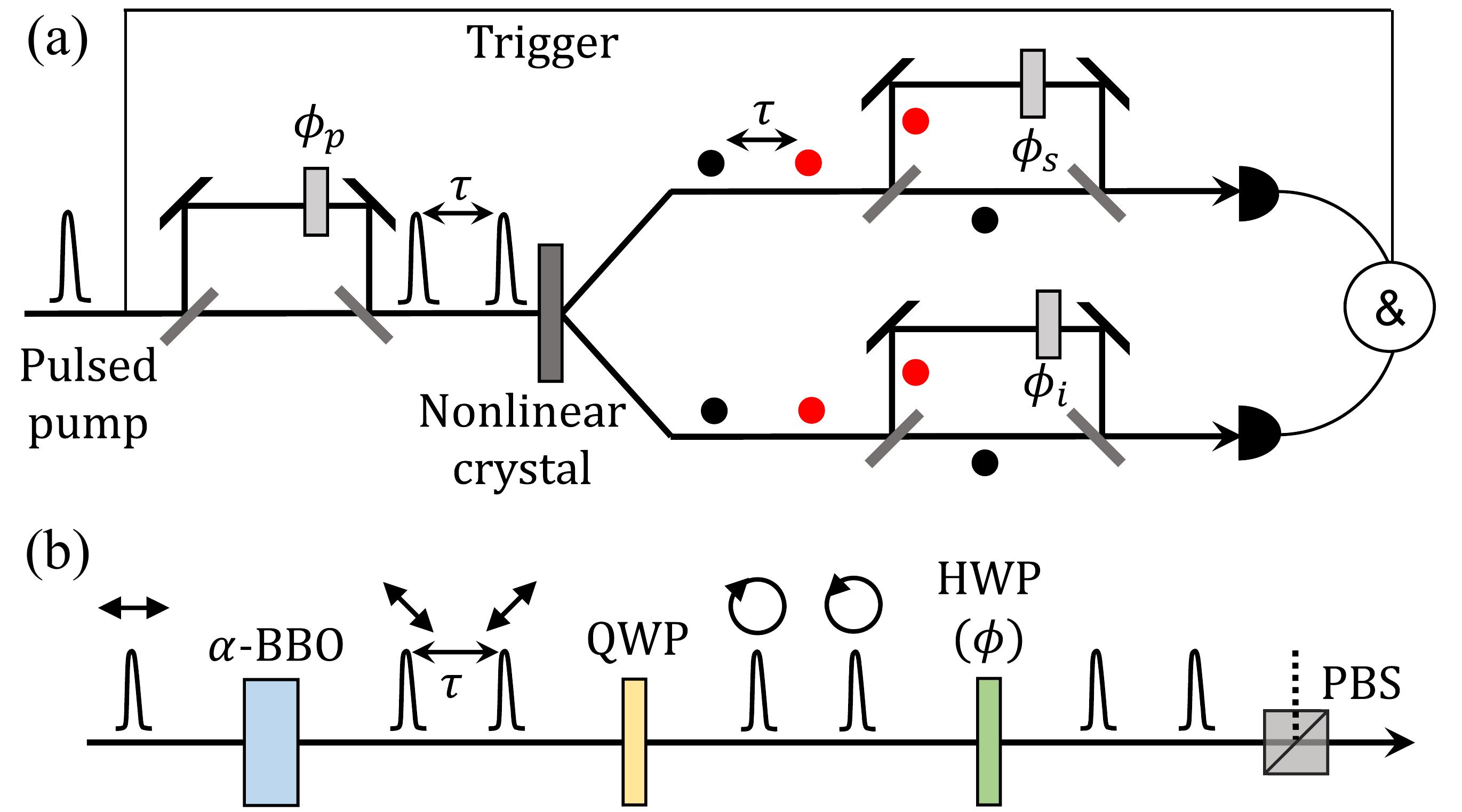}
	\caption{(a) Experimental schematic for generation of time-bin entangled photons with pulsed pump \cite{brendel1999pulsed}. A pump pulse is split into two pulses with a time delay, upon passing through the interferometer. The 'early' and 'late' pulses coming after the interferometer constitutes the state $\alpha\ket{\text{early}}_{\text{pump}}+\beta\ket{\text{late}}_{\text{pump}}$. The photon-pair from an early pump pulse traversing longer arms of the interferometers are indistiguishable with the pair from a late pump pulse traversing shorter arms. So, the joint state is $\alpha\ket{\text{early}}_s\ket{\text{early}}_i+\beta\ket{\text{late}}_s\ket{\text{late}}_i$, which is similar to the energy-time correlations obtained from Franson's experiment \cite{franson1989bell}. The red and black circles denote the photons generated from 'early' and 'late' pump pulses respectively \cite{maclean2018ultrafast}. (b) Scheme for single path Franson interferometer using $\alpha$-BBO, quarter-wave plate (QWP) and a half-wave plate (HWP). $\alpha$-BBO splits the initial horizontally polarized photon to diagonal and delayed anti-diagonal pulses. The QWP converts them to right and left circular polarizations. The HWP introduces a phase between the circularly polarized photons. Both polarizations are projected to the horizontal state upon transmission through the PBS.}
	\label{time-bin-schemes}
\end{figure}

Similar to energy-time entanglement, Franson interferometers are also used to generate time-bin entanglement but with a pulsed pump \cite{brendel1999pulsed,marcikic2002time,marcikic2004distribution}. The experimental scheme is shown in Fig. \ref{time-bin-schemes}(a). Entanglement is characterized by three-fold coincidence measurements in gated mode with the pump as the trigger. Unlike energy-time entanglement, this design does not need information about the coherence of the pump laser because the information of arrival times of the pump at the crystal is obtained directly from the pump superposition state in the bases $\ket{\text{early}}$ and $\ket{\text{late}}$. The use of all fiber Michelson interferometers along with Faraday mirrors in the pump as well as down-converted arms, makes the source compact and suitable for long distance fiber based quantum communication \cite{tittel1998experimental,tittel2000quantum}. Time-bin entangled qubits are also constructed using SPDC with a multi-mode continuous-wave pump \cite{kwon2013time} based on coherence revival of such lasers with short coherence length \cite{baek2007high,kwon2009coherence}. 

Instead of an unbalanced interferometer, an alternative method of giving temporal delay between pulses was implemented based on a combination of $\alpha$-BBO, quarter-wave plate (QWP), half-wave plate (HWP) and a polarizing beam splitter (PBS) \cite{maclean2018ultrafast} (Fig. \ref{time-bin-schemes}(b)). 
Another method is the generation of active time-bin states with a modified Franson interferometer after replacing the input beam-splitter by a balanced Mach-Zehnder interferometer (MZI) \cite{vedovato2018postselection}. The MZI acts as a fast optical switch that makes the 'early' and 'late' photon pulses in the measurement interferometer indistinguishable.

Another interesting configuration is the generation of hybrid-entangled states, such as polarization for one photon, and and time-bin encoding for the other. A straight forward method to implement such entangled states is to utilize a source for time-bin entangled photon-pairs, and effectively convert the time-bin state of one photon into a polarization state using a polarized unbalanced interferometer \cite{fujiwara:261103}.

\subsubsection{Frequency-bin entanglement}
Time-energy and time-bin entanglement has been successfully distributed in numerous experimental field trials and proof-of-concept experiments. While the generation of entanglement of this form  is relatively straightforwardly obtained via SPDC, the analysis of energy-time and time-bin superposition states typically requires actively stabilized unbalanced interferometers. Another promising manifestation of entanglement due to energy conservation is entanglement of discrete frequency bins \cite{rarity1990two,ou1988observation}, superposition of which may be analysed using microwave domain electro-optic modulation and optical waveshaping techniques \cite{lu2018quantum} or quantum interference on a beam splitter \cite{chen2020verification}. 

Frequency entangled states may be generated by adaptive approaches discussed in the context of polarization-entanglement. For example, in Ref. \cite{ramelow2009discrete} the authors used a modified Sagnac loop geometry with non-degenerate signal and idler wavelengths, to generate tunable entanglement of frequencies separated by up to 10 THz, without additional spectral filtering. Likewise, modified crossed-crystal configuration \cite{chen2019hong} and two-period quasi-phase-matched parametric down conversion \cite{kaneda2019direct} have been used to this end. 

Frequency-bin entanglement of many narrowband frequency modes – a biphoton frequency comb \cite{lu2003mode} –  may be achieved by engineering cavity-enhanced SPDC to exhibit multiple narrow-band resonances \cite{rielander2017frequency}, an approach that is also highly suited to photonic integration, in particular on-chip micro-ring resonators \cite{kues2019quantum}. Moreover, the approach has already been combined with entanglement in polarization \cite{xie2015harnessing} and time-bin \cite{reimer2019high} – a hyperentangled frequency comb – as a very promising platform for high-dimensional quantum information processing. 

\section{\label{applications}Applications}

Development of SPDC entangled photon sources using different generation techniques has motivated researchers for their use in major quantum based applications. For a polarization entangled source, the major performance parameters such as source brightness, pair-to-singles ratio and entanglement fidelity determines its suitability for different applications. Time-energy and time-bin entanglement have been well widely exploited for long-distance fiber communications. Spatial correlations and orbital angular momentum entanglement of photon-pairs are used in quantum metrology, imaging and sensing applications \cite{giovannetti2011advances,pirandola2018advances,polino2020photonicreview,moreau2019imaging,Maga_a_Loaiza_2019}. Some of the major applications of SPDC entangled photon sources are discussed below. Also, Table \ref{table4} gives the list of some applications of entangled photon sources and their practical realization using different source designs.

\subsection{Fundamental quantum physics}

Presence of non-local correlation of photon-pairs in different degrees of freedom makes SPDC sources a useful resource for testing foundations of quantum physics \cite{zeilinger1999experiment,shadbolt2014testing}. Non-collinear source of polarization entangled photon-pairs are widely used in experiments for testing foundations of quantum physics, such as complementarity principle \cite{weston2013experimental}, delayed-choice quantum eraser \cite{ma2016delayed}, non-locality \cite{weihs1998violation,christensen2013detection} and EPR steering \cite{saunders2010experimental}. Recently, similar experiments for testing fundamental quantum physics \cite{vermeyden2013experimental,xiao2017experimental,giustina2013bell} make use of sources with stable Sagnac interferometric designs \cite{anderson1994sagnac,kim2006phase,kuzucu2008pulsed,fedrizzi2007wavelength,predojevic2012pulsed,steinlechner2014efficient}, which ensures compactness for their use in both laboratory as well as outdoor environments. Non-collinear sources with overlapping SPDC cones are a useful resources for multi-path experiments involving two or more entangled photon-pairs from different parts of the cone \cite{pan2019direct}.

\subsection{Multi-photon experiments}

Correlation and entanglement among three or more photons are experimentally realized using photon-pair sources by entangling two independent photons each from separate sources, based on quantum interference \cite{zukowski1995entangling}. This typically involves  the use of ultra-fast pumped SPDC and the technique of forcing indistiguishability \cite{Zukowski93a} of photons from photon-pairs through narrow-band filtering, which was used first used in quatnum teleportation experiments \cite{bouwmeester1997experimental}. Multiple non-collinear sources of polarization entangled photon-pairs in crossed-crystal and parallel crystal designs are used to demonstrate six- \cite{zhang2015experimental}, eight- \cite{yao2012observation} and ten-photon \cite{wang2016experimental} entanglement. Another method is group-velocity matching of pump and signal (or idler) pulses, to achieve high-quality multi-pair interference. The group-velocity matching can be achieved through careful design and selection of the SPDC phase matching and configuration to generate bi-photon states that leave the photons spectral uncorrelated states. For instance, by using 5 mm thick KDP, pumped with 415nm, 50 fs pulses Mosley et al. \cite{mosley:133601} where able to achieve two-photon interference for the 830~nm photons generated from two different SPDC processes with  around 90\%, without applying the filtering technique. Another approach is to match the group-velocities through geometric configuration of the emission cones. For instance, 1550nm photon-pairs produced from type-II SPDC in BBO, will match the group velocity of the pump event at modest emission angles of $3^\circ$ \cite{Lutz:14}, allowing to control the spectral correlations with optimal beam waists (ca. $100 \mu$m) and pump pulse length (90~fs). 

\subsection{Quantum communication}

Entangled photon sources are advantageous in QKD schemes \cite{ekert1991quantum,bennett1992quantum} because the inherent randomness in the generation of photon-pairs leads to pure random key generation. Energy-time and time-bin qubits are extensively used for demonstrating quantum communication over large distances through optical fiber \cite{jennewein2000quantum,ribordy2000long,naik2000entangled,tittel2000quantum,ribordy2000long,marcikic2003long,fasel2004quantum,poppe2004practical,ursin2004quantum,honjo2007differential,honjo2008long,zhang2008distribution,dynes2009efficient,takesue2010long,inagaki2013entanglement,takesue2015quantum,sun2017entanglement,huo2018deterministic,wengerowsky2019entanglement,wengerowsky2020passively}. Recently, distribution of multidimensional entanglement in spatial degrees of freedom through long fibers was demonstrated \cite{ikuta2018four,liu2020multidimensional,cao2020distribution}.

While several ground-based QKD or Bell-test experiments were performed using non-collinear polarization entangled photon-pairs \cite{aspelmeyer2003long,resch2005distributing,marcikic2006free,jin2010experimental,yin2012quantum}, some realizations used the brighter and more stable Sagnac sources to reach record distances of up to 144~km on ground \cite{ursin2007entanglement,ma2012quantum}. Free-space distribution of high dimensional entanglement using spatial modes of light was demonstrated  \cite{vallone2014free,krenn2015twisted,ecker2019overcoming}. Recently, entanglement distribution was accomplished using flying drones, constructing a scalable airborne system for multi-node connection leading towards mobile quantum networks \cite{liu2021optical}.

\begin{figure}[h]
	\centering
	\includegraphics[width=0.48\textwidth]{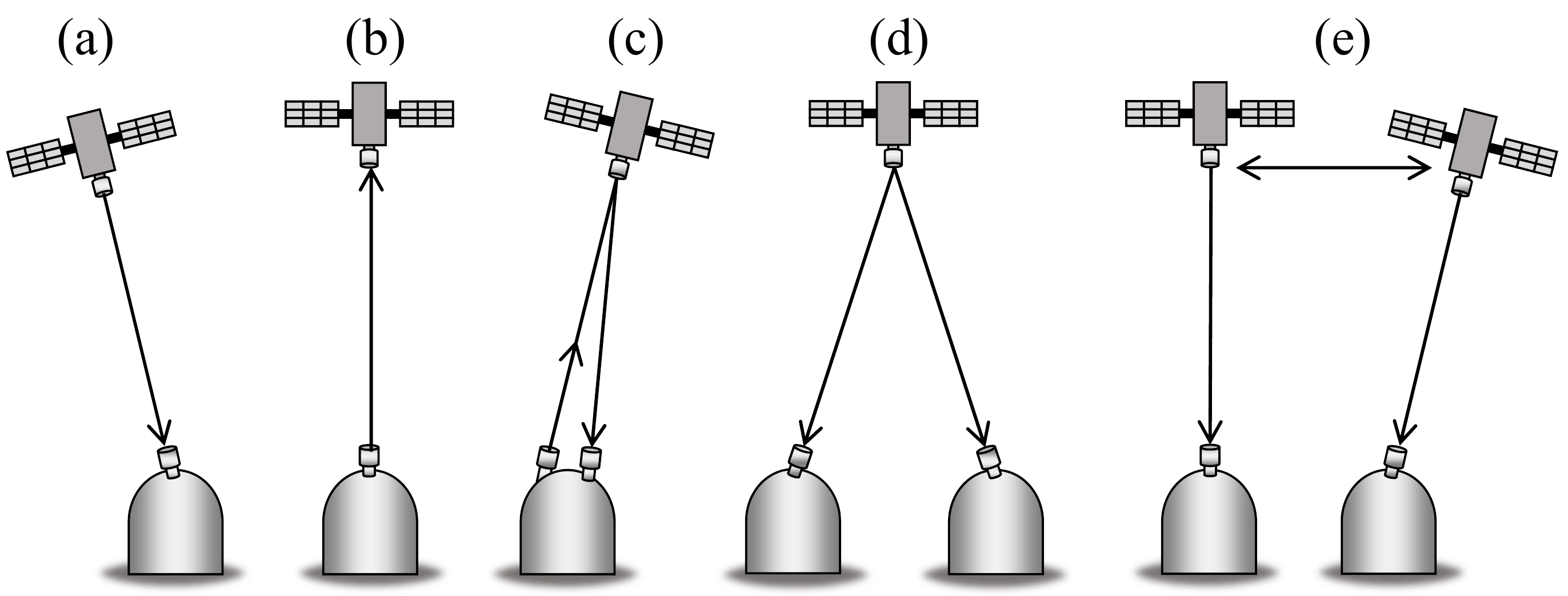}
	\caption{Illustration of different schemes for performing satellite QKD. Scenarios (a) and (b) depict a downlink and an uplink, respectively, while in scenario (c) a downlink is simulated by using a retro-reflector on board the satellite. In (d), pairs of entangled photons are being transmitted to Earth so that two ground stations can share entangled states. Finally, scenario (e) illustrates how inter-satellite links can allow more complex satellite QKD networks.}
	\label{satelliteQKD}
\end{figure}

Quantum communication based on satellites can benefit from a quadratic loss scaling, and therefore overcome the exponential scaling of losses which is a major drawback of ground-based QKD systems \cite{bedington2017progress}. Different schemes for performing satellite QKD are depicted in Fig.\ref{satelliteQKD}.
Micius - the first satellite that successfully demonstrated the distribution of polarization entangled photon-pairs over 1200 km from a Sagnac SPDC source \cite{yin2017satellite}. Following this, intercontinental communication networks was established \cite{yin2017satellitetoground,ren2017ground,liao2018satellite}. 

Because the satellite-ground link is inherently variable due to satellite motion, the optimal pair-production rate for entangled photons will also vary. This is due to the increase of multi-photon emissions at higher pump powers. In 2013, Holloway et al. \cite{Holloway:2013} modelled satellite-QKD links and found that the optimal pair production for SPDC entanglement sources could vary between <1 MHz up to 500 MHz, depending on the link losses, detector timing performance and background noise. And, during the pass of a low-earth-orbit satellite the optimal SPDC pair rate would even have to be adjusted to match the instantaneous link conditions. In particular, the uplink scenario depicted in Fig. \ref{satelliteQKD}(b) could make use of various quantum sources at the ground station \cite{jennewein2014b}, and  implementing a high-performance, tunable-rate entangled photon source is certainly feasible. 

The development of low-cost nanosatellites carrying compact entangled photon sources has become popular in the quantum space race \cite{jennewein2014nanoqey,oi2017nanosatellites,lee2019updated}. Iterations of small photon entangling quantum system (SPEQS series) aimed on achieving optimal performance of source as required for entanglement distribution and satellite QKD \cite{bedington2016nanosatellite}. The robustness of such sources under harsh environments was verified \cite{tang2014near,tang2016generation,tang2016photon}. An all-in-line design of polarization entangled photon source \cite{villar2018experimental} was used in a quantum nanosatellite which successfully demonstrated polarization entanglement distribution to the ground \cite{villar2020entanglement}. Future mission aims at deployment and testing of high performance entanglement-QKD system in space with the use of photon-pair source based on linear displacement interferometer \cite{lohrmann2019PAPA}. Recently, a drone-based entanglement distribution using unmanned aerial vehicle (UAV) was introduced, which offers a full-time all-location multi-weather coverage in a cost-effective way \cite{liu2020drone}. 

Apart from a free-space atmospheric channel, quantum communication in a water environment has attracted researchers due to the fact that major part of Earth's area comprises of water. Recent theoretical studies \cite{gariano2019theoretical} and experimental demonstrations \cite{ji2017towards,bouchard2018quantum} with entangled photon-pair sources show promising avenues for underwater quantum networking. 

\subsection{Quantum sensing, metrology and imaging}

SPDC sources generating entangled photons are found to be useful in the broad field of quantum sensing \cite{pirandola2018advances}, which includes the enhancement of quantum resources for the detection of faint objects, extraction of information from optical memories and optical resolution of extremely close point-like sources. Photon-pair sources are effectively used for quantum sensing \cite{lopaeva2013experimental,zhang2015entanglement} and imaging \cite{gregory2020imaging} protocols based on the idea of 'Quantum illumination' \cite{tan2008quantum,lloyd2008enhanced}, which show significant improvement in noise resilience over other classical schemes. With the use of ultra-high efficient photon source and detectors, performing an unconditional entanglement-enhanced photonic interferometry demonstrated a precision beyond the shot-noise limit \cite{slussarenko2017unconditional}.

SPDC sources of entangled photon-pairs are extensively used in quantum metrology, which aims at beating standard bounds on estimation precision by exploiting quantum probes \cite{braunstein1992quantum,braunstein1994statistical,giovannetti2004quantum,giovannetti2006quantum,paris2009quantum,giovannetti2011advances,demkowicz2015quantum}. Entanglement-assisted quantum metrology protocol was experimentally demonstrated in a noisy environment \cite{wang2018entanglement}. Orbital angular momentum entanglement between photon-pairs in SPDC is well exploited for performing ultra-sensitive angular measurements \cite{courtial1998measurement,lavery2013detection,jha2011supersensitive,fickler2012quantum,d2013photonic,zhang2019quantum}.

Spatial correlation of photon-pairs are utilized for quantum imaging \cite{gilaberte2019perspectives}. This was practically realized using different techniques \cite{pittman1995optical,simon2012two,fickler2013real}. Quantum imaging experiments such as sub-shot noise imaging \cite{blanchet2008measurement,brida2010experimental,toninelli2017sub} and quantum ghost imaging \cite{strekalov1995observation,morris2015imaging,padgett2017introduction} overcome the spatial resolution beyond classical bound. Entanglement of photon-pairs has found to have a key role in various imaging experiments beating classical limits \cite{boto2000quantum,tsang2009quantum,shin2011quantum,rozema2014scalable,xu2015experimental,hong2017heisenberg}. Entangled photon-pairs are also used for practical realization of various quantum-enhanced sensing and imaging technologies such as quantum lithography \cite{boto2000quantum,bjork2001entangled,d2001two,kothe2011efficiency}, quantum ellipsometry \cite{abouraddy2002entangled,graham2006ellipsometry}, quantum optical coherence tomography \cite{nasr2003demonstration,booth2011polarization,mazurek2013dispersion,graciano2019interference}, clock synchronization \cite{valencia2004distant} and optical gyroscope \cite{fink2019entanglement}.

\section{Summary}

In this article, we have discussed various designs of entangled photon sources using bulk nonlinear crystals and waveguides. These bulk sources have been used to demonstrate many quantum technology applications and will continue to play a role in the practical realization of quantum information. Some novel technological possibilities in source design like domain engineering of QPM crystals that are yet unexplored further, may add further scope in building compact sources for future technologies.



\section*{data availability statement}

All the available data will be shared upon reasonable request.


\begin{acknowledgments}
FS acknowledges financial support by the Fraunhofer Internal Programs under Grant No. Attract 066-604178. This research is supported by the Research Centres of Excellence programme supported by the National Research Foundation (NRF) Singapore and the Ministry of Education, Singapore.
\end{acknowledgments}

\appendix*

\section{Performance parameters of polarization entangled photon-pair sources}

This section gives comparison tables of the performance parameters of major sources for polarization entangled photon-pairs. Table \ref{table1} \& \ref{table2} lists the sources in infrared and telecom wavelengths respectively. Table \ref{table3} and \ref{table4} list the sources based on cavity-enhanced designs and waveguide (WG) devices respectively. Entangled photon-pair source based applications are listed in Table \ref{table5} with examples. $\dagger$ denotes values estimated from published data and $\ddagger$ denotes values corresponding to pair production. The spectral characteristics listed are central wavelengths of the signal and idler photons ($\lambda_s$, $\lambda_i$) and the full width at half maximum bandwidth ($\Delta\lambda$). In order to facilitate a comparison of the heralding efficiency for different detector efficiencies, both the uncorrected experimental values ($\eta_{s,i}$) as well as the values corrected for the estimated detector efficiency ($\eta_{s,i}/\eta_{det}$) are listed. The Bell-state fidelity is that obtained for low
pump powers. In particular, entanglement fidelity is approximated from reported visibilities as $F=1-(1-V)/2$ where $V$ is the highest reported average visibility in rectilinear and diagonal bases without subtracting accidental coincidences. SB - Spectral Brightness; SCSP - Single Crystal Single Pass; DCSP - Double Crystal Single Pass;; SCDP - Single Crystal Double Pass;; DCDP - Double Crystal Double Pass; TES - Transition Edge Sensor; SPCM - Single Photon Counting Module; APD - Avalanche Photodiode; 

\begin{widetext}

\begin{sidewaystable}[h]
\footnotesize\addtolength{\tabcolsep}{0.05cm}
\renewcommand{\arraystretch}{1.4}
    \caption{List of polarization entangled photon-pair sources at near infrared wavelengths with their corresponding performance parameters. For details, see the text.}
    \label{table1}
 \begin{tabular}{!{\vrule width 1pt}p{0.3cm}!{\vrule width 1pt}p{0.8cm}!{\vrule width 1pt}p{2.7cm}p{1.1cm}p{4cm}p{1.4cm}p{1.5cm}p{2.6cm}p{1.15cm}p{1.15cm}p{0.9cm}p{2.7cm}!{\vrule width 1pt}}
\noalign{\hrule height 1pt}
& & Process (Material) & Detector used & Comments & Brightness (cps/mW) & SB (cps/mW/nm) & $\lambda_s$, $\lambda_i$, ($\Delta\lambda$) (in nm) & $\eta_s$,($\frac{\eta_s}{\eta_{det}}$) (in \%) & $\eta_i$,($\frac{\eta_i}{\eta_{det}}$) (in \%) & Fidelity (in \%) & Reference \\
\noalign{\hrule height 1pt}

\parbox[t]{3mm}{\multirow{12}{*}{\rotatebox[origin=c]{90}{\large Non-collinear}}} & \parbox[t]{9mm}{\multirow{5}{*}{\rotatebox[origin=c]{0}{SCSP}}} & type-II CPM (BBO) & - & Overlapping cones & $1.0\times10^1$ & $2.0\times10^0$ & 702, 702, (5) & >10 & >10 & 98.4 & Kwiat 1995 \cite{kwiattype2} \\  \cline{3-12} 
& & type-II QPM (PPKTP) & SPCM & Overlapping cones & $8.2\times10^2$ & $8.2\times10^2$ & 797, 797, (1) & - & - & $96.5^\dagger$ & Fiorentino 2005 \cite{fiorentino2005source} \\ \cline{3-12}
& & type-II CPM (BiBO) & - & Universal Bell-state synthesizer & $9.4\times10^{1\dagger}$ & $3.1\times10^{1\dagger}$ & 797, 797, (3) & 8.7 & 8.7 & 94 & Halevy 2011 \cite{halevy2011biaxial} \\ \cline{3-12}
& & type-II QPM (PPKTP) & SPCM & Single-mode pump & $4.2\times10^3$ & $7.4\times10^{3\dagger}$ & 812.4, 812.4, (0.55) & 0.4 & 0.4 & $98.8^\dagger$ & Lee 2016 \cite{lee2016polarization} \\ \cline{3-12}
& & type-II QPM (PPKTP) & - & Universal Bell-state synthesizer & $7.0\times10^3$ & - & 812, 812, (8.7,5.8) & - & - & 99.2 & Jeong 2016 \cite{jeong2016bright} \\ \cline{2-12}
& \parbox[t]{9mm}{\multirow{7}{*}{\rotatebox[origin=c]{0}{DCSP}}} & type-I CPM (BBO) & SPCM & crossed crystal & $1.4\times10^2$ & $2.8\times10^{1\dagger}$ & 701, 701, (5) & 10, (15) & 10, (15) & $97.5^\dagger$ & Kwiat 1999 \cite{kwiattype1dual} \\ \cline{3-12}
& & type-I CPM (BBO) & SPCM & crossed crystal,phase compensated & $3.6\times10^{4\dagger}$ & $1.4\times10^{2\dagger}$ & 701, 701, (25) & 30, (46) & 30, (46) & 97.7 & Altepeter 2005 \cite{altepeter2005phase} \\ \cline{3-12}
& & type-II CPM (BBO) & SPCM & flipped crystal, pulsed pump & $3.0\times10^{2\dagger}$ & - & - & - & - & 97.4 & Niu 2008  \cite{niu2008beamlike} \\ \cline{3-12}
& & type-I CPM (BBO) & - & crossed crystal, pulsed pump & $5.4\times10^3$ & $5.4\times10^{2\dagger}$ & 810, 810, (10) & - & - & 98.9 & Rangarajan 2009 \cite{rangarajan2009optimizing} \\ \cline{3-12}
& & type-I CPM (BiBO) & - & crossed crystal, pulsed pump & $1.3\times10^4$ & $1.3\times10^{3\dagger}$ & 810, 810, (10) & - & - & - & Rangarajan 2009 \cite{rangarajan2009optimizing} \\ \cline{3-12}
& & type-I CPM (BiBO) & TES & crossed crystal,pulsed pump, SMF & - & - & 710, 710, (20) & 75  & 75 & $99.7^\dagger$ & Christensen 2013 \cite{christensen2013detection} \\ \cline{3-12}
& & type-II CPM (BBO) & Si-APD & flipped crystal, pulsed pump & $2.0\times10^{3\dagger}$ & - & - & - & - & 99 & Zhang 2015 \cite{zhang2015experimental} \\ \cline{2-12}
& \parbox[t]{9mm}{\multirow{2}{*}{\rotatebox[origin=c]{0}{SCDP}}} & type-II CPM (BBO) & Si-APD & rail-cross, pulsed pump & $1.5\times10^{2\dagger}$ & $2.9\times10^{1\dagger}$ & 780, 780, (5) & - & - & 98.1 & Hodelin 2006 \cite{hodelin2006optimal} \\ \cline{3-12} 
& & type-0 QPM (PPKTP) & SPCM & Sagnac loop,overlapping cones & - & - & 810, 810, (2,10) & - & - & 97.5 & Jabir 2017 \cite{jabir2017robust} \\
\noalign{\hrule height 1pt}

\parbox[t]{2mm}{\multirow{23}{*}{\rotatebox[origin=c]{90}{\large Collinear}}} & \parbox[t]{9mm}{\multirow{2}{*}{\rotatebox[origin=c]{0}{SCSP}}}  & type-II QPM (PPKTP) & Si-APD & beam splitter post-selection & $3.0\times10^2$ & $3.0\times10^2$ & 795, 795, (1) & 10, (20) & - & 99.3 & Kuklewicz 2004 \cite{kuklewicz2004high} \\\cline{3-12} 
& & type-0 QPM (PPKTP) & Si-APD & position correlation based & $1.2\times10^5$ & - & 792, 829 & 18 & 15 & 99.1 & Perumangatt 2019 \cite{perumangatt2019experimental} \\ \cline{2-12}
& \parbox[t]{9mm}{\multirow{5}{*}{\rotatebox[origin=c]{0}{DCSP}}} & type-I CPM (BBO) & Si-APD & crossed-crystal, broadband pump & $2.7\times10^5$ & $1.8\times10^3$ & 763, 850, (14.5, 15.4) & 38, (75) & - & 99.4 & Trojek 2008 \cite{trojek2008collinear} \\ \cline{3-12} 
& & type-0 QPM (PPKTP) & Si-SPAD & crossed-crystal, SMF & $6.4\times10^5$ & $2.8\times10^5$ & 783, 839, (2.3) & 18, (36)  & 9, (18) & 98.3 & Steinlechner 2012 \cite{steinlechner2012high} \\ \cline{3-12}
& & type-II QPM (PPKTP) & Si-APD & crossed-crystal, pulsed pump & $5.0\times10^{2\dagger}$ & $1.7\times10^2$ & 760, 810, (3) & 15 & 15 & 95.8 & Scheidl 2014 \cite{scheidl2014crossed} \\ \cline{3-12}
& & type-I CPM (BBO) & Si-APD & parallel crystal & $6.5\times10^4$ & - & 776, 847 & 27 & 22 & 99.6 & Villar 2018 \cite{villar2018experimental} \\ \cline{3-12}
& & type-I CPM (BBO) & Si-APD & parallel crystal, circular field stop & $1.0\times10^5$ & $6.7\times10^{3\dagger}$ & 785, 837, (15) & 18.7 & 20 & 99.5 & Lohrmann 2018 \cite{lohrmann2018high} \\ \cline{2-12}
&  \parbox[t]{9mm}{\multirow{15}{*}{\rotatebox[origin=c]{0}{SCDP}}} & type-II QPM (PPKTP) & SPCM & bi-directional Mach-Zehnder & $1.2\times10^4$ & $4.0\times10^3$ & 797, 797, (3) & $18^\dagger$, ($36^\dagger$) & $18^\dagger$, ($36^\dagger$) & $92^\dagger$ & Fiorentino 2004 \cite{fiorentino2004generation} \\ \cline{3-12} 
&& type-II QPM (PPKTP) & SPCM & Sagnac loop & $5.0\times10^3$ & $5.0\times10^3$ & 810, 810, (1) & $15^\dagger$ & $15^\dagger$ & $98.4^\dagger$ & Kim 2006 \cite{kim2006phase} \\ \cline{3-12}
& & type-II QPM (PPKTP) & SPCM & Sagnac loop, tunable wavelength & $8.2\times10^4$ & $2.7\times10^5$ & 810$\pm$26, (0.3) & 20, (71) & - & 99.6 & Fedrizzi 2007 \cite{fedrizzi2007wavelength} \\ \cline{3-12}

& & type-II QPM (PPKTP) & Si-APD & Sagnac loop, pulsed pump & $4.0\times10^3$ & $1.5\times10^3$ & 808, 808, (1.8, 2.7) & - & - & 98.2 & Predojevic 2012 \cite{predojevic2012pulsed} \\ \cline{3-12}
& & type-II QPM (PPKTP) & TES & Sagnac loop, SMF & $4.0\times10^3$ & $8.0\times10^3$ & 820, 820, (0.5) & 60, (80) & 60, (80) & 98 & Smith 2012 \cite{smith2012conclusive} \\ \cline{3-12}
& & type-II QPM (PPKTP) & TES & Sagnac loop, SMF. herlading & $1.0\times10^4$ & $2.0\times10^4$ & 810, 810, (0.5) & 12 & 83 & 92 & Ramelow 2013 \cite{ramelow2013highly} \\ \cline{3-12}
& & type-0 QPM (PPKTP) & SPAD & folded sandwich, SMF & $1.1\times10^6$ & $3.9\times10^5$ & 783, 839, (2.9) & 19, (35) & 13, (25) & 99.1 & Steinlechner 2013 \cite{steinlechner2013phase} \\ \cline{3-12}
& & type-II QPM (PPKTP) & SPAD & Sagnac loop, SMF & $8.0\times10^3$ & $2.7\times10^4$ & 810, 810, (0.3) & 45, (83) & 39, (78) & 99.4 & Steinlechner 2014 \cite{steinlechner2014efficient} \\ \cline{3-12}
& & type-0 QPM (PPKTP) & SPAD & Sagnac loop, SMF & $1.0\times10^6$ & $4.0\times10^5$ & 783, 839, (2.3) & 45, (83) & 35, (78) & 99.1 & Steinlechner 2014 \cite{steinlechner2014efficient} \\ \cline{3-12}
& & type-0 QPM (PPKTP) & SPAD & folded sandwich, SMF coupled & $3.0\times10^6$ & $1.0\times10^6$ & 783, 839, (3) & 18, (32) & 9, (17) & 98.3 & Steinlechner 2014 \cite{steinlechner2015sources} \\ \cline{3-12}
& & type-II QPM (PPKTP) & TES & Sagnac loop, SMF & $8.0\times10^3$ & $2.0\times10^4$ & 810, 810, (0.4) & 81.4, (86) & 69.6, (86) & 99.8 & Joshi 2014 \cite{joshi2014entangled} \\ \cline{3-12}
& & type-0 QPM (PPKTP) & SPCM & Sagnac loop, two pump pass & $9.1\times10^2$ & $3.0\times10^{2\dagger}$ & 810, 810, (3) & - & - & 87.3 & Terashima 2018 \cite{terashima2018quantum} \\ \cline{3-12}
& & type-II QPM (PPKTP) & SNSPD & Sagnac loop & $4.7\times10^3$ & - & 807.8, 807.8, (3) & 12.5 & 12.5 & 99.7 & Sandra 2020 \cite{ausserlechner2020approaching} \\ \cline{3-12}
&& type-0 QPM (PPKTP) & Si-APD & beam displacer, broadband pump & $5.6\times10^5$ & $4.0\times10^{5\dagger}$ & 810, 810, (14) & 21 & 21 & $99.4\dagger$ & Lohrmann 2020 \cite{lohrmann2019PAPA} \\ \cline{3-12}
& & type-0 QPM (PPKTP) & Si-APD & beam displacer, narrowband pump & $1.2\times10^5$ & $3.0\times10^{5\dagger}$ & 810, 810, (14) & 21 & 21 & $99.4\dagger$ & Lohrmann 2020 \cite{lohrmann2019PAPA} \\ \cline{2-12}
& \parbox[t]{9mm}{\multirow{1}{*}{\rotatebox[origin=c]{0}{DCDP}}} & type-0 QPM (PPKTP) & Si-APD & crossed crystal, Sagnac loop & $1.6\times10^5$ & $5.3\times10^4$ & 810, 810, (3) & 18.5 & 18.5 & 99.2 & Chen 2018 \cite{chen2018polarization} \\ 
\noalign{\hrule height 1pt}
\end{tabular}
\end{sidewaystable}

\begin{sidewaystable}[h]
\footnotesize\addtolength{\tabcolsep}{0.03cm}
\renewcommand{\arraystretch}{1.4}
    \centering
    \caption{List of polarization entangled photon-pair sources at telecom wavelengths with their corresponding performance parameters. For details, see the text.}
    \label{table2}
        \begin{tabular}{!{\vrule width 1pt}p{4cm}p{2.2cm}p{3.5cm}p{1.3cm}p{1.6cm}p{2.5cm}p{1.3cm}p{1.3cm}p{1cm}p{2.5cm}!{\vrule width 1pt}}
    \noalign{\hrule height 1pt}
Process (Material) & Detector & Comments & Brightness (cps/mW)  & SB (cps/mW/nm) & $\lambda_s$, $\lambda_i$, ($\Delta\lambda$) (in nm) & $\eta_s$,($\frac{\eta_s}{\eta_{det}}$) (in \%) & $\eta_i$,($\frac{\eta_i}{\eta_{det}}$) (in \%) & Fidelity (in \%) & Reference \\
\noalign{\hrule height 1pt}

Collinear type-0 QPM (PPKTP) & Si-APD, InGaAs & Crossed crystal & $3.2\times10^3$ & $1.1\times10^{3\dagger}$ & 810, 1550, (3) & 21, ($36^\dagger$) & 7.5, ($94^\dagger$) & $91.6^\dagger$ & Pelton 2004 \cite{pelton2004bright} \\
\hline
Collinear type-0 QPM (PPLN) & Si-SPCM, InGaAs & Folded MZI, SMF & $4.5\times10^2$ & $3.0\times10^{2\dagger}$ & 795, 1609, (0.11) & - & 9.4, ($47.5^\dagger$) & 98 & K\"{o}nig  2004 \cite{konig2005efficient} \\ \hline
Non-collinear type-I QPM (PPLN) & Si-APD, InGaAs &  Single crystal, dual process & - & - & 810, 1550, (10) & - & - & $92.5\dagger$ & Hugues 2006 \cite{de2006non} \\ \hline
Non-collinear type-II CPM (BBO) & InGaAs/InP & Single crystal & - & - & 1550, 1550, (10) & - & - & $99\dagger$ & Noh  2007 \cite{noh2007efficient} \\ \hline
Collinear type-I QPM (PPKTP) & Si-APD, InGaAs & Crossed crystals & $3.0\times10^3$ & - & 810, 1550, (1) & - & 3 & $99.5\dagger$ & H\"{u}bel  2007 \cite{hubel2007high} \\ \hline
Collinear type-I QPM (PPLN) & Si-APD, InGaAs & Folded Mach-Zehnder & $2.5\times10^{5\dagger}$ & $3.7\times10^{5\dagger}$ & 810, 1550, (0.8) & 3, ($20^\dagger$) & - & $95.5\dagger$ & Sauge 2008 \cite{sauge2008single} \\ \hline
Collinear type-0 QPM (PPKTP) & Si-APD, InGaAs & Modified sagnac loop & $9.3\times10^{3\dagger}$ & $1.1\times10^{3\dagger}$ & 810, 1550, (0.4, 1.5) & 1.55, ($10^\dagger$) & 1.55, ($5.2^\dagger$) & 97.5 & Hentschel 2009 \cite{hentschel2009three} \\ \hline
Collinear type-II QPM (PPKTP) & InGaAs/InP & Pulsed, double displacement & $7.5\times10^{2\dagger}$ & - & 1552, 1552 & - & - & $97.3^\dagger$ & Evans 2010 \cite{evans2010bright} \\ \hline
Collinear type-II QPM (PPLN) & InGaAs & Pulsed, Two-period QPM & - & - & 1506, 1594, (1.4) & - & - & 94 & Ueno 2012 \cite{ueno2012entangled} \\ \hline
Collinear type-II QPM (PPKTP) & InGaAs & Single crystal, broadband & - & - & 1540-1600 & - & - & $85^\dagger$ & Zhou 2013 \cite{zhou2013ultra} \\ \hline
Collinear type-II QPM (PPKTP) & SNSPD & Pulsed, GVM based Sagnac & $2.0\times10^3$ & - & 792, 1584 & - & - & $98^\dagger$ & Jin 2014 \cite{jin2014pulsed} \\ \hline
Collinear type-II QPM (PPKTP) & InGaAs & Sagnac & - & $2.0\times10^0$ & 1550, 1550, (2.4) & - & - & 93.5 & Yan 2015 \cite{li2015cw} \\ \hline
Collinear type-II QPM (PPKTP) & SNSPD & Pulsed, Sagnac & $5.7\times10^2$ & - & 1570, 1570, (8) & 52 & 52 & 99 & Weston 2016 \cite{weston2016efficient} \\ \hline
Non-collinear type-0 QPM (PPLN) & InGaAs/InP & Sagnac loop, pulsed pump & - & $4.4\times10^1$ & 1550, 1550 & - & - & $98.5^\dagger$ & Kim 2019 \cite{kim2019pulsed} \\ \hline
Collinear type-0 QPM (PPLN) & - & MZI, double displacement & - & - & 1552, 1552 & - & - & 98 & Horn 2019 \cite{horn2019auto} \\ \hline
Collinear type-II QPM (PPLN) & SNSPD & Domain engineered & - & - & 1530, 1569 & - & - & $96.6^\dagger$ & Kuo 2020 \cite{kuo2020demonstration}
\\ \hline
Collinear type-0 QPM (PPLN) & SNSPD & Pulsed pump & - & - & 2090, 2090 & - & - & $78^\dagger$ & Prabhakar 2020 \cite{prabhakar2020two} \\ \hline
Collinear type-0 QPM (PPLN) & SNSPD & MgO doped & - & - & 1549, 1552 (0.6) & - & - & 93.5 & S\"{o}ren 2020 \cite{wengerowsky2020passively} \\

\noalign{\hrule height 1pt}
    
    \end{tabular}

\smallskip
\footnotesize\addtolength{\tabcolsep}{0.07cm}
\renewcommand{\arraystretch}{1.4}
    \centering
    \caption{List of polarization entangled photon-pair sources using cavity enhanced PDC process with their corresponding performance parameters. For details, see the text.}
    \label{table3}
    \begin{tabular}{!{\vrule width 1pt}p{3.9cm}p{1.2cm}p{3cm}p{1.4cm}p{1.6cm}p{2.3cm}p{1.1cm}p{1.1cm}p{0.9cm}p{2.9cm}!{\vrule width 1pt}}
    \noalign{\hrule height 1pt}
Process (Material) & Detector & Comments & Brightness (cps/mW)  & SB (cps/mW/nm) & $\lambda_s$, $\lambda_i$, ($\Delta\lambda$) (in nm) & $\eta_s$,($\frac{\eta_s}{\eta_{det}}$) (in \%) & $\eta_i$,($\frac{\eta_i}{\eta_{det}}$) (in \%) & Fidelity (in \%) & Reference \\
\noalign{\hrule height 1pt}

Non-collinear type-I CPM (BBO) & Si-APD & Linear cavity & $1.0\times 10^2$ & $2.1\times 10^{1^\dagger}$ & 702, 702, (5) & - & - & $97.4^\dagger$ & Oberparleiter 2000 \cite{oberparleiter2000cavity} \\
\hline
Collinear type-I CPM (KTP) & SPCM & Crossed-crystal, Z-cavity & $2.2\times10^4$ & $2.2\times10^{3\dagger}$ & 860, 860, (10) & - & - & $70.4^\dagger$ & Wang  2004 \cite{wang2004polarization} \\ \hline
Collinear type-II QPM (PPKTP) & - & Linear cavity & $6.6\times10^{1{\dagger,\ddagger}}$ & $6.0\times10^{3{\dagger,\ddagger}}$ & 780, 780 & - & - & 94.3 & Bao 2008 \cite{bao2008generation} \\ \hline
Collinear type-0 QPM (PPKTP) & - & Sagnac loop, Z-cavity & - & $9.0\times10^{6\ddagger}$ & 854, 854 & - & - & 99.2 & Arenskotter 2017 \cite{arenskotter2017polarization} \\ 

\noalign{\hrule height 1pt}
    
    \end{tabular}

\end{sidewaystable}

\begin{sidewaystable}[h]
\footnotesize\addtolength{\tabcolsep}{0.0006cm}
\renewcommand{\arraystretch}{1.4}
    \centering
    \caption{List of polarization entangled photon-pair sources using waveguides (WG) with their corresponding performance parameters. For details, see the text.}
    \label{table4}
\begin{tabular}{!{\vrule width 1pt}p{0.3cm}!{\vrule width 1pt}p{2.4cm}p{2cm}p{2.2cm}p{1.3cm}p{1.3cm}p{1.5cm}p{1.6cm}p{2.4cm}p{1.4cm}p{1.4cm}p{0.9cm}p{2.6cm}!{\vrule width 1pt}}
\noalign{\hrule height 1pt}
 & Process (Material) & Detector & Comments & Generated brightness (cps/mW) &  Detected brightness (cps/mW) & Generated SB (cps/mW/nm) & Detected \quad SB (cps/mW/nm) & $\lambda_s$,$\lambda_i$,($\Delta\lambda$) (in nm) & $\eta_s$, ($\eta_s/\eta_{det}$) (in \%) & $\eta_i$, ($\eta_i/\eta_{det}$) (in \%) & Fidelity (in \%) & Reference \\

\noalign{\hrule height 1pt}
\parbox[t]{3mm}{\multirow{14}{*}{\rotatebox[origin=c]{90}{CW-pumped}}} & type-I (PPLN) & - & Two WGs & $2.4\times10^{6\dagger}$ & $9.6\times10^2$ & - & - & 1550, 1550 & $2^\dagger$, ($10^\dagger$) & $2^\dagger$, ($12^\dagger$) & $97^\dagger$ & Yoshizawa 2003 \cite{yoshizawa2003generation} \\ \cline{2-13}
 & type-II (PPLN) & Ge-APD, InGaAs & Ti in-diffused WG &  & $1.8\times10^{2\dagger}$ & $3.0\times10^5$ & & 1310, 1310, (0.7) & $0.5^\dagger$, ($0.13^\dagger$) & $0.5^\dagger$, ($0.05^\dagger$) & $91.5^\dagger$ & Martin 2010 \cite{martin2010polarization} \\  \cline{2-13} 
 & type-II (PPKTP) & SNSPD & post-selection & - & $2.3\times10^{5\dagger}$ & - & $3.3\times10^{5\dagger}$ & 1316, 1316, (0.7) & - & - & $98.5^\dagger$ & Zhong 2010 \cite{zhong2010high} \\ \cline{2-13}
 & type-II (PPLN) & InGaAs & post-selection & $1.7\times10^{4\dagger}$ & $4.4\times10^{2\dagger}$ & $2.0\times10^4$ & - & 1540, 1540, (0.8) & - & - & $98.6^\dagger$ & Kaiser 2012 \cite{kaiser2012high} \\ \cline{2-13} 
 & type-II (PPLN) & InGaAs & double-poling & $1.0\times10^{6\dagger}$ & - & - & $7.0\times10^3$ & 1551, 1571, (0.7) & $2^\dagger$ & $2^\dagger$ & $97.5^\dagger$ & Herrmann 2013 \cite{herrmann2013post} \\ \cline{2-13} 
 & type-II (AlGaAs) & InGaAs & Bragg grating & - & - & - & - & 1537, 1575 & - & - & 83 & Horn 2013 \cite{horn2013inherent} \\ \cline{2-13}
 & type-0 (PPKTP) & InGaAs & MZI & $9.0\times10^6$ & $2.0\times10^3$ & $2.4\times10^6$ & - & 1560, 1560, (32) & - & - & $99.8^\dagger$ & Kaiser 2014 \cite{kaiser2014polarization} \\ \cline{2-13}
 & type-I (PPKTP) & SPCM, SNSPD & free-space MZI & - & - & $2.4\times10^6$ & - & 883, 1338, (2.1, 4.3) & 3.1 & 7.4 & $97.9^\dagger$ & Clausen 2014 \cite{clausen2014source} \\ \cline{2-13}
 & type-I (PPLN) & SPCM, SNSPD & free-space MZI & - & & $3.1\times10^6$ & - & 883, 1338, (1.1, 2.0) & 2.6 & 6.6 & $97.9^\dagger$ & Clausen 2014 \cite{clausen2014source} \\ \cline{2-13}
 & type-II (AlGaAs) & InGaAs & Bragg grating & - & & - & & 1560, 1560 & 1 & 1 & $93.4^\dagger$ & Autebert 2016 \cite{autebert2016multi} \\ \cline{2-13}
 & type-0 (MgO:PPLN) & InGaAs & fiber Sagnac & $2.0\times10^6$ & $1.8\times10^{3\dagger}$ & $2.0\times10^6$ & $1.8\times10^{3\dagger}$ & 1546, 1574, (1) & 3 & 3 & $92.6^\dagger$ & Vergyris 2017 \cite{vergyris2017fully} \\ \cline{2-13}
  & type-0 (PPLN) & InGaAs & fbr WG + pol MZI & $1.6\times10^8$ & $1.0\times10^{6\dagger}$ & - & - & 1530, 1570 & 0.05 & 0.12 & $90.0^\dagger$ & Shalm 2013 \cite{shalm2013three}(SI) \\ \cline{2-13}
 & type-0 (PPLN) & InGaAs & WG+BS, MZI & $2.2\times10^{9\dagger}$ & $3.5\times10^{2\dagger}$ & - & - & 1560, 1560 & 0.04 ($0.16^\dagger$) & 0.04 ($0.16^\dagger$) & $92.9^\dagger$ & Atzeni 2018 \cite{atzeni2018integrated} \\ \cline{2-13}
 & type-0 (PPKTP) & SNSPD & WG+Sagnac & $5.6\times10^6$ & - & - & - & 1550, 1550 & 27 & 27 & 98.8 & Meyer 2018 \cite{meyer2018high} \\ \cline{2-13}
 & type-0 (Ti:LiNbO$_3$) & InGaAs & WG+Sagnac & $2.8\times10^7$ & $1.5\times10^{6\dagger}$ & $1.2\times10^7$ & $6.4\times10^{5\dagger}$ &  1335, 1490, (6, 2) & 5.4 (36) & 5.4 (22) & 94.5 & Sun 2019 \cite{sun2019compact} \\
\noalign{\hrule height 1pt}

\parbox[t]{3mm}{\multirow{8}{*}{\rotatebox[origin=c]{90}{Pulse-pumped}}} & type-I (PPLN) & InGaAs & two WGs & $2.4\times10^{6\dagger}$ & $1.2\times10^{5\dagger}$ & $1.6\times10^{5\dagger}$ & $8.0\times10^{3\dagger}$ & 1431, 1611, (15) & 16.3 & 6.3 & $95.7^\dagger$ & Jiang 2007 \cite{jiang2007generation} \\ \cline{2-13}
 & type-I (PPLN) & InGaAs SPCM & fiber Sagnac & - & - & - & - & 1542, 1562, (9) & 0.6 & 0.6 & 96.8 & Lim 2008 \cite{lim2008stable} \\ \cline{2-13}
 & type-0 (PPLN) & InGaAs/InP & cascaded Sagnac & $5.0\times10^6$ & - & - & - & 1539, 1559, (36) & $0.6^\dagger$ & $0.6^\dagger$ & $99.4^\dagger$ & Arahira 2011 \cite{arahira2011generation} \\ \cline{2-13}
& type-0 (PPLN) & SNSPD & MZI & $9.6\times10^{5\dagger}$ & $1.0\times10^{0\dagger}$ & $4.8\times10^6$ & $5.2\times10^{0\dagger}$ & 1554, 1554, (0.2) & 0.12 & 0.09 & 97.3 & Sansoni 2017 \cite{sansoni2017two} \\ \cline{2-13}
& type-I (silica fiber) & InGaAs & periodically poled & $1.4\times10^6$ & & $1.4\times10^4$ & & 1517, 1594, (0.2) & $2.5^\dagger$ & $2.5^\dagger$ & 98.9 & Chen 2017 \cite{chen2017compensation} \\ \cline{2-13}
& type-II (BRW) & InGaAs APD & Bragg grating & $1.9\times10^{5\dagger}$ & $3.0\times10^2$ & $4.7\times10^3$ & $7.6\times10^0$ & 1535, 1535, (40) & $0.4^\dagger$ & $0.4^\dagger$ & 93.2 & Schlager 2017 \cite{schlager2017temporally} \\ \cline{2-13}
& type-II (silica fiber) & InGaAs & periodically poled & - & $2.4\times10^{1\dagger}$ & - & $2.2\times10^{1\dagger}$ & 1554, 1577, (1.1) & - & - & 98.1 & Chen 2018 \cite{chen2018turn} \\
\noalign{\hrule height 1pt}

\end{tabular}

\smallskip
\renewcommand{\arraystretch}{1.4}
\centering
\caption{List of entangled photon-pair source-based applications and their examples in which different source designs are implemented.}
\label{table5}
\begin{tabular}{!{\vrule width 1pt}p{5.4cm}|p{6.4cm}|p{10.8cm}!{\vrule width 1pt}}
\noalign{\hrule height 1pt}
\textbf{Application} & \textbf{Key requirements} & \textbf{Examples of source designs} \\

\noalign{\hrule height 1pt}

Bell test, foundation of entanglement science, Multi-path experiments & Relatively compact and stable, but with good entanglement visibility and count rates & Sagnac \cite{vermeyden2013experimental,giustina2013bell,xiao2017experimental}, Standard BBO (or BiBO) \cite{herzog1995complementarity,weihs1998violation,saunders2010experimental,pan2019direct}, Crossed-crystal \cite{christensen2013detection,ono2013entanglement,weston2013experimental}, Review article \cite{ma2016delayed} \\
\hline
Bell test - loophole free & Very high coupling efficiency and very high visibility & Sagnac \cite{giustina2015significant}, Beam displacer \cite{shalm2015strong}. \\ 
\hline
Entanglement distribution and QKD - high rate & Compact, robust and movable, high count rate & Single-crystal (BBO) \cite{liu2016experimental,ursin2007entanglement,poppe2004practical}, Sagnac \cite{beckert2019space,liu2020drone,ji2017towards,horn2019auto}, Crossed-crystal \cite{trojek2008collinear,hubel2007high}, Beam displacer \cite{shi2020stable}  \\
\hline
Satellite entanglement distribution and QKD & Compact, very high count rate and mechanically stable & Parallel-crystal \cite{villar2020entanglement}, Sagnac \cite{beckert2019space,yin2017satellite,yin2017satellitetoground} \\ \hline
Teleportation/Swapping & Spectrally separable biphoton states, group-velocity matching & BBO (or BiBO) \cite{ibarra2020experimental}, BBO (BiBO) with PBS\cite{Kim:2002aa,yin2012quantum,ma2012quantum} which also works well at telecom\cite{Lutz:14}, QPM crystal \cite{yepiz2020spectrally},  GVM based Sagnac \cite{Jin:2015aa}, Sagnac \cite{ren2017ground}, Review article \cite{pirandola2015advances} \\
\hline
Multi-photon entanglement & Generate multiple photon-pairs with high count rates,  heralded photons for multi-photon interference &  Crossed-crystal \cite{wang2016experimental}, Cascaded SPDC of triplets \cite{Huebel2010,Hamel:2014qf}, multi-SPDC interference \cite{yao2012observation,wang2016experimental} based on BBO with PBS \cite{Kim:2002aa} \\
\hline
HOM interference, Optical coherence tomography, Quantum sensing, metrology and imaging & High count rate, spectrally factorizable & Single-crystal \cite{lopez2012quantum,slussarenko2017unconditional,fink2019entanglement}, Group velocity matched crystals\cite{mosley:133601,Gerrits:2015aa}, Review articles \cite{pirandola2018advances,Maga_a_Loaiza_2019,moreau2019imaging,polino2020photonicreview,gregory2020imaging}.\\
\hline
\noalign{\hrule height 1pt}
    
    \end{tabular}
\end{sidewaystable}

\end{widetext}

\bibliography{aipsamp}

\providecommand{\noopsort}[1]{}\providecommand{\singleletter}[1]{#1}%
\begin{thebibliography}{398}%
\makeatletter
\providecommand \@ifxundefined [1]{%
 \@ifx{#1\undefined}
}%
\providecommand \@ifnum [1]{%
 \ifnum #1\expandafter \@firstoftwo
 \else \expandafter \@secondoftwo
 \fi
}%
\providecommand \@ifx [1]{%
 \ifx #1\expandafter \@firstoftwo
 \else \expandafter \@secondoftwo
 \fi
}%
\providecommand \natexlab [1]{#1}%
\providecommand \enquote  [1]{``#1''}%
\providecommand \bibnamefont  [1]{#1}%
\providecommand \bibfnamefont [1]{#1}%
\providecommand \citenamefont [1]{#1}%
\providecommand \href@noop [0]{\@secondoftwo}%
\providecommand \href [0]{\begingroup \@sanitize@url \@href}%
\providecommand \@href[1]{\@@startlink{#1}\@@href}%
\providecommand \@@href[1]{\endgroup#1\@@endlink}%
\providecommand \@sanitize@url [0]{\catcode `\\12\catcode `\$12\catcode
  `\&12\catcode `\#12\catcode `\^12\catcode `\_12\catcode `\%12\relax}%
\providecommand \@@startlink[1]{}%
\providecommand \@@endlink[0]{}%
\providecommand \url  [0]{\begingroup\@sanitize@url \@url }%
\providecommand \@url [1]{\endgroup\@href {#1}{\urlprefix }}%
\providecommand \urlprefix  [0]{URL }%
\providecommand \Eprint [0]{\href }%
\providecommand \doibase [0]{http://dx.doi.org/}%
\providecommand \selectlanguage [0]{\@gobble}%
\providecommand \bibinfo  [0]{\@secondoftwo}%
\providecommand \bibfield  [0]{\@secondoftwo}%
\providecommand \translation [1]{[#1]}%
\providecommand \BibitemOpen [0]{}%
\providecommand \bibitemStop [0]{}%
\providecommand \bibitemNoStop [0]{.\EOS\space}%
\providecommand \EOS [0]{\spacefactor3000\relax}%
\providecommand \BibitemShut  [1]{\csname bibitem#1\endcsname}%
\let\auto@bib@innerbib\@empty
\bibitem [{\citenamefont {Corona}\ \emph {et~al.}(2011)\citenamefont {Corona},
  \citenamefont {Garay-Palmett},\ and\ \citenamefont {U'Ren}}]{Corona:11}%
  \BibitemOpen
  \bibfield  {author} {\bibinfo {author} {\bibfnamefont {M.}~\bibnamefont
  {Corona}}, \bibinfo {author} {\bibfnamefont {K.}~\bibnamefont
  {Garay-Palmett}}, \ and\ \bibinfo {author} {\bibfnamefont {A.~B.}\
  \bibnamefont {U'Ren}},\ }\href {\doibase 10.1364/OL.36.000190} {\bibfield
  {journal} {\bibinfo  {journal} {Opt. Lett.}\ }\textbf {\bibinfo {volume}
  {36}},\ \bibinfo {pages} {190} (\bibinfo {year} {2011})}\BibitemShut
  {NoStop}%
\bibitem [{\citenamefont {Sandbo~Chang}\ \emph {et~al.}(2018)\citenamefont
  {Sandbo~Chang}, \citenamefont {Simoen}, \citenamefont {Aumentado},
  \citenamefont {Sab\'{\i}n}, \citenamefont {Forn-D\'{\i}az}, \citenamefont
  {Vadiraj}, \citenamefont {Quijandr\'{\i}a}, \citenamefont {Johansson},
  \citenamefont {Fuentes},\ and\ \citenamefont
  {Wilson}}]{PhysRevApplied.10.044019}%
  \BibitemOpen
  \bibfield  {author} {\bibinfo {author} {\bibfnamefont {C.~W.}\ \bibnamefont
  {Sandbo~Chang}}, \bibinfo {author} {\bibfnamefont {M.}~\bibnamefont
  {Simoen}}, \bibinfo {author} {\bibfnamefont {J.}~\bibnamefont {Aumentado}},
  \bibinfo {author} {\bibfnamefont {C.}~\bibnamefont {Sab\'{\i}n}}, \bibinfo
  {author} {\bibfnamefont {P.}~\bibnamefont {Forn-D\'{\i}az}}, \bibinfo
  {author} {\bibfnamefont {A.~M.}\ \bibnamefont {Vadiraj}}, \bibinfo {author}
  {\bibfnamefont {F.}~\bibnamefont {Quijandr\'{\i}a}}, \bibinfo {author}
  {\bibfnamefont {G.}~\bibnamefont {Johansson}}, \bibinfo {author}
  {\bibfnamefont {I.}~\bibnamefont {Fuentes}}, \ and\ \bibinfo {author}
  {\bibfnamefont {C.~M.}\ \bibnamefont {Wilson}},\ }\href {\doibase
  10.1103/PhysRevApplied.10.044019} {\bibfield  {journal} {\bibinfo  {journal}
  {Phys. Rev. Appl.}\ }\textbf {\bibinfo {volume} {10}},\ \bibinfo {pages}
  {044019} (\bibinfo {year} {2018})}\BibitemShut {NoStop}%
\bibitem [{\citenamefont {Huebel}\ \emph {et~al.}(2010)\citenamefont {Huebel},
  \citenamefont {Hamel}, \citenamefont {Fedrizzi}, \citenamefont {Ramelow},
  \citenamefont {Resch},\ and\ \citenamefont {Jennewein}}]{Huebel2010}%
  \BibitemOpen
  \bibfield  {author} {\bibinfo {author} {\bibfnamefont {H.}~\bibnamefont
  {Huebel}}, \bibinfo {author} {\bibfnamefont {D.~R.}\ \bibnamefont {Hamel}},
  \bibinfo {author} {\bibfnamefont {A.}~\bibnamefont {Fedrizzi}}, \bibinfo
  {author} {\bibfnamefont {S.}~\bibnamefont {Ramelow}}, \bibinfo {author}
  {\bibfnamefont {K.~J.}\ \bibnamefont {Resch}}, \ and\ \bibinfo {author}
  {\bibfnamefont {T.}~\bibnamefont {Jennewein}},\ }\href {\doibase
  10.1038/nature09175} {\bibfield  {journal} {\bibinfo  {journal} {Nature}\
  }\textbf {\bibinfo {volume} {466}},\ \bibinfo {pages} {601} (\bibinfo {year}
  {2010})}\BibitemShut {NoStop}%
\bibitem [{\citenamefont {Fulconis}\ \emph {et~al.}(2005)\citenamefont
  {Fulconis}, \citenamefont {Alibart}, \citenamefont {Wadsworth}, \citenamefont
  {Russell},\ and\ \citenamefont {Rarity}}]{fulconis2005high}%
  \BibitemOpen
  \bibfield  {author} {\bibinfo {author} {\bibfnamefont {J.}~\bibnamefont
  {Fulconis}}, \bibinfo {author} {\bibfnamefont {O.}~\bibnamefont {Alibart}},
  \bibinfo {author} {\bibfnamefont {W.~J.}\ \bibnamefont {Wadsworth}}, \bibinfo
  {author} {\bibfnamefont {P.~S.~J.}\ \bibnamefont {Russell}}, \ and\ \bibinfo
  {author} {\bibfnamefont {J.~G.}\ \bibnamefont {Rarity}},\ }\href@noop {}
  {\bibfield  {journal} {\bibinfo  {journal} {Opt. Express}\ }\textbf {\bibinfo
  {volume} {13}},\ \bibinfo {pages} {7572} (\bibinfo {year}
  {2005})}\BibitemShut {NoStop}%
\bibitem [{\citenamefont {Kwiat}\ \emph {et~al.}(1995)\citenamefont {Kwiat},
  \citenamefont {Mattle}, \citenamefont {Weinfurter}, \citenamefont
  {Zeilinger}, \citenamefont {Sergienko},\ and\ \citenamefont
  {Shih}}]{kwiattype2}%
  \BibitemOpen
  \bibfield  {author} {\bibinfo {author} {\bibfnamefont {P.~G.}\ \bibnamefont
  {Kwiat}}, \bibinfo {author} {\bibfnamefont {K.}~\bibnamefont {Mattle}},
  \bibinfo {author} {\bibfnamefont {H.}~\bibnamefont {Weinfurter}}, \bibinfo
  {author} {\bibfnamefont {A.}~\bibnamefont {Zeilinger}}, \bibinfo {author}
  {\bibfnamefont {A.~V.}\ \bibnamefont {Sergienko}}, \ and\ \bibinfo {author}
  {\bibfnamefont {Y.}~\bibnamefont {Shih}},\ }\href {\doibase
  10.1103/PhysRevLett.75.4337} {\bibfield  {journal} {\bibinfo  {journal}
  {Phys. Rev. Lett.}\ }\textbf {\bibinfo {volume} {75}},\ \bibinfo {pages}
  {4337} (\bibinfo {year} {1995})}\BibitemShut {NoStop}%
\bibitem [{\citenamefont {Kwiat}\ \emph {et~al.}(1999)\citenamefont {Kwiat},
  \citenamefont {Waks}, \citenamefont {White}, \citenamefont {Appelbaum},\ and\
  \citenamefont {Eberhard}}]{kwiattype1dual}%
  \BibitemOpen
  \bibfield  {author} {\bibinfo {author} {\bibfnamefont {P.~G.}\ \bibnamefont
  {Kwiat}}, \bibinfo {author} {\bibfnamefont {E.}~\bibnamefont {Waks}},
  \bibinfo {author} {\bibfnamefont {A.~G.}\ \bibnamefont {White}}, \bibinfo
  {author} {\bibfnamefont {I.}~\bibnamefont {Appelbaum}}, \ and\ \bibinfo
  {author} {\bibfnamefont {P.~H.}\ \bibnamefont {Eberhard}},\ }\href@noop {}
  {\bibfield  {journal} {\bibinfo  {journal} {Phys. Rev. A}\ }\textbf {\bibinfo
  {volume} {60}},\ \bibinfo {pages} {R773} (\bibinfo {year}
  {1999})}\BibitemShut {NoStop}%
\bibitem [{\citenamefont {Krenn}\ \emph {et~al.}(2017)\citenamefont {Krenn},
  \citenamefont {Malik}, \citenamefont {Erhard},\ and\ \citenamefont
  {Zeilinger}}]{krenn2017orbital}%
  \BibitemOpen
  \bibfield  {author} {\bibinfo {author} {\bibfnamefont {M.}~\bibnamefont
  {Krenn}}, \bibinfo {author} {\bibfnamefont {M.}~\bibnamefont {Malik}},
  \bibinfo {author} {\bibfnamefont {M.}~\bibnamefont {Erhard}}, \ and\ \bibinfo
  {author} {\bibfnamefont {A.}~\bibnamefont {Zeilinger}},\ }\href@noop {}
  {\bibfield  {journal} {\bibinfo  {journal} {Phil. Trans. Roy. Soc. A: Math.
  Phys. Engg. Sci.}\ }\textbf {\bibinfo {volume} {375}},\ \bibinfo {pages}
  {20150442} (\bibinfo {year} {2017})}\BibitemShut {NoStop}%
\bibitem [{\citenamefont {Forbes}\ and\ \citenamefont
  {Nape}(2019)}]{forbes2019quantum}%
  \BibitemOpen
  \bibfield  {author} {\bibinfo {author} {\bibfnamefont {A.}~\bibnamefont
  {Forbes}}\ and\ \bibinfo {author} {\bibfnamefont {I.}~\bibnamefont {Nape}},\
  }\href@noop {} {\bibfield  {journal} {\bibinfo  {journal} {AVS Quant. Sci.}\
  }\textbf {\bibinfo {volume} {1}},\ \bibinfo {pages} {011701} (\bibinfo {year}
  {2019})}\BibitemShut {NoStop}%
\bibitem [{\citenamefont {Brendel}\ \emph {et~al.}(1999)\citenamefont
  {Brendel}, \citenamefont {Gisin}, \citenamefont {Tittel},\ and\ \citenamefont
  {Zbinden}}]{brendel1999pulsed}%
  \BibitemOpen
  \bibfield  {author} {\bibinfo {author} {\bibfnamefont {J.}~\bibnamefont
  {Brendel}}, \bibinfo {author} {\bibfnamefont {N.}~\bibnamefont {Gisin}},
  \bibinfo {author} {\bibfnamefont {W.}~\bibnamefont {Tittel}}, \ and\ \bibinfo
  {author} {\bibfnamefont {H.}~\bibnamefont {Zbinden}},\ }\href@noop {}
  {\bibfield  {journal} {\bibinfo  {journal} {Phys. Rev. Lett.}\ }\textbf
  {\bibinfo {volume} {82}},\ \bibinfo {pages} {2594} (\bibinfo {year}
  {1999})}\BibitemShut {NoStop}%
\bibitem [{\citenamefont {Avenhaus}\ \emph {et~al.}(2009)\citenamefont
  {Avenhaus}, \citenamefont {Chekhova}, \citenamefont {Krivitsky},
  \citenamefont {Leuchs},\ and\ \citenamefont
  {Silberhorn}}]{avenhaus2009experimental}%
  \BibitemOpen
  \bibfield  {author} {\bibinfo {author} {\bibfnamefont {M.}~\bibnamefont
  {Avenhaus}}, \bibinfo {author} {\bibfnamefont {M.~V.}\ \bibnamefont
  {Chekhova}}, \bibinfo {author} {\bibfnamefont {L.~A.}\ \bibnamefont
  {Krivitsky}}, \bibinfo {author} {\bibfnamefont {G.}~\bibnamefont {Leuchs}}, \
  and\ \bibinfo {author} {\bibfnamefont {C.}~\bibnamefont {Silberhorn}},\
  }\href@noop {} {\bibfield  {journal} {\bibinfo  {journal} {Phys. Rev. A}\
  }\textbf {\bibinfo {volume} {79}},\ \bibinfo {pages} {043836} (\bibinfo
  {year} {2009})}\BibitemShut {NoStop}%
\bibitem [{\citenamefont {Fujiwara}\ \emph {et~al.}(2009)\citenamefont
  {Fujiwara}, \citenamefont {Toyoshima}, \citenamefont {Sasaki}, \citenamefont
  {Yoshino}, \citenamefont {Nambu},\ and\ \citenamefont
  {Tomita}}]{fujiwara:261103}%
  \BibitemOpen
  \bibfield  {author} {\bibinfo {author} {\bibfnamefont {M.}~\bibnamefont
  {Fujiwara}}, \bibinfo {author} {\bibfnamefont {M.}~\bibnamefont {Toyoshima}},
  \bibinfo {author} {\bibfnamefont {M.}~\bibnamefont {Sasaki}}, \bibinfo
  {author} {\bibfnamefont {K.}~\bibnamefont {Yoshino}}, \bibinfo {author}
  {\bibfnamefont {Y.}~\bibnamefont {Nambu}}, \ and\ \bibinfo {author}
  {\bibfnamefont {A.}~\bibnamefont {Tomita}},\ }\href {\doibase
  10.1063/1.3276559} {\bibfield  {journal} {\bibinfo  {journal} {Appl. Phys.
  Lett.}\ }\textbf {\bibinfo {volume} {95}},\ \bibinfo {eid} {261103} (\bibinfo
  {year} {2009})}\BibitemShut {NoStop}%
\bibitem [{\citenamefont {Ma}\ \emph {et~al.}(2009)\citenamefont {Ma},
  \citenamefont {Qarry}, \citenamefont {Kofler}, \citenamefont {Jennewein},\
  and\ \citenamefont {Zeilinger}}]{Ma:2009}%
  \BibitemOpen
  \bibfield  {author} {\bibinfo {author} {\bibfnamefont {X.-s.}\ \bibnamefont
  {Ma}}, \bibinfo {author} {\bibfnamefont {A.}~\bibnamefont {Qarry}}, \bibinfo
  {author} {\bibfnamefont {J.}~\bibnamefont {Kofler}}, \bibinfo {author}
  {\bibfnamefont {T.}~\bibnamefont {Jennewein}}, \ and\ \bibinfo {author}
  {\bibfnamefont {A.}~\bibnamefont {Zeilinger}},\ }\href@noop {} {\bibfield
  {journal} {\bibinfo  {journal} {Phys. Rev. A}\ }\textbf {\bibinfo {volume}
  {79}},\ \bibinfo {pages} {042101} (\bibinfo {year} {2009})}\BibitemShut
  {NoStop}%
\bibitem [{\citenamefont {Meyer-Scott}\ \emph {et~al.}(2018)\citenamefont
  {Meyer-Scott}, \citenamefont {Prasannan}, \citenamefont {Eigner},
  \citenamefont {Quiring}, \citenamefont {Donohue}, \citenamefont {Barkhofen},\
  and\ \citenamefont {Silberhorn}}]{meyer2018high}%
  \BibitemOpen
  \bibfield  {author} {\bibinfo {author} {\bibfnamefont {E.}~\bibnamefont
  {Meyer-Scott}}, \bibinfo {author} {\bibfnamefont {N.}~\bibnamefont
  {Prasannan}}, \bibinfo {author} {\bibfnamefont {C.}~\bibnamefont {Eigner}},
  \bibinfo {author} {\bibfnamefont {V.}~\bibnamefont {Quiring}}, \bibinfo
  {author} {\bibfnamefont {J.~M.}\ \bibnamefont {Donohue}}, \bibinfo {author}
  {\bibfnamefont {S.}~\bibnamefont {Barkhofen}}, \ and\ \bibinfo {author}
  {\bibfnamefont {C.}~\bibnamefont {Silberhorn}},\ }\href@noop {} {\bibfield
  {journal} {\bibinfo  {journal} {Opt. Express}\ }\textbf {\bibinfo {volume}
  {26}},\ \bibinfo {pages} {32475} (\bibinfo {year} {2018})}\BibitemShut
  {NoStop}%
\bibitem [{\citenamefont {Orieux}\ \emph {et~al.}(2017)\citenamefont {Orieux},
  \citenamefont {Versteegh}, \citenamefont {J{\"o}ns},\ and\ \citenamefont
  {Ducci}}]{orieux2017semiconductor}%
  \BibitemOpen
  \bibfield  {author} {\bibinfo {author} {\bibfnamefont {A.}~\bibnamefont
  {Orieux}}, \bibinfo {author} {\bibfnamefont {M.~A.~M.}\ \bibnamefont
  {Versteegh}}, \bibinfo {author} {\bibfnamefont {K.~D.}\ \bibnamefont
  {J{\"o}ns}}, \ and\ \bibinfo {author} {\bibfnamefont {S.}~\bibnamefont
  {Ducci}},\ }\href@noop {} {\bibfield  {journal} {\bibinfo  {journal} {Rep.
  Prog. Phys.}\ }\textbf {\bibinfo {volume} {80}},\ \bibinfo {pages} {076001}
  (\bibinfo {year} {2017})}\BibitemShut {NoStop}%
\bibitem [{\citenamefont {Huber}\ \emph {et~al.}(2018)\citenamefont {Huber},
  \citenamefont {Reindl}, \citenamefont {Aberl}, \citenamefont {Rastelli},\
  and\ \citenamefont {Trotta}}]{huber2018semiconductor}%
  \BibitemOpen
  \bibfield  {author} {\bibinfo {author} {\bibfnamefont {D.}~\bibnamefont
  {Huber}}, \bibinfo {author} {\bibfnamefont {M.}~\bibnamefont {Reindl}},
  \bibinfo {author} {\bibfnamefont {J.}~\bibnamefont {Aberl}}, \bibinfo
  {author} {\bibfnamefont {A.}~\bibnamefont {Rastelli}}, \ and\ \bibinfo
  {author} {\bibfnamefont {R.}~\bibnamefont {Trotta}},\ }\href@noop {}
  {\bibfield  {journal} {\bibinfo  {journal} {J. Opt.}\ }\textbf {\bibinfo
  {volume} {20}},\ \bibinfo {pages} {073002} (\bibinfo {year}
  {2018})}\BibitemShut {NoStop}%
\bibitem [{\citenamefont {Vergyris}\ \emph {et~al.}(2017)\citenamefont
  {Vergyris}, \citenamefont {Kaiser}, \citenamefont {Gouzien}, \citenamefont
  {Sauder}, \citenamefont {Lunghi},\ and\ \citenamefont
  {Tanzilli}}]{vergyris2017fully}%
  \BibitemOpen
  \bibfield  {author} {\bibinfo {author} {\bibfnamefont {P.}~\bibnamefont
  {Vergyris}}, \bibinfo {author} {\bibfnamefont {F.}~\bibnamefont {Kaiser}},
  \bibinfo {author} {\bibfnamefont {E.}~\bibnamefont {Gouzien}}, \bibinfo
  {author} {\bibfnamefont {G.}~\bibnamefont {Sauder}}, \bibinfo {author}
  {\bibfnamefont {T.}~\bibnamefont {Lunghi}}, \ and\ \bibinfo {author}
  {\bibfnamefont {S.}~\bibnamefont {Tanzilli}},\ }\href@noop {} {\bibfield
  {journal} {\bibinfo  {journal} {Quant. Sci. Tech.}\ }\textbf {\bibinfo
  {volume} {2}},\ \bibinfo {pages} {024007} (\bibinfo {year}
  {2017})}\BibitemShut {NoStop}%
\bibitem [{\citenamefont {\ifmmode~\dot{Z}\else \.{Z}\fi{}ukowski}\ \emph
  {et~al.}(1993)\citenamefont {\ifmmode~\dot{Z}\else \.{Z}\fi{}ukowski},
  \citenamefont {Zeilinger}, \citenamefont {Horne},\ and\ \citenamefont
  {Ekert}}]{Zukowski93a}%
  \BibitemOpen
  \bibfield  {author} {\bibinfo {author} {\bibfnamefont {M.}~\bibnamefont
  {\ifmmode~\dot{Z}\else \.{Z}\fi{}ukowski}}, \bibinfo {author} {\bibfnamefont
  {A.}~\bibnamefont {Zeilinger}}, \bibinfo {author} {\bibfnamefont {M.~A.}\
  \bibnamefont {Horne}}, \ and\ \bibinfo {author} {\bibfnamefont {A.~K.}\
  \bibnamefont {Ekert}},\ }\href {\doibase 10.1103/PhysRevLett.71.4287}
  {\bibfield  {journal} {\bibinfo  {journal} {Phys. Rev. Lett.}\ }\textbf
  {\bibinfo {volume} {71}},\ \bibinfo {pages} {4287} (\bibinfo {year}
  {1993})}\BibitemShut {NoStop}%
\bibitem [{\citenamefont {Pan}\ \emph {et~al.}(2001)\citenamefont {Pan},
  \citenamefont {Daniell}, \citenamefont {Gasparoni}, \citenamefont {Weihs},\
  and\ \citenamefont {Zeilinger}}]{Pan01b}%
  \BibitemOpen
  \bibfield  {author} {\bibinfo {author} {\bibfnamefont {J.-W.}\ \bibnamefont
  {Pan}}, \bibinfo {author} {\bibfnamefont {M.}~\bibnamefont {Daniell}},
  \bibinfo {author} {\bibfnamefont {S.}~\bibnamefont {Gasparoni}}, \bibinfo
  {author} {\bibfnamefont {G.}~\bibnamefont {Weihs}}, \ and\ \bibinfo {author}
  {\bibfnamefont {A.}~\bibnamefont {Zeilinger}},\ }\href@noop {} {\bibfield
  {journal} {\bibinfo  {journal} {Phys. Rev. Lett.}\ }\textbf {\bibinfo
  {volume} {86}},\ \bibinfo {pages} {4435} (\bibinfo {year}
  {2001})}\BibitemShut {NoStop}%
\bibitem [{\citenamefont {Bouwmeester}\ \emph {et~al.}(1997)\citenamefont
  {Bouwmeester}, \citenamefont {Pan}, \citenamefont {Mattle}, \citenamefont
  {Eibl}, \citenamefont {Weinfurter},\ and\ \citenamefont
  {Zeilinger}}]{bouwmeester1997experimental}%
  \BibitemOpen
  \bibfield  {author} {\bibinfo {author} {\bibfnamefont {D.}~\bibnamefont
  {Bouwmeester}}, \bibinfo {author} {\bibfnamefont {J.-W.}\ \bibnamefont
  {Pan}}, \bibinfo {author} {\bibfnamefont {K.}~\bibnamefont {Mattle}},
  \bibinfo {author} {\bibfnamefont {M.}~\bibnamefont {Eibl}}, \bibinfo {author}
  {\bibfnamefont {H.}~\bibnamefont {Weinfurter}}, \ and\ \bibinfo {author}
  {\bibfnamefont {A.}~\bibnamefont {Zeilinger}},\ }\href@noop {} {\bibfield
  {journal} {\bibinfo  {journal} {Nature}\ }\textbf {\bibinfo {volume} {390}},\
  \bibinfo {pages} {575} (\bibinfo {year} {1997})}\BibitemShut {NoStop}%
\bibitem [{\citenamefont {Pan}\ \emph {et~al.}(2012)\citenamefont {Pan},
  \citenamefont {Chen}, \citenamefont {Lu}, \citenamefont {Weinfurter},
  \citenamefont {Zeilinger},\ and\ \citenamefont
  {{\.Z}ukowski}}]{pan2012multiphoton}%
  \BibitemOpen
  \bibfield  {author} {\bibinfo {author} {\bibfnamefont {J.-W.}\ \bibnamefont
  {Pan}}, \bibinfo {author} {\bibfnamefont {Z.-B.}\ \bibnamefont {Chen}},
  \bibinfo {author} {\bibfnamefont {C.-Y.}\ \bibnamefont {Lu}}, \bibinfo
  {author} {\bibfnamefont {H.}~\bibnamefont {Weinfurter}}, \bibinfo {author}
  {\bibfnamefont {A.}~\bibnamefont {Zeilinger}}, \ and\ \bibinfo {author}
  {\bibfnamefont {M.}~\bibnamefont {{\.Z}ukowski}},\ }\href@noop {} {\bibfield
  {journal} {\bibinfo  {journal} {Rev. Mod. Phys.}\ }\textbf {\bibinfo {volume}
  {84}},\ \bibinfo {pages} {777} (\bibinfo {year} {2012})}\BibitemShut
  {NoStop}%
\bibitem [{\citenamefont {Mandel}\ and\ \citenamefont
  {Wolf}(1995)}]{mandel1995optical}%
  \BibitemOpen
  \bibfield  {author} {\bibinfo {author} {\bibfnamefont {L.}~\bibnamefont
  {Mandel}}\ and\ \bibinfo {author} {\bibfnamefont {E.}~\bibnamefont {Wolf}},\
  }\href@noop {} {\emph {\bibinfo {title} {Optical coherence and quantum
  optics}}}\ (\bibinfo  {publisher} {Cambridge university press},\ \bibinfo
  {year} {1995})\BibitemShut {NoStop}%
\bibitem [{\citenamefont {Glauber}(2007)}]{glauber2007quantum}%
  \BibitemOpen
  \bibfield  {author} {\bibinfo {author} {\bibfnamefont {R.~J.}\ \bibnamefont
  {Glauber}},\ }\href@noop {} {\emph {\bibinfo {title} {Quantum theory of
  optical coherence: selected papers and lectures}}}\ (\bibinfo  {publisher}
  {John Wiley \& Sons},\ \bibinfo {year} {2007})\BibitemShut {NoStop}%
\bibitem [{\citenamefont {Ou}(2007)}]{ou2007multi}%
  \BibitemOpen
  \bibfield  {author} {\bibinfo {author} {\bibfnamefont {Z.-Y.~J.}\
  \bibnamefont {Ou}},\ }\href@noop {} {\emph {\bibinfo {title} {Multi-photon
  quantum interference}}},\ Vol.~\bibinfo {volume} {43}\ (\bibinfo  {publisher}
  {Springer},\ \bibinfo {year} {2007})\BibitemShut {NoStop}%
\bibitem [{\citenamefont {Torres}\ \emph {et~al.}(2011)\citenamefont {Torres},
  \citenamefont {Banaszek},\ and\ \citenamefont
  {Walmsley}}]{torres2011engineering}%
  \BibitemOpen
  \bibfield  {author} {\bibinfo {author} {\bibfnamefont {J.~P.}\ \bibnamefont
  {Torres}}, \bibinfo {author} {\bibfnamefont {K.}~\bibnamefont {Banaszek}}, \
  and\ \bibinfo {author} {\bibfnamefont {I.}~\bibnamefont {Walmsley}},\ }in\
  \href@noop {} {\emph {\bibinfo {booktitle} {Progress in Optics}}},\
  Vol.~\bibinfo {volume} {56}\ (\bibinfo  {publisher} {Elsevier},\ \bibinfo
  {year} {2011})\ pp.\ \bibinfo {pages} {227--331}\BibitemShut {NoStop}%
\bibitem [{\citenamefont {Loudon}(1983)}]{Loudon83a}%
  \BibitemOpen
  \bibfield  {author} {\bibinfo {author} {\bibfnamefont {R.}~\bibnamefont
  {Loudon}},\ }\href@noop {} {\emph {\bibinfo {title} {The Quantum Theory of
  Light}}},\ \bibinfo {edition} {2nd}\ ed.\ (\bibinfo  {publisher} {Oxford
  University Press},\ \bibinfo {address} {Oxford},\ \bibinfo {year}
  {1983})\BibitemShut {NoStop}%
\bibitem [{\citenamefont {Takesue}\ and\ \citenamefont
  {Shimizu}(2010)}]{takesue2010effects}%
  \BibitemOpen
  \bibfield  {author} {\bibinfo {author} {\bibfnamefont {H.}~\bibnamefont
  {Takesue}}\ and\ \bibinfo {author} {\bibfnamefont {K.}~\bibnamefont
  {Shimizu}},\ }\href@noop {} {\bibfield  {journal} {\bibinfo  {journal} {Opt.
  Commun.}\ }\textbf {\bibinfo {volume} {283}},\ \bibinfo {pages} {276}
  (\bibinfo {year} {2010})}\BibitemShut {NoStop}%
\bibitem [{\citenamefont {Takeoka}\ \emph {et~al.}(2015)\citenamefont
  {Takeoka}, \citenamefont {Jin},\ and\ \citenamefont
  {Sasaki}}]{takeoka2015full}%
  \BibitemOpen
  \bibfield  {author} {\bibinfo {author} {\bibfnamefont {M.}~\bibnamefont
  {Takeoka}}, \bibinfo {author} {\bibfnamefont {R.-B.}\ \bibnamefont {Jin}}, \
  and\ \bibinfo {author} {\bibfnamefont {M.}~\bibnamefont {Sasaki}},\
  }\href@noop {} {\bibfield  {journal} {\bibinfo  {journal} {New J. Phys.}\
  }\textbf {\bibinfo {volume} {17}},\ \bibinfo {pages} {043030} (\bibinfo
  {year} {2015})}\BibitemShut {NoStop}%
\bibitem [{\citenamefont {Tsujimoto}\ \emph
  {et~al.}(2018{\natexlab{a}})\citenamefont {Tsujimoto}, \citenamefont {Wakui},
  \citenamefont {Fujiwara}, \citenamefont {Hayasaka}, \citenamefont {Miki},
  \citenamefont {Terai}, \citenamefont {Sasaki},\ and\ \citenamefont
  {Takeoka}}]{tsujimoto2018optimal}%
  \BibitemOpen
  \bibfield  {author} {\bibinfo {author} {\bibfnamefont {Y.}~\bibnamefont
  {Tsujimoto}}, \bibinfo {author} {\bibfnamefont {K.}~\bibnamefont {Wakui}},
  \bibinfo {author} {\bibfnamefont {M.}~\bibnamefont {Fujiwara}}, \bibinfo
  {author} {\bibfnamefont {K.}~\bibnamefont {Hayasaka}}, \bibinfo {author}
  {\bibfnamefont {S.}~\bibnamefont {Miki}}, \bibinfo {author} {\bibfnamefont
  {H.}~\bibnamefont {Terai}}, \bibinfo {author} {\bibfnamefont
  {M.}~\bibnamefont {Sasaki}}, \ and\ \bibinfo {author} {\bibfnamefont
  {M.}~\bibnamefont {Takeoka}},\ }\href@noop {} {\bibfield  {journal} {\bibinfo
   {journal} {Phys. Rev. A}\ }\textbf {\bibinfo {volume} {98}},\ \bibinfo
  {pages} {063842} (\bibinfo {year} {2018}{\natexlab{a}})}\BibitemShut
  {NoStop}%
\bibitem [{\citenamefont {Braunstein}\ and\ \citenamefont {van
  Loocuk}(2005)}]{RevModPhys.77.513}%
  \BibitemOpen
  \bibfield  {author} {\bibinfo {author} {\bibfnamefont {S.~L.}\ \bibnamefont
  {Braunstein}}\ and\ \bibinfo {author} {\bibfnamefont {P.}~\bibnamefont {van
  Loocuk}},\ }\href {\doibase 10.1103/RevModPhys.77.513} {\bibfield  {journal}
  {\bibinfo  {journal} {Rev. Mod. Phys.}\ }\textbf {\bibinfo {volume} {77}},\
  \bibinfo {pages} {513} (\bibinfo {year} {2005})}\BibitemShut {NoStop}%
\bibitem [{\citenamefont {Ling}\ \emph {et~al.}(2008)\citenamefont {Ling},
  \citenamefont {Lamas-Linares},\ and\ \citenamefont
  {Kurtsiefer}}]{ling2008absolute}%
  \BibitemOpen
  \bibfield  {author} {\bibinfo {author} {\bibfnamefont {A.}~\bibnamefont
  {Ling}}, \bibinfo {author} {\bibfnamefont {A.}~\bibnamefont {Lamas-Linares}},
  \ and\ \bibinfo {author} {\bibfnamefont {C.}~\bibnamefont {Kurtsiefer}},\
  }\href@noop {} {\bibfield  {journal} {\bibinfo  {journal} {Phys. Rev. A}\
  }\textbf {\bibinfo {volume} {77}},\ \bibinfo {pages} {043834} (\bibinfo
  {year} {2008})}\BibitemShut {NoStop}%
\bibitem [{\citenamefont {Bennink}(2010)}]{bennink2010optimal}%
  \BibitemOpen
  \bibfield  {author} {\bibinfo {author} {\bibfnamefont {R.~S.}\ \bibnamefont
  {Bennink}},\ }\href@noop {} {\bibfield  {journal} {\bibinfo  {journal} {Phys.
  Rev. A}\ }\textbf {\bibinfo {volume} {81}},\ \bibinfo {pages} {053805}
  (\bibinfo {year} {2010})}\BibitemShut {NoStop}%
\bibitem [{\citenamefont {Grice}\ and\ \citenamefont
  {Walmsley}(1997)}]{grice1997spectral}%
  \BibitemOpen
  \bibfield  {author} {\bibinfo {author} {\bibfnamefont {W.~P.}\ \bibnamefont
  {Grice}}\ and\ \bibinfo {author} {\bibfnamefont {I.~A.}\ \bibnamefont
  {Walmsley}},\ }\href@noop {} {\bibfield  {journal} {\bibinfo  {journal}
  {Phys. Rev. A}\ }\textbf {\bibinfo {volume} {56}},\ \bibinfo {pages} {1627}
  (\bibinfo {year} {1997})}\BibitemShut {NoStop}%
\bibitem [{\citenamefont {Boeuf}\ \emph {et~al.}(2000)\citenamefont {Boeuf},
  \citenamefont {Branning}, \citenamefont {Chaperot}, \citenamefont {Dauler},
  \citenamefont {Guerin}, \citenamefont {Jaeger}, \citenamefont {Muller},\ and\
  \citenamefont {Migdall}}]{boeuf2000calculating}%
  \BibitemOpen
  \bibfield  {author} {\bibinfo {author} {\bibfnamefont {N.}~\bibnamefont
  {Boeuf}}, \bibinfo {author} {\bibfnamefont {D.}~\bibnamefont {Branning}},
  \bibinfo {author} {\bibfnamefont {I.}~\bibnamefont {Chaperot}}, \bibinfo
  {author} {\bibfnamefont {E.}~\bibnamefont {Dauler}}, \bibinfo {author}
  {\bibfnamefont {S.}~\bibnamefont {Guerin}}, \bibinfo {author} {\bibfnamefont
  {G.}~\bibnamefont {Jaeger}}, \bibinfo {author} {\bibfnamefont
  {A.}~\bibnamefont {Muller}}, \ and\ \bibinfo {author} {\bibfnamefont
  {A.}~\bibnamefont {Migdall}},\ }\href@noop {} {\bibfield  {journal} {\bibinfo
   {journal} {OPTICAL ENGINEERING-BELLINGHAM-INTERNATIONAL SOCIETY FOR OPTICAL
  ENGINEERING-}\ }\textbf {\bibinfo {volume} {39}},\ \bibinfo {pages} {1016}
  (\bibinfo {year} {2000})}\BibitemShut {NoStop}%
\bibitem [{\citenamefont {Fejer}\ \emph {et~al.}(1992)\citenamefont {Fejer},
  \citenamefont {Magel}, \citenamefont {Jundt},\ and\ \citenamefont
  {Byer}}]{fejer1992quasi}%
  \BibitemOpen
  \bibfield  {author} {\bibinfo {author} {\bibfnamefont {M.~M.}\ \bibnamefont
  {Fejer}}, \bibinfo {author} {\bibfnamefont {G.~A.}\ \bibnamefont {Magel}},
  \bibinfo {author} {\bibfnamefont {D.~H.}\ \bibnamefont {Jundt}}, \ and\
  \bibinfo {author} {\bibfnamefont {R.~L.}\ \bibnamefont {Byer}},\ }\href@noop
  {} {\bibfield  {journal} {\bibinfo  {journal} {IEEE J. Quant. Electron.}\
  }\textbf {\bibinfo {volume} {28}},\ \bibinfo {pages} {2631} (\bibinfo {year}
  {1992})}\BibitemShut {NoStop}%
\bibitem [{\citenamefont {Shichijyo}\ \emph {et~al.}(2004)\citenamefont
  {Shichijyo}, \citenamefont {Hirohashi}, \citenamefont {Kamio},\ and\
  \citenamefont {Yamada}}]{shichijyo2004total}%
  \BibitemOpen
  \bibfield  {author} {\bibinfo {author} {\bibfnamefont {S.}~\bibnamefont
  {Shichijyo}}, \bibinfo {author} {\bibfnamefont {J.}~\bibnamefont
  {Hirohashi}}, \bibinfo {author} {\bibfnamefont {H.}~\bibnamefont {Kamio}}, \
  and\ \bibinfo {author} {\bibfnamefont {K.}~\bibnamefont {Yamada}},\
  }\href@noop {} {\bibfield  {journal} {\bibinfo  {journal} {Jap. J. Appl.
  Phys.}\ }\textbf {\bibinfo {volume} {43}},\ \bibinfo {pages} {3413} (\bibinfo
  {year} {2004})}\BibitemShut {NoStop}%
\bibitem [{Note1()}]{Note1}%
  \BibitemOpen
  \bibinfo {note} {In the context of heralded single photon generation, the
  heralding efficiency is typically normalized with respect to detection
  efficiency.}\BibitemShut {Stop}%
\bibitem [{Note2()}]{Note2}%
  \BibitemOpen
  \bibinfo {note} {In the context of characterizing the spatial mode
  distribution in the SPDC process, the concept of conditional spatial mode
  overlap, may also be considered equivalent to the heralding
  efficiency.}\BibitemShut {Stop}%
\bibitem [{\citenamefont {Grieve}\ \emph {et~al.}(2016)\citenamefont {Grieve},
  \citenamefont {Chandrasekara}, \citenamefont {Tang}, \citenamefont {Cheng},\
  and\ \citenamefont {Ling}}]{grieve2016correcting}%
  \BibitemOpen
  \bibfield  {author} {\bibinfo {author} {\bibfnamefont {J.~A.}\ \bibnamefont
  {Grieve}}, \bibinfo {author} {\bibfnamefont {R.}~\bibnamefont
  {Chandrasekara}}, \bibinfo {author} {\bibfnamefont {Z.}~\bibnamefont {Tang}},
  \bibinfo {author} {\bibfnamefont {C.}~\bibnamefont {Cheng}}, \ and\ \bibinfo
  {author} {\bibfnamefont {A.}~\bibnamefont {Ling}},\ }\href@noop {} {\bibfield
   {journal} {\bibinfo  {journal} {Opt. Express}\ }\textbf {\bibinfo {volume}
  {24}},\ \bibinfo {pages} {3592} (\bibinfo {year} {2016})}\BibitemShut
  {NoStop}%
\bibitem [{\citenamefont {Villar}\ \emph {et~al.}(2018)\citenamefont {Villar},
  \citenamefont {Lohrmann},\ and\ \citenamefont
  {Ling}}]{villar2018experimental}%
  \BibitemOpen
  \bibfield  {author} {\bibinfo {author} {\bibfnamefont {A.}~\bibnamefont
  {Villar}}, \bibinfo {author} {\bibfnamefont {A.}~\bibnamefont {Lohrmann}}, \
  and\ \bibinfo {author} {\bibfnamefont {A.}~\bibnamefont {Ling}},\ }\href@noop
  {} {\bibfield  {journal} {\bibinfo  {journal} {Opt. Express}\ }\textbf
  {\bibinfo {volume} {26}},\ \bibinfo {pages} {12396} (\bibinfo {year}
  {2018})}\BibitemShut {NoStop}%
\bibitem [{\citenamefont {Steinlechner}\ \emph {et~al.}(2017)\citenamefont
  {Steinlechner}, \citenamefont {Ecker}, \citenamefont {Fink}, \citenamefont
  {Liu}, \citenamefont {Bavaresco}, \citenamefont {Huber}, \citenamefont
  {Scheidl},\ and\ \citenamefont {Ursin}}]{steinlechner2017distribution}%
  \BibitemOpen
  \bibfield  {author} {\bibinfo {author} {\bibfnamefont {F.}~\bibnamefont
  {Steinlechner}}, \bibinfo {author} {\bibfnamefont {S.}~\bibnamefont {Ecker}},
  \bibinfo {author} {\bibfnamefont {M.}~\bibnamefont {Fink}}, \bibinfo {author}
  {\bibfnamefont {B.}~\bibnamefont {Liu}}, \bibinfo {author} {\bibfnamefont
  {J.}~\bibnamefont {Bavaresco}}, \bibinfo {author} {\bibfnamefont
  {M.}~\bibnamefont {Huber}}, \bibinfo {author} {\bibfnamefont
  {T.}~\bibnamefont {Scheidl}}, \ and\ \bibinfo {author} {\bibfnamefont
  {R.}~\bibnamefont {Ursin}},\ }\href@noop {} {\bibfield  {journal} {\bibinfo
  {journal} {Nat. Commun.}\ }\textbf {\bibinfo {volume} {8}},\ \bibinfo {pages}
  {15971} (\bibinfo {year} {2017})}\BibitemShut {NoStop}%
\bibitem [{\citenamefont {Clauser}\ \emph {et~al.}(1969)\citenamefont
  {Clauser}, \citenamefont {Horne}, \citenamefont {Shimony},\ and\
  \citenamefont {Holt}}]{clauser1969proposed}%
  \BibitemOpen
  \bibfield  {author} {\bibinfo {author} {\bibfnamefont {J.~F.}\ \bibnamefont
  {Clauser}}, \bibinfo {author} {\bibfnamefont {M.~A.}\ \bibnamefont {Horne}},
  \bibinfo {author} {\bibfnamefont {A.}~\bibnamefont {Shimony}}, \ and\
  \bibinfo {author} {\bibfnamefont {R.~A.}\ \bibnamefont {Holt}},\ }\href@noop
  {} {\bibfield  {journal} {\bibinfo  {journal} {Phys. Rev. Lett.}\ }\textbf
  {\bibinfo {volume} {23}},\ \bibinfo {pages} {880} (\bibinfo {year}
  {1969})}\BibitemShut {NoStop}%
\bibitem [{\citenamefont {Aspect}\ \emph {et~al.}(1982)\citenamefont {Aspect},
  \citenamefont {Grangier},\ and\ \citenamefont
  {Roger}}]{aspect1982aexperimental}%
  \BibitemOpen
  \bibfield  {author} {\bibinfo {author} {\bibfnamefont {A.}~\bibnamefont
  {Aspect}}, \bibinfo {author} {\bibfnamefont {P.}~\bibnamefont {Grangier}}, \
  and\ \bibinfo {author} {\bibfnamefont {G.}~\bibnamefont {Roger}},\
  }\href@noop {} {\bibfield  {journal} {\bibinfo  {journal} {Phys. Rev. Lett.}\
  }\textbf {\bibinfo {volume} {49}},\ \bibinfo {pages} {91} (\bibinfo {year}
  {1982})}\BibitemShut {NoStop}%
\bibitem [{\citenamefont {Banaszek}\ \emph {et~al.}(1999)\citenamefont
  {Banaszek}, \citenamefont {D’ariano}, \citenamefont {Paris},\ and\
  \citenamefont {Sacchi}}]{banaszek1999maximum}%
  \BibitemOpen
  \bibfield  {author} {\bibinfo {author} {\bibfnamefont {K.}~\bibnamefont
  {Banaszek}}, \bibinfo {author} {\bibfnamefont {G.~M.}\ \bibnamefont
  {D’ariano}}, \bibinfo {author} {\bibfnamefont {M.~G.~A.}\ \bibnamefont
  {Paris}}, \ and\ \bibinfo {author} {\bibfnamefont {M.~F.}\ \bibnamefont
  {Sacchi}},\ }\href@noop {} {\bibfield  {journal} {\bibinfo  {journal} {Phys.
  Rev. A}\ }\textbf {\bibinfo {volume} {61}},\ \bibinfo {pages} {010304}
  (\bibinfo {year} {1999})}\BibitemShut {NoStop}%
\bibitem [{\citenamefont {James}\ \emph {et~al.}(2001)\citenamefont {James},
  \citenamefont {Kwiat}, \citenamefont {Munro},\ and\ \citenamefont
  {White}}]{kwiat2001measure}%
  \BibitemOpen
  \bibfield  {author} {\bibinfo {author} {\bibfnamefont {D.~F.~V.}\
  \bibnamefont {James}}, \bibinfo {author} {\bibfnamefont {P.~G.}\ \bibnamefont
  {Kwiat}}, \bibinfo {author} {\bibfnamefont {W.~J.}\ \bibnamefont {Munro}}, \
  and\ \bibinfo {author} {\bibfnamefont {A.~G.}\ \bibnamefont {White}},\ }\href
  {\doibase 10.1103/PhysRevA.64.052312} {\bibfield  {journal} {\bibinfo
  {journal} {Phys. Rev. A}\ }\textbf {\bibinfo {volume} {64}},\ \bibinfo
  {pages} {052312} (\bibinfo {year} {2001})}\BibitemShut {NoStop}%
\bibitem [{\citenamefont {Altepeter}\ \emph
  {et~al.}(2005{\natexlab{a}})\citenamefont {Altepeter}, \citenamefont
  {Jeffrey},\ and\ \citenamefont {Kwiat}}]{altepeter2005photonic}%
  \BibitemOpen
  \bibfield  {author} {\bibinfo {author} {\bibfnamefont {J.~B.}\ \bibnamefont
  {Altepeter}}, \bibinfo {author} {\bibfnamefont {E.~R.}\ \bibnamefont
  {Jeffrey}}, \ and\ \bibinfo {author} {\bibfnamefont {P.~G.}\ \bibnamefont
  {Kwiat}},\ }\href@noop {} {\bibfield  {journal} {\bibinfo  {journal} {Adv.
  At. Mol. Opt. Phys.}\ }\textbf {\bibinfo {volume} {52}},\ \bibinfo {pages}
  {105} (\bibinfo {year} {2005}{\natexlab{a}})}\BibitemShut {NoStop}%
\bibitem [{\citenamefont {Jennewein}\ \emph {et~al.}(2009)\citenamefont
  {Jennewein}, \citenamefont {Ursin}, \citenamefont {Aspelmeyer},\ and\
  \citenamefont {Zeilinger}}]{jennewein2009performing}%
  \BibitemOpen
  \bibfield  {author} {\bibinfo {author} {\bibfnamefont {T.}~\bibnamefont
  {Jennewein}}, \bibinfo {author} {\bibfnamefont {R.}~\bibnamefont {Ursin}},
  \bibinfo {author} {\bibfnamefont {M.}~\bibnamefont {Aspelmeyer}}, \ and\
  \bibinfo {author} {\bibfnamefont {A.}~\bibnamefont {Zeilinger}},\ }\href@noop
  {} {\bibfield  {journal} {\bibinfo  {journal} {J Phys. B: At. Mol. Opt.
  Phys.}\ }\textbf {\bibinfo {volume} {42}},\ \bibinfo {pages} {114008}
  (\bibinfo {year} {2009})}\BibitemShut {NoStop}%
\bibitem [{\citenamefont {G{\"u}hne}\ and\ \citenamefont
  {T{\'o}th}(2009)}]{guhne2009entanglement}%
  \BibitemOpen
  \bibfield  {author} {\bibinfo {author} {\bibfnamefont {O.}~\bibnamefont
  {G{\"u}hne}}\ and\ \bibinfo {author} {\bibfnamefont {G.}~\bibnamefont
  {T{\'o}th}},\ }\href@noop {} {\bibfield  {journal} {\bibinfo  {journal}
  {Phys. Rep.}\ }\textbf {\bibinfo {volume} {474}},\ \bibinfo {pages} {1}
  (\bibinfo {year} {2009})}\BibitemShut {NoStop}%
\bibitem [{\citenamefont {Coffman}\ \emph {et~al.}(2000)\citenamefont
  {Coffman}, \citenamefont {Kundu},\ and\ \citenamefont
  {Wootters}}]{coffman2000distributed}%
  \BibitemOpen
  \bibfield  {author} {\bibinfo {author} {\bibfnamefont {V.}~\bibnamefont
  {Coffman}}, \bibinfo {author} {\bibfnamefont {J.}~\bibnamefont {Kundu}}, \
  and\ \bibinfo {author} {\bibfnamefont {W.~K.}\ \bibnamefont {Wootters}},\
  }\href@noop {} {\bibfield  {journal} {\bibinfo  {journal} {Phys. Rev. A}\
  }\textbf {\bibinfo {volume} {61}},\ \bibinfo {pages} {052306} (\bibinfo
  {year} {2000})}\BibitemShut {NoStop}%
\bibitem [{\citenamefont {Wootters}(1998)}]{wootters1998entanglement}%
  \BibitemOpen
  \bibfield  {author} {\bibinfo {author} {\bibfnamefont {W.~K.}\ \bibnamefont
  {Wootters}},\ }\href@noop {} {\bibfield  {journal} {\bibinfo  {journal}
  {Phys. Rev. Lett.}\ }\textbf {\bibinfo {volume} {80}},\ \bibinfo {pages}
  {2245} (\bibinfo {year} {1998})}\BibitemShut {NoStop}%
\bibitem [{\citenamefont {Horodecki}\ \emph {et~al.}(2009)\citenamefont
  {Horodecki}, \citenamefont {Horodecki}, \citenamefont {Horodecki},\ and\
  \citenamefont {Horodecki}}]{horodecki2009quantum}%
  \BibitemOpen
  \bibfield  {author} {\bibinfo {author} {\bibfnamefont {R.}~\bibnamefont
  {Horodecki}}, \bibinfo {author} {\bibfnamefont {P.}~\bibnamefont
  {Horodecki}}, \bibinfo {author} {\bibfnamefont {M.}~\bibnamefont
  {Horodecki}}, \ and\ \bibinfo {author} {\bibfnamefont {K.}~\bibnamefont
  {Horodecki}},\ }\href@noop {} {\bibfield  {journal} {\bibinfo  {journal}
  {Rev. Mod. Phys.}\ }\textbf {\bibinfo {volume} {81}},\ \bibinfo {pages} {865}
  (\bibinfo {year} {2009})}\BibitemShut {NoStop}%
\bibitem [{\citenamefont {Bavaresco}\ \emph {et~al.}(2018)\citenamefont
  {Bavaresco}, \citenamefont {Valencia}, \citenamefont {Kl{\"o}ckl},
  \citenamefont {Pivoluska}, \citenamefont {Erker}, \citenamefont {Friis},
  \citenamefont {Malik},\ and\ \citenamefont
  {Huber}}]{bavaresco2018measurements}%
  \BibitemOpen
  \bibfield  {author} {\bibinfo {author} {\bibfnamefont {J.}~\bibnamefont
  {Bavaresco}}, \bibinfo {author} {\bibfnamefont {N.~H.}\ \bibnamefont
  {Valencia}}, \bibinfo {author} {\bibfnamefont {C.}~\bibnamefont
  {Kl{\"o}ckl}}, \bibinfo {author} {\bibfnamefont {M.}~\bibnamefont
  {Pivoluska}}, \bibinfo {author} {\bibfnamefont {P.}~\bibnamefont {Erker}},
  \bibinfo {author} {\bibfnamefont {N.}~\bibnamefont {Friis}}, \bibinfo
  {author} {\bibfnamefont {M.}~\bibnamefont {Malik}}, \ and\ \bibinfo {author}
  {\bibfnamefont {M.}~\bibnamefont {Huber}},\ }\href@noop {} {\bibfield
  {journal} {\bibinfo  {journal} {Nat. Phys.}\ }\textbf {\bibinfo {volume}
  {14}},\ \bibinfo {pages} {1032} (\bibinfo {year} {2018})}\BibitemShut
  {NoStop}%
\bibitem [{\citenamefont {Scully}\ and\ \citenamefont
  {Dr{\"u}hl}(1982)}]{scully1982quantum}%
  \BibitemOpen
  \bibfield  {author} {\bibinfo {author} {\bibfnamefont {M.~O.}\ \bibnamefont
  {Scully}}\ and\ \bibinfo {author} {\bibfnamefont {K.}~\bibnamefont
  {Dr{\"u}hl}},\ }\href@noop {} {\bibfield  {journal} {\bibinfo  {journal}
  {Phys. Rev. A}\ }\textbf {\bibinfo {volume} {25}},\ \bibinfo {pages} {2208}
  (\bibinfo {year} {1982})}\BibitemShut {NoStop}%
\bibitem [{\citenamefont {Scully}\ \emph {et~al.}(1991)\citenamefont {Scully},
  \citenamefont {Englert},\ and\ \citenamefont {Walther}}]{scully1991quantum}%
  \BibitemOpen
  \bibfield  {author} {\bibinfo {author} {\bibfnamefont {M.~O.}\ \bibnamefont
  {Scully}}, \bibinfo {author} {\bibfnamefont {B.-G.}\ \bibnamefont {Englert}},
  \ and\ \bibinfo {author} {\bibfnamefont {H.}~\bibnamefont {Walther}},\
  }\href@noop {} {\bibfield  {journal} {\bibinfo  {journal} {Nature}\ }\textbf
  {\bibinfo {volume} {351}},\ \bibinfo {pages} {111} (\bibinfo {year}
  {1991})}\BibitemShut {NoStop}%
\bibitem [{\citenamefont {Aharonov}\ and\ \citenamefont
  {Zubairy}(2005)}]{aharonov2005time}%
  \BibitemOpen
  \bibfield  {author} {\bibinfo {author} {\bibfnamefont {Y.}~\bibnamefont
  {Aharonov}}\ and\ \bibinfo {author} {\bibfnamefont {M.~S.}\ \bibnamefont
  {Zubairy}},\ }\href@noop {} {\bibfield  {journal} {\bibinfo  {journal}
  {Science}\ }\textbf {\bibinfo {volume} {307}},\ \bibinfo {pages} {875}
  (\bibinfo {year} {2005})}\BibitemShut {NoStop}%
\bibitem [{\citenamefont {Ou}\ and\ \citenamefont
  {Mandel}(1988{\natexlab{a}})}]{Ou88a}%
  \BibitemOpen
  \bibfield  {author} {\bibinfo {author} {\bibfnamefont {Z.}~\bibnamefont
  {Ou}}\ and\ \bibinfo {author} {\bibfnamefont {L.}~\bibnamefont {Mandel}},\
  }\href@noop {} {\bibfield  {journal} {\bibinfo  {journal} {Phys. Rev. Lett.}\
  }\textbf {\bibinfo {volume} {61}},\ \bibinfo {pages} {50} (\bibinfo {year}
  {1988}{\natexlab{a}})}\BibitemShut {NoStop}%
\bibitem [{\citenamefont {Rarity}\ and\ \citenamefont
  {Tapster}(1990)}]{Rarity90b}%
  \BibitemOpen
  \bibfield  {author} {\bibinfo {author} {\bibfnamefont {J.}~\bibnamefont
  {Rarity}}\ and\ \bibinfo {author} {\bibfnamefont {P.}~\bibnamefont
  {Tapster}},\ }\href@noop {} {\bibfield  {journal} {\bibinfo  {journal} {Phys.
  Rev. Lett.}\ }\textbf {\bibinfo {volume} {64}},\ \bibinfo {pages} {2495}
  (\bibinfo {year} {1990})}\BibitemShut {NoStop}%
\bibitem [{\citenamefont {Chekhova}\ and\ \citenamefont
  {Ou}(2016)}]{chekhova2016nonlinear}%
  \BibitemOpen
  \bibfield  {author} {\bibinfo {author} {\bibfnamefont {M.~V.}\ \bibnamefont
  {Chekhova}}\ and\ \bibinfo {author} {\bibfnamefont {Z.~Y.}\ \bibnamefont
  {Ou}},\ }\href@noop {} {\bibfield  {journal} {\bibinfo  {journal} {Adv. Opt.
  Photon.}\ }\textbf {\bibinfo {volume} {8}},\ \bibinfo {pages} {104} (\bibinfo
  {year} {2016})}\BibitemShut {NoStop}%
\bibitem [{\citenamefont {Kim}\ \emph {et~al.}(2006)\citenamefont {Kim},
  \citenamefont {Fiorentino},\ and\ \citenamefont {Wong}}]{kim2006phase}%
  \BibitemOpen
  \bibfield  {author} {\bibinfo {author} {\bibfnamefont {T.}~\bibnamefont
  {Kim}}, \bibinfo {author} {\bibfnamefont {M.}~\bibnamefont {Fiorentino}}, \
  and\ \bibinfo {author} {\bibfnamefont {F.~N.~C.}\ \bibnamefont {Wong}},\
  }\href@noop {} {\bibfield  {journal} {\bibinfo  {journal} {Phys. Rev. A}\
  }\textbf {\bibinfo {volume} {73}},\ \bibinfo {pages} {012316} (\bibinfo
  {year} {2006})}\BibitemShut {NoStop}%
\bibitem [{\citenamefont {Hentschel}\ \emph {et~al.}(2009)\citenamefont
  {Hentschel}, \citenamefont {H{\"u}bel}, \citenamefont {Poppe},\ and\
  \citenamefont {Zeilinger}}]{hentschel2009three}%
  \BibitemOpen
  \bibfield  {author} {\bibinfo {author} {\bibfnamefont {M.}~\bibnamefont
  {Hentschel}}, \bibinfo {author} {\bibfnamefont {H.}~\bibnamefont
  {H{\"u}bel}}, \bibinfo {author} {\bibfnamefont {A.}~\bibnamefont {Poppe}}, \
  and\ \bibinfo {author} {\bibfnamefont {A.}~\bibnamefont {Zeilinger}},\
  }\href@noop {} {\bibfield  {journal} {\bibinfo  {journal} {Opt. Express}\
  }\textbf {\bibinfo {volume} {17}},\ \bibinfo {pages} {23153} (\bibinfo {year}
  {2009})}\BibitemShut {NoStop}%
\bibitem [{\citenamefont {Perumangatt}\ \emph {et~al.}(2020)\citenamefont
  {Perumangatt}, \citenamefont {Lohrmann},\ and\ \citenamefont
  {Ling}}]{perumangatt2019experimental}%
  \BibitemOpen
  \bibfield  {author} {\bibinfo {author} {\bibfnamefont {C.}~\bibnamefont
  {Perumangatt}}, \bibinfo {author} {\bibfnamefont {A.}~\bibnamefont
  {Lohrmann}}, \ and\ \bibinfo {author} {\bibfnamefont {A.}~\bibnamefont
  {Ling}},\ }\href@noop {} {\bibfield  {journal} {\bibinfo  {journal} {Phys.
  Rev. A}\ }\textbf {\bibinfo {volume} {102}},\ \bibinfo {pages} {012404}
  (\bibinfo {year} {2020})}\BibitemShut {NoStop}%
\bibitem [{\citenamefont {Lohrmann}\ \emph {et~al.}(2020)\citenamefont
  {Lohrmann}, \citenamefont {Perumangatt}, \citenamefont {Villar},\ and\
  \citenamefont {Ling}}]{lohrmann2019PAPA}%
  \BibitemOpen
  \bibfield  {author} {\bibinfo {author} {\bibfnamefont {A.}~\bibnamefont
  {Lohrmann}}, \bibinfo {author} {\bibfnamefont {C.}~\bibnamefont
  {Perumangatt}}, \bibinfo {author} {\bibfnamefont {A.}~\bibnamefont {Villar}},
  \ and\ \bibinfo {author} {\bibfnamefont {A.}~\bibnamefont {Ling}},\
  }\href@noop {} {\bibfield  {journal} {\bibinfo  {journal} {Appl. Phys.
  Lett.}\ }\textbf {\bibinfo {volume} {116}},\ \bibinfo {pages} {021101}
  (\bibinfo {year} {2020})}\BibitemShut {NoStop}%
\bibitem [{\citenamefont {Trojek}\ and\ \citenamefont
  {Weinfurter}(2008)}]{trojek2008collinear}%
  \BibitemOpen
  \bibfield  {author} {\bibinfo {author} {\bibfnamefont {P.}~\bibnamefont
  {Trojek}}\ and\ \bibinfo {author} {\bibfnamefont {H.}~\bibnamefont
  {Weinfurter}},\ }\href@noop {} {\bibfield  {journal} {\bibinfo  {journal}
  {Appl. Phys. Lett.}\ }\textbf {\bibinfo {volume} {92}},\ \bibinfo {pages}
  {211103} (\bibinfo {year} {2008})}\BibitemShut {NoStop}%
\bibitem [{\citenamefont {Steinlechner}\ \emph {et~al.}(2013)\citenamefont
  {Steinlechner}, \citenamefont {Ramelow}, \citenamefont {Jofre}, \citenamefont
  {Gilaberte}, \citenamefont {Jennewein}, \citenamefont {Torres}, \citenamefont
  {Mitchell},\ and\ \citenamefont {Pruneri}}]{steinlechner2013phase}%
  \BibitemOpen
  \bibfield  {author} {\bibinfo {author} {\bibfnamefont {F.}~\bibnamefont
  {Steinlechner}}, \bibinfo {author} {\bibfnamefont {S.}~\bibnamefont
  {Ramelow}}, \bibinfo {author} {\bibfnamefont {M.}~\bibnamefont {Jofre}},
  \bibinfo {author} {\bibfnamefont {M.}~\bibnamefont {Gilaberte}}, \bibinfo
  {author} {\bibfnamefont {T.}~\bibnamefont {Jennewein}}, \bibinfo {author}
  {\bibfnamefont {J.~P.}\ \bibnamefont {Torres}}, \bibinfo {author}
  {\bibfnamefont {M.~W.}\ \bibnamefont {Mitchell}}, \ and\ \bibinfo {author}
  {\bibfnamefont {V.}~\bibnamefont {Pruneri}},\ }\href@noop {} {\bibfield
  {journal} {\bibinfo  {journal} {Opt. Express}\ }\textbf {\bibinfo {volume}
  {21}},\ \bibinfo {pages} {11943} (\bibinfo {year} {2013})}\BibitemShut
  {NoStop}%
\bibitem [{\citenamefont {Trojek}\ \emph {et~al.}(2004)\citenamefont {Trojek},
  \citenamefont {Schmid}, \citenamefont {Bourennane}, \citenamefont
  {Weinfurter},\ and\ \citenamefont {Kurtsiefer}}]{trojek2004compact}%
  \BibitemOpen
  \bibfield  {author} {\bibinfo {author} {\bibfnamefont {P.}~\bibnamefont
  {Trojek}}, \bibinfo {author} {\bibfnamefont {C.}~\bibnamefont {Schmid}},
  \bibinfo {author} {\bibfnamefont {M.}~\bibnamefont {Bourennane}}, \bibinfo
  {author} {\bibfnamefont {H.}~\bibnamefont {Weinfurter}}, \ and\ \bibinfo
  {author} {\bibfnamefont {C.}~\bibnamefont {Kurtsiefer}},\ }\href@noop {}
  {\bibfield  {journal} {\bibinfo  {journal} {Opt. Express}\ }\textbf {\bibinfo
  {volume} {12}},\ \bibinfo {pages} {276} (\bibinfo {year} {2004})}\BibitemShut
  {NoStop}%
\bibitem [{\citenamefont {Fiorentino}\ \emph {et~al.}(2005)\citenamefont
  {Fiorentino}, \citenamefont {Kuklewicz},\ and\ \citenamefont
  {Wong}}]{fiorentino2005source}%
  \BibitemOpen
  \bibfield  {author} {\bibinfo {author} {\bibfnamefont {M.}~\bibnamefont
  {Fiorentino}}, \bibinfo {author} {\bibfnamefont {C.~E.}\ \bibnamefont
  {Kuklewicz}}, \ and\ \bibinfo {author} {\bibfnamefont {F.~N.~C.}\
  \bibnamefont {Wong}},\ }\href@noop {} {\bibfield  {journal} {\bibinfo
  {journal} {Opt. Express}\ }\textbf {\bibinfo {volume} {13}},\ \bibinfo
  {pages} {127} (\bibinfo {year} {2005})}\BibitemShut {NoStop}%
\bibitem [{\citenamefont {Lee}\ \emph {et~al.}(2016)\citenamefont {Lee},
  \citenamefont {Kim}, \citenamefont {Cha},\ and\ \citenamefont
  {Moon}}]{lee2016polarization}%
  \BibitemOpen
  \bibfield  {author} {\bibinfo {author} {\bibfnamefont {S.~M.}\ \bibnamefont
  {Lee}}, \bibinfo {author} {\bibfnamefont {H.}~\bibnamefont {Kim}}, \bibinfo
  {author} {\bibfnamefont {M.}~\bibnamefont {Cha}}, \ and\ \bibinfo {author}
  {\bibfnamefont {H.~S.}\ \bibnamefont {Moon}},\ }\href@noop {} {\bibfield
  {journal} {\bibinfo  {journal} {Opt. Express}\ }\textbf {\bibinfo {volume}
  {24}},\ \bibinfo {pages} {2941} (\bibinfo {year} {2016})}\BibitemShut
  {NoStop}%
\bibitem [{\citenamefont {de~Chatellus}\ \emph {et~al.}(2006)\citenamefont
  {de~Chatellus}, \citenamefont {Sergienko}, \citenamefont {Saleh},
  \citenamefont {Teich},\ and\ \citenamefont {Di~Giuseppe}}]{de2006non}%
  \BibitemOpen
  \bibfield  {author} {\bibinfo {author} {\bibfnamefont {H.~G.}\ \bibnamefont
  {de~Chatellus}}, \bibinfo {author} {\bibfnamefont {A.~V.}\ \bibnamefont
  {Sergienko}}, \bibinfo {author} {\bibfnamefont {B.~E.~A.}\ \bibnamefont
  {Saleh}}, \bibinfo {author} {\bibfnamefont {M.~C.}\ \bibnamefont {Teich}}, \
  and\ \bibinfo {author} {\bibfnamefont {G.}~\bibnamefont {Di~Giuseppe}},\
  }\href@noop {} {\bibfield  {journal} {\bibinfo  {journal} {Opt. Express}\
  }\textbf {\bibinfo {volume} {14}},\ \bibinfo {pages} {10060} (\bibinfo {year}
  {2006})}\BibitemShut {NoStop}%
\bibitem [{\citenamefont {Kim}\ \emph {et~al.}(2003)\citenamefont {Kim},
  \citenamefont {Kulik}, \citenamefont {Chekhova}, \citenamefont {Grice},\ and\
  \citenamefont {Shih}}]{kim2003experimental}%
  \BibitemOpen
  \bibfield  {author} {\bibinfo {author} {\bibfnamefont {Y.-H.}\ \bibnamefont
  {Kim}}, \bibinfo {author} {\bibfnamefont {S.~P.}\ \bibnamefont {Kulik}},
  \bibinfo {author} {\bibfnamefont {M.~V.}\ \bibnamefont {Chekhova}}, \bibinfo
  {author} {\bibfnamefont {W.~P.}\ \bibnamefont {Grice}}, \ and\ \bibinfo
  {author} {\bibfnamefont {Y.}~\bibnamefont {Shih}},\ }\href@noop {} {\bibfield
   {journal} {\bibinfo  {journal} {Phys. Rev. A}\ }\textbf {\bibinfo {volume}
  {67}},\ \bibinfo {pages} {010301} (\bibinfo {year} {2003})}\BibitemShut
  {NoStop}%
\bibitem [{\citenamefont {Lee}\ \emph {et~al.}(2005)\citenamefont {Lee},
  \citenamefont {Van~Exter},\ and\ \citenamefont {Woerdman}}]{lee2005focused}%
  \BibitemOpen
  \bibfield  {author} {\bibinfo {author} {\bibfnamefont {P.~S.~K.}\
  \bibnamefont {Lee}}, \bibinfo {author} {\bibfnamefont {M.~P.}\ \bibnamefont
  {Van~Exter}}, \ and\ \bibinfo {author} {\bibfnamefont {J.~P.}\ \bibnamefont
  {Woerdman}},\ }\href@noop {} {\bibfield  {journal} {\bibinfo  {journal}
  {Phys. Rev. A}\ }\textbf {\bibinfo {volume} {72}},\ \bibinfo {pages} {033803}
  (\bibinfo {year} {2005})}\BibitemShut {NoStop}%
\bibitem [{\citenamefont {Halevy}\ \emph {et~al.}(2011)\citenamefont {Halevy},
  \citenamefont {Megidish}, \citenamefont {Dovrat}, \citenamefont {Eisenberg},
  \citenamefont {Becker},\ and\ \citenamefont
  {Bohat{\`y}}}]{halevy2011biaxial}%
  \BibitemOpen
  \bibfield  {author} {\bibinfo {author} {\bibfnamefont {A.}~\bibnamefont
  {Halevy}}, \bibinfo {author} {\bibfnamefont {E.}~\bibnamefont {Megidish}},
  \bibinfo {author} {\bibfnamefont {L.}~\bibnamefont {Dovrat}}, \bibinfo
  {author} {\bibfnamefont {H.~S.}\ \bibnamefont {Eisenberg}}, \bibinfo {author}
  {\bibfnamefont {P.}~\bibnamefont {Becker}}, \ and\ \bibinfo {author}
  {\bibfnamefont {L.}~\bibnamefont {Bohat{\`y}}},\ }\href@noop {} {\bibfield
  {journal} {\bibinfo  {journal} {Opt. Express}\ }\textbf {\bibinfo {volume}
  {19}},\ \bibinfo {pages} {20420} (\bibinfo {year} {2011})}\BibitemShut
  {NoStop}%
\bibitem [{\citenamefont {White}\ \emph {et~al.}(1999)\citenamefont {White},
  \citenamefont {James}, \citenamefont {Eberhard},\ and\ \citenamefont
  {Kwiat}}]{white1999nonmaximally}%
  \BibitemOpen
  \bibfield  {author} {\bibinfo {author} {\bibfnamefont {A.~G.}\ \bibnamefont
  {White}}, \bibinfo {author} {\bibfnamefont {D.~F.}\ \bibnamefont {James}},
  \bibinfo {author} {\bibfnamefont {P.~H.}\ \bibnamefont {Eberhard}}, \ and\
  \bibinfo {author} {\bibfnamefont {P.~G.}\ \bibnamefont {Kwiat}},\ }\href@noop
  {} {\bibfield  {journal} {\bibinfo  {journal} {Phys. Rev. Lett.}\ }\textbf
  {\bibinfo {volume} {83}},\ \bibinfo {pages} {3103} (\bibinfo {year}
  {1999})}\BibitemShut {NoStop}%
\bibitem [{\citenamefont {Migdall}(1997)}]{migdall1997polarization}%
  \BibitemOpen
  \bibfield  {author} {\bibinfo {author} {\bibfnamefont {A.}~\bibnamefont
  {Migdall}},\ }\href@noop {} {\bibfield  {journal} {\bibinfo  {journal} {J.
  Opt. Soc. Am. B}\ }\textbf {\bibinfo {volume} {14}},\ \bibinfo {pages} {1093}
  (\bibinfo {year} {1997})}\BibitemShut {NoStop}%
\bibitem [{\citenamefont {Rangarajan}\ \emph
  {et~al.}(2011{\natexlab{a}})\citenamefont {Rangarajan}, \citenamefont
  {U’Ren},\ and\ \citenamefont {Kwiat}}]{rangarajan2011migdall}%
  \BibitemOpen
  \bibfield  {author} {\bibinfo {author} {\bibfnamefont {R.}~\bibnamefont
  {Rangarajan}}, \bibinfo {author} {\bibfnamefont {A.~B.}\ \bibnamefont
  {U’Ren}}, \ and\ \bibinfo {author} {\bibfnamefont {P.~G.}\ \bibnamefont
  {Kwiat}},\ }\href@noop {} {\bibfield  {journal} {\bibinfo  {journal} {J. Mod.
  Opt.}\ }\textbf {\bibinfo {volume} {58}},\ \bibinfo {pages} {312} (\bibinfo
  {year} {2011}{\natexlab{a}})}\BibitemShut {NoStop}%
\bibitem [{\citenamefont {Nambu}\ \emph {et~al.}(2002)\citenamefont {Nambu},
  \citenamefont {Usami}, \citenamefont {Tsuda}, \citenamefont {Matsumoto},\
  and\ \citenamefont {Nakamura}}]{nambu2002generation}%
  \BibitemOpen
  \bibfield  {author} {\bibinfo {author} {\bibfnamefont {Y.}~\bibnamefont
  {Nambu}}, \bibinfo {author} {\bibfnamefont {K.}~\bibnamefont {Usami}},
  \bibinfo {author} {\bibfnamefont {Y.}~\bibnamefont {Tsuda}}, \bibinfo
  {author} {\bibfnamefont {K.}~\bibnamefont {Matsumoto}}, \ and\ \bibinfo
  {author} {\bibfnamefont {K.}~\bibnamefont {Nakamura}},\ }\href@noop {}
  {\bibfield  {journal} {\bibinfo  {journal} {Phys. Rev. A}\ }\textbf {\bibinfo
  {volume} {66}},\ \bibinfo {pages} {033816} (\bibinfo {year}
  {2002})}\BibitemShut {NoStop}%
\bibitem [{\citenamefont {Rangarajan}\ \emph {et~al.}(2009)\citenamefont
  {Rangarajan}, \citenamefont {Goggin},\ and\ \citenamefont
  {Kwiat}}]{rangarajan2009optimizing}%
  \BibitemOpen
  \bibfield  {author} {\bibinfo {author} {\bibfnamefont {R.}~\bibnamefont
  {Rangarajan}}, \bibinfo {author} {\bibfnamefont {M.}~\bibnamefont {Goggin}},
  \ and\ \bibinfo {author} {\bibfnamefont {P.}~\bibnamefont {Kwiat}},\
  }\href@noop {} {\bibfield  {journal} {\bibinfo  {journal} {Opt. Express}\
  }\textbf {\bibinfo {volume} {17}},\ \bibinfo {pages} {18920} (\bibinfo {year}
  {2009})}\BibitemShut {NoStop}%
\bibitem [{\citenamefont {Rangarajan}\ \emph
  {et~al.}(2011{\natexlab{b}})\citenamefont {Rangarajan}, \citenamefont
  {Vicent}, \citenamefont {U’Ren},\ and\ \citenamefont
  {Kwiat}}]{rangarajan2011engineering}%
  \BibitemOpen
  \bibfield  {author} {\bibinfo {author} {\bibfnamefont {R.}~\bibnamefont
  {Rangarajan}}, \bibinfo {author} {\bibfnamefont {L.~E.}\ \bibnamefont
  {Vicent}}, \bibinfo {author} {\bibfnamefont {A.~B.}\ \bibnamefont {U’Ren}},
  \ and\ \bibinfo {author} {\bibfnamefont {P.~G.}\ \bibnamefont {Kwiat}},\
  }\href@noop {} {\bibfield  {journal} {\bibinfo  {journal} {J. Mod. Opt.}\
  }\textbf {\bibinfo {volume} {58}},\ \bibinfo {pages} {318} (\bibinfo {year}
  {2011}{\natexlab{b}})}\BibitemShut {NoStop}%
\bibitem [{\citenamefont {Bitton}\ \emph {et~al.}(2002)\citenamefont {Bitton},
  \citenamefont {Grice}, \citenamefont {Moreau},\ and\ \citenamefont
  {Zhang}}]{bitton2002cascaded}%
  \BibitemOpen
  \bibfield  {author} {\bibinfo {author} {\bibfnamefont {G.}~\bibnamefont
  {Bitton}}, \bibinfo {author} {\bibfnamefont {W.}~\bibnamefont {Grice}},
  \bibinfo {author} {\bibfnamefont {J.}~\bibnamefont {Moreau}}, \ and\ \bibinfo
  {author} {\bibfnamefont {L.}~\bibnamefont {Zhang}},\ }\href@noop {}
  {\bibfield  {journal} {\bibinfo  {journal} {Phys. Rev. A}\ }\textbf {\bibinfo
  {volume} {65}},\ \bibinfo {pages} {063805} (\bibinfo {year}
  {2002})}\BibitemShut {NoStop}%
\bibitem [{\citenamefont {Niu}\ \emph {et~al.}(2008)\citenamefont {Niu},
  \citenamefont {Huang}, \citenamefont {Xiang}, \citenamefont {Guo},\ and\
  \citenamefont {Ou}}]{niu2008beamlike}%
  \BibitemOpen
  \bibfield  {author} {\bibinfo {author} {\bibfnamefont {X.-L.}\ \bibnamefont
  {Niu}}, \bibinfo {author} {\bibfnamefont {Y.-F.}\ \bibnamefont {Huang}},
  \bibinfo {author} {\bibfnamefont {G.-Y.}\ \bibnamefont {Xiang}}, \bibinfo
  {author} {\bibfnamefont {G.-C.}\ \bibnamefont {Guo}}, \ and\ \bibinfo
  {author} {\bibfnamefont {Z.}~\bibnamefont {Ou}},\ }\href@noop {} {\bibfield
  {journal} {\bibinfo  {journal} {Opt. Lett.}\ }\textbf {\bibinfo {volume}
  {33}},\ \bibinfo {pages} {968} (\bibinfo {year} {2008})}\BibitemShut
  {NoStop}%
\bibitem [{\citenamefont {Zhang}\ \emph
  {et~al.}(2015{\natexlab{a}})\citenamefont {Zhang}, \citenamefont {Huang},
  \citenamefont {Wang}, \citenamefont {Liu}, \citenamefont {Li},\ and\
  \citenamefont {Guo}}]{zhang2015experimental}%
  \BibitemOpen
  \bibfield  {author} {\bibinfo {author} {\bibfnamefont {C.}~\bibnamefont
  {Zhang}}, \bibinfo {author} {\bibfnamefont {Y.-F.}\ \bibnamefont {Huang}},
  \bibinfo {author} {\bibfnamefont {Z.}~\bibnamefont {Wang}}, \bibinfo {author}
  {\bibfnamefont {B.-H.}\ \bibnamefont {Liu}}, \bibinfo {author} {\bibfnamefont
  {C.-F.}\ \bibnamefont {Li}}, \ and\ \bibinfo {author} {\bibfnamefont {G.-C.}\
  \bibnamefont {Guo}},\ }\href@noop {} {\bibfield  {journal} {\bibinfo
  {journal} {Phys. Rev. Lett.}\ }\textbf {\bibinfo {volume} {115}},\ \bibinfo
  {pages} {260402} (\bibinfo {year} {2015}{\natexlab{a}})}\BibitemShut
  {NoStop}%
\bibitem [{\citenamefont {Herzog}\ \emph {et~al.}(1994)\citenamefont {Herzog},
  \citenamefont {Rarity}, \citenamefont {Weinfurter},\ and\ \citenamefont
  {Zeilinger}}]{herzog1994frustrated}%
  \BibitemOpen
  \bibfield  {author} {\bibinfo {author} {\bibfnamefont {T.}~\bibnamefont
  {Herzog}}, \bibinfo {author} {\bibfnamefont {J.}~\bibnamefont {Rarity}},
  \bibinfo {author} {\bibfnamefont {H.}~\bibnamefont {Weinfurter}}, \ and\
  \bibinfo {author} {\bibfnamefont {A.}~\bibnamefont {Zeilinger}},\ }\href@noop
  {} {\bibfield  {journal} {\bibinfo  {journal} {Phys. Rev. Lett.}\ }\textbf
  {\bibinfo {volume} {72}},\ \bibinfo {pages} {629} (\bibinfo {year}
  {1994})}\BibitemShut {NoStop}%
\bibitem [{\citenamefont {Herzog}\ \emph {et~al.}(1995)\citenamefont {Herzog},
  \citenamefont {Kwiat}, \citenamefont {Weinfurter},\ and\ \citenamefont
  {Zeilinger}}]{herzog1995complementarity}%
  \BibitemOpen
  \bibfield  {author} {\bibinfo {author} {\bibfnamefont {T.~J.}\ \bibnamefont
  {Herzog}}, \bibinfo {author} {\bibfnamefont {P.~G.}\ \bibnamefont {Kwiat}},
  \bibinfo {author} {\bibfnamefont {H.}~\bibnamefont {Weinfurter}}, \ and\
  \bibinfo {author} {\bibfnamefont {A.}~\bibnamefont {Zeilinger}},\ }\href@noop
  {} {\bibfield  {journal} {\bibinfo  {journal} {Phys. Rev. Lett.}\ }\textbf
  {\bibinfo {volume} {75}},\ \bibinfo {pages} {3034} (\bibinfo {year}
  {1995})}\BibitemShut {NoStop}%
\bibitem [{\citenamefont {Vallone}\ \emph {et~al.}(2009)\citenamefont
  {Vallone}, \citenamefont {Donati}, \citenamefont {De~Martini},\ and\
  \citenamefont {Mataloni}}]{vallone2009polarization}%
  \BibitemOpen
  \bibfield  {author} {\bibinfo {author} {\bibfnamefont {G.}~\bibnamefont
  {Vallone}}, \bibinfo {author} {\bibfnamefont {G.}~\bibnamefont {Donati}},
  \bibinfo {author} {\bibfnamefont {F.}~\bibnamefont {De~Martini}}, \ and\
  \bibinfo {author} {\bibfnamefont {P.}~\bibnamefont {Mataloni}},\ }\href@noop
  {} {\bibfield  {journal} {\bibinfo  {journal} {Appl. Phys. Lett.}\ }\textbf
  {\bibinfo {volume} {95}},\ \bibinfo {pages} {181110} (\bibinfo {year}
  {2009})}\BibitemShut {NoStop}%
\bibitem [{\citenamefont {Hodelin}\ \emph {et~al.}(2006)\citenamefont
  {Hodelin}, \citenamefont {Khoury},\ and\ \citenamefont
  {Bouwmeester}}]{hodelin2006optimal}%
  \BibitemOpen
  \bibfield  {author} {\bibinfo {author} {\bibfnamefont {J.~F.}\ \bibnamefont
  {Hodelin}}, \bibinfo {author} {\bibfnamefont {G.}~\bibnamefont {Khoury}}, \
  and\ \bibinfo {author} {\bibfnamefont {D.}~\bibnamefont {Bouwmeester}},\
  }\href@noop {} {\bibfield  {journal} {\bibinfo  {journal} {Phys. Rev. A}\
  }\textbf {\bibinfo {volume} {74}},\ \bibinfo {pages} {013802} (\bibinfo
  {year} {2006})}\BibitemShut {NoStop}%
\bibitem [{\citenamefont {Jabir}\ and\ \citenamefont
  {Samanta}(2017)}]{jabir2017robust}%
  \BibitemOpen
  \bibfield  {author} {\bibinfo {author} {\bibfnamefont {M.}~\bibnamefont
  {Jabir}}\ and\ \bibinfo {author} {\bibfnamefont {G.}~\bibnamefont
  {Samanta}},\ }\href@noop {} {\bibfield  {journal} {\bibinfo  {journal} {Sci.
  Rep.}\ }\textbf {\bibinfo {volume} {7}},\ \bibinfo {pages} {12613} (\bibinfo
  {year} {2017})}\BibitemShut {NoStop}%
\bibitem [{\citenamefont {Kim}\ \emph {et~al.}(2019)\citenamefont {Kim},
  \citenamefont {Kwon},\ and\ \citenamefont {Moon}}]{kim2019pulsed}%
  \BibitemOpen
  \bibfield  {author} {\bibinfo {author} {\bibfnamefont {H.}~\bibnamefont
  {Kim}}, \bibinfo {author} {\bibfnamefont {O.}~\bibnamefont {Kwon}}, \ and\
  \bibinfo {author} {\bibfnamefont {H.~S.}\ \bibnamefont {Moon}},\ }\href@noop
  {} {\bibfield  {journal} {\bibinfo  {journal} {Sci. Rep.}\ }\textbf {\bibinfo
  {volume} {9}},\ \bibinfo {pages} {5031} (\bibinfo {year} {2019})}\BibitemShut
  {NoStop}%
\bibitem [{\citenamefont {Slussarenko}\ \emph {et~al.}(2010)\citenamefont
  {Slussarenko}, \citenamefont {D’Ambrosio}, \citenamefont {Piccirillo},
  \citenamefont {Marrucci},\ and\ \citenamefont
  {Santamato}}]{slussarenko2010polarizing}%
  \BibitemOpen
  \bibfield  {author} {\bibinfo {author} {\bibfnamefont {S.}~\bibnamefont
  {Slussarenko}}, \bibinfo {author} {\bibfnamefont {V.}~\bibnamefont
  {D’Ambrosio}}, \bibinfo {author} {\bibfnamefont {B.}~\bibnamefont
  {Piccirillo}}, \bibinfo {author} {\bibfnamefont {L.}~\bibnamefont
  {Marrucci}}, \ and\ \bibinfo {author} {\bibfnamefont {E.}~\bibnamefont
  {Santamato}},\ }\href@noop {} {\bibfield  {journal} {\bibinfo  {journal}
  {Opt. Express}\ }\textbf {\bibinfo {volume} {18}},\ \bibinfo {pages} {27205}
  (\bibinfo {year} {2010})}\BibitemShut {NoStop}%
\bibitem [{\citenamefont {Kuklewicz}\ \emph {et~al.}(2004)\citenamefont
  {Kuklewicz}, \citenamefont {Fiorentino}, \citenamefont {Messin},
  \citenamefont {Wong},\ and\ \citenamefont {Shapiro}}]{kuklewicz2004high}%
  \BibitemOpen
  \bibfield  {author} {\bibinfo {author} {\bibfnamefont {C.~E.}\ \bibnamefont
  {Kuklewicz}}, \bibinfo {author} {\bibfnamefont {M.}~\bibnamefont
  {Fiorentino}}, \bibinfo {author} {\bibfnamefont {G.}~\bibnamefont {Messin}},
  \bibinfo {author} {\bibfnamefont {F.~N.}\ \bibnamefont {Wong}}, \ and\
  \bibinfo {author} {\bibfnamefont {J.~H.}\ \bibnamefont {Shapiro}},\
  }\href@noop {} {\bibfield  {journal} {\bibinfo  {journal} {Phys. Rev. A}\
  }\textbf {\bibinfo {volume} {69}},\ \bibinfo {pages} {013807} (\bibinfo
  {year} {2004})}\BibitemShut {NoStop}%
\bibitem [{\citenamefont {Zhou}\ \emph {et~al.}(2013)\citenamefont {Zhou},
  \citenamefont {Jiang}, \citenamefont {Ding},\ and\ \citenamefont
  {Shi}}]{zhou2013ultra}%
  \BibitemOpen
  \bibfield  {author} {\bibinfo {author} {\bibfnamefont {Z.-Y.}\ \bibnamefont
  {Zhou}}, \bibinfo {author} {\bibfnamefont {Y.-K.}\ \bibnamefont {Jiang}},
  \bibinfo {author} {\bibfnamefont {D.-S.}\ \bibnamefont {Ding}}, \ and\
  \bibinfo {author} {\bibfnamefont {B.-S.}\ \bibnamefont {Shi}},\ }\href@noop
  {} {\bibfield  {journal} {\bibinfo  {journal} {J. Mod. Opt.}\ }\textbf
  {\bibinfo {volume} {60}},\ \bibinfo {pages} {720} (\bibinfo {year}
  {2013})}\BibitemShut {NoStop}%
\bibitem [{\citenamefont {Ueno}\ \emph {et~al.}(2012)\citenamefont {Ueno},
  \citenamefont {Kaneda}, \citenamefont {Suzuki}, \citenamefont {Nagano},
  \citenamefont {Syouji}, \citenamefont {Shimizu}, \citenamefont {Suizu},\ and\
  \citenamefont {Edamatsu}}]{ueno2012entangled}%
  \BibitemOpen
  \bibfield  {author} {\bibinfo {author} {\bibfnamefont {W.}~\bibnamefont
  {Ueno}}, \bibinfo {author} {\bibfnamefont {F.}~\bibnamefont {Kaneda}},
  \bibinfo {author} {\bibfnamefont {H.}~\bibnamefont {Suzuki}}, \bibinfo
  {author} {\bibfnamefont {S.}~\bibnamefont {Nagano}}, \bibinfo {author}
  {\bibfnamefont {A.}~\bibnamefont {Syouji}}, \bibinfo {author} {\bibfnamefont
  {R.}~\bibnamefont {Shimizu}}, \bibinfo {author} {\bibfnamefont
  {K.}~\bibnamefont {Suizu}}, \ and\ \bibinfo {author} {\bibfnamefont
  {K.}~\bibnamefont {Edamatsu}},\ }\href@noop {} {\bibfield  {journal}
  {\bibinfo  {journal} {Opt. Express}\ }\textbf {\bibinfo {volume} {20}},\
  \bibinfo {pages} {5508} (\bibinfo {year} {2012})}\BibitemShut {NoStop}%
\bibitem [{\citenamefont {Gong}\ \emph {et~al.}(2011)\citenamefont {Gong},
  \citenamefont {Xie}, \citenamefont {Xu}, \citenamefont {Yu}, \citenamefont
  {Xue},\ and\ \citenamefont {Zhu}}]{gong2011compact}%
  \BibitemOpen
  \bibfield  {author} {\bibinfo {author} {\bibfnamefont {Y.-X.}\ \bibnamefont
  {Gong}}, \bibinfo {author} {\bibfnamefont {Z.-D.}\ \bibnamefont {Xie}},
  \bibinfo {author} {\bibfnamefont {P.}~\bibnamefont {Xu}}, \bibinfo {author}
  {\bibfnamefont {X.-Q.}\ \bibnamefont {Yu}}, \bibinfo {author} {\bibfnamefont
  {P.}~\bibnamefont {Xue}}, \ and\ \bibinfo {author} {\bibfnamefont {S.-N.}\
  \bibnamefont {Zhu}},\ }\href@noop {} {\bibfield  {journal} {\bibinfo
  {journal} {Phys. Rev. A}\ }\textbf {\bibinfo {volume} {84}},\ \bibinfo
  {pages} {053825} (\bibinfo {year} {2011})}\BibitemShut {NoStop}%
\bibitem [{\citenamefont {Laudenbach}\ \emph {et~al.}(2017)\citenamefont
  {Laudenbach}, \citenamefont {Kalista}, \citenamefont {Hentschel},
  \citenamefont {Walther},\ and\ \citenamefont
  {H{\"u}bel}}]{laudenbach2017novel}%
  \BibitemOpen
  \bibfield  {author} {\bibinfo {author} {\bibfnamefont {F.}~\bibnamefont
  {Laudenbach}}, \bibinfo {author} {\bibfnamefont {S.}~\bibnamefont {Kalista}},
  \bibinfo {author} {\bibfnamefont {M.}~\bibnamefont {Hentschel}}, \bibinfo
  {author} {\bibfnamefont {P.}~\bibnamefont {Walther}}, \ and\ \bibinfo
  {author} {\bibfnamefont {H.}~\bibnamefont {H{\"u}bel}},\ }\href@noop {}
  {\bibfield  {journal} {\bibinfo  {journal} {Sci. Rep.}\ }\textbf {\bibinfo
  {volume} {7}},\ \bibinfo {pages} {1} (\bibinfo {year} {2017})}\BibitemShut
  {NoStop}%
\bibitem [{\citenamefont {Kuo}\ \emph {et~al.}(2020)\citenamefont {Kuo},
  \citenamefont {Verma},\ and\ \citenamefont {Nam}}]{kuo2020demonstration}%
  \BibitemOpen
  \bibfield  {author} {\bibinfo {author} {\bibfnamefont {P.~S.}\ \bibnamefont
  {Kuo}}, \bibinfo {author} {\bibfnamefont {V.~B.}\ \bibnamefont {Verma}}, \
  and\ \bibinfo {author} {\bibfnamefont {S.~W.}\ \bibnamefont {Nam}},\
  }\href@noop {} {\bibfield  {journal} {\bibinfo  {journal} {OSA Cont.}\
  }\textbf {\bibinfo {volume} {3}},\ \bibinfo {pages} {295} (\bibinfo {year}
  {2020})}\BibitemShut {NoStop}%
\bibitem [{\citenamefont {Prabhakar}\ \emph {et~al.}(2020)\citenamefont
  {Prabhakar}, \citenamefont {Shields}, \citenamefont {Dada}, \citenamefont
  {Ebrahim}, \citenamefont {Taylor}, \citenamefont {Morozov}, \citenamefont
  {Erotokritou}, \citenamefont {Miki}, \citenamefont {Yabuno}, \citenamefont
  {Terai} \emph {et~al.}}]{prabhakar2020two}%
  \BibitemOpen
  \bibfield  {author} {\bibinfo {author} {\bibfnamefont {S.}~\bibnamefont
  {Prabhakar}}, \bibinfo {author} {\bibfnamefont {T.}~\bibnamefont {Shields}},
  \bibinfo {author} {\bibfnamefont {A.~C.}\ \bibnamefont {Dada}}, \bibinfo
  {author} {\bibfnamefont {M.}~\bibnamefont {Ebrahim}}, \bibinfo {author}
  {\bibfnamefont {G.~G.}\ \bibnamefont {Taylor}}, \bibinfo {author}
  {\bibfnamefont {D.}~\bibnamefont {Morozov}}, \bibinfo {author} {\bibfnamefont
  {K.}~\bibnamefont {Erotokritou}}, \bibinfo {author} {\bibfnamefont
  {S.}~\bibnamefont {Miki}}, \bibinfo {author} {\bibfnamefont {M.}~\bibnamefont
  {Yabuno}}, \bibinfo {author} {\bibfnamefont {H.}~\bibnamefont {Terai}},
  \emph {et~al.},\ }\href@noop {} {\bibfield  {journal} {\bibinfo  {journal}
  {Sci. Adv.}\ }\textbf {\bibinfo {volume} {6}},\ \bibinfo {pages} {eaay5195}
  (\bibinfo {year} {2020})}\BibitemShut {NoStop}%
\bibitem [{\citenamefont {Shimizu}\ \emph {et~al.}(2008)\citenamefont
  {Shimizu}, \citenamefont {Yamaguchi}, \citenamefont {Mitsumori},
  \citenamefont {Kosaka},\ and\ \citenamefont
  {Edamatsu}}]{shimizu2008generation}%
  \BibitemOpen
  \bibfield  {author} {\bibinfo {author} {\bibfnamefont {R.}~\bibnamefont
  {Shimizu}}, \bibinfo {author} {\bibfnamefont {T.}~\bibnamefont {Yamaguchi}},
  \bibinfo {author} {\bibfnamefont {Y.}~\bibnamefont {Mitsumori}}, \bibinfo
  {author} {\bibfnamefont {H.}~\bibnamefont {Kosaka}}, \ and\ \bibinfo {author}
  {\bibfnamefont {K.}~\bibnamefont {Edamatsu}},\ }\href@noop {} {\bibfield
  {journal} {\bibinfo  {journal} {Phys. Rev. A}\ }\textbf {\bibinfo {volume}
  {77}},\ \bibinfo {pages} {032338} (\bibinfo {year} {2008})}\BibitemShut
  {NoStop}%
\bibitem [{\citenamefont {Kim}\ \emph {et~al.}(2001)\citenamefont {Kim},
  \citenamefont {Chekhova}, \citenamefont {Kulik}, \citenamefont {Rubin},\ and\
  \citenamefont {Shih}}]{kim2001interferometric}%
  \BibitemOpen
  \bibfield  {author} {\bibinfo {author} {\bibfnamefont {Y.-H.}\ \bibnamefont
  {Kim}}, \bibinfo {author} {\bibfnamefont {M.~V.}\ \bibnamefont {Chekhova}},
  \bibinfo {author} {\bibfnamefont {S.~P.}\ \bibnamefont {Kulik}}, \bibinfo
  {author} {\bibfnamefont {M.~H.}\ \bibnamefont {Rubin}}, \ and\ \bibinfo
  {author} {\bibfnamefont {Y.}~\bibnamefont {Shih}},\ }\href@noop {} {\bibfield
   {journal} {\bibinfo  {journal} {Phys. Rev. A}\ }\textbf {\bibinfo {volume}
  {63}},\ \bibinfo {pages} {062301} (\bibinfo {year} {2001})}\BibitemShut
  {NoStop}%
\bibitem [{\citenamefont {Kwiat}\ \emph {et~al.}(1994)\citenamefont {Kwiat},
  \citenamefont {Eberhard}, \citenamefont {Steinberg},\ and\ \citenamefont
  {Chiao}}]{kwiat1994proposal}%
  \BibitemOpen
  \bibfield  {author} {\bibinfo {author} {\bibfnamefont {P.~G.}\ \bibnamefont
  {Kwiat}}, \bibinfo {author} {\bibfnamefont {P.~H.}\ \bibnamefont {Eberhard}},
  \bibinfo {author} {\bibfnamefont {A.~M.}\ \bibnamefont {Steinberg}}, \ and\
  \bibinfo {author} {\bibfnamefont {R.~Y.}\ \bibnamefont {Chiao}},\ }\href@noop
  {} {\bibfield  {journal} {\bibinfo  {journal} {Phys. Rev. A}\ }\textbf
  {\bibinfo {volume} {49}},\ \bibinfo {pages} {3209} (\bibinfo {year}
  {1994})}\BibitemShut {NoStop}%
\bibitem [{\citenamefont {Shapiro}\ and\ \citenamefont
  {Wong}(2000)}]{shapiro2000ultrabright}%
  \BibitemOpen
  \bibfield  {author} {\bibinfo {author} {\bibfnamefont {J.~H.}\ \bibnamefont
  {Shapiro}}\ and\ \bibinfo {author} {\bibfnamefont {N.}~\bibnamefont {Wong}},\
  }\href@noop {} {\bibfield  {journal} {\bibinfo  {journal} {J. Opt. B: Quant.
  Semiclass. Opt.}\ }\textbf {\bibinfo {volume} {2}},\ \bibinfo {pages} {L1}
  (\bibinfo {year} {2000})}\BibitemShut {NoStop}%
\bibitem [{\citenamefont {Shi}\ and\ \citenamefont
  {Tomita}(2003)}]{shi2003generation}%
  \BibitemOpen
  \bibfield  {author} {\bibinfo {author} {\bibfnamefont {B.-S.}\ \bibnamefont
  {Shi}}\ and\ \bibinfo {author} {\bibfnamefont {A.}~\bibnamefont {Tomita}},\
  }\href@noop {} {\bibfield  {journal} {\bibinfo  {journal} {Phys. Rev. A}\
  }\textbf {\bibinfo {volume} {67}},\ \bibinfo {pages} {043804} (\bibinfo
  {year} {2003})}\BibitemShut {NoStop}%
\bibitem [{\citenamefont {Septriani}\ \emph {et~al.}(2016)\citenamefont
  {Septriani}, \citenamefont {Grieve}, \citenamefont {Durak},\ and\
  \citenamefont {Ling}}]{septriani2016thick}%
  \BibitemOpen
  \bibfield  {author} {\bibinfo {author} {\bibfnamefont {B.}~\bibnamefont
  {Septriani}}, \bibinfo {author} {\bibfnamefont {J.~A.}\ \bibnamefont
  {Grieve}}, \bibinfo {author} {\bibfnamefont {K.}~\bibnamefont {Durak}}, \
  and\ \bibinfo {author} {\bibfnamefont {A.}~\bibnamefont {Ling}},\ }\href@noop
  {} {\bibfield  {journal} {\bibinfo  {journal} {Optica}\ }\textbf {\bibinfo
  {volume} {3}},\ \bibinfo {pages} {347} (\bibinfo {year} {2016})}\BibitemShut
  {NoStop}%
\bibitem [{\citenamefont {Pelton}\ \emph {et~al.}(2004)\citenamefont {Pelton},
  \citenamefont {Marsden}, \citenamefont {Ljunggren}, \citenamefont {Tengner},
  \citenamefont {Karlsson}, \citenamefont {Fragemann}, \citenamefont
  {Canalias},\ and\ \citenamefont {Laurell}}]{pelton2004bright}%
  \BibitemOpen
  \bibfield  {author} {\bibinfo {author} {\bibfnamefont {M.}~\bibnamefont
  {Pelton}}, \bibinfo {author} {\bibfnamefont {P.}~\bibnamefont {Marsden}},
  \bibinfo {author} {\bibfnamefont {D.}~\bibnamefont {Ljunggren}}, \bibinfo
  {author} {\bibfnamefont {M.}~\bibnamefont {Tengner}}, \bibinfo {author}
  {\bibfnamefont {A.}~\bibnamefont {Karlsson}}, \bibinfo {author}
  {\bibfnamefont {A.}~\bibnamefont {Fragemann}}, \bibinfo {author}
  {\bibfnamefont {C.}~\bibnamefont {Canalias}}, \ and\ \bibinfo {author}
  {\bibfnamefont {F.}~\bibnamefont {Laurell}},\ }\href@noop {} {\bibfield
  {journal} {\bibinfo  {journal} {Opt. Express}\ }\textbf {\bibinfo {volume}
  {12}},\ \bibinfo {pages} {3573} (\bibinfo {year} {2004})}\BibitemShut
  {NoStop}%
\bibitem [{\citenamefont {Ljunggren}\ \emph {et~al.}(2006)\citenamefont
  {Ljunggren}, \citenamefont {Tengner}, \citenamefont {Marsden},\ and\
  \citenamefont {Pelton}}]{ljunggren2006theory}%
  \BibitemOpen
  \bibfield  {author} {\bibinfo {author} {\bibfnamefont {D.}~\bibnamefont
  {Ljunggren}}, \bibinfo {author} {\bibfnamefont {M.}~\bibnamefont {Tengner}},
  \bibinfo {author} {\bibfnamefont {P.}~\bibnamefont {Marsden}}, \ and\
  \bibinfo {author} {\bibfnamefont {M.}~\bibnamefont {Pelton}},\ }\href@noop {}
  {\bibfield  {journal} {\bibinfo  {journal} {Phys. Rev. A}\ }\textbf {\bibinfo
  {volume} {73}},\ \bibinfo {pages} {032326} (\bibinfo {year}
  {2006})}\BibitemShut {NoStop}%
\bibitem [{\citenamefont {H{\"u}bel}\ \emph {et~al.}(2007)\citenamefont
  {H{\"u}bel}, \citenamefont {Vanner}, \citenamefont {Lederer}, \citenamefont
  {Blauensteiner}, \citenamefont {Lor{\"u}nser}, \citenamefont {Poppe},\ and\
  \citenamefont {Zeilinger}}]{hubel2007high}%
  \BibitemOpen
  \bibfield  {author} {\bibinfo {author} {\bibfnamefont {H.}~\bibnamefont
  {H{\"u}bel}}, \bibinfo {author} {\bibfnamefont {M.~R.}\ \bibnamefont
  {Vanner}}, \bibinfo {author} {\bibfnamefont {T.}~\bibnamefont {Lederer}},
  \bibinfo {author} {\bibfnamefont {B.}~\bibnamefont {Blauensteiner}}, \bibinfo
  {author} {\bibfnamefont {T.}~\bibnamefont {Lor{\"u}nser}}, \bibinfo {author}
  {\bibfnamefont {A.}~\bibnamefont {Poppe}}, \ and\ \bibinfo {author}
  {\bibfnamefont {A.}~\bibnamefont {Zeilinger}},\ }\href@noop {} {\bibfield
  {journal} {\bibinfo  {journal} {Opt. Express}\ }\textbf {\bibinfo {volume}
  {15}},\ \bibinfo {pages} {7853} (\bibinfo {year} {2007})}\BibitemShut
  {NoStop}%
\bibitem [{\citenamefont {Steinlechner}\ \emph {et~al.}(2012)\citenamefont
  {Steinlechner}, \citenamefont {Trojek}, \citenamefont {Jofre}, \citenamefont
  {Weier}, \citenamefont {Perez}, \citenamefont {Jennewein}, \citenamefont
  {Ursin}, \citenamefont {Rarity}, \citenamefont {Mitchell}, \citenamefont
  {Torres} \emph {et~al.}}]{steinlechner2012high}%
  \BibitemOpen
  \bibfield  {author} {\bibinfo {author} {\bibfnamefont {F.}~\bibnamefont
  {Steinlechner}}, \bibinfo {author} {\bibfnamefont {P.}~\bibnamefont
  {Trojek}}, \bibinfo {author} {\bibfnamefont {M.}~\bibnamefont {Jofre}},
  \bibinfo {author} {\bibfnamefont {H.}~\bibnamefont {Weier}}, \bibinfo
  {author} {\bibfnamefont {D.}~\bibnamefont {Perez}}, \bibinfo {author}
  {\bibfnamefont {T.}~\bibnamefont {Jennewein}}, \bibinfo {author}
  {\bibfnamefont {R.}~\bibnamefont {Ursin}}, \bibinfo {author} {\bibfnamefont
  {J.}~\bibnamefont {Rarity}}, \bibinfo {author} {\bibfnamefont {M.~W.}\
  \bibnamefont {Mitchell}}, \bibinfo {author} {\bibfnamefont {J.~P.}\
  \bibnamefont {Torres}},  \emph {et~al.},\ }\href@noop {} {\bibfield
  {journal} {\bibinfo  {journal} {Opt. Express}\ }\textbf {\bibinfo {volume}
  {20}},\ \bibinfo {pages} {9640} (\bibinfo {year} {2012})}\BibitemShut
  {NoStop}%
\bibitem [{\citenamefont {Scheidl}\ \emph {et~al.}(2014)\citenamefont
  {Scheidl}, \citenamefont {Tiefenbacher}, \citenamefont {Prevedel},
  \citenamefont {Steinlechner}, \citenamefont {Ursin},\ and\ \citenamefont
  {Zeilinger}}]{scheidl2014crossed}%
  \BibitemOpen
  \bibfield  {author} {\bibinfo {author} {\bibfnamefont {T.}~\bibnamefont
  {Scheidl}}, \bibinfo {author} {\bibfnamefont {F.}~\bibnamefont
  {Tiefenbacher}}, \bibinfo {author} {\bibfnamefont {R.}~\bibnamefont
  {Prevedel}}, \bibinfo {author} {\bibfnamefont {F.}~\bibnamefont
  {Steinlechner}}, \bibinfo {author} {\bibfnamefont {R.}~\bibnamefont {Ursin}},
  \ and\ \bibinfo {author} {\bibfnamefont {A.}~\bibnamefont {Zeilinger}},\
  }\href@noop {} {\bibfield  {journal} {\bibinfo  {journal} {Phys. Rev. A}\
  }\textbf {\bibinfo {volume} {89}},\ \bibinfo {pages} {042324} (\bibinfo
  {year} {2014})}\BibitemShut {NoStop}%
\bibitem [{\citenamefont {Fiorentino}\ \emph {et~al.}(2004)\citenamefont
  {Fiorentino}, \citenamefont {Messin}, \citenamefont {Kuklewicz},
  \citenamefont {Wong},\ and\ \citenamefont
  {Shapiro}}]{fiorentino2004generation}%
  \BibitemOpen
  \bibfield  {author} {\bibinfo {author} {\bibfnamefont {M.}~\bibnamefont
  {Fiorentino}}, \bibinfo {author} {\bibfnamefont {G.}~\bibnamefont {Messin}},
  \bibinfo {author} {\bibfnamefont {C.~E.}\ \bibnamefont {Kuklewicz}}, \bibinfo
  {author} {\bibfnamefont {F.~N.}\ \bibnamefont {Wong}}, \ and\ \bibinfo
  {author} {\bibfnamefont {J.~H.}\ \bibnamefont {Shapiro}},\ }\href@noop {}
  {\bibfield  {journal} {\bibinfo  {journal} {Phys. Rev. A}\ }\textbf {\bibinfo
  {volume} {69}},\ \bibinfo {pages} {041801} (\bibinfo {year}
  {2004})}\BibitemShut {NoStop}%
\bibitem [{\citenamefont {K{\"o}nig}\ \emph {et~al.}(2005)\citenamefont
  {K{\"o}nig}, \citenamefont {Mason}, \citenamefont {Wong},\ and\ \citenamefont
  {Albota}}]{konig2005efficient}%
  \BibitemOpen
  \bibfield  {author} {\bibinfo {author} {\bibfnamefont {F.}~\bibnamefont
  {K{\"o}nig}}, \bibinfo {author} {\bibfnamefont {E.~J.}\ \bibnamefont
  {Mason}}, \bibinfo {author} {\bibfnamefont {F.~N.}\ \bibnamefont {Wong}}, \
  and\ \bibinfo {author} {\bibfnamefont {M.~A.}\ \bibnamefont {Albota}},\
  }\href@noop {} {\bibfield  {journal} {\bibinfo  {journal} {Phys. Rev. A}\
  }\textbf {\bibinfo {volume} {71}},\ \bibinfo {pages} {033805} (\bibinfo
  {year} {2005})}\BibitemShut {NoStop}%
\bibitem [{\citenamefont {Anderson}\ \emph {et~al.}(1994)\citenamefont
  {Anderson}, \citenamefont {Bilger},\ and\ \citenamefont
  {Stedman}}]{anderson1994sagnac}%
  \BibitemOpen
  \bibfield  {author} {\bibinfo {author} {\bibfnamefont {R.}~\bibnamefont
  {Anderson}}, \bibinfo {author} {\bibfnamefont {H.}~\bibnamefont {Bilger}}, \
  and\ \bibinfo {author} {\bibfnamefont {G.~E.}\ \bibnamefont {Stedman}},\
  }\href@noop {} {\bibfield  {journal} {\bibinfo  {journal} {Am. J. Phys.}\
  }\textbf {\bibinfo {volume} {62}},\ \bibinfo {pages} {975} (\bibinfo {year}
  {1994})}\BibitemShut {NoStop}%
\bibitem [{\citenamefont {Kuzucu}\ and\ \citenamefont
  {Wong}(2008)}]{kuzucu2008pulsed}%
  \BibitemOpen
  \bibfield  {author} {\bibinfo {author} {\bibfnamefont {O.}~\bibnamefont
  {Kuzucu}}\ and\ \bibinfo {author} {\bibfnamefont {F.~N.}\ \bibnamefont
  {Wong}},\ }\href@noop {} {\bibfield  {journal} {\bibinfo  {journal} {Phys.
  Rev. A}\ }\textbf {\bibinfo {volume} {77}},\ \bibinfo {pages} {032314}
  (\bibinfo {year} {2008})}\BibitemShut {NoStop}%
\bibitem [{\citenamefont {Fedrizzi}\ \emph {et~al.}(2007)\citenamefont
  {Fedrizzi}, \citenamefont {Herbst}, \citenamefont {Poppe}, \citenamefont
  {Jennewein},\ and\ \citenamefont {Zeilinger}}]{fedrizzi2007wavelength}%
  \BibitemOpen
  \bibfield  {author} {\bibinfo {author} {\bibfnamefont {A.}~\bibnamefont
  {Fedrizzi}}, \bibinfo {author} {\bibfnamefont {T.}~\bibnamefont {Herbst}},
  \bibinfo {author} {\bibfnamefont {A.}~\bibnamefont {Poppe}}, \bibinfo
  {author} {\bibfnamefont {T.}~\bibnamefont {Jennewein}}, \ and\ \bibinfo
  {author} {\bibfnamefont {A.}~\bibnamefont {Zeilinger}},\ }\href@noop {}
  {\bibfield  {journal} {\bibinfo  {journal} {Opt. Express}\ }\textbf {\bibinfo
  {volume} {15}},\ \bibinfo {pages} {15377} (\bibinfo {year}
  {2007})}\BibitemShut {NoStop}%
\bibitem [{\citenamefont {Predojevi{\'c}}\ \emph {et~al.}(2012)\citenamefont
  {Predojevi{\'c}}, \citenamefont {Grabher},\ and\ \citenamefont
  {Weihs}}]{predojevic2012pulsed}%
  \BibitemOpen
  \bibfield  {author} {\bibinfo {author} {\bibfnamefont {A.}~\bibnamefont
  {Predojevi{\'c}}}, \bibinfo {author} {\bibfnamefont {S.}~\bibnamefont
  {Grabher}}, \ and\ \bibinfo {author} {\bibfnamefont {G.}~\bibnamefont
  {Weihs}},\ }\href@noop {} {\bibfield  {journal} {\bibinfo  {journal} {Opt.
  Express}\ }\textbf {\bibinfo {volume} {20}},\ \bibinfo {pages} {25022}
  (\bibinfo {year} {2012})}\BibitemShut {NoStop}%
\bibitem [{\citenamefont {Steinlechner}\ \emph {et~al.}(2014)\citenamefont
  {Steinlechner}, \citenamefont {Gilaberte}, \citenamefont {Jofre},
  \citenamefont {Scheidl}, \citenamefont {Torres}, \citenamefont {Pruneri},\
  and\ \citenamefont {Ursin}}]{steinlechner2014efficient}%
  \BibitemOpen
  \bibfield  {author} {\bibinfo {author} {\bibfnamefont {F.}~\bibnamefont
  {Steinlechner}}, \bibinfo {author} {\bibfnamefont {M.}~\bibnamefont
  {Gilaberte}}, \bibinfo {author} {\bibfnamefont {M.}~\bibnamefont {Jofre}},
  \bibinfo {author} {\bibfnamefont {T.}~\bibnamefont {Scheidl}}, \bibinfo
  {author} {\bibfnamefont {J.~P.}\ \bibnamefont {Torres}}, \bibinfo {author}
  {\bibfnamefont {V.}~\bibnamefont {Pruneri}}, \ and\ \bibinfo {author}
  {\bibfnamefont {R.}~\bibnamefont {Ursin}},\ }\href@noop {} {\bibfield
  {journal} {\bibinfo  {journal} {J. Opt. Soc. Am. B}\ }\textbf {\bibinfo
  {volume} {31}},\ \bibinfo {pages} {2068} (\bibinfo {year}
  {2014})}\BibitemShut {NoStop}%
\bibitem [{\citenamefont {Shi}\ and\ \citenamefont
  {Tomita}(2004)}]{shi2004generation}%
  \BibitemOpen
  \bibfield  {author} {\bibinfo {author} {\bibfnamefont {B.-S.}\ \bibnamefont
  {Shi}}\ and\ \bibinfo {author} {\bibfnamefont {A.}~\bibnamefont {Tomita}},\
  }\href@noop {} {\bibfield  {journal} {\bibinfo  {journal} {Phys. Rev. A}\
  }\textbf {\bibinfo {volume} {69}},\ \bibinfo {pages} {013803} (\bibinfo
  {year} {2004})}\BibitemShut {NoStop}%
\bibitem [{\citenamefont {Beckert}\ \emph {et~al.}(2019)\citenamefont
  {Beckert}, \citenamefont {de~Vries}, \citenamefont {Ursin}, \citenamefont
  {Steinlechner}, \citenamefont {Gr{\"a}fe},\ and\ \citenamefont
  {Basset}}]{beckert2019space}%
  \BibitemOpen
  \bibfield  {author} {\bibinfo {author} {\bibfnamefont {E.}~\bibnamefont
  {Beckert}}, \bibinfo {author} {\bibfnamefont {O.}~\bibnamefont {de~Vries}},
  \bibinfo {author} {\bibfnamefont {R.}~\bibnamefont {Ursin}}, \bibinfo
  {author} {\bibfnamefont {F.-O.}\ \bibnamefont {Steinlechner}}, \bibinfo
  {author} {\bibfnamefont {M.}~\bibnamefont {Gr{\"a}fe}}, \ and\ \bibinfo
  {author} {\bibfnamefont {M.~G.}\ \bibnamefont {Basset}},\ }in\ \href@noop {}
  {\emph {\bibinfo {booktitle} {Free-Space Laser Communications XXXI}}},\ Vol.\
  \bibinfo {volume} {10910}\ (\bibinfo {organization} {International Society
  for Optics and Photonics},\ \bibinfo {year} {2019})\ p.\ \bibinfo {pages}
  {1091016}\BibitemShut {NoStop}%
\bibitem [{\citenamefont {Yin}\ \emph {et~al.}(2017{\natexlab{a}})\citenamefont
  {Yin}, \citenamefont {Cao}, \citenamefont {Li}, \citenamefont {Liao},
  \citenamefont {Zhang}, \citenamefont {Ren}, \citenamefont {Cai},
  \citenamefont {Liu}, \citenamefont {Li}, \citenamefont {Dai} \emph
  {et~al.}}]{yin2017satellite}%
  \BibitemOpen
  \bibfield  {author} {\bibinfo {author} {\bibfnamefont {J.}~\bibnamefont
  {Yin}}, \bibinfo {author} {\bibfnamefont {Y.}~\bibnamefont {Cao}}, \bibinfo
  {author} {\bibfnamefont {Y.-H.}\ \bibnamefont {Li}}, \bibinfo {author}
  {\bibfnamefont {S.-K.}\ \bibnamefont {Liao}}, \bibinfo {author}
  {\bibfnamefont {L.}~\bibnamefont {Zhang}}, \bibinfo {author} {\bibfnamefont
  {J.-G.}\ \bibnamefont {Ren}}, \bibinfo {author} {\bibfnamefont {W.-Q.}\
  \bibnamefont {Cai}}, \bibinfo {author} {\bibfnamefont {W.-Y.}\ \bibnamefont
  {Liu}}, \bibinfo {author} {\bibfnamefont {B.}~\bibnamefont {Li}}, \bibinfo
  {author} {\bibfnamefont {H.}~\bibnamefont {Dai}},  \emph {et~al.},\
  }\href@noop {} {\bibfield  {journal} {\bibinfo  {journal} {Science}\ }\textbf
  {\bibinfo {volume} {356}},\ \bibinfo {pages} {1140} (\bibinfo {year}
  {2017}{\natexlab{a}})}\BibitemShut {NoStop}%
\bibitem [{\citenamefont {Steinlechner}(2015)}]{steinlechner2015sources}%
  \BibitemOpen
  \bibfield  {author} {\bibinfo {author} {\bibfnamefont {F.}~\bibnamefont
  {Steinlechner}},\ }\emph {\bibinfo {title} {Sources of photonic entanglement
  for applications in space}},\ \href@noop {} {Ph.D. thesis},\ \bibinfo
  {address} {Universitat Polit{\`e}cnica de Catalunya} (\bibinfo {year}
  {2015})\BibitemShut {NoStop}%
\bibitem [{\citenamefont {Terashima}\ \emph {et~al.}(2018)\citenamefont
  {Terashima}, \citenamefont {Kobayashi}, \citenamefont {Tsubakiyama},\ and\
  \citenamefont {Sanaka}}]{terashima2018quantum}%
  \BibitemOpen
  \bibfield  {author} {\bibinfo {author} {\bibfnamefont {H.}~\bibnamefont
  {Terashima}}, \bibinfo {author} {\bibfnamefont {S.}~\bibnamefont
  {Kobayashi}}, \bibinfo {author} {\bibfnamefont {T.}~\bibnamefont
  {Tsubakiyama}}, \ and\ \bibinfo {author} {\bibfnamefont {K.}~\bibnamefont
  {Sanaka}},\ }\href@noop {} {\bibfield  {journal} {\bibinfo  {journal} {Sci.
  Rep.}\ }\textbf {\bibinfo {volume} {8}},\ \bibinfo {pages} {15733} (\bibinfo
  {year} {2018})}\BibitemShut {NoStop}%
\bibitem [{\citenamefont {Fiorentino}\ and\ \citenamefont
  {Beausoleil}(2008)}]{fiorentino2008compact}%
  \BibitemOpen
  \bibfield  {author} {\bibinfo {author} {\bibfnamefont {M.}~\bibnamefont
  {Fiorentino}}\ and\ \bibinfo {author} {\bibfnamefont {R.~G.}\ \bibnamefont
  {Beausoleil}},\ }\href@noop {} {\bibfield  {journal} {\bibinfo  {journal}
  {Opt. Express}\ }\textbf {\bibinfo {volume} {16}},\ \bibinfo {pages} {20149}
  (\bibinfo {year} {2008})}\BibitemShut {NoStop}%
\bibitem [{\citenamefont {Evans}\ \emph {et~al.}(2010)\citenamefont {Evans},
  \citenamefont {Bennink}, \citenamefont {Grice}, \citenamefont {Humble},\ and\
  \citenamefont {Schaake}}]{evans2010bright}%
  \BibitemOpen
  \bibfield  {author} {\bibinfo {author} {\bibfnamefont {P.~G.}\ \bibnamefont
  {Evans}}, \bibinfo {author} {\bibfnamefont {R.~S.}\ \bibnamefont {Bennink}},
  \bibinfo {author} {\bibfnamefont {W.~P.}\ \bibnamefont {Grice}}, \bibinfo
  {author} {\bibfnamefont {T.~S.}\ \bibnamefont {Humble}}, \ and\ \bibinfo
  {author} {\bibfnamefont {J.}~\bibnamefont {Schaake}},\ }\href@noop {}
  {\bibfield  {journal} {\bibinfo  {journal} {Phys. Rev. Lett.}\ }\textbf
  {\bibinfo {volume} {105}},\ \bibinfo {pages} {253601} (\bibinfo {year}
  {2010})}\BibitemShut {NoStop}%
\bibitem [{\citenamefont {Chen}\ \emph {et~al.}(2007)\citenamefont {Chen},
  \citenamefont {Lee},\ and\ \citenamefont {Kumar}}]{chen2007deterministic}%
  \BibitemOpen
  \bibfield  {author} {\bibinfo {author} {\bibfnamefont {J.}~\bibnamefont
  {Chen}}, \bibinfo {author} {\bibfnamefont {K.~F.}\ \bibnamefont {Lee}}, \
  and\ \bibinfo {author} {\bibfnamefont {P.}~\bibnamefont {Kumar}},\
  }\href@noop {} {\bibfield  {journal} {\bibinfo  {journal} {Phys. Rev. A}\
  }\textbf {\bibinfo {volume} {76}},\ \bibinfo {pages} {031804} (\bibinfo
  {year} {2007})}\BibitemShut {NoStop}%
\bibitem [{\citenamefont {Chen}\ \emph
  {et~al.}(2018{\natexlab{a}})\citenamefont {Chen}, \citenamefont {Ecker},
  \citenamefont {Wengerowsky}, \citenamefont {Bulla}, \citenamefont {Joshi},
  \citenamefont {Steinlechner},\ and\ \citenamefont
  {Ursin}}]{chen2018polarization}%
  \BibitemOpen
  \bibfield  {author} {\bibinfo {author} {\bibfnamefont {Y.}~\bibnamefont
  {Chen}}, \bibinfo {author} {\bibfnamefont {S.}~\bibnamefont {Ecker}},
  \bibinfo {author} {\bibfnamefont {S.}~\bibnamefont {Wengerowsky}}, \bibinfo
  {author} {\bibfnamefont {L.}~\bibnamefont {Bulla}}, \bibinfo {author}
  {\bibfnamefont {S.~K.}\ \bibnamefont {Joshi}}, \bibinfo {author}
  {\bibfnamefont {F.}~\bibnamefont {Steinlechner}}, \ and\ \bibinfo {author}
  {\bibfnamefont {R.}~\bibnamefont {Ursin}},\ }\href@noop {} {\bibfield
  {journal} {\bibinfo  {journal} {Phys. Rev. Lett.}\ }\textbf {\bibinfo
  {volume} {121}},\ \bibinfo {pages} {200502} (\bibinfo {year}
  {2018}{\natexlab{a}})}\BibitemShut {NoStop}%
\bibitem [{\citenamefont {Horn}\ and\ \citenamefont
  {Jennewein}(2019)}]{horn2019auto}%
  \BibitemOpen
  \bibfield  {author} {\bibinfo {author} {\bibfnamefont {R.}~\bibnamefont
  {Horn}}\ and\ \bibinfo {author} {\bibfnamefont {T.}~\bibnamefont
  {Jennewein}},\ }\href@noop {} {\bibfield  {journal} {\bibinfo  {journal}
  {Opt. Express}\ }\textbf {\bibinfo {volume} {27}},\ \bibinfo {pages} {17369}
  (\bibinfo {year} {2019})}\BibitemShut {NoStop}%
\bibitem [{\citenamefont {Ou}\ and\ \citenamefont {Lu}(1999)}]{ou1999cavity}%
  \BibitemOpen
  \bibfield  {author} {\bibinfo {author} {\bibfnamefont {Z.}~\bibnamefont
  {Ou}}\ and\ \bibinfo {author} {\bibfnamefont {Y.}~\bibnamefont {Lu}},\
  }\href@noop {} {\bibfield  {journal} {\bibinfo  {journal} {Phys. Rev. Lett.}\
  }\textbf {\bibinfo {volume} {83}},\ \bibinfo {pages} {2556} (\bibinfo {year}
  {1999})}\BibitemShut {NoStop}%
\bibitem [{\citenamefont {Slattery}\ \emph {et~al.}(2019)\citenamefont
  {Slattery}, \citenamefont {Ma}, \citenamefont {Zong},\ and\ \citenamefont
  {Tang}}]{slattery2019background}%
  \BibitemOpen
  \bibfield  {author} {\bibinfo {author} {\bibfnamefont {O.}~\bibnamefont
  {Slattery}}, \bibinfo {author} {\bibfnamefont {L.}~\bibnamefont {Ma}},
  \bibinfo {author} {\bibfnamefont {K.}~\bibnamefont {Zong}}, \ and\ \bibinfo
  {author} {\bibfnamefont {X.}~\bibnamefont {Tang}},\ }\href@noop {} {\bibfield
   {journal} {\bibinfo  {journal} {J. Res. Nat. Inst. Stand. Tech.}\ }\textbf
  {\bibinfo {volume} {124}},\ \bibinfo {pages} {1} (\bibinfo {year}
  {2019})}\BibitemShut {NoStop}%
\bibitem [{\citenamefont {Oberparleiter}\ and\ \citenamefont
  {Weinfurter}(2000)}]{oberparleiter2000cavity}%
  \BibitemOpen
  \bibfield  {author} {\bibinfo {author} {\bibfnamefont {M.}~\bibnamefont
  {Oberparleiter}}\ and\ \bibinfo {author} {\bibfnamefont {H.}~\bibnamefont
  {Weinfurter}},\ }\href@noop {} {\bibfield  {journal} {\bibinfo  {journal}
  {Opt. Commun.}\ }\textbf {\bibinfo {volume} {183}},\ \bibinfo {pages} {133}
  (\bibinfo {year} {2000})}\BibitemShut {NoStop}%
\bibitem [{\citenamefont {Kuklewicz}\ \emph {et~al.}(2002)\citenamefont
  {Kuklewicz}, \citenamefont {Keskiner}, \citenamefont {Wong},\ and\
  \citenamefont {Shapiro}}]{Kuklewicz_2002}%
  \BibitemOpen
  \bibfield  {author} {\bibinfo {author} {\bibfnamefont {C.~E.}\ \bibnamefont
  {Kuklewicz}}, \bibinfo {author} {\bibfnamefont {E.}~\bibnamefont {Keskiner}},
  \bibinfo {author} {\bibfnamefont {F.~N.~C.}\ \bibnamefont {Wong}}, \ and\
  \bibinfo {author} {\bibfnamefont {J.~H.}\ \bibnamefont {Shapiro}},\ }\href
  {\doibase 10.1088/1464-4266/4/3/370} {\bibfield  {journal} {\bibinfo
  {journal} {J. Opt. B: Quant. Semiclass. Opt.}\ }\textbf {\bibinfo {volume}
  {4}},\ \bibinfo {pages} {S162} (\bibinfo {year} {2002})}\BibitemShut
  {NoStop}%
\bibitem [{\citenamefont {Kuklewicz}\ \emph {et~al.}(2006)\citenamefont
  {Kuklewicz}, \citenamefont {Wong},\ and\ \citenamefont
  {Shapiro}}]{Kuklewicz2006}%
  \BibitemOpen
  \bibfield  {author} {\bibinfo {author} {\bibfnamefont {C.~E.}\ \bibnamefont
  {Kuklewicz}}, \bibinfo {author} {\bibfnamefont {F.~N.~C.}\ \bibnamefont
  {Wong}}, \ and\ \bibinfo {author} {\bibfnamefont {J.~H.}\ \bibnamefont
  {Shapiro}},\ }\href {\doibase 10.1103/PhysRevLett.97.223601} {\bibfield
  {journal} {\bibinfo  {journal} {Phys. Rev. Lett.}\ }\textbf {\bibinfo
  {volume} {97}},\ \bibinfo {pages} {223601} (\bibinfo {year}
  {2006})}\BibitemShut {NoStop}%
\bibitem [{\citenamefont {Wolfgramm}\ \emph {et~al.}(2010)\citenamefont
  {Wolfgramm}, \citenamefont {Cere},\ and\ \citenamefont
  {Mitchell}}]{wolfgramm2010noon}%
  \BibitemOpen
  \bibfield  {author} {\bibinfo {author} {\bibfnamefont {F.}~\bibnamefont
  {Wolfgramm}}, \bibinfo {author} {\bibfnamefont {A.}~\bibnamefont {Cere}}, \
  and\ \bibinfo {author} {\bibfnamefont {M.~W.}\ \bibnamefont {Mitchell}},\
  }\href@noop {} {\bibfield  {journal} {\bibinfo  {journal} {J Opt. Soc. Am.
  B}\ }\textbf {\bibinfo {volume} {27}},\ \bibinfo {pages} {A25} (\bibinfo
  {year} {2010})}\BibitemShut {NoStop}%
\bibitem [{\citenamefont {Zhang}\ \emph {et~al.}(2011)\citenamefont {Zhang},
  \citenamefont {Jin}, \citenamefont {Yang}, \citenamefont {Dai}, \citenamefont
  {Yang}, \citenamefont {Zhao}, \citenamefont {Rui}, \citenamefont {He},
  \citenamefont {Jiang}, \citenamefont {Yang} \emph
  {et~al.}}]{zhang2011preparation}%
  \BibitemOpen
  \bibfield  {author} {\bibinfo {author} {\bibfnamefont {H.}~\bibnamefont
  {Zhang}}, \bibinfo {author} {\bibfnamefont {X.-M.}\ \bibnamefont {Jin}},
  \bibinfo {author} {\bibfnamefont {J.}~\bibnamefont {Yang}}, \bibinfo {author}
  {\bibfnamefont {H.-N.}\ \bibnamefont {Dai}}, \bibinfo {author} {\bibfnamefont
  {S.-J.}\ \bibnamefont {Yang}}, \bibinfo {author} {\bibfnamefont {T.-M.}\
  \bibnamefont {Zhao}}, \bibinfo {author} {\bibfnamefont {J.}~\bibnamefont
  {Rui}}, \bibinfo {author} {\bibfnamefont {Y.}~\bibnamefont {He}}, \bibinfo
  {author} {\bibfnamefont {X.}~\bibnamefont {Jiang}}, \bibinfo {author}
  {\bibfnamefont {F.}~\bibnamefont {Yang}},  \emph {et~al.},\ }\href@noop {}
  {\bibfield  {journal} {\bibinfo  {journal} {Nat. Photonics}\ }\textbf
  {\bibinfo {volume} {5}},\ \bibinfo {pages} {628} (\bibinfo {year}
  {2011})}\BibitemShut {NoStop}%
\bibitem [{\citenamefont {Bao}\ \emph {et~al.}(2008)\citenamefont {Bao},
  \citenamefont {Qian}, \citenamefont {Yang}, \citenamefont {Zhang},
  \citenamefont {Chen}, \citenamefont {Yang},\ and\ \citenamefont
  {Pan}}]{bao2008generation}%
  \BibitemOpen
  \bibfield  {author} {\bibinfo {author} {\bibfnamefont {X.-H.}\ \bibnamefont
  {Bao}}, \bibinfo {author} {\bibfnamefont {Y.}~\bibnamefont {Qian}}, \bibinfo
  {author} {\bibfnamefont {J.}~\bibnamefont {Yang}}, \bibinfo {author}
  {\bibfnamefont {H.}~\bibnamefont {Zhang}}, \bibinfo {author} {\bibfnamefont
  {Z.-B.}\ \bibnamefont {Chen}}, \bibinfo {author} {\bibfnamefont
  {T.}~\bibnamefont {Yang}}, \ and\ \bibinfo {author} {\bibfnamefont {J.-W.}\
  \bibnamefont {Pan}},\ }\href@noop {} {\bibfield  {journal} {\bibinfo
  {journal} {Phys. Rev. Lett.}\ }\textbf {\bibinfo {volume} {101}},\ \bibinfo
  {pages} {190501} (\bibinfo {year} {2008})}\BibitemShut {NoStop}%
\bibitem [{\citenamefont {Arenskotter}\ \emph {et~al.}(2017)\citenamefont
  {Arenskotter}, \citenamefont {Kucera},\ and\ \citenamefont
  {Eschner}}]{arenskotter2017polarization}%
  \BibitemOpen
  \bibfield  {author} {\bibinfo {author} {\bibfnamefont {J.}~\bibnamefont
  {Arenskotter}}, \bibinfo {author} {\bibfnamefont {S.}~\bibnamefont {Kucera}},
  \ and\ \bibinfo {author} {\bibfnamefont {J.}~\bibnamefont {Eschner}},\ }in\
  \href@noop {} {\emph {\bibinfo {booktitle} {Proc. CLEO/Europe-EQEC}}}\
  (\bibinfo {organization} {IEEE},\ \bibinfo {year} {2017})\ pp.\ \bibinfo
  {pages} {1--1}\BibitemShut {NoStop}%
\bibitem [{\citenamefont {Lu}\ \emph {et~al.}(2003)\citenamefont {Lu},
  \citenamefont {Campbell},\ and\ \citenamefont {Ou}}]{lu2003mode}%
  \BibitemOpen
  \bibfield  {author} {\bibinfo {author} {\bibfnamefont {Y.~J.}\ \bibnamefont
  {Lu}}, \bibinfo {author} {\bibfnamefont {R.~L.}\ \bibnamefont {Campbell}}, \
  and\ \bibinfo {author} {\bibfnamefont {Z.~Y.}\ \bibnamefont {Ou}},\
  }\href@noop {} {\bibfield  {journal} {\bibinfo  {journal} {Phys. Rev. Lett.}\
  }\textbf {\bibinfo {volume} {91}},\ \bibinfo {pages} {163602} (\bibinfo
  {year} {2003})}\BibitemShut {NoStop}%
\bibitem [{\citenamefont {Zavatta}\ \emph {et~al.}(2008)\citenamefont
  {Zavatta}, \citenamefont {Parigi},\ and\ \citenamefont
  {Bellini}}]{zavatta2008toward}%
  \BibitemOpen
  \bibfield  {author} {\bibinfo {author} {\bibfnamefont {A.}~\bibnamefont
  {Zavatta}}, \bibinfo {author} {\bibfnamefont {V.}~\bibnamefont {Parigi}}, \
  and\ \bibinfo {author} {\bibfnamefont {M.}~\bibnamefont {Bellini}},\
  }\href@noop {} {\bibfield  {journal} {\bibinfo  {journal} {Phys. Rev. A}\
  }\textbf {\bibinfo {volume} {78}},\ \bibinfo {pages} {033809} (\bibinfo
  {year} {2008})}\BibitemShut {NoStop}%
\bibitem [{\citenamefont {Xie}\ \emph {et~al.}(2015)\citenamefont {Xie},
  \citenamefont {Zhong}, \citenamefont {Shrestha}, \citenamefont {Xu},
  \citenamefont {Liang}, \citenamefont {Gong}, \citenamefont {Bienfang},
  \citenamefont {Restelli}, \citenamefont {Shapiro}, \citenamefont {Wong} \emph
  {et~al.}}]{xie2015harnessing}%
  \BibitemOpen
  \bibfield  {author} {\bibinfo {author} {\bibfnamefont {Z.}~\bibnamefont
  {Xie}}, \bibinfo {author} {\bibfnamefont {T.}~\bibnamefont {Zhong}}, \bibinfo
  {author} {\bibfnamefont {S.}~\bibnamefont {Shrestha}}, \bibinfo {author}
  {\bibfnamefont {X.}~\bibnamefont {Xu}}, \bibinfo {author} {\bibfnamefont
  {J.}~\bibnamefont {Liang}}, \bibinfo {author} {\bibfnamefont {Y.-X.}\
  \bibnamefont {Gong}}, \bibinfo {author} {\bibfnamefont {J.~C.}\ \bibnamefont
  {Bienfang}}, \bibinfo {author} {\bibfnamefont {A.}~\bibnamefont {Restelli}},
  \bibinfo {author} {\bibfnamefont {J.~H.}\ \bibnamefont {Shapiro}}, \bibinfo
  {author} {\bibfnamefont {F.~N.}\ \bibnamefont {Wong}},  \emph {et~al.},\
  }\href@noop {} {\bibfield  {journal} {\bibinfo  {journal} {Nat. Photonics}\
  }\textbf {\bibinfo {volume} {9}},\ \bibinfo {pages} {536} (\bibinfo {year}
  {2015})}\BibitemShut {NoStop}%
\bibitem [{\citenamefont {Riel{\"a}nder}\ \emph {et~al.}(2017)\citenamefont
  {Riel{\"a}nder}, \citenamefont {Lenhard}, \citenamefont {Far{\`\i}as},
  \citenamefont {M{\'a}ttar}, \citenamefont {Cavalcanti}, \citenamefont
  {Mazzera}, \citenamefont {Ac{\'\i}n},\ and\ \citenamefont
  {de~Riedmatten}}]{rielander2017frequency}%
  \BibitemOpen
  \bibfield  {author} {\bibinfo {author} {\bibfnamefont {D.}~\bibnamefont
  {Riel{\"a}nder}}, \bibinfo {author} {\bibfnamefont {A.}~\bibnamefont
  {Lenhard}}, \bibinfo {author} {\bibfnamefont {O.~J.}\ \bibnamefont
  {Far{\`\i}as}}, \bibinfo {author} {\bibfnamefont {A.}~\bibnamefont
  {M{\'a}ttar}}, \bibinfo {author} {\bibfnamefont {D.}~\bibnamefont
  {Cavalcanti}}, \bibinfo {author} {\bibfnamefont {M.}~\bibnamefont {Mazzera}},
  \bibinfo {author} {\bibfnamefont {A.}~\bibnamefont {Ac{\'\i}n}}, \ and\
  \bibinfo {author} {\bibfnamefont {H.}~\bibnamefont {de~Riedmatten}},\
  }\href@noop {} {\bibfield  {journal} {\bibinfo  {journal} {Quant. Sci.
  Tech.}\ }\textbf {\bibinfo {volume} {3}},\ \bibinfo {pages} {014007}
  (\bibinfo {year} {2017})}\BibitemShut {NoStop}%
\bibitem [{\citenamefont {Caspani}\ \emph {et~al.}(2017)\citenamefont
  {Caspani}, \citenamefont {Xiong}, \citenamefont {Eggleton}, \citenamefont
  {Bajoni}, \citenamefont {Liscidini}, \citenamefont {Galli}, \citenamefont
  {Morandotti},\ and\ \citenamefont {Moss}}]{caspani2017integrated}%
  \BibitemOpen
  \bibfield  {author} {\bibinfo {author} {\bibfnamefont {L.}~\bibnamefont
  {Caspani}}, \bibinfo {author} {\bibfnamefont {C.}~\bibnamefont {Xiong}},
  \bibinfo {author} {\bibfnamefont {B.~J.}\ \bibnamefont {Eggleton}}, \bibinfo
  {author} {\bibfnamefont {D.}~\bibnamefont {Bajoni}}, \bibinfo {author}
  {\bibfnamefont {M.}~\bibnamefont {Liscidini}}, \bibinfo {author}
  {\bibfnamefont {M.}~\bibnamefont {Galli}}, \bibinfo {author} {\bibfnamefont
  {R.}~\bibnamefont {Morandotti}}, \ and\ \bibinfo {author} {\bibfnamefont
  {D.~J.}\ \bibnamefont {Moss}},\ }\href@noop {} {\bibfield  {journal}
  {\bibinfo  {journal} {Light: Sci. Appl.}\ }\textbf {\bibinfo {volume} {6}},\
  \bibinfo {pages} {e17100} (\bibinfo {year} {2017})}\BibitemShut {NoStop}%
\bibitem [{\citenamefont {Wang}\ \emph {et~al.}(2020)\citenamefont {Wang},
  \citenamefont {Sciarrino}, \citenamefont {Laing},\ and\ \citenamefont
  {Thompson}}]{wang2020integrated}%
  \BibitemOpen
  \bibfield  {author} {\bibinfo {author} {\bibfnamefont {J.}~\bibnamefont
  {Wang}}, \bibinfo {author} {\bibfnamefont {F.}~\bibnamefont {Sciarrino}},
  \bibinfo {author} {\bibfnamefont {A.}~\bibnamefont {Laing}}, \ and\ \bibinfo
  {author} {\bibfnamefont {M.~G.}\ \bibnamefont {Thompson}},\ }\href@noop {}
  {\bibfield  {journal} {\bibinfo  {journal} {Nat. Photonics}\ }\textbf
  {\bibinfo {volume} {14}},\ \bibinfo {pages} {273} (\bibinfo {year}
  {2020})}\BibitemShut {NoStop}%
\bibitem [{\citenamefont {Vall{\'e}s}\ \emph {et~al.}(2013)\citenamefont
  {Vall{\'e}s}, \citenamefont {Hendrych}, \citenamefont {Svozilik},
  \citenamefont {Machulka}, \citenamefont {Abolghasem}, \citenamefont {Kang},
  \citenamefont {Bijlani}, \citenamefont {Helmy},\ and\ \citenamefont
  {Torres}}]{valles2013generation}%
  \BibitemOpen
  \bibfield  {author} {\bibinfo {author} {\bibfnamefont {A.}~\bibnamefont
  {Vall{\'e}s}}, \bibinfo {author} {\bibfnamefont {M.}~\bibnamefont
  {Hendrych}}, \bibinfo {author} {\bibfnamefont {J.}~\bibnamefont {Svozilik}},
  \bibinfo {author} {\bibfnamefont {R.}~\bibnamefont {Machulka}}, \bibinfo
  {author} {\bibfnamefont {P.}~\bibnamefont {Abolghasem}}, \bibinfo {author}
  {\bibfnamefont {D.}~\bibnamefont {Kang}}, \bibinfo {author} {\bibfnamefont
  {B.}~\bibnamefont {Bijlani}}, \bibinfo {author} {\bibfnamefont
  {A.}~\bibnamefont {Helmy}}, \ and\ \bibinfo {author} {\bibfnamefont
  {J.}~\bibnamefont {Torres}},\ }\href@noop {} {\bibfield  {journal} {\bibinfo
  {journal} {Opt. Express}\ }\textbf {\bibinfo {volume} {21}},\ \bibinfo
  {pages} {10841} (\bibinfo {year} {2013})}\BibitemShut {NoStop}%
\bibitem [{\citenamefont {Helt}\ \emph {et~al.}(2012)\citenamefont {Helt},
  \citenamefont {Liscidini},\ and\ \citenamefont {Sipe}}]{helt2012does}%
  \BibitemOpen
  \bibfield  {author} {\bibinfo {author} {\bibfnamefont {L.~G.}\ \bibnamefont
  {Helt}}, \bibinfo {author} {\bibfnamefont {M.}~\bibnamefont {Liscidini}}, \
  and\ \bibinfo {author} {\bibfnamefont {J.~E.}\ \bibnamefont {Sipe}},\
  }\href@noop {} {\bibfield  {journal} {\bibinfo  {journal} {J. Opt. Soc. Am.
  B}\ }\textbf {\bibinfo {volume} {29}},\ \bibinfo {pages} {2199} (\bibinfo
  {year} {2012})}\BibitemShut {NoStop}%
\bibitem [{\citenamefont {Christ}\ \emph {et~al.}(2009)\citenamefont {Christ},
  \citenamefont {Laiho}, \citenamefont {Eckstein}, \citenamefont {Lauckner},
  \citenamefont {Mosley},\ and\ \citenamefont {Silberhorn}}]{Christ:2009aa}%
  \BibitemOpen
  \bibfield  {author} {\bibinfo {author} {\bibfnamefont {A.}~\bibnamefont
  {Christ}}, \bibinfo {author} {\bibfnamefont {K.}~\bibnamefont {Laiho}},
  \bibinfo {author} {\bibfnamefont {A.}~\bibnamefont {Eckstein}}, \bibinfo
  {author} {\bibfnamefont {T.}~\bibnamefont {Lauckner}}, \bibinfo {author}
  {\bibfnamefont {P.~J.}\ \bibnamefont {Mosley}}, \ and\ \bibinfo {author}
  {\bibfnamefont {C.}~\bibnamefont {Silberhorn}},\ }\href {\doibase
  10.1103/PhysRevA.80.033829} {\bibfield  {journal} {\bibinfo  {journal} {Phys.
  Rev. A}\ }\textbf {\bibinfo {volume} {80}},\ \bibinfo {pages} {033829}
  (\bibinfo {year} {2009})}\BibitemShut {NoStop}%
\bibitem [{\citenamefont {Horn}\ \emph {et~al.}(2013)\citenamefont {Horn},
  \citenamefont {Kolenderski}, \citenamefont {Kang}, \citenamefont
  {Abolghasem}, \citenamefont {Scarcella}, \citenamefont {Della~Frera},
  \citenamefont {Tosi}, \citenamefont {Helt}, \citenamefont {Zhukovsky},
  \citenamefont {Sipe} \emph {et~al.}}]{horn2013inherent}%
  \BibitemOpen
  \bibfield  {author} {\bibinfo {author} {\bibfnamefont {R.~T.}\ \bibnamefont
  {Horn}}, \bibinfo {author} {\bibfnamefont {P.}~\bibnamefont {Kolenderski}},
  \bibinfo {author} {\bibfnamefont {D.}~\bibnamefont {Kang}}, \bibinfo {author}
  {\bibfnamefont {P.}~\bibnamefont {Abolghasem}}, \bibinfo {author}
  {\bibfnamefont {C.}~\bibnamefont {Scarcella}}, \bibinfo {author}
  {\bibfnamefont {A.}~\bibnamefont {Della~Frera}}, \bibinfo {author}
  {\bibfnamefont {A.}~\bibnamefont {Tosi}}, \bibinfo {author} {\bibfnamefont
  {L.~G.}\ \bibnamefont {Helt}}, \bibinfo {author} {\bibfnamefont {S.~V.}\
  \bibnamefont {Zhukovsky}}, \bibinfo {author} {\bibfnamefont {J.~E.}\
  \bibnamefont {Sipe}},  \emph {et~al.},\ }\href@noop {} {\bibfield  {journal}
  {\bibinfo  {journal} {Sci. Rep.}\ }\textbf {\bibinfo {volume} {3}},\ \bibinfo
  {pages} {2314} (\bibinfo {year} {2013})}\BibitemShut {NoStop}%
\bibitem [{\citenamefont {Yoshizawa}\ \emph {et~al.}(2003)\citenamefont
  {Yoshizawa}, \citenamefont {Kaji},\ and\ \citenamefont
  {Tsuchida}}]{yoshizawa2003generation}%
  \BibitemOpen
  \bibfield  {author} {\bibinfo {author} {\bibfnamefont {A.}~\bibnamefont
  {Yoshizawa}}, \bibinfo {author} {\bibfnamefont {R.}~\bibnamefont {Kaji}}, \
  and\ \bibinfo {author} {\bibfnamefont {H.}~\bibnamefont {Tsuchida}},\
  }\href@noop {} {\bibfield  {journal} {\bibinfo  {journal} {Electron. Lett.}\
  }\textbf {\bibinfo {volume} {39}},\ \bibinfo {pages} {621} (\bibinfo {year}
  {2003})}\BibitemShut {NoStop}%
\bibitem [{\citenamefont {Herrmann}\ \emph {et~al.}(2013)\citenamefont
  {Herrmann}, \citenamefont {Yang}, \citenamefont {Thomas}, \citenamefont
  {Poppe}, \citenamefont {Sohler},\ and\ \citenamefont
  {Silberhorn}}]{herrmann2013post}%
  \BibitemOpen
  \bibfield  {author} {\bibinfo {author} {\bibfnamefont {H.}~\bibnamefont
  {Herrmann}}, \bibinfo {author} {\bibfnamefont {X.}~\bibnamefont {Yang}},
  \bibinfo {author} {\bibfnamefont {A.}~\bibnamefont {Thomas}}, \bibinfo
  {author} {\bibfnamefont {A.}~\bibnamefont {Poppe}}, \bibinfo {author}
  {\bibfnamefont {W.}~\bibnamefont {Sohler}}, \ and\ \bibinfo {author}
  {\bibfnamefont {C.}~\bibnamefont {Silberhorn}},\ }\href@noop {} {\bibfield
  {journal} {\bibinfo  {journal} {Opt. Express}\ }\textbf {\bibinfo {volume}
  {21}},\ \bibinfo {pages} {27981} (\bibinfo {year} {2013})}\BibitemShut
  {NoStop}%
\bibitem [{\citenamefont {Sun}\ \emph {et~al.}(2019)\citenamefont {Sun},
  \citenamefont {Wu}, \citenamefont {Duan}, \citenamefont {Zhou}, \citenamefont
  {Xia}, \citenamefont {Xu}, \citenamefont {Xie}, \citenamefont {Gong},\ and\
  \citenamefont {Zhu}}]{sun2019compact}%
  \BibitemOpen
  \bibfield  {author} {\bibinfo {author} {\bibfnamefont {C.-W.}\ \bibnamefont
  {Sun}}, \bibinfo {author} {\bibfnamefont {S.-H.}\ \bibnamefont {Wu}},
  \bibinfo {author} {\bibfnamefont {J.-C.}\ \bibnamefont {Duan}}, \bibinfo
  {author} {\bibfnamefont {J.-W.}\ \bibnamefont {Zhou}}, \bibinfo {author}
  {\bibfnamefont {J.-L.}\ \bibnamefont {Xia}}, \bibinfo {author} {\bibfnamefont
  {P.}~\bibnamefont {Xu}}, \bibinfo {author} {\bibfnamefont {Z.}~\bibnamefont
  {Xie}}, \bibinfo {author} {\bibfnamefont {Y.-X.}\ \bibnamefont {Gong}}, \
  and\ \bibinfo {author} {\bibfnamefont {S.-N.}\ \bibnamefont {Zhu}},\
  }\href@noop {} {\bibfield  {journal} {\bibinfo  {journal} {Opt. Lett.}\
  }\textbf {\bibinfo {volume} {44}},\ \bibinfo {pages} {5598} (\bibinfo {year}
  {2019})}\BibitemShut {NoStop}%
\bibitem [{\citenamefont {Kaiser}\ \emph {et~al.}(2012)\citenamefont {Kaiser},
  \citenamefont {Issautier}, \citenamefont {Ngah}, \citenamefont
  {D{\u{a}}nil{\u{a}}}, \citenamefont {Herrmann}, \citenamefont {Sohler},
  \citenamefont {Martin},\ and\ \citenamefont {Tanzilli}}]{kaiser2012high}%
  \BibitemOpen
  \bibfield  {author} {\bibinfo {author} {\bibfnamefont {F.}~\bibnamefont
  {Kaiser}}, \bibinfo {author} {\bibfnamefont {A.}~\bibnamefont {Issautier}},
  \bibinfo {author} {\bibfnamefont {L.~A.}\ \bibnamefont {Ngah}}, \bibinfo
  {author} {\bibfnamefont {O.}~\bibnamefont {D{\u{a}}nil{\u{a}}}}, \bibinfo
  {author} {\bibfnamefont {H.}~\bibnamefont {Herrmann}}, \bibinfo {author}
  {\bibfnamefont {W.}~\bibnamefont {Sohler}}, \bibinfo {author} {\bibfnamefont
  {A.}~\bibnamefont {Martin}}, \ and\ \bibinfo {author} {\bibfnamefont
  {S.}~\bibnamefont {Tanzilli}},\ }\href@noop {} {\bibfield  {journal}
  {\bibinfo  {journal} {New J. Phys.}\ }\textbf {\bibinfo {volume} {14}},\
  \bibinfo {pages} {085015} (\bibinfo {year} {2012})}\BibitemShut {NoStop}%
\bibitem [{\citenamefont {Kaiser}\ \emph {et~al.}(2014)\citenamefont {Kaiser},
  \citenamefont {Ngah}, \citenamefont {Issautier}, \citenamefont {Delord},
  \citenamefont {Aktas}, \citenamefont {De~Micheli}, \citenamefont {Kastberg},
  \citenamefont {Labont{\'e}}, \citenamefont {Alibart}, \citenamefont {Martin}
  \emph {et~al.}}]{kaiser2014polarization}%
  \BibitemOpen
  \bibfield  {author} {\bibinfo {author} {\bibfnamefont {F.}~\bibnamefont
  {Kaiser}}, \bibinfo {author} {\bibfnamefont {L.~A.}\ \bibnamefont {Ngah}},
  \bibinfo {author} {\bibfnamefont {A.}~\bibnamefont {Issautier}}, \bibinfo
  {author} {\bibfnamefont {T.}~\bibnamefont {Delord}}, \bibinfo {author}
  {\bibfnamefont {D.}~\bibnamefont {Aktas}}, \bibinfo {author} {\bibfnamefont
  {M.}~\bibnamefont {De~Micheli}}, \bibinfo {author} {\bibfnamefont
  {A.}~\bibnamefont {Kastberg}}, \bibinfo {author} {\bibfnamefont
  {L.}~\bibnamefont {Labont{\'e}}}, \bibinfo {author} {\bibfnamefont
  {O.}~\bibnamefont {Alibart}}, \bibinfo {author} {\bibfnamefont
  {A.}~\bibnamefont {Martin}},  \emph {et~al.},\ }\href@noop {} {\bibfield
  {journal} {\bibinfo  {journal} {Opt. Commun.}\ }\textbf {\bibinfo {volume}
  {327}},\ \bibinfo {pages} {7} (\bibinfo {year} {2014})}\BibitemShut {NoStop}%
\bibitem [{\citenamefont {Bonfrate}\ \emph {et~al.}(1999)\citenamefont
  {Bonfrate}, \citenamefont {Pruneri}, \citenamefont {Kazansky}, \citenamefont
  {Tapster},\ and\ \citenamefont {Rarity}}]{bonfrate1999parametric}%
  \BibitemOpen
  \bibfield  {author} {\bibinfo {author} {\bibfnamefont {G.}~\bibnamefont
  {Bonfrate}}, \bibinfo {author} {\bibfnamefont {V.}~\bibnamefont {Pruneri}},
  \bibinfo {author} {\bibfnamefont {P.}~\bibnamefont {Kazansky}}, \bibinfo
  {author} {\bibfnamefont {P.}~\bibnamefont {Tapster}}, \ and\ \bibinfo
  {author} {\bibfnamefont {J.}~\bibnamefont {Rarity}},\ }\href@noop {}
  {\bibfield  {journal} {\bibinfo  {journal} {Appl. Phys. Lett.}\ }\textbf
  {\bibinfo {volume} {75}},\ \bibinfo {pages} {2356} (\bibinfo {year}
  {1999})}\BibitemShut {NoStop}%
\bibitem [{\citenamefont {Zhu}\ \emph {et~al.}(2013)\citenamefont {Zhu},
  \citenamefont {Tang}, \citenamefont {Qian}, \citenamefont {Helt},
  \citenamefont {Liscidini}, \citenamefont {Sipe}, \citenamefont {Corbari},
  \citenamefont {Canagasabey}, \citenamefont {Ibsen},\ and\ \citenamefont
  {Kazansky}}]{zhu2013poled}%
  \BibitemOpen
  \bibfield  {author} {\bibinfo {author} {\bibfnamefont {E.}~\bibnamefont
  {Zhu}}, \bibinfo {author} {\bibfnamefont {Z.}~\bibnamefont {Tang}}, \bibinfo
  {author} {\bibfnamefont {L.}~\bibnamefont {Qian}}, \bibinfo {author}
  {\bibfnamefont {L.}~\bibnamefont {Helt}}, \bibinfo {author} {\bibfnamefont
  {M.}~\bibnamefont {Liscidini}}, \bibinfo {author} {\bibfnamefont
  {J.}~\bibnamefont {Sipe}}, \bibinfo {author} {\bibfnamefont {C.}~\bibnamefont
  {Corbari}}, \bibinfo {author} {\bibfnamefont {A.}~\bibnamefont
  {Canagasabey}}, \bibinfo {author} {\bibfnamefont {M.}~\bibnamefont {Ibsen}},
  \ and\ \bibinfo {author} {\bibfnamefont {P.}~\bibnamefont {Kazansky}},\
  }\href@noop {} {\bibfield  {journal} {\bibinfo  {journal} {Opt. Lett.}\
  }\textbf {\bibinfo {volume} {38}},\ \bibinfo {pages} {4397} (\bibinfo {year}
  {2013})}\BibitemShut {NoStop}%
\bibitem [{\citenamefont {Chen}\ \emph {et~al.}(2017)\citenamefont {Chen},
  \citenamefont {Zhu}, \citenamefont {Riazi}, \citenamefont {Gladyshev},
  \citenamefont {Corbari}, \citenamefont {Ibsen}, \citenamefont {Kazansky},\
  and\ \citenamefont {Qian}}]{chen2017compensation}%
  \BibitemOpen
  \bibfield  {author} {\bibinfo {author} {\bibfnamefont {C.}~\bibnamefont
  {Chen}}, \bibinfo {author} {\bibfnamefont {E.~Y.}\ \bibnamefont {Zhu}},
  \bibinfo {author} {\bibfnamefont {A.}~\bibnamefont {Riazi}}, \bibinfo
  {author} {\bibfnamefont {A.~V.}\ \bibnamefont {Gladyshev}}, \bibinfo {author}
  {\bibfnamefont {C.}~\bibnamefont {Corbari}}, \bibinfo {author} {\bibfnamefont
  {M.}~\bibnamefont {Ibsen}}, \bibinfo {author} {\bibfnamefont {P.~G.}\
  \bibnamefont {Kazansky}}, \ and\ \bibinfo {author} {\bibfnamefont
  {L.}~\bibnamefont {Qian}},\ }\href@noop {} {\bibfield  {journal} {\bibinfo
  {journal} {Opt. Express}\ }\textbf {\bibinfo {volume} {25}},\ \bibinfo
  {pages} {22667} (\bibinfo {year} {2017})}\BibitemShut {NoStop}%
\bibitem [{\citenamefont {Shalm}\ \emph {et~al.}(2015)\citenamefont {Shalm},
  \citenamefont {Meyer-Scott}, \citenamefont {Christensen}, \citenamefont
  {Bierhorst}, \citenamefont {Wayne}, \citenamefont {Stevens}, \citenamefont
  {Gerrits}, \citenamefont {Glancy}, \citenamefont {Hamel}, \citenamefont
  {Allman} \emph {et~al.}}]{shalm2015strong}%
  \BibitemOpen
  \bibfield  {author} {\bibinfo {author} {\bibfnamefont {L.~K.}\ \bibnamefont
  {Shalm}}, \bibinfo {author} {\bibfnamefont {E.}~\bibnamefont {Meyer-Scott}},
  \bibinfo {author} {\bibfnamefont {B.~G.}\ \bibnamefont {Christensen}},
  \bibinfo {author} {\bibfnamefont {P.}~\bibnamefont {Bierhorst}}, \bibinfo
  {author} {\bibfnamefont {M.~A.}\ \bibnamefont {Wayne}}, \bibinfo {author}
  {\bibfnamefont {M.~J.}\ \bibnamefont {Stevens}}, \bibinfo {author}
  {\bibfnamefont {T.}~\bibnamefont {Gerrits}}, \bibinfo {author} {\bibfnamefont
  {S.}~\bibnamefont {Glancy}}, \bibinfo {author} {\bibfnamefont {D.~R.}\
  \bibnamefont {Hamel}}, \bibinfo {author} {\bibfnamefont {M.~S.}\ \bibnamefont
  {Allman}},  \emph {et~al.},\ }\href@noop {} {\bibfield  {journal} {\bibinfo
  {journal} {Phys. Rev. Lett.}\ }\textbf {\bibinfo {volume} {115}},\ \bibinfo
  {pages} {250402} (\bibinfo {year} {2015})}\BibitemShut {NoStop}%
\bibitem [{\citenamefont {Giustina}\ \emph {et~al.}(2015)\citenamefont
  {Giustina}, \citenamefont {Versteegh}, \citenamefont {Wengerowsky},
  \citenamefont {Handsteiner}, \citenamefont {Hochrainer}, \citenamefont
  {Phelan}, \citenamefont {Steinlechner}, \citenamefont {Kofler}, \citenamefont
  {Larsson}, \citenamefont {Abell{\'a}n} \emph
  {et~al.}}]{giustina2015significant}%
  \BibitemOpen
  \bibfield  {author} {\bibinfo {author} {\bibfnamefont {M.}~\bibnamefont
  {Giustina}}, \bibinfo {author} {\bibfnamefont {M.~A.}\ \bibnamefont
  {Versteegh}}, \bibinfo {author} {\bibfnamefont {S.}~\bibnamefont
  {Wengerowsky}}, \bibinfo {author} {\bibfnamefont {J.}~\bibnamefont
  {Handsteiner}}, \bibinfo {author} {\bibfnamefont {A.}~\bibnamefont
  {Hochrainer}}, \bibinfo {author} {\bibfnamefont {K.}~\bibnamefont {Phelan}},
  \bibinfo {author} {\bibfnamefont {F.}~\bibnamefont {Steinlechner}}, \bibinfo
  {author} {\bibfnamefont {J.}~\bibnamefont {Kofler}}, \bibinfo {author}
  {\bibfnamefont {J.-{\AA}.}\ \bibnamefont {Larsson}}, \bibinfo {author}
  {\bibfnamefont {C.}~\bibnamefont {Abell{\'a}n}},  \emph {et~al.},\
  }\href@noop {} {\bibfield  {journal} {\bibinfo  {journal} {Phys. Rev. Lett.}\
  }\textbf {\bibinfo {volume} {115}},\ \bibinfo {pages} {250401} (\bibinfo
  {year} {2015})}\BibitemShut {NoStop}%
\bibitem [{\citenamefont {Poh}\ \emph {et~al.}(2015)\citenamefont {Poh},
  \citenamefont {Joshi}, \citenamefont {Cere}, \citenamefont {Cabello},\ and\
  \citenamefont {Kurtsiefer}}]{poh2015approaching}%
  \BibitemOpen
  \bibfield  {author} {\bibinfo {author} {\bibfnamefont {H.~S.}\ \bibnamefont
  {Poh}}, \bibinfo {author} {\bibfnamefont {S.~K.}\ \bibnamefont {Joshi}},
  \bibinfo {author} {\bibfnamefont {A.}~\bibnamefont {Cere}}, \bibinfo {author}
  {\bibfnamefont {A.}~\bibnamefont {Cabello}}, \ and\ \bibinfo {author}
  {\bibfnamefont {C.}~\bibnamefont {Kurtsiefer}},\ }\href@noop {} {\bibfield
  {journal} {\bibinfo  {journal} {Phys. Rev. Lett.}\ }\textbf {\bibinfo
  {volume} {115}},\ \bibinfo {pages} {180408} (\bibinfo {year}
  {2015})}\BibitemShut {NoStop}%
\bibitem [{\citenamefont {Tsujimoto}\ \emph
  {et~al.}(2018{\natexlab{b}})\citenamefont {Tsujimoto}, \citenamefont
  {Tanaka}, \citenamefont {Iwasaki}, \citenamefont {Ikuta}, \citenamefont
  {Miki}, \citenamefont {Yamashita}, \citenamefont {Terai}, \citenamefont
  {Yamamoto}, \citenamefont {Koashi},\ and\ \citenamefont
  {Imoto}}]{tsujimoto2018high}%
  \BibitemOpen
  \bibfield  {author} {\bibinfo {author} {\bibfnamefont {Y.}~\bibnamefont
  {Tsujimoto}}, \bibinfo {author} {\bibfnamefont {M.}~\bibnamefont {Tanaka}},
  \bibinfo {author} {\bibfnamefont {N.}~\bibnamefont {Iwasaki}}, \bibinfo
  {author} {\bibfnamefont {R.}~\bibnamefont {Ikuta}}, \bibinfo {author}
  {\bibfnamefont {S.}~\bibnamefont {Miki}}, \bibinfo {author} {\bibfnamefont
  {T.}~\bibnamefont {Yamashita}}, \bibinfo {author} {\bibfnamefont
  {H.}~\bibnamefont {Terai}}, \bibinfo {author} {\bibfnamefont
  {T.}~\bibnamefont {Yamamoto}}, \bibinfo {author} {\bibfnamefont
  {M.}~\bibnamefont {Koashi}}, \ and\ \bibinfo {author} {\bibfnamefont
  {N.}~\bibnamefont {Imoto}},\ }\href@noop {} {\bibfield  {journal} {\bibinfo
  {journal} {Sci. Rep.}\ }\textbf {\bibinfo {volume} {8}},\ \bibinfo {pages}
  {1} (\bibinfo {year} {2018}{\natexlab{b}})}\BibitemShut {NoStop}%
\bibitem [{\citenamefont {Tanzilli}\ \emph {et~al.}(2012)\citenamefont
  {Tanzilli}, \citenamefont {Martin}, \citenamefont {Kaiser}, \citenamefont
  {De~Micheli}, \citenamefont {Alibart},\ and\ \citenamefont
  {Ostrowsky}}]{tanzilli2012genesis}%
  \BibitemOpen
  \bibfield  {author} {\bibinfo {author} {\bibfnamefont {S.}~\bibnamefont
  {Tanzilli}}, \bibinfo {author} {\bibfnamefont {A.}~\bibnamefont {Martin}},
  \bibinfo {author} {\bibfnamefont {F.}~\bibnamefont {Kaiser}}, \bibinfo
  {author} {\bibfnamefont {M.~P.}\ \bibnamefont {De~Micheli}}, \bibinfo
  {author} {\bibfnamefont {O.}~\bibnamefont {Alibart}}, \ and\ \bibinfo
  {author} {\bibfnamefont {D.~B.}\ \bibnamefont {Ostrowsky}},\ }\href@noop {}
  {\bibfield  {journal} {\bibinfo  {journal} {Laser Photon. Rev.}\ }\textbf
  {\bibinfo {volume} {6}},\ \bibinfo {pages} {115} (\bibinfo {year}
  {2012})}\BibitemShut {NoStop}%
\bibitem [{\citenamefont {Alibart}\ \emph {et~al.}(2016)\citenamefont
  {Alibart}, \citenamefont {D’Auria}, \citenamefont {De~Micheli},
  \citenamefont {Doutre}, \citenamefont {Kaiser}, \citenamefont {Labont{\'e}},
  \citenamefont {Lunghi}, \citenamefont {Picholle},\ and\ \citenamefont
  {Tanzilli}}]{alibart2016quantum}%
  \BibitemOpen
  \bibfield  {author} {\bibinfo {author} {\bibfnamefont {O.}~\bibnamefont
  {Alibart}}, \bibinfo {author} {\bibfnamefont {V.}~\bibnamefont {D’Auria}},
  \bibinfo {author} {\bibfnamefont {M.}~\bibnamefont {De~Micheli}}, \bibinfo
  {author} {\bibfnamefont {F.}~\bibnamefont {Doutre}}, \bibinfo {author}
  {\bibfnamefont {F.}~\bibnamefont {Kaiser}}, \bibinfo {author} {\bibfnamefont
  {L.}~\bibnamefont {Labont{\'e}}}, \bibinfo {author} {\bibfnamefont
  {T.}~\bibnamefont {Lunghi}}, \bibinfo {author} {\bibfnamefont
  {{\'E}.}~\bibnamefont {Picholle}}, \ and\ \bibinfo {author} {\bibfnamefont
  {S.}~\bibnamefont {Tanzilli}},\ }\href@noop {} {\bibfield  {journal}
  {\bibinfo  {journal} {J. Opt.}\ }\textbf {\bibinfo {volume} {18}},\ \bibinfo
  {pages} {104001} (\bibinfo {year} {2016})}\BibitemShut {NoStop}%
\bibitem [{\citenamefont {Kwiat}(1997)}]{kwiat1997hyper}%
  \BibitemOpen
  \bibfield  {author} {\bibinfo {author} {\bibfnamefont {P.~G.}\ \bibnamefont
  {Kwiat}},\ }\href@noop {} {\bibfield  {journal} {\bibinfo  {journal} {J. Mod.
  Opt.}\ }\textbf {\bibinfo {volume} {44}},\ \bibinfo {pages} {2173} (\bibinfo
  {year} {1997})}\BibitemShut {NoStop}%
\bibitem [{\citenamefont {Nagali}\ and\ \citenamefont
  {Sciarrino}(2010)}]{nagali2010generation}%
  \BibitemOpen
  \bibfield  {author} {\bibinfo {author} {\bibfnamefont {E.}~\bibnamefont
  {Nagali}}\ and\ \bibinfo {author} {\bibfnamefont {F.}~\bibnamefont
  {Sciarrino}},\ }\href@noop {} {\bibfield  {journal} {\bibinfo  {journal}
  {Opt. Express}\ }\textbf {\bibinfo {volume} {18}},\ \bibinfo {pages} {18243}
  (\bibinfo {year} {2010})}\BibitemShut {NoStop}%
\bibitem [{\citenamefont {Walborn}\ \emph {et~al.}(2010)\citenamefont
  {Walborn}, \citenamefont {Monken}, \citenamefont {P{\'a}dua},\ and\
  \citenamefont {Ribeiro}}]{walborn2010spatial}%
  \BibitemOpen
  \bibfield  {author} {\bibinfo {author} {\bibfnamefont {S.~P.}\ \bibnamefont
  {Walborn}}, \bibinfo {author} {\bibfnamefont {C.}~\bibnamefont {Monken}},
  \bibinfo {author} {\bibfnamefont {S.}~\bibnamefont {P{\'a}dua}}, \ and\
  \bibinfo {author} {\bibfnamefont {P.~S.}\ \bibnamefont {Ribeiro}},\
  }\href@noop {} {\bibfield  {journal} {\bibinfo  {journal} {Phys. Rep.}\
  }\textbf {\bibinfo {volume} {495}},\ \bibinfo {pages} {87} (\bibinfo {year}
  {2010})}\BibitemShut {NoStop}%
\bibitem [{\citenamefont {Howell}\ \emph {et~al.}(2004)\citenamefont {Howell},
  \citenamefont {Bennink}, \citenamefont {Bentley},\ and\ \citenamefont
  {Boyd}}]{howell2004realization}%
  \BibitemOpen
  \bibfield  {author} {\bibinfo {author} {\bibfnamefont {J.~C.}\ \bibnamefont
  {Howell}}, \bibinfo {author} {\bibfnamefont {R.~S.}\ \bibnamefont {Bennink}},
  \bibinfo {author} {\bibfnamefont {S.~J.}\ \bibnamefont {Bentley}}, \ and\
  \bibinfo {author} {\bibfnamefont {R.}~\bibnamefont {Boyd}},\ }\href@noop {}
  {\bibfield  {journal} {\bibinfo  {journal} {Phys. Rev. Lett.}\ }\textbf
  {\bibinfo {volume} {92}},\ \bibinfo {pages} {210403} (\bibinfo {year}
  {2004})}\BibitemShut {NoStop}%
\bibitem [{\citenamefont {Chen}\ \emph
  {et~al.}(2019{\natexlab{a}})\citenamefont {Chen}, \citenamefont {Ma},
  \citenamefont {Qiu}, \citenamefont {Zhang}, \citenamefont {Zhang},\ and\
  \citenamefont {Boyd}}]{chen2019realization}%
  \BibitemOpen
  \bibfield  {author} {\bibinfo {author} {\bibfnamefont {L.}~\bibnamefont
  {Chen}}, \bibinfo {author} {\bibfnamefont {T.}~\bibnamefont {Ma}}, \bibinfo
  {author} {\bibfnamefont {X.}~\bibnamefont {Qiu}}, \bibinfo {author}
  {\bibfnamefont {D.}~\bibnamefont {Zhang}}, \bibinfo {author} {\bibfnamefont
  {W.}~\bibnamefont {Zhang}}, \ and\ \bibinfo {author} {\bibfnamefont {R.~W.}\
  \bibnamefont {Boyd}},\ }\href@noop {} {\bibfield  {journal} {\bibinfo
  {journal} {Phys. Rev. Lett.}\ }\textbf {\bibinfo {volume} {123}},\ \bibinfo
  {pages} {060403} (\bibinfo {year} {2019}{\natexlab{a}})}\BibitemShut
  {NoStop}%
\bibitem [{\citenamefont {Neves}\ \emph {et~al.}(2007)\citenamefont {Neves},
  \citenamefont {Lima}, \citenamefont {Fonseca}, \citenamefont {Davidovich},\
  and\ \citenamefont {P{\'a}dua}}]{neves2007characterizing}%
  \BibitemOpen
  \bibfield  {author} {\bibinfo {author} {\bibfnamefont {L.}~\bibnamefont
  {Neves}}, \bibinfo {author} {\bibfnamefont {G.}~\bibnamefont {Lima}},
  \bibinfo {author} {\bibfnamefont {E.}~\bibnamefont {Fonseca}}, \bibinfo
  {author} {\bibfnamefont {L.}~\bibnamefont {Davidovich}}, \ and\ \bibinfo
  {author} {\bibfnamefont {S.}~\bibnamefont {P{\'a}dua}},\ }\href@noop {}
  {\bibfield  {journal} {\bibinfo  {journal} {Phys. Rev. A}\ }\textbf {\bibinfo
  {volume} {76}},\ \bibinfo {pages} {032314} (\bibinfo {year}
  {2007})}\BibitemShut {NoStop}%
\bibitem [{\citenamefont {Rossi}\ \emph {et~al.}(2009)\citenamefont {Rossi},
  \citenamefont {Vallone}, \citenamefont {Chiuri}, \citenamefont {De~Martini},\
  and\ \citenamefont {Mataloni}}]{rossi2009multipath}%
  \BibitemOpen
  \bibfield  {author} {\bibinfo {author} {\bibfnamefont {A.}~\bibnamefont
  {Rossi}}, \bibinfo {author} {\bibfnamefont {G.}~\bibnamefont {Vallone}},
  \bibinfo {author} {\bibfnamefont {A.}~\bibnamefont {Chiuri}}, \bibinfo
  {author} {\bibfnamefont {F.}~\bibnamefont {De~Martini}}, \ and\ \bibinfo
  {author} {\bibfnamefont {P.}~\bibnamefont {Mataloni}},\ }\href@noop {}
  {\bibfield  {journal} {\bibinfo  {journal} {Phys. Rev. Lett.}\ }\textbf
  {\bibinfo {volume} {102}},\ \bibinfo {pages} {153902} (\bibinfo {year}
  {2009})}\BibitemShut {NoStop}%
\bibitem [{\citenamefont {Law}\ and\ \citenamefont
  {Eberly}(2004)}]{law2004analysis}%
  \BibitemOpen
  \bibfield  {author} {\bibinfo {author} {\bibfnamefont {C.}~\bibnamefont
  {Law}}\ and\ \bibinfo {author} {\bibfnamefont {J.}~\bibnamefont {Eberly}},\
  }\href@noop {} {\bibfield  {journal} {\bibinfo  {journal} {Phys. Rev. Lett.}\
  }\textbf {\bibinfo {volume} {92}},\ \bibinfo {pages} {127903} (\bibinfo
  {year} {2004})}\BibitemShut {NoStop}%
\bibitem [{\citenamefont {Just}\ \emph {et~al.}(2013)\citenamefont {Just},
  \citenamefont {Cavanna}, \citenamefont {Chekhova},\ and\ \citenamefont
  {Leuchs}}]{just2013transverse}%
  \BibitemOpen
  \bibfield  {author} {\bibinfo {author} {\bibfnamefont {F.}~\bibnamefont
  {Just}}, \bibinfo {author} {\bibfnamefont {A.}~\bibnamefont {Cavanna}},
  \bibinfo {author} {\bibfnamefont {M.~V.}\ \bibnamefont {Chekhova}}, \ and\
  \bibinfo {author} {\bibfnamefont {G.}~\bibnamefont {Leuchs}},\ }\href@noop {}
  {\bibfield  {journal} {\bibinfo  {journal} {New J. Phys.}\ }\textbf {\bibinfo
  {volume} {15}},\ \bibinfo {pages} {083015} (\bibinfo {year}
  {2013})}\BibitemShut {NoStop}%
\bibitem [{\citenamefont {Neves}\ \emph {et~al.}(2005)\citenamefont {Neves},
  \citenamefont {Lima}, \citenamefont {G{\'o}mez}, \citenamefont {Monken},
  \citenamefont {Saavedra},\ and\ \citenamefont
  {P{\'a}dua}}]{neves2005generation}%
  \BibitemOpen
  \bibfield  {author} {\bibinfo {author} {\bibfnamefont {L.}~\bibnamefont
  {Neves}}, \bibinfo {author} {\bibfnamefont {G.}~\bibnamefont {Lima}},
  \bibinfo {author} {\bibfnamefont {J.~A.}\ \bibnamefont {G{\'o}mez}}, \bibinfo
  {author} {\bibfnamefont {C.}~\bibnamefont {Monken}}, \bibinfo {author}
  {\bibfnamefont {C.}~\bibnamefont {Saavedra}}, \ and\ \bibinfo {author}
  {\bibfnamefont {S.}~\bibnamefont {P{\'a}dua}},\ }\href@noop {} {\bibfield
  {journal} {\bibinfo  {journal} {Phys. Rev. Lett.}\ }\textbf {\bibinfo
  {volume} {94}},\ \bibinfo {pages} {100501} (\bibinfo {year}
  {2005})}\BibitemShut {NoStop}%
\bibitem [{\citenamefont {O’Sullivan-Hale}\ \emph {et~al.}(2005)\citenamefont
  {O’Sullivan-Hale}, \citenamefont {Khan}, \citenamefont {Boyd},\ and\
  \citenamefont {Howell}}]{o2005pixel}%
  \BibitemOpen
  \bibfield  {author} {\bibinfo {author} {\bibfnamefont {M.~N.}\ \bibnamefont
  {O’Sullivan-Hale}}, \bibinfo {author} {\bibfnamefont {I.~A.}\ \bibnamefont
  {Khan}}, \bibinfo {author} {\bibfnamefont {R.~W.}\ \bibnamefont {Boyd}}, \
  and\ \bibinfo {author} {\bibfnamefont {J.~C.}\ \bibnamefont {Howell}},\
  }\href@noop {} {\bibfield  {journal} {\bibinfo  {journal} {Phys. Rev. Lett.}\
  }\textbf {\bibinfo {volume} {94}},\ \bibinfo {pages} {220501} (\bibinfo
  {year} {2005})}\BibitemShut {NoStop}%
\bibitem [{\citenamefont {Lima}\ \emph {et~al.}(2009)\citenamefont {Lima},
  \citenamefont {Vargas}, \citenamefont {Neves}, \citenamefont {Guzm{\'a}n},\
  and\ \citenamefont {Saavedra}}]{lima2009manipulating}%
  \BibitemOpen
  \bibfield  {author} {\bibinfo {author} {\bibfnamefont {G.}~\bibnamefont
  {Lima}}, \bibinfo {author} {\bibfnamefont {A.}~\bibnamefont {Vargas}},
  \bibinfo {author} {\bibfnamefont {L.}~\bibnamefont {Neves}}, \bibinfo
  {author} {\bibfnamefont {R.}~\bibnamefont {Guzm{\'a}n}}, \ and\ \bibinfo
  {author} {\bibfnamefont {C.}~\bibnamefont {Saavedra}},\ }\href@noop {}
  {\bibfield  {journal} {\bibinfo  {journal} {Opt. Express}\ }\textbf {\bibinfo
  {volume} {17}},\ \bibinfo {pages} {10688} (\bibinfo {year}
  {2009})}\BibitemShut {NoStop}%
\bibitem [{\citenamefont {Ostermeyer}\ \emph {et~al.}(2009)\citenamefont
  {Ostermeyer}, \citenamefont {Korn}, \citenamefont {Puhlmann}, \citenamefont
  {Henkel},\ and\ \citenamefont {Eisert}}]{ostermeyer2009two}%
  \BibitemOpen
  \bibfield  {author} {\bibinfo {author} {\bibfnamefont {M.}~\bibnamefont
  {Ostermeyer}}, \bibinfo {author} {\bibfnamefont {D.}~\bibnamefont {Korn}},
  \bibinfo {author} {\bibfnamefont {D.}~\bibnamefont {Puhlmann}}, \bibinfo
  {author} {\bibfnamefont {C.}~\bibnamefont {Henkel}}, \ and\ \bibinfo {author}
  {\bibfnamefont {J.}~\bibnamefont {Eisert}},\ }\href@noop {} {\bibfield
  {journal} {\bibinfo  {journal} {J. Mod. Opt.}\ }\textbf {\bibinfo {volume}
  {56}},\ \bibinfo {pages} {1829} (\bibinfo {year} {2009})}\BibitemShut
  {NoStop}%
\bibitem [{\citenamefont {Edgar}\ \emph {et~al.}(2012)\citenamefont {Edgar},
  \citenamefont {Tasca}, \citenamefont {Izdebski}, \citenamefont {Warburton},
  \citenamefont {Leach}, \citenamefont {Agnew}, \citenamefont {Buller},
  \citenamefont {Boyd},\ and\ \citenamefont {Padgett}}]{edgar2012imaging}%
  \BibitemOpen
  \bibfield  {author} {\bibinfo {author} {\bibfnamefont {M.~P.}\ \bibnamefont
  {Edgar}}, \bibinfo {author} {\bibfnamefont {D.~S.}\ \bibnamefont {Tasca}},
  \bibinfo {author} {\bibfnamefont {F.}~\bibnamefont {Izdebski}}, \bibinfo
  {author} {\bibfnamefont {R.~E.}\ \bibnamefont {Warburton}}, \bibinfo {author}
  {\bibfnamefont {J.}~\bibnamefont {Leach}}, \bibinfo {author} {\bibfnamefont
  {M.}~\bibnamefont {Agnew}}, \bibinfo {author} {\bibfnamefont {G.~S.}\
  \bibnamefont {Buller}}, \bibinfo {author} {\bibfnamefont {R.~W.}\
  \bibnamefont {Boyd}}, \ and\ \bibinfo {author} {\bibfnamefont {M.~J.}\
  \bibnamefont {Padgett}},\ }\href@noop {} {\bibfield  {journal} {\bibinfo
  {journal} {Nat. Commun.}\ }\textbf {\bibinfo {volume} {3}},\ \bibinfo {pages}
  {984} (\bibinfo {year} {2012})}\BibitemShut {NoStop}%
\bibitem [{\citenamefont {Yu}\ \emph {et~al.}(2008)\citenamefont {Yu},
  \citenamefont {Xu}, \citenamefont {Xie}, \citenamefont {Wang}, \citenamefont
  {Leng}, \citenamefont {Zhao}, \citenamefont {Zhu},\ and\ \citenamefont
  {Ming}}]{yu2008transforming}%
  \BibitemOpen
  \bibfield  {author} {\bibinfo {author} {\bibfnamefont {X.}~\bibnamefont
  {Yu}}, \bibinfo {author} {\bibfnamefont {P.}~\bibnamefont {Xu}}, \bibinfo
  {author} {\bibfnamefont {Z.}~\bibnamefont {Xie}}, \bibinfo {author}
  {\bibfnamefont {J.}~\bibnamefont {Wang}}, \bibinfo {author} {\bibfnamefont
  {H.}~\bibnamefont {Leng}}, \bibinfo {author} {\bibfnamefont {J.}~\bibnamefont
  {Zhao}}, \bibinfo {author} {\bibfnamefont {S.}~\bibnamefont {Zhu}}, \ and\
  \bibinfo {author} {\bibfnamefont {N.}~\bibnamefont {Ming}},\ }\href@noop {}
  {\bibfield  {journal} {\bibinfo  {journal} {Phys. Rev. Lett.}\ }\textbf
  {\bibinfo {volume} {101}},\ \bibinfo {pages} {233601} (\bibinfo {year}
  {2008})}\BibitemShut {NoStop}%
\bibitem [{\citenamefont {Svozil{\'\i}k}\ \emph {et~al.}(2012)\citenamefont
  {Svozil{\'\i}k}, \citenamefont {Pe{\v{r}}ina~Jr},\ and\ \citenamefont
  {Torres}}]{svozilik2012high}%
  \BibitemOpen
  \bibfield  {author} {\bibinfo {author} {\bibfnamefont {J.}~\bibnamefont
  {Svozil{\'\i}k}}, \bibinfo {author} {\bibfnamefont {J.}~\bibnamefont
  {Pe{\v{r}}ina~Jr}}, \ and\ \bibinfo {author} {\bibfnamefont {J.~P.}\
  \bibnamefont {Torres}},\ }\href@noop {} {\bibfield  {journal} {\bibinfo
  {journal} {Phys. Rev. A}\ }\textbf {\bibinfo {volume} {86}},\ \bibinfo
  {pages} {052318} (\bibinfo {year} {2012})}\BibitemShut {NoStop}%
\bibitem [{\citenamefont {Gomes}\ \emph {et~al.}(2009)\citenamefont {Gomes},
  \citenamefont {Salles}, \citenamefont {Toscano}, \citenamefont {Ribeiro},\
  and\ \citenamefont {Walborn}}]{gomes2009observation}%
  \BibitemOpen
  \bibfield  {author} {\bibinfo {author} {\bibfnamefont {R.}~\bibnamefont
  {Gomes}}, \bibinfo {author} {\bibfnamefont {A.}~\bibnamefont {Salles}},
  \bibinfo {author} {\bibfnamefont {F.}~\bibnamefont {Toscano}}, \bibinfo
  {author} {\bibfnamefont {P.~S.}\ \bibnamefont {Ribeiro}}, \ and\ \bibinfo
  {author} {\bibfnamefont {S.}~\bibnamefont {Walborn}},\ }\href@noop {}
  {\bibfield  {journal} {\bibinfo  {journal} {Phys. Rev. Lett.}\ }\textbf
  {\bibinfo {volume} {103}},\ \bibinfo {pages} {033602} (\bibinfo {year}
  {2009})}\BibitemShut {NoStop}%
\bibitem [{\citenamefont {Tasca}\ \emph {et~al.}(2011)\citenamefont {Tasca},
  \citenamefont {Gomes}, \citenamefont {Toscano}, \citenamefont {Ribeiro},\
  and\ \citenamefont {Walborn}}]{tasca2011continuous}%
  \BibitemOpen
  \bibfield  {author} {\bibinfo {author} {\bibfnamefont {D.}~\bibnamefont
  {Tasca}}, \bibinfo {author} {\bibfnamefont {R.}~\bibnamefont {Gomes}},
  \bibinfo {author} {\bibfnamefont {F.}~\bibnamefont {Toscano}}, \bibinfo
  {author} {\bibfnamefont {P.~S.}\ \bibnamefont {Ribeiro}}, \ and\ \bibinfo
  {author} {\bibfnamefont {S.}~\bibnamefont {Walborn}},\ }\href@noop {}
  {\bibfield  {journal} {\bibinfo  {journal} {Phys. Rev. A}\ }\textbf {\bibinfo
  {volume} {83}},\ \bibinfo {pages} {052325} (\bibinfo {year}
  {2011})}\BibitemShut {NoStop}%
\bibitem [{\citenamefont {Zhang}\ \emph
  {et~al.}(2008{\natexlab{a}})\citenamefont {Zhang}, \citenamefont
  {Silberhorn},\ and\ \citenamefont {Walmsley}}]{zhang2008secure}%
  \BibitemOpen
  \bibfield  {author} {\bibinfo {author} {\bibfnamefont {L.}~\bibnamefont
  {Zhang}}, \bibinfo {author} {\bibfnamefont {C.}~\bibnamefont {Silberhorn}}, \
  and\ \bibinfo {author} {\bibfnamefont {I.~A.}\ \bibnamefont {Walmsley}},\
  }\href@noop {} {\bibfield  {journal} {\bibinfo  {journal} {Phys. Rev. Lett.}\
  }\textbf {\bibinfo {volume} {100}},\ \bibinfo {pages} {110504} (\bibinfo
  {year} {2008}{\natexlab{a}})}\BibitemShut {NoStop}%
\bibitem [{\citenamefont {Padgett}(2017)}]{padgett2017orbital}%
  \BibitemOpen
  \bibfield  {author} {\bibinfo {author} {\bibfnamefont {M.~J.}\ \bibnamefont
  {Padgett}},\ }\href@noop {} {\bibfield  {journal} {\bibinfo  {journal} {Opt.
  Express}\ }\textbf {\bibinfo {volume} {25}},\ \bibinfo {pages} {11265}
  (\bibinfo {year} {2017})}\BibitemShut {NoStop}%
\bibitem [{\citenamefont {Allen}\ \emph {et~al.}(1992)\citenamefont {Allen},
  \citenamefont {Beijersbergen}, \citenamefont {Spreeuw},\ and\ \citenamefont
  {Woerdman}}]{allen1992orbital}%
  \BibitemOpen
  \bibfield  {author} {\bibinfo {author} {\bibfnamefont {L.}~\bibnamefont
  {Allen}}, \bibinfo {author} {\bibfnamefont {M.~W.}\ \bibnamefont
  {Beijersbergen}}, \bibinfo {author} {\bibfnamefont {R.}~\bibnamefont
  {Spreeuw}}, \ and\ \bibinfo {author} {\bibfnamefont {J.}~\bibnamefont
  {Woerdman}},\ }\href@noop {} {\bibfield  {journal} {\bibinfo  {journal}
  {Phys. Rev. A}\ }\textbf {\bibinfo {volume} {45}},\ \bibinfo {pages} {8185}
  (\bibinfo {year} {1992})}\BibitemShut {NoStop}%
\bibitem [{\citenamefont {Beijersbergen}\ \emph {et~al.}(1993)\citenamefont
  {Beijersbergen}, \citenamefont {Allen}, \citenamefont {Van~der Veen},\ and\
  \citenamefont {Woerdman}}]{beijersbergen1993astigmatic}%
  \BibitemOpen
  \bibfield  {author} {\bibinfo {author} {\bibfnamefont {M.~W.}\ \bibnamefont
  {Beijersbergen}}, \bibinfo {author} {\bibfnamefont {L.}~\bibnamefont
  {Allen}}, \bibinfo {author} {\bibfnamefont {H.}~\bibnamefont {Van~der Veen}},
  \ and\ \bibinfo {author} {\bibfnamefont {J.}~\bibnamefont {Woerdman}},\
  }\href@noop {} {\bibfield  {journal} {\bibinfo  {journal} {Opt. Commun.}\
  }\textbf {\bibinfo {volume} {96}},\ \bibinfo {pages} {123} (\bibinfo {year}
  {1993})}\BibitemShut {NoStop}%
\bibitem [{\citenamefont {Mair}\ \emph {et~al.}(2001)\citenamefont {Mair},
  \citenamefont {Vaziri}, \citenamefont {Weihs},\ and\ \citenamefont
  {Zeilinger}}]{mair}%
  \BibitemOpen
  \bibfield  {author} {\bibinfo {author} {\bibfnamefont {A.}~\bibnamefont
  {Mair}}, \bibinfo {author} {\bibfnamefont {A.}~\bibnamefont {Vaziri}},
  \bibinfo {author} {\bibfnamefont {G.}~\bibnamefont {Weihs}}, \ and\ \bibinfo
  {author} {\bibfnamefont {A.}~\bibnamefont {Zeilinger}},\ }\href@noop {}
  {\bibfield  {journal} {\bibinfo  {journal} {Nature}\ }\textbf {\bibinfo
  {volume} {412}},\ \bibinfo {pages} {313} (\bibinfo {year}
  {2001})}\BibitemShut {NoStop}%
\bibitem [{\citenamefont {Walborn}\ \emph {et~al.}(2004)\citenamefont
  {Walborn}, \citenamefont {De~Oliveira}, \citenamefont {Thebaldi},\ and\
  \citenamefont {Monken}}]{walborn2004entanglement}%
  \BibitemOpen
  \bibfield  {author} {\bibinfo {author} {\bibfnamefont {S.}~\bibnamefont
  {Walborn}}, \bibinfo {author} {\bibfnamefont {A.}~\bibnamefont
  {De~Oliveira}}, \bibinfo {author} {\bibfnamefont {R.}~\bibnamefont
  {Thebaldi}}, \ and\ \bibinfo {author} {\bibfnamefont {C.}~\bibnamefont
  {Monken}},\ }\href@noop {} {\bibfield  {journal} {\bibinfo  {journal} {Phys.
  Rev. A}\ }\textbf {\bibinfo {volume} {69}},\ \bibinfo {pages} {023811}
  (\bibinfo {year} {2004})}\BibitemShut {NoStop}%
\bibitem [{\citenamefont {Oemrawsingh}\ \emph {et~al.}(2005)\citenamefont
  {Oemrawsingh}, \citenamefont {Ma}, \citenamefont {Voigt}, \citenamefont
  {Aiello}, \citenamefont {Eliel}, \citenamefont {Woerdman} \emph
  {et~al.}}]{oemrawsingh2005experimental}%
  \BibitemOpen
  \bibfield  {author} {\bibinfo {author} {\bibfnamefont {S.}~\bibnamefont
  {Oemrawsingh}}, \bibinfo {author} {\bibfnamefont {X.}~\bibnamefont {Ma}},
  \bibinfo {author} {\bibfnamefont {D.}~\bibnamefont {Voigt}}, \bibinfo
  {author} {\bibfnamefont {A.}~\bibnamefont {Aiello}}, \bibinfo {author}
  {\bibfnamefont {E.~t.}\ \bibnamefont {Eliel}}, \bibinfo {author}
  {\bibfnamefont {J.}~\bibnamefont {Woerdman}},  \emph {et~al.},\ }\href@noop
  {} {\bibfield  {journal} {\bibinfo  {journal} {Phys. Rev. Lett.}\ }\textbf
  {\bibinfo {volume} {95}},\ \bibinfo {pages} {240501} (\bibinfo {year}
  {2005})}\BibitemShut {NoStop}%
\bibitem [{\citenamefont {Oemrawsingh}\ \emph {et~al.}(2006)\citenamefont
  {Oemrawsingh}, \citenamefont {De~Jong}, \citenamefont {Ma}, \citenamefont
  {Aiello}, \citenamefont {Eliel}, \citenamefont {Woerdman} \emph
  {et~al.}}]{oemrawsingh2006high}%
  \BibitemOpen
  \bibfield  {author} {\bibinfo {author} {\bibfnamefont {S.}~\bibnamefont
  {Oemrawsingh}}, \bibinfo {author} {\bibfnamefont {J.}~\bibnamefont
  {De~Jong}}, \bibinfo {author} {\bibfnamefont {X.}~\bibnamefont {Ma}},
  \bibinfo {author} {\bibfnamefont {A.}~\bibnamefont {Aiello}}, \bibinfo
  {author} {\bibfnamefont {E.}~\bibnamefont {Eliel}}, \bibinfo {author}
  {\bibfnamefont {J.}~\bibnamefont {Woerdman}},  \emph {et~al.},\ }\href@noop
  {} {\bibfield  {journal} {\bibinfo  {journal} {Phys. Rev. A}\ }\textbf
  {\bibinfo {volume} {73}},\ \bibinfo {pages} {032339} (\bibinfo {year}
  {2006})}\BibitemShut {NoStop}%
\bibitem [{\citenamefont {Pors}\ \emph {et~al.}(2008)\citenamefont {Pors},
  \citenamefont {Oemrawsingh}, \citenamefont {Aiello}, \citenamefont
  {Van~Exter}, \citenamefont {Eliel}, \citenamefont {Woerdman} \emph
  {et~al.}}]{pors2008shannon}%
  \BibitemOpen
  \bibfield  {author} {\bibinfo {author} {\bibfnamefont {J.}~\bibnamefont
  {Pors}}, \bibinfo {author} {\bibfnamefont {S.}~\bibnamefont {Oemrawsingh}},
  \bibinfo {author} {\bibfnamefont {A.}~\bibnamefont {Aiello}}, \bibinfo
  {author} {\bibfnamefont {M.}~\bibnamefont {Van~Exter}}, \bibinfo {author}
  {\bibfnamefont {E.}~\bibnamefont {Eliel}}, \bibinfo {author} {\bibfnamefont
  {J.}~\bibnamefont {Woerdman}},  \emph {et~al.},\ }\href@noop {} {\bibfield
  {journal} {\bibinfo  {journal} {Phys. Rev. Lett.}\ }\textbf {\bibinfo
  {volume} {101}},\ \bibinfo {pages} {120502} (\bibinfo {year}
  {2008})}\BibitemShut {NoStop}%
\bibitem [{\citenamefont {Pors}\ \emph {et~al.}(2011)\citenamefont {Pors},
  \citenamefont {Miatto}, \citenamefont {Eliel}, \citenamefont {Woerdman} \emph
  {et~al.}}]{pors2011high}%
  \BibitemOpen
  \bibfield  {author} {\bibinfo {author} {\bibfnamefont {B.-J.}\ \bibnamefont
  {Pors}}, \bibinfo {author} {\bibfnamefont {F.}~\bibnamefont {Miatto}},
  \bibinfo {author} {\bibfnamefont {E.}~\bibnamefont {Eliel}}, \bibinfo
  {author} {\bibfnamefont {J.}~\bibnamefont {Woerdman}},  \emph {et~al.},\
  }\href@noop {} {\bibfield  {journal} {\bibinfo  {journal} {J. Opt.}\ }\textbf
  {\bibinfo {volume} {13}},\ \bibinfo {pages} {064008} (\bibinfo {year}
  {2011})}\BibitemShut {NoStop}%
\bibitem [{\citenamefont {Franke-Arnold}\ \emph {et~al.}(2002)\citenamefont
  {Franke-Arnold}, \citenamefont {Barnett}, \citenamefont {Padgett},\ and\
  \citenamefont {Allen}}]{franke2002two}%
  \BibitemOpen
  \bibfield  {author} {\bibinfo {author} {\bibfnamefont {S.}~\bibnamefont
  {Franke-Arnold}}, \bibinfo {author} {\bibfnamefont {S.~M.}\ \bibnamefont
  {Barnett}}, \bibinfo {author} {\bibfnamefont {M.~J.}\ \bibnamefont
  {Padgett}}, \ and\ \bibinfo {author} {\bibfnamefont {L.}~\bibnamefont
  {Allen}},\ }\href@noop {} {\bibfield  {journal} {\bibinfo  {journal} {Phys.
  Rev. A}\ }\textbf {\bibinfo {volume} {65}},\ \bibinfo {pages} {033823}
  (\bibinfo {year} {2002})}\BibitemShut {NoStop}%
\bibitem [{\citenamefont {Leach}\ \emph {et~al.}(2009)\citenamefont {Leach},
  \citenamefont {Jack}, \citenamefont {Romero}, \citenamefont {Ritsch-Marte},
  \citenamefont {Boyd}, \citenamefont {Jha}, \citenamefont {Barnett},
  \citenamefont {Franke-Arnold},\ and\ \citenamefont
  {Padgett}}]{leach2009violation}%
  \BibitemOpen
  \bibfield  {author} {\bibinfo {author} {\bibfnamefont {J.}~\bibnamefont
  {Leach}}, \bibinfo {author} {\bibfnamefont {B.}~\bibnamefont {Jack}},
  \bibinfo {author} {\bibfnamefont {J.}~\bibnamefont {Romero}}, \bibinfo
  {author} {\bibfnamefont {M.}~\bibnamefont {Ritsch-Marte}}, \bibinfo {author}
  {\bibfnamefont {R.}~\bibnamefont {Boyd}}, \bibinfo {author} {\bibfnamefont
  {A.}~\bibnamefont {Jha}}, \bibinfo {author} {\bibfnamefont {S.}~\bibnamefont
  {Barnett}}, \bibinfo {author} {\bibfnamefont {S.}~\bibnamefont
  {Franke-Arnold}}, \ and\ \bibinfo {author} {\bibfnamefont {M.}~\bibnamefont
  {Padgett}},\ }\href@noop {} {\bibfield  {journal} {\bibinfo  {journal} {Opt.
  Express}\ }\textbf {\bibinfo {volume} {17}},\ \bibinfo {pages} {8287}
  (\bibinfo {year} {2009})}\BibitemShut {NoStop}%
\bibitem [{\citenamefont {Dada}\ \emph {et~al.}(2011)\citenamefont {Dada},
  \citenamefont {Leach}, \citenamefont {Buller}, \citenamefont {Padgett},\ and\
  \citenamefont {Andersson}}]{dada2011experimental}%
  \BibitemOpen
  \bibfield  {author} {\bibinfo {author} {\bibfnamefont {A.~C.}\ \bibnamefont
  {Dada}}, \bibinfo {author} {\bibfnamefont {J.}~\bibnamefont {Leach}},
  \bibinfo {author} {\bibfnamefont {G.~S.}\ \bibnamefont {Buller}}, \bibinfo
  {author} {\bibfnamefont {M.~J.}\ \bibnamefont {Padgett}}, \ and\ \bibinfo
  {author} {\bibfnamefont {E.}~\bibnamefont {Andersson}},\ }\href@noop {}
  {\bibfield  {journal} {\bibinfo  {journal} {Nat. Phys.}\ }\textbf {\bibinfo
  {volume} {7}},\ \bibinfo {pages} {677} (\bibinfo {year} {2011})}\BibitemShut
  {NoStop}%
\bibitem [{\citenamefont {Vaziri}\ \emph {et~al.}(2002)\citenamefont {Vaziri},
  \citenamefont {Weihs},\ and\ \citenamefont
  {Zeilinger}}]{vaziri2002experimental}%
  \BibitemOpen
  \bibfield  {author} {\bibinfo {author} {\bibfnamefont {A.}~\bibnamefont
  {Vaziri}}, \bibinfo {author} {\bibfnamefont {G.}~\bibnamefont {Weihs}}, \
  and\ \bibinfo {author} {\bibfnamefont {A.}~\bibnamefont {Zeilinger}},\
  }\href@noop {} {\bibfield  {journal} {\bibinfo  {journal} {Phys. Rev. Lett.}\
  }\textbf {\bibinfo {volume} {89}},\ \bibinfo {pages} {240401} (\bibinfo
  {year} {2002})}\BibitemShut {NoStop}%
\bibitem [{\citenamefont {Torres}\ \emph
  {et~al.}(2003{\natexlab{a}})\citenamefont {Torres}, \citenamefont {Deyanova},
  \citenamefont {Torner},\ and\ \citenamefont
  {Molina-Terriza}}]{torres2003preparation}%
  \BibitemOpen
  \bibfield  {author} {\bibinfo {author} {\bibfnamefont {J.~P.}\ \bibnamefont
  {Torres}}, \bibinfo {author} {\bibfnamefont {Y.}~\bibnamefont {Deyanova}},
  \bibinfo {author} {\bibfnamefont {L.}~\bibnamefont {Torner}}, \ and\ \bibinfo
  {author} {\bibfnamefont {G.}~\bibnamefont {Molina-Terriza}},\ }\href@noop {}
  {\bibfield  {journal} {\bibinfo  {journal} {Phys. Rev. A}\ }\textbf {\bibinfo
  {volume} {67}},\ \bibinfo {pages} {052313} (\bibinfo {year}
  {2003}{\natexlab{a}})}\BibitemShut {NoStop}%
\bibitem [{\citenamefont {Vaziri}\ \emph {et~al.}(2003)\citenamefont {Vaziri},
  \citenamefont {Pan}, \citenamefont {Jennewein}, \citenamefont {Weihs},\ and\
  \citenamefont {Zeilinger}}]{vaziri2003concentration}%
  \BibitemOpen
  \bibfield  {author} {\bibinfo {author} {\bibfnamefont {A.}~\bibnamefont
  {Vaziri}}, \bibinfo {author} {\bibfnamefont {J.-W.}\ \bibnamefont {Pan}},
  \bibinfo {author} {\bibfnamefont {T.}~\bibnamefont {Jennewein}}, \bibinfo
  {author} {\bibfnamefont {G.}~\bibnamefont {Weihs}}, \ and\ \bibinfo {author}
  {\bibfnamefont {A.}~\bibnamefont {Zeilinger}},\ }\href@noop {} {\bibfield
  {journal} {\bibinfo  {journal} {Phys. Rev. Lett.}\ }\textbf {\bibinfo
  {volume} {91}},\ \bibinfo {pages} {227902} (\bibinfo {year}
  {2003})}\BibitemShut {NoStop}%
\bibitem [{\citenamefont {Malik}\ \emph {et~al.}(2016)\citenamefont {Malik},
  \citenamefont {Erhard}, \citenamefont {Huber}, \citenamefont {Krenn},
  \citenamefont {Fickler},\ and\ \citenamefont {Zeilinger}}]{malik2016multi}%
  \BibitemOpen
  \bibfield  {author} {\bibinfo {author} {\bibfnamefont {M.}~\bibnamefont
  {Malik}}, \bibinfo {author} {\bibfnamefont {M.}~\bibnamefont {Erhard}},
  \bibinfo {author} {\bibfnamefont {M.}~\bibnamefont {Huber}}, \bibinfo
  {author} {\bibfnamefont {M.}~\bibnamefont {Krenn}}, \bibinfo {author}
  {\bibfnamefont {R.}~\bibnamefont {Fickler}}, \ and\ \bibinfo {author}
  {\bibfnamefont {A.}~\bibnamefont {Zeilinger}},\ }\href@noop {} {\bibfield
  {journal} {\bibinfo  {journal} {Nat. Photonics}\ }\textbf {\bibinfo {volume}
  {10}},\ \bibinfo {pages} {248} (\bibinfo {year} {2016})}\BibitemShut
  {NoStop}%
\bibitem [{\citenamefont {Perumangatt}\ \emph {et~al.}(2017)\citenamefont
  {Perumangatt}, \citenamefont {Lal}, \citenamefont {Anwar}, \citenamefont
  {Reddy},\ and\ \citenamefont {Singh}}]{perumangatt2017quantum}%
  \BibitemOpen
  \bibfield  {author} {\bibinfo {author} {\bibfnamefont {C.}~\bibnamefont
  {Perumangatt}}, \bibinfo {author} {\bibfnamefont {N.}~\bibnamefont {Lal}},
  \bibinfo {author} {\bibfnamefont {A.}~\bibnamefont {Anwar}}, \bibinfo
  {author} {\bibfnamefont {S.~G.}\ \bibnamefont {Reddy}}, \ and\ \bibinfo
  {author} {\bibfnamefont {R.}~\bibnamefont {Singh}},\ }\href@noop {}
  {\bibfield  {journal} {\bibinfo  {journal} {Phys. Lett. A}\ }\textbf
  {\bibinfo {volume} {381}},\ \bibinfo {pages} {1858} (\bibinfo {year}
  {2017})}\BibitemShut {NoStop}%
\bibitem [{\citenamefont {Romero}\ \emph {et~al.}(2012)\citenamefont {Romero},
  \citenamefont {Giovannini}, \citenamefont {McLaren}, \citenamefont {Galvez},
  \citenamefont {Forbes},\ and\ \citenamefont {Padgett}}]{romero2012orbital}%
  \BibitemOpen
  \bibfield  {author} {\bibinfo {author} {\bibfnamefont {J.}~\bibnamefont
  {Romero}}, \bibinfo {author} {\bibfnamefont {D.}~\bibnamefont {Giovannini}},
  \bibinfo {author} {\bibfnamefont {M.}~\bibnamefont {McLaren}}, \bibinfo
  {author} {\bibfnamefont {E.}~\bibnamefont {Galvez}}, \bibinfo {author}
  {\bibfnamefont {A.}~\bibnamefont {Forbes}}, \ and\ \bibinfo {author}
  {\bibfnamefont {M.}~\bibnamefont {Padgett}},\ }\href@noop {} {\bibfield
  {journal} {\bibinfo  {journal} {J. Opt.}\ }\textbf {\bibinfo {volume} {14}},\
  \bibinfo {pages} {085401} (\bibinfo {year} {2012})}\BibitemShut {NoStop}%
\bibitem [{\citenamefont {Kovlakov}\ \emph {et~al.}(2018)\citenamefont
  {Kovlakov}, \citenamefont {Straupe},\ and\ \citenamefont
  {Kulik}}]{kovlakov2018quantum}%
  \BibitemOpen
  \bibfield  {author} {\bibinfo {author} {\bibfnamefont {E.}~\bibnamefont
  {Kovlakov}}, \bibinfo {author} {\bibfnamefont {S.}~\bibnamefont {Straupe}}, \
  and\ \bibinfo {author} {\bibfnamefont {S.}~\bibnamefont {Kulik}},\
  }\href@noop {} {\bibfield  {journal} {\bibinfo  {journal} {Phys. Rev. A}\
  }\textbf {\bibinfo {volume} {98}},\ \bibinfo {pages} {060301} (\bibinfo
  {year} {2018})}\BibitemShut {NoStop}%
\bibitem [{\citenamefont {Liu}\ \emph {et~al.}(2018)\citenamefont {Liu},
  \citenamefont {Zhou}, \citenamefont {Liu}, \citenamefont {Li}, \citenamefont
  {Li}, \citenamefont {Yang}, \citenamefont {Xu}, \citenamefont {Liu},
  \citenamefont {Guo},\ and\ \citenamefont {Shi}}]{liu2018coherent}%
  \BibitemOpen
  \bibfield  {author} {\bibinfo {author} {\bibfnamefont {S.}~\bibnamefont
  {Liu}}, \bibinfo {author} {\bibfnamefont {Z.}~\bibnamefont {Zhou}}, \bibinfo
  {author} {\bibfnamefont {S.}~\bibnamefont {Liu}}, \bibinfo {author}
  {\bibfnamefont {Y.}~\bibnamefont {Li}}, \bibinfo {author} {\bibfnamefont
  {Y.}~\bibnamefont {Li}}, \bibinfo {author} {\bibfnamefont {C.}~\bibnamefont
  {Yang}}, \bibinfo {author} {\bibfnamefont {Z.}~\bibnamefont {Xu}}, \bibinfo
  {author} {\bibfnamefont {Z.}~\bibnamefont {Liu}}, \bibinfo {author}
  {\bibfnamefont {G.}~\bibnamefont {Guo}}, \ and\ \bibinfo {author}
  {\bibfnamefont {B.}~\bibnamefont {Shi}},\ }\href@noop {} {\bibfield
  {journal} {\bibinfo  {journal} {Phys. Rev. A}\ }\textbf {\bibinfo {volume}
  {98}},\ \bibinfo {pages} {062316} (\bibinfo {year} {2018})}\BibitemShut
  {NoStop}%
\bibitem [{\citenamefont {Anwar}\ \emph {et~al.}(2020)\citenamefont {Anwar},
  \citenamefont {Lal}, \citenamefont {Prabhakar},\ and\ \citenamefont
  {Singh}}]{anwar2020selective}%
  \BibitemOpen
  \bibfield  {author} {\bibinfo {author} {\bibfnamefont {A.}~\bibnamefont
  {Anwar}}, \bibinfo {author} {\bibfnamefont {N.}~\bibnamefont {Lal}}, \bibinfo
  {author} {\bibfnamefont {S.}~\bibnamefont {Prabhakar}}, \ and\ \bibinfo
  {author} {\bibfnamefont {R.}~\bibnamefont {Singh}},\ }\href@noop {}
  {\bibfield  {journal} {\bibinfo  {journal} {New J. Phys.}\ }\textbf {\bibinfo
  {volume} {22}},\ \bibinfo {pages} {113020} (\bibinfo {year}
  {2020})}\BibitemShut {NoStop}%
\bibitem [{\citenamefont {Torres}\ \emph
  {et~al.}(2003{\natexlab{b}})\citenamefont {Torres}, \citenamefont
  {Alexandrescu},\ and\ \citenamefont {Torner}}]{torres2003quantum}%
  \BibitemOpen
  \bibfield  {author} {\bibinfo {author} {\bibfnamefont {J.}~\bibnamefont
  {Torres}}, \bibinfo {author} {\bibfnamefont {A.}~\bibnamefont
  {Alexandrescu}}, \ and\ \bibinfo {author} {\bibfnamefont {L.}~\bibnamefont
  {Torner}},\ }\href@noop {} {\bibfield  {journal} {\bibinfo  {journal} {Phys.
  Rev. A}\ }\textbf {\bibinfo {volume} {68}},\ \bibinfo {pages} {050301}
  (\bibinfo {year} {2003}{\natexlab{b}})}\BibitemShut {NoStop}%
\bibitem [{\citenamefont {Qassim}\ \emph {et~al.}(2014)\citenamefont {Qassim},
  \citenamefont {Miatto}, \citenamefont {Torres}, \citenamefont {Padgett},
  \citenamefont {Karimi},\ and\ \citenamefont {Boyd}}]{qassim2014limitations}%
  \BibitemOpen
  \bibfield  {author} {\bibinfo {author} {\bibfnamefont {H.}~\bibnamefont
  {Qassim}}, \bibinfo {author} {\bibfnamefont {F.~M.}\ \bibnamefont {Miatto}},
  \bibinfo {author} {\bibfnamefont {J.~P.}\ \bibnamefont {Torres}}, \bibinfo
  {author} {\bibfnamefont {M.~J.}\ \bibnamefont {Padgett}}, \bibinfo {author}
  {\bibfnamefont {E.}~\bibnamefont {Karimi}}, \ and\ \bibinfo {author}
  {\bibfnamefont {R.~W.}\ \bibnamefont {Boyd}},\ }\href@noop {} {\bibfield
  {journal} {\bibinfo  {journal} {J. Opt. Soc. Am. B}\ }\textbf {\bibinfo
  {volume} {31}},\ \bibinfo {pages} {A20} (\bibinfo {year} {2014})}\BibitemShut
  {NoStop}%
\bibitem [{\citenamefont {Tyler}\ and\ \citenamefont
  {Boyd}(2009)}]{tyler2009influence}%
  \BibitemOpen
  \bibfield  {author} {\bibinfo {author} {\bibfnamefont {G.~A.}\ \bibnamefont
  {Tyler}}\ and\ \bibinfo {author} {\bibfnamefont {R.~W.}\ \bibnamefont
  {Boyd}},\ }\href@noop {} {\bibfield  {journal} {\bibinfo  {journal} {Opt.
  Lett.}\ }\textbf {\bibinfo {volume} {34}},\ \bibinfo {pages} {142} (\bibinfo
  {year} {2009})}\BibitemShut {NoStop}%
\bibitem [{\citenamefont {Ibrahim}\ \emph {et~al.}(2013)\citenamefont
  {Ibrahim}, \citenamefont {Roux}, \citenamefont {McLaren}, \citenamefont
  {Konrad},\ and\ \citenamefont {Forbes}}]{ibrahim2013orbital}%
  \BibitemOpen
  \bibfield  {author} {\bibinfo {author} {\bibfnamefont {A.~H.}\ \bibnamefont
  {Ibrahim}}, \bibinfo {author} {\bibfnamefont {F.~S.}\ \bibnamefont {Roux}},
  \bibinfo {author} {\bibfnamefont {M.}~\bibnamefont {McLaren}}, \bibinfo
  {author} {\bibfnamefont {T.}~\bibnamefont {Konrad}}, \ and\ \bibinfo {author}
  {\bibfnamefont {A.}~\bibnamefont {Forbes}},\ }\href@noop {} {\bibfield
  {journal} {\bibinfo  {journal} {Phys. Rev. A}\ }\textbf {\bibinfo {volume}
  {88}},\ \bibinfo {pages} {012312} (\bibinfo {year} {2013})}\BibitemShut
  {NoStop}%
\bibitem [{\citenamefont {Zhang}\ \emph {et~al.}(2016)\citenamefont {Zhang},
  \citenamefont {Prabhakar}, \citenamefont {Roux}, \citenamefont {Forbes},
  \citenamefont {Konrad} \emph {et~al.}}]{zhang2016experimentally}%
  \BibitemOpen
  \bibfield  {author} {\bibinfo {author} {\bibfnamefont {Y.}~\bibnamefont
  {Zhang}}, \bibinfo {author} {\bibfnamefont {S.}~\bibnamefont {Prabhakar}},
  \bibinfo {author} {\bibfnamefont {F.~S.}\ \bibnamefont {Roux}}, \bibinfo
  {author} {\bibfnamefont {A.}~\bibnamefont {Forbes}}, \bibinfo {author}
  {\bibfnamefont {T.}~\bibnamefont {Konrad}},  \emph {et~al.},\ }\href@noop {}
  {\bibfield  {journal} {\bibinfo  {journal} {Phys. Rev. A}\ }\textbf {\bibinfo
  {volume} {94}},\ \bibinfo {pages} {032310} (\bibinfo {year}
  {2016})}\BibitemShut {NoStop}%
\bibitem [{\citenamefont {Kovlakov}\ \emph {et~al.}(2017)\citenamefont
  {Kovlakov}, \citenamefont {Bobrov}, \citenamefont {Straupe},\ and\
  \citenamefont {Kulik}}]{kovlakov2017spatial}%
  \BibitemOpen
  \bibfield  {author} {\bibinfo {author} {\bibfnamefont {E.}~\bibnamefont
  {Kovlakov}}, \bibinfo {author} {\bibfnamefont {I.}~\bibnamefont {Bobrov}},
  \bibinfo {author} {\bibfnamefont {S.}~\bibnamefont {Straupe}}, \ and\
  \bibinfo {author} {\bibfnamefont {S.}~\bibnamefont {Kulik}},\ }\href@noop {}
  {\bibfield  {journal} {\bibinfo  {journal} {Phys. Rev. Lett.}\ }\textbf
  {\bibinfo {volume} {118}},\ \bibinfo {pages} {030503} (\bibinfo {year}
  {2017})}\BibitemShut {NoStop}%
\bibitem [{\citenamefont {McLaren}\ \emph {et~al.}(2012)\citenamefont
  {McLaren}, \citenamefont {Agnew}, \citenamefont {Leach}, \citenamefont
  {Roux}, \citenamefont {Padgett}, \citenamefont {Boyd},\ and\ \citenamefont
  {Forbes}}]{mclaren2012entangled}%
  \BibitemOpen
  \bibfield  {author} {\bibinfo {author} {\bibfnamefont {M.}~\bibnamefont
  {McLaren}}, \bibinfo {author} {\bibfnamefont {M.}~\bibnamefont {Agnew}},
  \bibinfo {author} {\bibfnamefont {J.}~\bibnamefont {Leach}}, \bibinfo
  {author} {\bibfnamefont {F.~S.}\ \bibnamefont {Roux}}, \bibinfo {author}
  {\bibfnamefont {M.~J.}\ \bibnamefont {Padgett}}, \bibinfo {author}
  {\bibfnamefont {R.~W.}\ \bibnamefont {Boyd}}, \ and\ \bibinfo {author}
  {\bibfnamefont {A.}~\bibnamefont {Forbes}},\ }\href@noop {} {\bibfield
  {journal} {\bibinfo  {journal} {Opt. Express}\ }\textbf {\bibinfo {volume}
  {20}},\ \bibinfo {pages} {23589} (\bibinfo {year} {2012})}\BibitemShut
  {NoStop}%
\bibitem [{\citenamefont {McLaren}\ \emph {et~al.}(2013)\citenamefont
  {McLaren}, \citenamefont {Romero}, \citenamefont {Padgett}, \citenamefont
  {Roux},\ and\ \citenamefont {Forbes}}]{mclaren2013two}%
  \BibitemOpen
  \bibfield  {author} {\bibinfo {author} {\bibfnamefont {M.}~\bibnamefont
  {McLaren}}, \bibinfo {author} {\bibfnamefont {J.}~\bibnamefont {Romero}},
  \bibinfo {author} {\bibfnamefont {M.~J.}\ \bibnamefont {Padgett}}, \bibinfo
  {author} {\bibfnamefont {F.~S.}\ \bibnamefont {Roux}}, \ and\ \bibinfo
  {author} {\bibfnamefont {A.}~\bibnamefont {Forbes}},\ }\href@noop {}
  {\bibfield  {journal} {\bibinfo  {journal} {Phys. Rev. A}\ }\textbf {\bibinfo
  {volume} {88}},\ \bibinfo {pages} {033818} (\bibinfo {year}
  {2013})}\BibitemShut {NoStop}%
\bibitem [{\citenamefont {McLaren}\ \emph {et~al.}(2014)\citenamefont
  {McLaren}, \citenamefont {Mhlanga}, \citenamefont {Padgett}, \citenamefont
  {Roux},\ and\ \citenamefont {Forbes}}]{mclaren2014self}%
  \BibitemOpen
  \bibfield  {author} {\bibinfo {author} {\bibfnamefont {M.}~\bibnamefont
  {McLaren}}, \bibinfo {author} {\bibfnamefont {T.}~\bibnamefont {Mhlanga}},
  \bibinfo {author} {\bibfnamefont {M.~J.}\ \bibnamefont {Padgett}}, \bibinfo
  {author} {\bibfnamefont {F.~S.}\ \bibnamefont {Roux}}, \ and\ \bibinfo
  {author} {\bibfnamefont {A.}~\bibnamefont {Forbes}},\ }\href@noop {}
  {\bibfield  {journal} {\bibinfo  {journal} {Nat. Commun.}\ }\textbf {\bibinfo
  {volume} {5}},\ \bibinfo {pages} {3248} (\bibinfo {year} {2014})}\BibitemShut
  {NoStop}%
\bibitem [{\citenamefont {Krenn}\ \emph {et~al.}(2013)\citenamefont {Krenn},
  \citenamefont {Fickler}, \citenamefont {Huber}, \citenamefont {Lapkiewicz},
  \citenamefont {Plick}, \citenamefont {Ramelow},\ and\ \citenamefont
  {Zeilinger}}]{krenn2013entangled}%
  \BibitemOpen
  \bibfield  {author} {\bibinfo {author} {\bibfnamefont {M.}~\bibnamefont
  {Krenn}}, \bibinfo {author} {\bibfnamefont {R.}~\bibnamefont {Fickler}},
  \bibinfo {author} {\bibfnamefont {M.}~\bibnamefont {Huber}}, \bibinfo
  {author} {\bibfnamefont {R.}~\bibnamefont {Lapkiewicz}}, \bibinfo {author}
  {\bibfnamefont {W.}~\bibnamefont {Plick}}, \bibinfo {author} {\bibfnamefont
  {S.}~\bibnamefont {Ramelow}}, \ and\ \bibinfo {author} {\bibfnamefont
  {A.}~\bibnamefont {Zeilinger}},\ }\href@noop {} {\bibfield  {journal}
  {\bibinfo  {journal} {Phys. Rev. A}\ }\textbf {\bibinfo {volume} {87}},\
  \bibinfo {pages} {012326} (\bibinfo {year} {2013})}\BibitemShut {NoStop}%
\bibitem [{\citenamefont {Rarity}\ \emph {et~al.}(1990)\citenamefont {Rarity},
  \citenamefont {Tapster}, \citenamefont {Jakeman}, \citenamefont {Larchuk},
  \citenamefont {Campos}, \citenamefont {Teich},\ and\ \citenamefont
  {Saleh}}]{rarity1990two}%
  \BibitemOpen
  \bibfield  {author} {\bibinfo {author} {\bibfnamefont {J.~G.}\ \bibnamefont
  {Rarity}}, \bibinfo {author} {\bibfnamefont {P.~R.}\ \bibnamefont {Tapster}},
  \bibinfo {author} {\bibfnamefont {E.}~\bibnamefont {Jakeman}}, \bibinfo
  {author} {\bibfnamefont {T.}~\bibnamefont {Larchuk}}, \bibinfo {author}
  {\bibfnamefont {R.~A.}\ \bibnamefont {Campos}}, \bibinfo {author}
  {\bibfnamefont {M.~C.}\ \bibnamefont {Teich}}, \ and\ \bibinfo {author}
  {\bibfnamefont {B.~E.~A.}\ \bibnamefont {Saleh}},\ }\href@noop {} {\bibfield
  {journal} {\bibinfo  {journal} {Phys. Rev. Lett.}\ }\textbf {\bibinfo
  {volume} {65}},\ \bibinfo {pages} {1348} (\bibinfo {year}
  {1990})}\BibitemShut {NoStop}%
\bibitem [{\citenamefont {Kwiat}\ \emph {et~al.}(1990)\citenamefont {Kwiat},
  \citenamefont {Vareka}, \citenamefont {Hong}, \citenamefont {Nathel},\ and\
  \citenamefont {Chiao}}]{kwiat1990correlated}%
  \BibitemOpen
  \bibfield  {author} {\bibinfo {author} {\bibfnamefont {P.~G.}\ \bibnamefont
  {Kwiat}}, \bibinfo {author} {\bibfnamefont {W.~A.}\ \bibnamefont {Vareka}},
  \bibinfo {author} {\bibfnamefont {C.~K.}\ \bibnamefont {Hong}}, \bibinfo
  {author} {\bibfnamefont {H.}~\bibnamefont {Nathel}}, \ and\ \bibinfo {author}
  {\bibfnamefont {R.}~\bibnamefont {Chiao}},\ }\href@noop {} {\bibfield
  {journal} {\bibinfo  {journal} {Phys. Rev. A}\ }\textbf {\bibinfo {volume}
  {41}},\ \bibinfo {pages} {2910} (\bibinfo {year} {1990})}\BibitemShut
  {NoStop}%
\bibitem [{\citenamefont {Ou}\ \emph {et~al.}(1990)\citenamefont {Ou},
  \citenamefont {Zou}, \citenamefont {Wang},\ and\ \citenamefont
  {Mandel}}]{ou1990observation}%
  \BibitemOpen
  \bibfield  {author} {\bibinfo {author} {\bibfnamefont {Z.~Y.}\ \bibnamefont
  {Ou}}, \bibinfo {author} {\bibfnamefont {X.~Y.}\ \bibnamefont {Zou}},
  \bibinfo {author} {\bibfnamefont {L.~J.}\ \bibnamefont {Wang}}, \ and\
  \bibinfo {author} {\bibfnamefont {L.}~\bibnamefont {Mandel}},\ }\href@noop {}
  {\bibfield  {journal} {\bibinfo  {journal} {Phys. Rev. Lett.}\ }\textbf
  {\bibinfo {volume} {65}},\ \bibinfo {pages} {321} (\bibinfo {year}
  {1990})}\BibitemShut {NoStop}%
\bibitem [{\citenamefont {Brendel}\ \emph {et~al.}(1992)\citenamefont
  {Brendel}, \citenamefont {Mohler},\ and\ \citenamefont
  {Martienssen}}]{brendel1992experimental}%
  \BibitemOpen
  \bibfield  {author} {\bibinfo {author} {\bibfnamefont {J.}~\bibnamefont
  {Brendel}}, \bibinfo {author} {\bibfnamefont {E.}~\bibnamefont {Mohler}}, \
  and\ \bibinfo {author} {\bibfnamefont {W.}~\bibnamefont {Martienssen}},\
  }\href@noop {} {\bibfield  {journal} {\bibinfo  {journal} {Europhys. Lett.}\
  }\textbf {\bibinfo {volume} {20}},\ \bibinfo {pages} {575} (\bibinfo {year}
  {1992})}\BibitemShut {NoStop}%
\bibitem [{\citenamefont {Chiao}\ \emph {et~al.}(1995)\citenamefont {Chiao},
  \citenamefont {Kwiat},\ and\ \citenamefont {Steinberg}}]{chiao1995quantum}%
  \BibitemOpen
  \bibfield  {author} {\bibinfo {author} {\bibfnamefont {R.~Y.}\ \bibnamefont
  {Chiao}}, \bibinfo {author} {\bibfnamefont {P.}~\bibnamefont {Kwiat}}, \ and\
  \bibinfo {author} {\bibfnamefont {A.~M.}\ \bibnamefont {Steinberg}},\
  }\href@noop {} {\bibfield  {journal} {\bibinfo  {journal} {Quant. Semiclass.
  Opt.: J. Eur. Opt. Soc. B}\ }\textbf {\bibinfo {volume} {7}},\ \bibinfo
  {pages} {259} (\bibinfo {year} {1995})}\BibitemShut {NoStop}%
\bibitem [{\citenamefont {Franson}(1989)}]{franson1989bell}%
  \BibitemOpen
  \bibfield  {author} {\bibinfo {author} {\bibfnamefont {J.~D.}\ \bibnamefont
  {Franson}},\ }\href@noop {} {\bibfield  {journal} {\bibinfo  {journal} {Phys.
  Rev. Lett.}\ }\textbf {\bibinfo {volume} {62}},\ \bibinfo {pages} {2205}
  (\bibinfo {year} {1989})}\BibitemShut {NoStop}%
\bibitem [{\citenamefont {Weinfurter}(1994)}]{weinfurter1994experimental}%
  \BibitemOpen
  \bibfield  {author} {\bibinfo {author} {\bibfnamefont {H.}~\bibnamefont
  {Weinfurter}},\ }\href@noop {} {\bibfield  {journal} {\bibinfo  {journal}
  {Europhys. Lett.}\ }\textbf {\bibinfo {volume} {25}},\ \bibinfo {pages} {559}
  (\bibinfo {year} {1994})}\BibitemShut {NoStop}%
\bibitem [{\citenamefont {Cabello}\ \emph {et~al.}(2009)\citenamefont
  {Cabello}, \citenamefont {Rossi}, \citenamefont {Vallone}, \citenamefont
  {De~Martini},\ and\ \citenamefont {Mataloni}}]{cabello2009proposed}%
  \BibitemOpen
  \bibfield  {author} {\bibinfo {author} {\bibfnamefont {A.}~\bibnamefont
  {Cabello}}, \bibinfo {author} {\bibfnamefont {A.}~\bibnamefont {Rossi}},
  \bibinfo {author} {\bibfnamefont {G.}~\bibnamefont {Vallone}}, \bibinfo
  {author} {\bibfnamefont {F.}~\bibnamefont {De~Martini}}, \ and\ \bibinfo
  {author} {\bibfnamefont {P.}~\bibnamefont {Mataloni}},\ }\href@noop {}
  {\bibfield  {journal} {\bibinfo  {journal} {Phys. Rev. Lett.}\ }\textbf
  {\bibinfo {volume} {102}},\ \bibinfo {pages} {040401} (\bibinfo {year}
  {2009})}\BibitemShut {NoStop}%
\bibitem [{\citenamefont {Kwiat}\ \emph {et~al.}(1993)\citenamefont {Kwiat},
  \citenamefont {Steinberg},\ and\ \citenamefont {Chiao}}]{kwiat1993high}%
  \BibitemOpen
  \bibfield  {author} {\bibinfo {author} {\bibfnamefont {P.~G.}\ \bibnamefont
  {Kwiat}}, \bibinfo {author} {\bibfnamefont {A.~M.}\ \bibnamefont
  {Steinberg}}, \ and\ \bibinfo {author} {\bibfnamefont {R.~Y.}\ \bibnamefont
  {Chiao}},\ }\href@noop {} {\bibfield  {journal} {\bibinfo  {journal} {Phys.
  Rev. A}\ }\textbf {\bibinfo {volume} {47}},\ \bibinfo {pages} {R2472}
  (\bibinfo {year} {1993})}\BibitemShut {NoStop}%
\bibitem [{\citenamefont {Strekalov}\ \emph {et~al.}(1996)\citenamefont
  {Strekalov}, \citenamefont {Pittman}, \citenamefont {Sergienko},
  \citenamefont {Shih},\ and\ \citenamefont
  {Kwiat}}]{strekalov1996postselection}%
  \BibitemOpen
  \bibfield  {author} {\bibinfo {author} {\bibfnamefont {D.}~\bibnamefont
  {Strekalov}}, \bibinfo {author} {\bibfnamefont {T.}~\bibnamefont {Pittman}},
  \bibinfo {author} {\bibfnamefont {A.}~\bibnamefont {Sergienko}}, \bibinfo
  {author} {\bibfnamefont {Y.}~\bibnamefont {Shih}}, \ and\ \bibinfo {author}
  {\bibfnamefont {P.~G.}\ \bibnamefont {Kwiat}},\ }\href@noop {} {\bibfield
  {journal} {\bibinfo  {journal} {Phys. Rev. A}\ }\textbf {\bibinfo {volume}
  {54}},\ \bibinfo {pages} {R1} (\bibinfo {year} {1996})}\BibitemShut {NoStop}%
\bibitem [{\citenamefont {Khan}\ and\ \citenamefont
  {Howell}(2006)}]{khan2006experimental}%
  \BibitemOpen
  \bibfield  {author} {\bibinfo {author} {\bibfnamefont {I.~A.}\ \bibnamefont
  {Khan}}\ and\ \bibinfo {author} {\bibfnamefont {J.~C.}\ \bibnamefont
  {Howell}},\ }\href@noop {} {\bibfield  {journal} {\bibinfo  {journal} {Phys.
  Rev. A}\ }\textbf {\bibinfo {volume} {73}},\ \bibinfo {pages} {031801}
  (\bibinfo {year} {2006})}\BibitemShut {NoStop}%
\bibitem [{\citenamefont {Ali-Khan}\ \emph {et~al.}(2007)\citenamefont
  {Ali-Khan}, \citenamefont {Broadbent},\ and\ \citenamefont
  {Howell}}]{ali2007large}%
  \BibitemOpen
  \bibfield  {author} {\bibinfo {author} {\bibfnamefont {I.}~\bibnamefont
  {Ali-Khan}}, \bibinfo {author} {\bibfnamefont {C.~J.}\ \bibnamefont
  {Broadbent}}, \ and\ \bibinfo {author} {\bibfnamefont {J.~C.}\ \bibnamefont
  {Howell}},\ }\href@noop {} {\bibfield  {journal} {\bibinfo  {journal} {Phys.
  Rev. Lett.}\ }\textbf {\bibinfo {volume} {98}},\ \bibinfo {pages} {060503}
  (\bibinfo {year} {2007})}\BibitemShut {NoStop}%
\bibitem [{\citenamefont {Shalm}\ \emph {et~al.}(2013)\citenamefont {Shalm},
  \citenamefont {Hamel}, \citenamefont {Yan}, \citenamefont {Simon},
  \citenamefont {Resch},\ and\ \citenamefont {Jennewein}}]{shalm2013three}%
  \BibitemOpen
  \bibfield  {author} {\bibinfo {author} {\bibfnamefont {L.~K.}\ \bibnamefont
  {Shalm}}, \bibinfo {author} {\bibfnamefont {D.~R.}\ \bibnamefont {Hamel}},
  \bibinfo {author} {\bibfnamefont {Z.}~\bibnamefont {Yan}}, \bibinfo {author}
  {\bibfnamefont {C.}~\bibnamefont {Simon}}, \bibinfo {author} {\bibfnamefont
  {K.~J.}\ \bibnamefont {Resch}}, \ and\ \bibinfo {author} {\bibfnamefont
  {T.}~\bibnamefont {Jennewein}},\ }\href@noop {} {\bibfield  {journal}
  {\bibinfo  {journal} {Nat. Phys.}\ }\textbf {\bibinfo {volume} {9}},\
  \bibinfo {pages} {19} (\bibinfo {year} {2013})}\BibitemShut {NoStop}%
\bibitem [{\citenamefont {Schwarz}\ \emph {et~al.}(2015)\citenamefont
  {Schwarz}, \citenamefont {Bessire},\ and\ \citenamefont
  {Stefanov}}]{schwarz2015experimental}%
  \BibitemOpen
  \bibfield  {author} {\bibinfo {author} {\bibfnamefont {S.}~\bibnamefont
  {Schwarz}}, \bibinfo {author} {\bibfnamefont {B.}~\bibnamefont {Bessire}}, \
  and\ \bibinfo {author} {\bibfnamefont {A.}~\bibnamefont {Stefanov}},\
  }\href@noop {} {\bibfield  {journal} {\bibinfo  {journal} {Intl. J. Quant.
  Inf.}\ }\textbf {\bibinfo {volume} {12}},\ \bibinfo {pages} {1560026}
  (\bibinfo {year} {2015})}\BibitemShut {NoStop}%
\bibitem [{\citenamefont {Bessire}\ \emph {et~al.}(2014)\citenamefont
  {Bessire}, \citenamefont {Bernhard}, \citenamefont {Feurer},\ and\
  \citenamefont {Stefanov}}]{bessire2014versatile}%
  \BibitemOpen
  \bibfield  {author} {\bibinfo {author} {\bibfnamefont {B.}~\bibnamefont
  {Bessire}}, \bibinfo {author} {\bibfnamefont {C.}~\bibnamefont {Bernhard}},
  \bibinfo {author} {\bibfnamefont {T.}~\bibnamefont {Feurer}}, \ and\ \bibinfo
  {author} {\bibfnamefont {A.}~\bibnamefont {Stefanov}},\ }\href@noop {}
  {\bibfield  {journal} {\bibinfo  {journal} {New J. Phys.}\ }\textbf {\bibinfo
  {volume} {16}},\ \bibinfo {pages} {033017} (\bibinfo {year}
  {2014})}\BibitemShut {NoStop}%
\bibitem [{\citenamefont {Brougham}\ \emph {et~al.}(2013)\citenamefont
  {Brougham}, \citenamefont {Barnett}, \citenamefont {McCusker}, \citenamefont
  {Kwiat},\ and\ \citenamefont {Gauthier}}]{brougham2013security}%
  \BibitemOpen
  \bibfield  {author} {\bibinfo {author} {\bibfnamefont {T.}~\bibnamefont
  {Brougham}}, \bibinfo {author} {\bibfnamefont {S.~M.}\ \bibnamefont
  {Barnett}}, \bibinfo {author} {\bibfnamefont {K.~T.}\ \bibnamefont
  {McCusker}}, \bibinfo {author} {\bibfnamefont {P.~G.}\ \bibnamefont {Kwiat}},
  \ and\ \bibinfo {author} {\bibfnamefont {D.~J.}\ \bibnamefont {Gauthier}},\
  }\href@noop {} {\bibfield  {journal} {\bibinfo  {journal} {J. Phys. B: At.
  Mol. Opt. Phys.}\ }\textbf {\bibinfo {volume} {46}},\ \bibinfo {pages}
  {104010} (\bibinfo {year} {2013})}\BibitemShut {NoStop}%
\bibitem [{\citenamefont {Aerts}\ \emph {et~al.}(1999)\citenamefont {Aerts},
  \citenamefont {Kwiat}, \citenamefont {Larsson},\ and\ \citenamefont
  {Zukowski}}]{aerts1999two}%
  \BibitemOpen
  \bibfield  {author} {\bibinfo {author} {\bibfnamefont {S.}~\bibnamefont
  {Aerts}}, \bibinfo {author} {\bibfnamefont {P.}~\bibnamefont {Kwiat}},
  \bibinfo {author} {\bibfnamefont {J.-{\AA}.}\ \bibnamefont {Larsson}}, \ and\
  \bibinfo {author} {\bibfnamefont {M.}~\bibnamefont {Zukowski}},\ }\href@noop
  {} {\bibfield  {journal} {\bibinfo  {journal} {Phys. Rev. Lett.}\ }\textbf
  {\bibinfo {volume} {83}},\ \bibinfo {pages} {2872} (\bibinfo {year}
  {1999})}\BibitemShut {NoStop}%
\bibitem [{\citenamefont {Larsson}(2002)}]{larson2002practical}%
  \BibitemOpen
  \bibfield  {author} {\bibinfo {author} {\bibfnamefont {J.}~\bibnamefont
  {Larsson}},\ }\href {http://portal.acm.org/citation.cfm?id=2011494}
  {\bibfield  {journal} {\bibinfo  {journal} {Quant.Inf. Comput.}\ }\textbf
  {\bibinfo {volume} {2}},\ \bibinfo {pages} {434} (\bibinfo {year}
  {2002})}\BibitemShut {NoStop}%
\bibitem [{\citenamefont {Barreiro}\ \emph {et~al.}(2005)\citenamefont
  {Barreiro}, \citenamefont {Langford}, \citenamefont {Peters},\ and\
  \citenamefont {Kwiat}}]{barreiro2005generation}%
  \BibitemOpen
  \bibfield  {author} {\bibinfo {author} {\bibfnamefont {J.~T.}\ \bibnamefont
  {Barreiro}}, \bibinfo {author} {\bibfnamefont {N.~K.}\ \bibnamefont
  {Langford}}, \bibinfo {author} {\bibfnamefont {N.~A.}\ \bibnamefont
  {Peters}}, \ and\ \bibinfo {author} {\bibfnamefont {P.~G.}\ \bibnamefont
  {Kwiat}},\ }\href@noop {} {\bibfield  {journal} {\bibinfo  {journal} {Phys.
  Rev. Lett.}\ }\textbf {\bibinfo {volume} {95}},\ \bibinfo {pages} {260501}
  (\bibinfo {year} {2005})}\BibitemShut {NoStop}%
\bibitem [{\citenamefont {Lima}\ \emph {et~al.}(2010)\citenamefont {Lima},
  \citenamefont {Vallone}, \citenamefont {Chiuri}, \citenamefont {Cabello},\
  and\ \citenamefont {Mataloni}}]{lima2010experimental}%
  \BibitemOpen
  \bibfield  {author} {\bibinfo {author} {\bibfnamefont {G.}~\bibnamefont
  {Lima}}, \bibinfo {author} {\bibfnamefont {G.}~\bibnamefont {Vallone}},
  \bibinfo {author} {\bibfnamefont {A.}~\bibnamefont {Chiuri}}, \bibinfo
  {author} {\bibfnamefont {A.}~\bibnamefont {Cabello}}, \ and\ \bibinfo
  {author} {\bibfnamefont {P.}~\bibnamefont {Mataloni}},\ }\href@noop {}
  {\bibfield  {journal} {\bibinfo  {journal} {Phys. Rev. A}\ }\textbf {\bibinfo
  {volume} {81}},\ \bibinfo {pages} {040101} (\bibinfo {year}
  {2010})}\BibitemShut {NoStop}%
\bibitem [{\citenamefont {Vallone}\ \emph {et~al.}(2011)\citenamefont
  {Vallone}, \citenamefont {Gianani}, \citenamefont {Inostroza}, \citenamefont
  {Saavedra}, \citenamefont {Lima}, \citenamefont {Cabello},\ and\
  \citenamefont {Mataloni}}]{vallone2011testing}%
  \BibitemOpen
  \bibfield  {author} {\bibinfo {author} {\bibfnamefont {G.}~\bibnamefont
  {Vallone}}, \bibinfo {author} {\bibfnamefont {I.}~\bibnamefont {Gianani}},
  \bibinfo {author} {\bibfnamefont {E.~B.}\ \bibnamefont {Inostroza}}, \bibinfo
  {author} {\bibfnamefont {C.}~\bibnamefont {Saavedra}}, \bibinfo {author}
  {\bibfnamefont {G.}~\bibnamefont {Lima}}, \bibinfo {author} {\bibfnamefont
  {A.}~\bibnamefont {Cabello}}, \ and\ \bibinfo {author} {\bibfnamefont
  {P.}~\bibnamefont {Mataloni}},\ }\href@noop {} {\bibfield  {journal}
  {\bibinfo  {journal} {Phys. Rev. A}\ }\textbf {\bibinfo {volume} {83}},\
  \bibinfo {pages} {042105} (\bibinfo {year} {2011})}\BibitemShut {NoStop}%
\bibitem [{\citenamefont {Cuevas}\ \emph {et~al.}(2013)\citenamefont {Cuevas},
  \citenamefont {Carvacho}, \citenamefont {Saavedra}, \citenamefont
  {Cari{\~n}e}, \citenamefont {Nogueira}, \citenamefont {Figueroa},
  \citenamefont {Cabello}, \citenamefont {Mataloni}, \citenamefont {Lima},\
  and\ \citenamefont {Xavier}}]{cuevas2013long}%
  \BibitemOpen
  \bibfield  {author} {\bibinfo {author} {\bibfnamefont {A.}~\bibnamefont
  {Cuevas}}, \bibinfo {author} {\bibfnamefont {G.}~\bibnamefont {Carvacho}},
  \bibinfo {author} {\bibfnamefont {G.}~\bibnamefont {Saavedra}}, \bibinfo
  {author} {\bibfnamefont {J.}~\bibnamefont {Cari{\~n}e}}, \bibinfo {author}
  {\bibfnamefont {W.}~\bibnamefont {Nogueira}}, \bibinfo {author}
  {\bibfnamefont {M.}~\bibnamefont {Figueroa}}, \bibinfo {author}
  {\bibfnamefont {A.}~\bibnamefont {Cabello}}, \bibinfo {author} {\bibfnamefont
  {P.}~\bibnamefont {Mataloni}}, \bibinfo {author} {\bibfnamefont
  {G.}~\bibnamefont {Lima}}, \ and\ \bibinfo {author} {\bibfnamefont
  {G.}~\bibnamefont {Xavier}},\ }\href@noop {} {\bibfield  {journal} {\bibinfo
  {journal} {Nat. Commun.}\ }\textbf {\bibinfo {volume} {4}},\ \bibinfo {pages}
  {1} (\bibinfo {year} {2013})}\BibitemShut {NoStop}%
\bibitem [{\citenamefont {Brecht}\ \emph {et~al.}(2015)\citenamefont {Brecht},
  \citenamefont {Reddy}, \citenamefont {Silberhorn},\ and\ \citenamefont
  {Raymer}}]{brecht2015photon}%
  \BibitemOpen
  \bibfield  {author} {\bibinfo {author} {\bibfnamefont {B.}~\bibnamefont
  {Brecht}}, \bibinfo {author} {\bibfnamefont {D.~V.}\ \bibnamefont {Reddy}},
  \bibinfo {author} {\bibfnamefont {C.}~\bibnamefont {Silberhorn}}, \ and\
  \bibinfo {author} {\bibfnamefont {M.}~\bibnamefont {Raymer}},\ }\href@noop {}
  {\bibfield  {journal} {\bibinfo  {journal} {Phys. Rev. X}\ }\textbf {\bibinfo
  {volume} {5}},\ \bibinfo {pages} {041017} (\bibinfo {year}
  {2015})}\BibitemShut {NoStop}%
\bibitem [{\citenamefont {Marcikic}\ \emph {et~al.}(2002)\citenamefont
  {Marcikic}, \citenamefont {de~Riedmatten}, \citenamefont {Tittel},
  \citenamefont {Scarani}, \citenamefont {Zbinden},\ and\ \citenamefont
  {Gisin}}]{marcikic2002time}%
  \BibitemOpen
  \bibfield  {author} {\bibinfo {author} {\bibfnamefont {I.}~\bibnamefont
  {Marcikic}}, \bibinfo {author} {\bibfnamefont {H.}~\bibnamefont
  {de~Riedmatten}}, \bibinfo {author} {\bibfnamefont {W.}~\bibnamefont
  {Tittel}}, \bibinfo {author} {\bibfnamefont {V.}~\bibnamefont {Scarani}},
  \bibinfo {author} {\bibfnamefont {H.}~\bibnamefont {Zbinden}}, \ and\
  \bibinfo {author} {\bibfnamefont {N.}~\bibnamefont {Gisin}},\ }\href@noop {}
  {\bibfield  {journal} {\bibinfo  {journal} {Phys. Rev. A}\ }\textbf {\bibinfo
  {volume} {66}},\ \bibinfo {pages} {062308} (\bibinfo {year}
  {2002})}\BibitemShut {NoStop}%
\bibitem [{\citenamefont {Marcikic}\ \emph {et~al.}(2004)\citenamefont
  {Marcikic}, \citenamefont {De~Riedmatten}, \citenamefont {Tittel},
  \citenamefont {Zbinden}, \citenamefont {Legr{\'e}},\ and\ \citenamefont
  {Gisin}}]{marcikic2004distribution}%
  \BibitemOpen
  \bibfield  {author} {\bibinfo {author} {\bibfnamefont {I.}~\bibnamefont
  {Marcikic}}, \bibinfo {author} {\bibfnamefont {H.}~\bibnamefont
  {De~Riedmatten}}, \bibinfo {author} {\bibfnamefont {W.}~\bibnamefont
  {Tittel}}, \bibinfo {author} {\bibfnamefont {H.}~\bibnamefont {Zbinden}},
  \bibinfo {author} {\bibfnamefont {M.}~\bibnamefont {Legr{\'e}}}, \ and\
  \bibinfo {author} {\bibfnamefont {N.}~\bibnamefont {Gisin}},\ }\href@noop {}
  {\bibfield  {journal} {\bibinfo  {journal} {Phys. Rev. Lett.}\ }\textbf
  {\bibinfo {volume} {93}},\ \bibinfo {pages} {180502} (\bibinfo {year}
  {2004})}\BibitemShut {NoStop}%
\bibitem [{\citenamefont {Tittel}\ \emph {et~al.}(1998)\citenamefont {Tittel},
  \citenamefont {Brendel}, \citenamefont {Gisin}, \citenamefont {Herzog},
  \citenamefont {Zbinden},\ and\ \citenamefont
  {Gisin}}]{tittel1998experimental}%
  \BibitemOpen
  \bibfield  {author} {\bibinfo {author} {\bibfnamefont {W.}~\bibnamefont
  {Tittel}}, \bibinfo {author} {\bibfnamefont {J.}~\bibnamefont {Brendel}},
  \bibinfo {author} {\bibfnamefont {B.}~\bibnamefont {Gisin}}, \bibinfo
  {author} {\bibfnamefont {T.}~\bibnamefont {Herzog}}, \bibinfo {author}
  {\bibfnamefont {H.}~\bibnamefont {Zbinden}}, \ and\ \bibinfo {author}
  {\bibfnamefont {N.}~\bibnamefont {Gisin}},\ }\href@noop {} {\bibfield
  {journal} {\bibinfo  {journal} {Phys. Rev. A}\ }\textbf {\bibinfo {volume}
  {57}},\ \bibinfo {pages} {3229} (\bibinfo {year} {1998})}\BibitemShut
  {NoStop}%
\bibitem [{\citenamefont {Tittel}\ \emph {et~al.}(2000)\citenamefont {Tittel},
  \citenamefont {Brendel}, \citenamefont {Zbinden},\ and\ \citenamefont
  {Gisin}}]{tittel2000quantum}%
  \BibitemOpen
  \bibfield  {author} {\bibinfo {author} {\bibfnamefont {W.}~\bibnamefont
  {Tittel}}, \bibinfo {author} {\bibfnamefont {J.}~\bibnamefont {Brendel}},
  \bibinfo {author} {\bibfnamefont {H.}~\bibnamefont {Zbinden}}, \ and\
  \bibinfo {author} {\bibfnamefont {N.}~\bibnamefont {Gisin}},\ }\href@noop {}
  {\bibfield  {journal} {\bibinfo  {journal} {Phys. Rev. Lett.}\ }\textbf
  {\bibinfo {volume} {84}},\ \bibinfo {pages} {4737} (\bibinfo {year}
  {2000})}\BibitemShut {NoStop}%
\bibitem [{\citenamefont {Kwon}\ \emph {et~al.}(2013)\citenamefont {Kwon},
  \citenamefont {Park}, \citenamefont {Ra}, \citenamefont {Kim},\ and\
  \citenamefont {Kim}}]{kwon2013time}%
  \BibitemOpen
  \bibfield  {author} {\bibinfo {author} {\bibfnamefont {O.}~\bibnamefont
  {Kwon}}, \bibinfo {author} {\bibfnamefont {K.-K.}\ \bibnamefont {Park}},
  \bibinfo {author} {\bibfnamefont {Y.-S.}\ \bibnamefont {Ra}}, \bibinfo
  {author} {\bibfnamefont {Y.-S.}\ \bibnamefont {Kim}}, \ and\ \bibinfo
  {author} {\bibfnamefont {Y.-H.}\ \bibnamefont {Kim}},\ }\href@noop {}
  {\bibfield  {journal} {\bibinfo  {journal} {Opt. Express}\ }\textbf {\bibinfo
  {volume} {21}},\ \bibinfo {pages} {25492} (\bibinfo {year}
  {2013})}\BibitemShut {NoStop}%
\bibitem [{\citenamefont {Baek}\ \emph {et~al.}(2007)\citenamefont {Baek},
  \citenamefont {Kwon},\ and\ \citenamefont {Kim}}]{baek2007high}%
  \BibitemOpen
  \bibfield  {author} {\bibinfo {author} {\bibfnamefont {S.-Y.}\ \bibnamefont
  {Baek}}, \bibinfo {author} {\bibfnamefont {O.}~\bibnamefont {Kwon}}, \ and\
  \bibinfo {author} {\bibfnamefont {Y.-H.}\ \bibnamefont {Kim}},\ }\href@noop
  {} {\bibfield  {journal} {\bibinfo  {journal} {Jap. J. Appl. Phys.}\ }\textbf
  {\bibinfo {volume} {46}},\ \bibinfo {pages} {7720} (\bibinfo {year}
  {2007})}\BibitemShut {NoStop}%
\bibitem [{\citenamefont {Kwon}\ \emph {et~al.}(2009)\citenamefont {Kwon},
  \citenamefont {Ra},\ and\ \citenamefont {Kim}}]{kwon2009coherence}%
  \BibitemOpen
  \bibfield  {author} {\bibinfo {author} {\bibfnamefont {O.}~\bibnamefont
  {Kwon}}, \bibinfo {author} {\bibfnamefont {Y.-S.}\ \bibnamefont {Ra}}, \ and\
  \bibinfo {author} {\bibfnamefont {Y.-H.}\ \bibnamefont {Kim}},\ }\href@noop
  {} {\bibfield  {journal} {\bibinfo  {journal} {Opt. Express}\ }\textbf
  {\bibinfo {volume} {17}},\ \bibinfo {pages} {13059} (\bibinfo {year}
  {2009})}\BibitemShut {NoStop}%
\bibitem [{\citenamefont {MacLean}\ \emph {et~al.}(2018)\citenamefont
  {MacLean}, \citenamefont {Donohue},\ and\ \citenamefont
  {Resch}}]{maclean2018ultrafast}%
  \BibitemOpen
  \bibfield  {author} {\bibinfo {author} {\bibfnamefont {J.-P.~W.}\
  \bibnamefont {MacLean}}, \bibinfo {author} {\bibfnamefont {J.~M.}\
  \bibnamefont {Donohue}}, \ and\ \bibinfo {author} {\bibfnamefont {K.~J.}\
  \bibnamefont {Resch}},\ }\href@noop {} {\bibfield  {journal} {\bibinfo
  {journal} {Phys. Rev. A}\ }\textbf {\bibinfo {volume} {97}},\ \bibinfo
  {pages} {063826} (\bibinfo {year} {2018})}\BibitemShut {NoStop}%
\bibitem [{\citenamefont {Vedovato}\ \emph {et~al.}(2018)\citenamefont
  {Vedovato}, \citenamefont {Agnesi}, \citenamefont {Tomasin}, \citenamefont
  {Avesani}, \citenamefont {Larsson}, \citenamefont {Vallone},\ and\
  \citenamefont {Villoresi}}]{vedovato2018postselection}%
  \BibitemOpen
  \bibfield  {author} {\bibinfo {author} {\bibfnamefont {F.}~\bibnamefont
  {Vedovato}}, \bibinfo {author} {\bibfnamefont {C.}~\bibnamefont {Agnesi}},
  \bibinfo {author} {\bibfnamefont {M.}~\bibnamefont {Tomasin}}, \bibinfo
  {author} {\bibfnamefont {M.}~\bibnamefont {Avesani}}, \bibinfo {author}
  {\bibfnamefont {J.-{\AA}.}\ \bibnamefont {Larsson}}, \bibinfo {author}
  {\bibfnamefont {G.}~\bibnamefont {Vallone}}, \ and\ \bibinfo {author}
  {\bibfnamefont {P.}~\bibnamefont {Villoresi}},\ }\href@noop {} {\bibfield
  {journal} {\bibinfo  {journal} {Phys. Rev. Lett.}\ }\textbf {\bibinfo
  {volume} {121}},\ \bibinfo {pages} {190401} (\bibinfo {year}
  {2018})}\BibitemShut {NoStop}%
\bibitem [{\citenamefont {Ou}\ and\ \citenamefont
  {Mandel}(1988{\natexlab{b}})}]{ou1988observation}%
  \BibitemOpen
  \bibfield  {author} {\bibinfo {author} {\bibfnamefont {Z.~Y.}\ \bibnamefont
  {Ou}}\ and\ \bibinfo {author} {\bibfnamefont {L.}~\bibnamefont {Mandel}},\
  }\href@noop {} {\bibfield  {journal} {\bibinfo  {journal} {Phys. Rev. Lett.}\
  }\textbf {\bibinfo {volume} {61}},\ \bibinfo {pages} {54} (\bibinfo {year}
  {1988}{\natexlab{b}})}\BibitemShut {NoStop}%
\bibitem [{\citenamefont {Lu}\ \emph {et~al.}(2018)\citenamefont {Lu},
  \citenamefont {Lukens}, \citenamefont {Peters}, \citenamefont {Williams},
  \citenamefont {Weiner},\ and\ \citenamefont {Lougovski}}]{lu2018quantum}%
  \BibitemOpen
  \bibfield  {author} {\bibinfo {author} {\bibfnamefont {H.-H.}\ \bibnamefont
  {Lu}}, \bibinfo {author} {\bibfnamefont {J.~M.}\ \bibnamefont {Lukens}},
  \bibinfo {author} {\bibfnamefont {N.~A.}\ \bibnamefont {Peters}}, \bibinfo
  {author} {\bibfnamefont {B.~P.}\ \bibnamefont {Williams}}, \bibinfo {author}
  {\bibfnamefont {A.~M.}\ \bibnamefont {Weiner}}, \ and\ \bibinfo {author}
  {\bibfnamefont {P.}~\bibnamefont {Lougovski}},\ }\href@noop {} {\bibfield
  {journal} {\bibinfo  {journal} {Optica}\ }\textbf {\bibinfo {volume} {5}},\
  \bibinfo {pages} {1455} (\bibinfo {year} {2018})}\BibitemShut {NoStop}%
\bibitem [{\citenamefont {Chen}\ \emph {et~al.}(2020)\citenamefont {Chen},
  \citenamefont {Ecker}, \citenamefont {Bavaresco}, \citenamefont {Scheidl},
  \citenamefont {Chen}, \citenamefont {Steinlechner}, \citenamefont {Huber},\
  and\ \citenamefont {Ursin}}]{chen2020verification}%
  \BibitemOpen
  \bibfield  {author} {\bibinfo {author} {\bibfnamefont {Y.}~\bibnamefont
  {Chen}}, \bibinfo {author} {\bibfnamefont {S.}~\bibnamefont {Ecker}},
  \bibinfo {author} {\bibfnamefont {J.}~\bibnamefont {Bavaresco}}, \bibinfo
  {author} {\bibfnamefont {T.}~\bibnamefont {Scheidl}}, \bibinfo {author}
  {\bibfnamefont {L.}~\bibnamefont {Chen}}, \bibinfo {author} {\bibfnamefont
  {F.}~\bibnamefont {Steinlechner}}, \bibinfo {author} {\bibfnamefont
  {M.}~\bibnamefont {Huber}}, \ and\ \bibinfo {author} {\bibfnamefont
  {R.}~\bibnamefont {Ursin}},\ }\href@noop {} {\bibfield  {journal} {\bibinfo
  {journal} {Phys. Rev. A}\ }\textbf {\bibinfo {volume} {101}},\ \bibinfo
  {pages} {032302} (\bibinfo {year} {2020})}\BibitemShut {NoStop}%
\bibitem [{\citenamefont {Ramelow}\ \emph {et~al.}(2009)\citenamefont
  {Ramelow}, \citenamefont {Ratschbacher}, \citenamefont {Fedrizzi},
  \citenamefont {Langford},\ and\ \citenamefont
  {Zeilinger}}]{ramelow2009discrete}%
  \BibitemOpen
  \bibfield  {author} {\bibinfo {author} {\bibfnamefont {S.}~\bibnamefont
  {Ramelow}}, \bibinfo {author} {\bibfnamefont {L.}~\bibnamefont
  {Ratschbacher}}, \bibinfo {author} {\bibfnamefont {A.}~\bibnamefont
  {Fedrizzi}}, \bibinfo {author} {\bibfnamefont {N.}~\bibnamefont {Langford}},
  \ and\ \bibinfo {author} {\bibfnamefont {A.}~\bibnamefont {Zeilinger}},\
  }\href@noop {} {\bibfield  {journal} {\bibinfo  {journal} {Phys. Rev. Lett.}\
  }\textbf {\bibinfo {volume} {103}},\ \bibinfo {pages} {253601} (\bibinfo
  {year} {2009})}\BibitemShut {NoStop}%
\bibitem [{\citenamefont {Chen}\ \emph
  {et~al.}(2019{\natexlab{b}})\citenamefont {Chen}, \citenamefont {Fink},
  \citenamefont {Steinlechner}, \citenamefont {Torres},\ and\ \citenamefont
  {Ursin}}]{chen2019hong}%
  \BibitemOpen
  \bibfield  {author} {\bibinfo {author} {\bibfnamefont {Y.}~\bibnamefont
  {Chen}}, \bibinfo {author} {\bibfnamefont {M.}~\bibnamefont {Fink}}, \bibinfo
  {author} {\bibfnamefont {F.}~\bibnamefont {Steinlechner}}, \bibinfo {author}
  {\bibfnamefont {J.~P.}\ \bibnamefont {Torres}}, \ and\ \bibinfo {author}
  {\bibfnamefont {R.}~\bibnamefont {Ursin}},\ }\href@noop {} {\bibfield
  {journal} {\bibinfo  {journal} {npj Quant. Inf.}\ }\textbf {\bibinfo {volume}
  {5}},\ \bibinfo {pages} {1} (\bibinfo {year}
  {2019}{\natexlab{b}})}\BibitemShut {NoStop}%
\bibitem [{\citenamefont {Kaneda}\ \emph {et~al.}(2019)\citenamefont {Kaneda},
  \citenamefont {Suzuki}, \citenamefont {Shimizu},\ and\ \citenamefont
  {Edamatsu}}]{kaneda2019direct}%
  \BibitemOpen
  \bibfield  {author} {\bibinfo {author} {\bibfnamefont {F.}~\bibnamefont
  {Kaneda}}, \bibinfo {author} {\bibfnamefont {H.}~\bibnamefont {Suzuki}},
  \bibinfo {author} {\bibfnamefont {R.}~\bibnamefont {Shimizu}}, \ and\
  \bibinfo {author} {\bibfnamefont {K.}~\bibnamefont {Edamatsu}},\ }\href@noop
  {} {\bibfield  {journal} {\bibinfo  {journal} {Opt. Express}\ }\textbf
  {\bibinfo {volume} {27}},\ \bibinfo {pages} {1416} (\bibinfo {year}
  {2019})}\BibitemShut {NoStop}%
\bibitem [{\citenamefont {Kues}\ \emph {et~al.}(2019)\citenamefont {Kues},
  \citenamefont {Reimer}, \citenamefont {Lukens}, \citenamefont {Munro},
  \citenamefont {Weiner}, \citenamefont {Moss},\ and\ \citenamefont
  {Morandotti}}]{kues2019quantum}%
  \BibitemOpen
  \bibfield  {author} {\bibinfo {author} {\bibfnamefont {M.}~\bibnamefont
  {Kues}}, \bibinfo {author} {\bibfnamefont {C.}~\bibnamefont {Reimer}},
  \bibinfo {author} {\bibfnamefont {J.~M.}\ \bibnamefont {Lukens}}, \bibinfo
  {author} {\bibfnamefont {W.~J.}\ \bibnamefont {Munro}}, \bibinfo {author}
  {\bibfnamefont {A.~M.}\ \bibnamefont {Weiner}}, \bibinfo {author}
  {\bibfnamefont {D.~J.}\ \bibnamefont {Moss}}, \ and\ \bibinfo {author}
  {\bibfnamefont {R.}~\bibnamefont {Morandotti}},\ }\href@noop {} {\bibfield
  {journal} {\bibinfo  {journal} {Nat. Photonics}\ }\textbf {\bibinfo {volume}
  {13}},\ \bibinfo {pages} {170} (\bibinfo {year} {2019})}\BibitemShut
  {NoStop}%
\bibitem [{\citenamefont {Reimer}\ \emph {et~al.}(2019)\citenamefont {Reimer},
  \citenamefont {Sciara}, \citenamefont {Roztocki}, \citenamefont {Islam},
  \citenamefont {Cort{\'e}s}, \citenamefont {Zhang}, \citenamefont {Fischer},
  \citenamefont {Loranger}, \citenamefont {Kashyap}, \citenamefont {Cino} \emph
  {et~al.}}]{reimer2019high}%
  \BibitemOpen
  \bibfield  {author} {\bibinfo {author} {\bibfnamefont {C.}~\bibnamefont
  {Reimer}}, \bibinfo {author} {\bibfnamefont {S.}~\bibnamefont {Sciara}},
  \bibinfo {author} {\bibfnamefont {P.}~\bibnamefont {Roztocki}}, \bibinfo
  {author} {\bibfnamefont {M.}~\bibnamefont {Islam}}, \bibinfo {author}
  {\bibfnamefont {L.~R.}\ \bibnamefont {Cort{\'e}s}}, \bibinfo {author}
  {\bibfnamefont {Y.}~\bibnamefont {Zhang}}, \bibinfo {author} {\bibfnamefont
  {B.}~\bibnamefont {Fischer}}, \bibinfo {author} {\bibfnamefont
  {S.}~\bibnamefont {Loranger}}, \bibinfo {author} {\bibfnamefont
  {R.}~\bibnamefont {Kashyap}}, \bibinfo {author} {\bibfnamefont
  {A.}~\bibnamefont {Cino}},  \emph {et~al.},\ }\href@noop {} {\bibfield
  {journal} {\bibinfo  {journal} {Nat. Phys.}\ }\textbf {\bibinfo {volume}
  {15}},\ \bibinfo {pages} {148} (\bibinfo {year} {2019})}\BibitemShut
  {NoStop}%
\bibitem [{\citenamefont {Giovannetti}\ \emph {et~al.}(2011)\citenamefont
  {Giovannetti}, \citenamefont {Lloyd},\ and\ \citenamefont
  {Maccone}}]{giovannetti2011advances}%
  \BibitemOpen
  \bibfield  {author} {\bibinfo {author} {\bibfnamefont {V.}~\bibnamefont
  {Giovannetti}}, \bibinfo {author} {\bibfnamefont {S.}~\bibnamefont {Lloyd}},
  \ and\ \bibinfo {author} {\bibfnamefont {L.}~\bibnamefont {Maccone}},\
  }\href@noop {} {\bibfield  {journal} {\bibinfo  {journal} {Nat. Photonics}\
  }\textbf {\bibinfo {volume} {5}},\ \bibinfo {pages} {222} (\bibinfo {year}
  {2011})}\BibitemShut {NoStop}%
\bibitem [{\citenamefont {Pirandola}\ \emph {et~al.}(2018)\citenamefont
  {Pirandola}, \citenamefont {Bardhan}, \citenamefont {Gehring}, \citenamefont
  {Weedbrook},\ and\ \citenamefont {Lloyd}}]{pirandola2018advances}%
  \BibitemOpen
  \bibfield  {author} {\bibinfo {author} {\bibfnamefont {S.}~\bibnamefont
  {Pirandola}}, \bibinfo {author} {\bibfnamefont {B.~R.}\ \bibnamefont
  {Bardhan}}, \bibinfo {author} {\bibfnamefont {T.}~\bibnamefont {Gehring}},
  \bibinfo {author} {\bibfnamefont {C.}~\bibnamefont {Weedbrook}}, \ and\
  \bibinfo {author} {\bibfnamefont {S.}~\bibnamefont {Lloyd}},\ }\href@noop {}
  {\bibfield  {journal} {\bibinfo  {journal} {Nat. Photonics}\ }\textbf
  {\bibinfo {volume} {12}},\ \bibinfo {pages} {724} (\bibinfo {year}
  {2018})}\BibitemShut {NoStop}%
\bibitem [{\citenamefont {Polino}\ \emph {et~al.}(2020)\citenamefont {Polino},
  \citenamefont {Valeri}, \citenamefont {Spagnolo},\ and\ \citenamefont
  {Sciarrino}}]{polino2020photonicreview}%
  \BibitemOpen
  \bibfield  {author} {\bibinfo {author} {\bibfnamefont {E.}~\bibnamefont
  {Polino}}, \bibinfo {author} {\bibfnamefont {M.}~\bibnamefont {Valeri}},
  \bibinfo {author} {\bibfnamefont {N.}~\bibnamefont {Spagnolo}}, \ and\
  \bibinfo {author} {\bibfnamefont {F.}~\bibnamefont {Sciarrino}},\ }\href
  {\doibase 10.1116/5.0007577} {\bibfield  {journal} {\bibinfo  {journal} {AVS
  Quant. Sci.}\ }\textbf {\bibinfo {volume} {2}},\ \bibinfo {pages} {024703}
  (\bibinfo {year} {2020})}\BibitemShut {NoStop}%
\bibitem [{\citenamefont {Moreau}\ \emph {et~al.}(2019)\citenamefont {Moreau},
  \citenamefont {Toninelli}, \citenamefont {Gregory},\ and\ \citenamefont
  {Padgett}}]{moreau2019imaging}%
  \BibitemOpen
  \bibfield  {author} {\bibinfo {author} {\bibfnamefont {P.-A.}\ \bibnamefont
  {Moreau}}, \bibinfo {author} {\bibfnamefont {E.}~\bibnamefont {Toninelli}},
  \bibinfo {author} {\bibfnamefont {T.}~\bibnamefont {Gregory}}, \ and\
  \bibinfo {author} {\bibfnamefont {M.~J.}\ \bibnamefont {Padgett}},\
  }\href@noop {} {\bibfield  {journal} {\bibinfo  {journal} {Nat. Rev. Phys.}\
  }\textbf {\bibinfo {volume} {1}},\ \bibinfo {pages} {367} (\bibinfo {year}
  {2019})}\BibitemShut {NoStop}%
\bibitem [{\citenamefont {Maga{\~{n}}a-Loaiza}\ and\ \citenamefont
  {Boyd}(2019)}]{Maga_a_Loaiza_2019}%
  \BibitemOpen
  \bibfield  {author} {\bibinfo {author} {\bibfnamefont {O.~S.}\ \bibnamefont
  {Maga{\~{n}}a-Loaiza}}\ and\ \bibinfo {author} {\bibfnamefont {R.~W.}\
  \bibnamefont {Boyd}},\ }\href {\doibase 10.1088/1361-6633/ab5005} {\bibfield
  {journal} {\bibinfo  {journal} {Rep. Prog. Phys.}\ }\textbf {\bibinfo
  {volume} {82}},\ \bibinfo {pages} {124401} (\bibinfo {year}
  {2019})}\BibitemShut {NoStop}%
\bibitem [{\citenamefont {Zeilinger}(1999)}]{zeilinger1999experiment}%
  \BibitemOpen
  \bibfield  {author} {\bibinfo {author} {\bibfnamefont {A.}~\bibnamefont
  {Zeilinger}},\ }in\ \href@noop {} {\emph {\bibinfo {booktitle} {More Things
  in Heaven and Earth}}}\ (\bibinfo  {publisher} {Springer},\ \bibinfo {year}
  {1999})\ pp.\ \bibinfo {pages} {482--498}\BibitemShut {NoStop}%
\bibitem [{\citenamefont {Shadbolt}\ \emph {et~al.}(2014)\citenamefont
  {Shadbolt}, \citenamefont {Mathews}, \citenamefont {Laing},\ and\
  \citenamefont {O'brien}}]{shadbolt2014testing}%
  \BibitemOpen
  \bibfield  {author} {\bibinfo {author} {\bibfnamefont {P.}~\bibnamefont
  {Shadbolt}}, \bibinfo {author} {\bibfnamefont {J.~C.}\ \bibnamefont
  {Mathews}}, \bibinfo {author} {\bibfnamefont {A.}~\bibnamefont {Laing}}, \
  and\ \bibinfo {author} {\bibfnamefont {J.~L.}\ \bibnamefont {O'brien}},\
  }\href@noop {} {\bibfield  {journal} {\bibinfo  {journal} {Nat. Phys.}\
  }\textbf {\bibinfo {volume} {10}},\ \bibinfo {pages} {278} (\bibinfo {year}
  {2014})}\BibitemShut {NoStop}%
\bibitem [{\citenamefont {Weston}\ \emph {et~al.}(2013)\citenamefont {Weston},
  \citenamefont {Hall}, \citenamefont {Palsson}, \citenamefont {Wiseman},\ and\
  \citenamefont {Pryde}}]{weston2013experimental}%
  \BibitemOpen
  \bibfield  {author} {\bibinfo {author} {\bibfnamefont {M.~M.}\ \bibnamefont
  {Weston}}, \bibinfo {author} {\bibfnamefont {M.~J.}\ \bibnamefont {Hall}},
  \bibinfo {author} {\bibfnamefont {M.~S.}\ \bibnamefont {Palsson}}, \bibinfo
  {author} {\bibfnamefont {H.~M.}\ \bibnamefont {Wiseman}}, \ and\ \bibinfo
  {author} {\bibfnamefont {G.~J.}\ \bibnamefont {Pryde}},\ }\href@noop {}
  {\bibfield  {journal} {\bibinfo  {journal} {Phys. Rev. Lett.}\ }\textbf
  {\bibinfo {volume} {110}},\ \bibinfo {pages} {220402} (\bibinfo {year}
  {2013})}\BibitemShut {NoStop}%
\bibitem [{\citenamefont {Ma}\ \emph {et~al.}(2016)\citenamefont {Ma},
  \citenamefont {Kofler},\ and\ \citenamefont {Zeilinger}}]{ma2016delayed}%
  \BibitemOpen
  \bibfield  {author} {\bibinfo {author} {\bibfnamefont {X.-s.}\ \bibnamefont
  {Ma}}, \bibinfo {author} {\bibfnamefont {J.}~\bibnamefont {Kofler}}, \ and\
  \bibinfo {author} {\bibfnamefont {A.}~\bibnamefont {Zeilinger}},\ }\href@noop
  {} {\bibfield  {journal} {\bibinfo  {journal} {Rev. Mod. Phys.}\ }\textbf
  {\bibinfo {volume} {88}},\ \bibinfo {pages} {015005} (\bibinfo {year}
  {2016})}\BibitemShut {NoStop}%
\bibitem [{\citenamefont {Weihs}\ \emph {et~al.}(1998)\citenamefont {Weihs},
  \citenamefont {Jennewein}, \citenamefont {Simon}, \citenamefont
  {Weinfurter},\ and\ \citenamefont {Zeilinger}}]{weihs1998violation}%
  \BibitemOpen
  \bibfield  {author} {\bibinfo {author} {\bibfnamefont {G.}~\bibnamefont
  {Weihs}}, \bibinfo {author} {\bibfnamefont {T.}~\bibnamefont {Jennewein}},
  \bibinfo {author} {\bibfnamefont {C.}~\bibnamefont {Simon}}, \bibinfo
  {author} {\bibfnamefont {H.}~\bibnamefont {Weinfurter}}, \ and\ \bibinfo
  {author} {\bibfnamefont {A.}~\bibnamefont {Zeilinger}},\ }\href@noop {}
  {\bibfield  {journal} {\bibinfo  {journal} {Phys. Rev. Lett.}\ }\textbf
  {\bibinfo {volume} {81}},\ \bibinfo {pages} {5039} (\bibinfo {year}
  {1998})}\BibitemShut {NoStop}%
\bibitem [{\citenamefont {Christensen}\ \emph {et~al.}(2013)\citenamefont
  {Christensen}, \citenamefont {McCusker}, \citenamefont {Altepeter},
  \citenamefont {Calkins}, \citenamefont {Gerrits}, \citenamefont {Lita},
  \citenamefont {Miller}, \citenamefont {Shalm}, \citenamefont {Zhang},
  \citenamefont {Nam} \emph {et~al.}}]{christensen2013detection}%
  \BibitemOpen
  \bibfield  {author} {\bibinfo {author} {\bibfnamefont {B.}~\bibnamefont
  {Christensen}}, \bibinfo {author} {\bibfnamefont {K.}~\bibnamefont
  {McCusker}}, \bibinfo {author} {\bibfnamefont {J.}~\bibnamefont {Altepeter}},
  \bibinfo {author} {\bibfnamefont {B.}~\bibnamefont {Calkins}}, \bibinfo
  {author} {\bibfnamefont {T.}~\bibnamefont {Gerrits}}, \bibinfo {author}
  {\bibfnamefont {A.~E.}\ \bibnamefont {Lita}}, \bibinfo {author}
  {\bibfnamefont {A.}~\bibnamefont {Miller}}, \bibinfo {author} {\bibfnamefont
  {L.~K.}\ \bibnamefont {Shalm}}, \bibinfo {author} {\bibfnamefont
  {Y.}~\bibnamefont {Zhang}}, \bibinfo {author} {\bibfnamefont
  {S.}~\bibnamefont {Nam}},  \emph {et~al.},\ }\href@noop {} {\bibfield
  {journal} {\bibinfo  {journal} {Phys. Rev. Lett.}\ }\textbf {\bibinfo
  {volume} {111}},\ \bibinfo {pages} {130406} (\bibinfo {year}
  {2013})}\BibitemShut {NoStop}%
\bibitem [{\citenamefont {Saunders}\ \emph {et~al.}(2010)\citenamefont
  {Saunders}, \citenamefont {Jones}, \citenamefont {Wiseman},\ and\
  \citenamefont {Pryde}}]{saunders2010experimental}%
  \BibitemOpen
  \bibfield  {author} {\bibinfo {author} {\bibfnamefont {D.~J.}\ \bibnamefont
  {Saunders}}, \bibinfo {author} {\bibfnamefont {S.~J.}\ \bibnamefont {Jones}},
  \bibinfo {author} {\bibfnamefont {H.~M.}\ \bibnamefont {Wiseman}}, \ and\
  \bibinfo {author} {\bibfnamefont {G.~J.}\ \bibnamefont {Pryde}},\ }\href@noop
  {} {\bibfield  {journal} {\bibinfo  {journal} {Nat. Phys.}\ }\textbf
  {\bibinfo {volume} {6}},\ \bibinfo {pages} {845} (\bibinfo {year}
  {2010})}\BibitemShut {NoStop}%
\bibitem [{\citenamefont {Vermeyden}\ \emph {et~al.}(2013)\citenamefont
  {Vermeyden}, \citenamefont {Bonsma}, \citenamefont {Noel}, \citenamefont
  {Donohue}, \citenamefont {Wolfe},\ and\ \citenamefont
  {Resch}}]{vermeyden2013experimental}%
  \BibitemOpen
  \bibfield  {author} {\bibinfo {author} {\bibfnamefont {L.}~\bibnamefont
  {Vermeyden}}, \bibinfo {author} {\bibfnamefont {M.}~\bibnamefont {Bonsma}},
  \bibinfo {author} {\bibfnamefont {C.}~\bibnamefont {Noel}}, \bibinfo {author}
  {\bibfnamefont {J.}~\bibnamefont {Donohue}}, \bibinfo {author} {\bibfnamefont
  {E.}~\bibnamefont {Wolfe}}, \ and\ \bibinfo {author} {\bibfnamefont
  {K.}~\bibnamefont {Resch}},\ }\href@noop {} {\bibfield  {journal} {\bibinfo
  {journal} {Phys. Rev. A}\ }\textbf {\bibinfo {volume} {87}},\ \bibinfo
  {pages} {032105} (\bibinfo {year} {2013})}\BibitemShut {NoStop}%
\bibitem [{\citenamefont {Xiao}\ \emph {et~al.}(2017)\citenamefont {Xiao},
  \citenamefont {Kedem}, \citenamefont {Xu}, \citenamefont {Li},\ and\
  \citenamefont {Guo}}]{xiao2017experimental}%
  \BibitemOpen
  \bibfield  {author} {\bibinfo {author} {\bibfnamefont {Y.}~\bibnamefont
  {Xiao}}, \bibinfo {author} {\bibfnamefont {Y.}~\bibnamefont {Kedem}},
  \bibinfo {author} {\bibfnamefont {J.-S.}\ \bibnamefont {Xu}}, \bibinfo
  {author} {\bibfnamefont {C.-F.}\ \bibnamefont {Li}}, \ and\ \bibinfo {author}
  {\bibfnamefont {G.-C.}\ \bibnamefont {Guo}},\ }\href@noop {} {\bibfield
  {journal} {\bibinfo  {journal} {Opt. Express}\ }\textbf {\bibinfo {volume}
  {25}},\ \bibinfo {pages} {14463} (\bibinfo {year} {2017})}\BibitemShut
  {NoStop}%
\bibitem [{\citenamefont {Giustina}\ \emph {et~al.}(2013)\citenamefont
  {Giustina}, \citenamefont {Mech}, \citenamefont {Ramelow}, \citenamefont
  {Wittmann}, \citenamefont {Kofler}, \citenamefont {Beyer}, \citenamefont
  {Lita}, \citenamefont {Calkins}, \citenamefont {Gerrits}, \citenamefont {Nam}
  \emph {et~al.}}]{giustina2013bell}%
  \BibitemOpen
  \bibfield  {author} {\bibinfo {author} {\bibfnamefont {M.}~\bibnamefont
  {Giustina}}, \bibinfo {author} {\bibfnamefont {A.}~\bibnamefont {Mech}},
  \bibinfo {author} {\bibfnamefont {S.}~\bibnamefont {Ramelow}}, \bibinfo
  {author} {\bibfnamefont {B.}~\bibnamefont {Wittmann}}, \bibinfo {author}
  {\bibfnamefont {J.}~\bibnamefont {Kofler}}, \bibinfo {author} {\bibfnamefont
  {J.}~\bibnamefont {Beyer}}, \bibinfo {author} {\bibfnamefont
  {A.}~\bibnamefont {Lita}}, \bibinfo {author} {\bibfnamefont {B.}~\bibnamefont
  {Calkins}}, \bibinfo {author} {\bibfnamefont {T.}~\bibnamefont {Gerrits}},
  \bibinfo {author} {\bibfnamefont {S.~W.}\ \bibnamefont {Nam}},  \emph
  {et~al.},\ }\href@noop {} {\bibfield  {journal} {\bibinfo  {journal}
  {Nature}\ }\textbf {\bibinfo {volume} {497}},\ \bibinfo {pages} {227}
  (\bibinfo {year} {2013})}\BibitemShut {NoStop}%
\bibitem [{\citenamefont {Pan}\ \emph {et~al.}(2019)\citenamefont {Pan},
  \citenamefont {Xu}, \citenamefont {Kedem}, \citenamefont {Wang},
  \citenamefont {Chen}, \citenamefont {Jan}, \citenamefont {Sun}, \citenamefont
  {Xu}, \citenamefont {Han}, \citenamefont {Li} \emph
  {et~al.}}]{pan2019direct}%
  \BibitemOpen
  \bibfield  {author} {\bibinfo {author} {\bibfnamefont {W.-W.}\ \bibnamefont
  {Pan}}, \bibinfo {author} {\bibfnamefont {X.-Y.}\ \bibnamefont {Xu}},
  \bibinfo {author} {\bibfnamefont {Y.}~\bibnamefont {Kedem}}, \bibinfo
  {author} {\bibfnamefont {Q.-Q.}\ \bibnamefont {Wang}}, \bibinfo {author}
  {\bibfnamefont {Z.}~\bibnamefont {Chen}}, \bibinfo {author} {\bibfnamefont
  {M.}~\bibnamefont {Jan}}, \bibinfo {author} {\bibfnamefont {K.}~\bibnamefont
  {Sun}}, \bibinfo {author} {\bibfnamefont {J.-S.}\ \bibnamefont {Xu}},
  \bibinfo {author} {\bibfnamefont {Y.-J.}\ \bibnamefont {Han}}, \bibinfo
  {author} {\bibfnamefont {C.-F.}\ \bibnamefont {Li}},  \emph {et~al.},\
  }\href@noop {} {\bibfield  {journal} {\bibinfo  {journal} {Phys. Rev. Lett.}\
  }\textbf {\bibinfo {volume} {123}},\ \bibinfo {pages} {150402} (\bibinfo
  {year} {2019})}\BibitemShut {NoStop}%
\bibitem [{\citenamefont {Zukowski}\ \emph {et~al.}(1995)\citenamefont
  {Zukowski}, \citenamefont {Zeilinger},\ and\ \citenamefont
  {Weinfurter}}]{zukowski1995entangling}%
  \BibitemOpen
  \bibfield  {author} {\bibinfo {author} {\bibfnamefont {M.}~\bibnamefont
  {Zukowski}}, \bibinfo {author} {\bibfnamefont {A.}~\bibnamefont {Zeilinger}},
  \ and\ \bibinfo {author} {\bibfnamefont {H.}~\bibnamefont {Weinfurter}},\
  }\href@noop {} {\bibfield  {journal} {\bibinfo  {journal} {Ann. New York
  Acad. Sci.}\ }\textbf {\bibinfo {volume} {755}},\ \bibinfo {pages} {91}
  (\bibinfo {year} {1995})}\BibitemShut {NoStop}%
\bibitem [{\citenamefont {Yao}\ \emph {et~al.}(2012)\citenamefont {Yao},
  \citenamefont {Wang}, \citenamefont {Xu}, \citenamefont {Lu}, \citenamefont
  {Pan}, \citenamefont {Bao}, \citenamefont {Peng}, \citenamefont {Lu},
  \citenamefont {Chen},\ and\ \citenamefont {Pan}}]{yao2012observation}%
  \BibitemOpen
  \bibfield  {author} {\bibinfo {author} {\bibfnamefont {X.-C.}\ \bibnamefont
  {Yao}}, \bibinfo {author} {\bibfnamefont {T.-X.}\ \bibnamefont {Wang}},
  \bibinfo {author} {\bibfnamefont {P.}~\bibnamefont {Xu}}, \bibinfo {author}
  {\bibfnamefont {H.}~\bibnamefont {Lu}}, \bibinfo {author} {\bibfnamefont
  {G.-S.}\ \bibnamefont {Pan}}, \bibinfo {author} {\bibfnamefont {X.-H.}\
  \bibnamefont {Bao}}, \bibinfo {author} {\bibfnamefont {C.-Z.}\ \bibnamefont
  {Peng}}, \bibinfo {author} {\bibfnamefont {C.-Y.}\ \bibnamefont {Lu}},
  \bibinfo {author} {\bibfnamefont {Y.-A.}\ \bibnamefont {Chen}}, \ and\
  \bibinfo {author} {\bibfnamefont {J.-W.}\ \bibnamefont {Pan}},\ }\href@noop
  {} {\bibfield  {journal} {\bibinfo  {journal} {Nat. Photonics}\ }\textbf
  {\bibinfo {volume} {6}},\ \bibinfo {pages} {225} (\bibinfo {year}
  {2012})}\BibitemShut {NoStop}%
\bibitem [{\citenamefont {Wang}\ \emph {et~al.}(2016)\citenamefont {Wang},
  \citenamefont {Chen}, \citenamefont {Li}, \citenamefont {Huang},
  \citenamefont {Liu}, \citenamefont {Chen}, \citenamefont {Luo}, \citenamefont
  {Su}, \citenamefont {Wu}, \citenamefont {Li} \emph
  {et~al.}}]{wang2016experimental}%
  \BibitemOpen
  \bibfield  {author} {\bibinfo {author} {\bibfnamefont {X.-L.}\ \bibnamefont
  {Wang}}, \bibinfo {author} {\bibfnamefont {L.-K.}\ \bibnamefont {Chen}},
  \bibinfo {author} {\bibfnamefont {W.}~\bibnamefont {Li}}, \bibinfo {author}
  {\bibfnamefont {H.-L.}\ \bibnamefont {Huang}}, \bibinfo {author}
  {\bibfnamefont {C.}~\bibnamefont {Liu}}, \bibinfo {author} {\bibfnamefont
  {C.}~\bibnamefont {Chen}}, \bibinfo {author} {\bibfnamefont {Y.-H.}\
  \bibnamefont {Luo}}, \bibinfo {author} {\bibfnamefont {Z.-E.}\ \bibnamefont
  {Su}}, \bibinfo {author} {\bibfnamefont {D.}~\bibnamefont {Wu}}, \bibinfo
  {author} {\bibfnamefont {Z.-D.}\ \bibnamefont {Li}},  \emph {et~al.},\
  }\href@noop {} {\bibfield  {journal} {\bibinfo  {journal} {Phys. Rev. Lett.}\
  }\textbf {\bibinfo {volume} {117}},\ \bibinfo {pages} {210502} (\bibinfo
  {year} {2016})}\BibitemShut {NoStop}%
\bibitem [{\citenamefont {Mosley}\ \emph {et~al.}(2008)\citenamefont {Mosley},
  \citenamefont {Lundeen}, \citenamefont {Smith}, \citenamefont {Wasylczyk},
  \citenamefont {U'Ren}, \citenamefont {Silberhorn},\ and\ \citenamefont
  {Walmsley}}]{mosley:133601}%
  \BibitemOpen
  \bibfield  {author} {\bibinfo {author} {\bibfnamefont {P.~J.}\ \bibnamefont
  {Mosley}}, \bibinfo {author} {\bibfnamefont {J.~S.}\ \bibnamefont {Lundeen}},
  \bibinfo {author} {\bibfnamefont {B.~J.}\ \bibnamefont {Smith}}, \bibinfo
  {author} {\bibfnamefont {P.}~\bibnamefont {Wasylczyk}}, \bibinfo {author}
  {\bibfnamefont {A.~B.}\ \bibnamefont {U'Ren}}, \bibinfo {author}
  {\bibfnamefont {C.}~\bibnamefont {Silberhorn}}, \ and\ \bibinfo {author}
  {\bibfnamefont {I.~A.}\ \bibnamefont {Walmsley}},\ }\href {\doibase
  10.1103/PhysRevLett.100.133601} {\bibfield  {journal} {\bibinfo  {journal}
  {Phys. Rev. Lett.}\ }\textbf {\bibinfo {volume} {100}},\ \bibinfo {eid}
  {133601} (\bibinfo {year} {2008})}\BibitemShut {NoStop}%
\bibitem [{\citenamefont {Lutz}\ \emph {et~al.}(2014)\citenamefont {Lutz},
  \citenamefont {Kolenderski},\ and\ \citenamefont {Jennewein}}]{Lutz:14}%
  \BibitemOpen
  \bibfield  {author} {\bibinfo {author} {\bibfnamefont {T.}~\bibnamefont
  {Lutz}}, \bibinfo {author} {\bibfnamefont {P.}~\bibnamefont {Kolenderski}}, \
  and\ \bibinfo {author} {\bibfnamefont {T.}~\bibnamefont {Jennewein}},\ }\href
  {\doibase 10.1364/OL.39.001481} {\bibfield  {journal} {\bibinfo  {journal}
  {Opt. Lett.}\ }\textbf {\bibinfo {volume} {39}},\ \bibinfo {pages} {1481}
  (\bibinfo {year} {2014})}\BibitemShut {NoStop}%
\bibitem [{\citenamefont {Ekert}(1991)}]{ekert1991quantum}%
  \BibitemOpen
  \bibfield  {author} {\bibinfo {author} {\bibfnamefont {A.~K.}\ \bibnamefont
  {Ekert}},\ }\href@noop {} {\bibfield  {journal} {\bibinfo  {journal} {Phys.
  Rev. Lett.}\ }\textbf {\bibinfo {volume} {67}},\ \bibinfo {pages} {661}
  (\bibinfo {year} {1991})}\BibitemShut {NoStop}%
\bibitem [{\citenamefont {Bennett}\ \emph {et~al.}(1992)\citenamefont
  {Bennett}, \citenamefont {Brassard},\ and\ \citenamefont
  {Mermin}}]{bennett1992quantum}%
  \BibitemOpen
  \bibfield  {author} {\bibinfo {author} {\bibfnamefont {C.~H.}\ \bibnamefont
  {Bennett}}, \bibinfo {author} {\bibfnamefont {G.}~\bibnamefont {Brassard}}, \
  and\ \bibinfo {author} {\bibfnamefont {N.~D.}\ \bibnamefont {Mermin}},\
  }\href@noop {} {\bibfield  {journal} {\bibinfo  {journal} {Phys. Rev. Lett.}\
  }\textbf {\bibinfo {volume} {68}},\ \bibinfo {pages} {557} (\bibinfo {year}
  {1992})}\BibitemShut {NoStop}%
\bibitem [{\citenamefont {Jennewein}\ \emph {et~al.}(2000)\citenamefont
  {Jennewein}, \citenamefont {Simon}, \citenamefont {Weihs}, \citenamefont
  {Weinfurter},\ and\ \citenamefont {Zeilinger}}]{jennewein2000quantum}%
  \BibitemOpen
  \bibfield  {author} {\bibinfo {author} {\bibfnamefont {T.}~\bibnamefont
  {Jennewein}}, \bibinfo {author} {\bibfnamefont {C.}~\bibnamefont {Simon}},
  \bibinfo {author} {\bibfnamefont {G.}~\bibnamefont {Weihs}}, \bibinfo
  {author} {\bibfnamefont {H.}~\bibnamefont {Weinfurter}}, \ and\ \bibinfo
  {author} {\bibfnamefont {A.}~\bibnamefont {Zeilinger}},\ }\href@noop {}
  {\bibfield  {journal} {\bibinfo  {journal} {Phys. Rev. Lett.}\ }\textbf
  {\bibinfo {volume} {84}},\ \bibinfo {pages} {4729} (\bibinfo {year}
  {2000})}\BibitemShut {NoStop}%
\bibitem [{\citenamefont {Ribordy}\ \emph {et~al.}(2000)\citenamefont
  {Ribordy}, \citenamefont {Brendel}, \citenamefont {Gautier}, \citenamefont
  {Gisin},\ and\ \citenamefont {Zbinden}}]{ribordy2000long}%
  \BibitemOpen
  \bibfield  {author} {\bibinfo {author} {\bibfnamefont {G.}~\bibnamefont
  {Ribordy}}, \bibinfo {author} {\bibfnamefont {J.}~\bibnamefont {Brendel}},
  \bibinfo {author} {\bibfnamefont {J.-D.}\ \bibnamefont {Gautier}}, \bibinfo
  {author} {\bibfnamefont {N.}~\bibnamefont {Gisin}}, \ and\ \bibinfo {author}
  {\bibfnamefont {H.}~\bibnamefont {Zbinden}},\ }\href@noop {} {\bibfield
  {journal} {\bibinfo  {journal} {Phys. Rev. A}\ }\textbf {\bibinfo {volume}
  {63}},\ \bibinfo {pages} {012309} (\bibinfo {year} {2000})}\BibitemShut
  {NoStop}%
\bibitem [{\citenamefont {Naik}\ \emph {et~al.}(2000)\citenamefont {Naik},
  \citenamefont {Peterson}, \citenamefont {White}, \citenamefont {Berglund},\
  and\ \citenamefont {Kwiat}}]{naik2000entangled}%
  \BibitemOpen
  \bibfield  {author} {\bibinfo {author} {\bibfnamefont {D.}~\bibnamefont
  {Naik}}, \bibinfo {author} {\bibfnamefont {C.}~\bibnamefont {Peterson}},
  \bibinfo {author} {\bibfnamefont {A.}~\bibnamefont {White}}, \bibinfo
  {author} {\bibfnamefont {A.}~\bibnamefont {Berglund}}, \ and\ \bibinfo
  {author} {\bibfnamefont {P.~G.}\ \bibnamefont {Kwiat}},\ }\href@noop {}
  {\bibfield  {journal} {\bibinfo  {journal} {Phys. Rev. Lett.}\ }\textbf
  {\bibinfo {volume} {84}},\ \bibinfo {pages} {4733} (\bibinfo {year}
  {2000})}\BibitemShut {NoStop}%
\bibitem [{\citenamefont {Marcikic}\ \emph {et~al.}(2003)\citenamefont
  {Marcikic}, \citenamefont {De~Riedmatten}, \citenamefont {Tittel},
  \citenamefont {Zbinden},\ and\ \citenamefont {Gisin}}]{marcikic2003long}%
  \BibitemOpen
  \bibfield  {author} {\bibinfo {author} {\bibfnamefont {I.}~\bibnamefont
  {Marcikic}}, \bibinfo {author} {\bibfnamefont {H.}~\bibnamefont
  {De~Riedmatten}}, \bibinfo {author} {\bibfnamefont {W.}~\bibnamefont
  {Tittel}}, \bibinfo {author} {\bibfnamefont {H.}~\bibnamefont {Zbinden}}, \
  and\ \bibinfo {author} {\bibfnamefont {N.}~\bibnamefont {Gisin}},\
  }\href@noop {} {\bibfield  {journal} {\bibinfo  {journal} {Nature}\ }\textbf
  {\bibinfo {volume} {421}},\ \bibinfo {pages} {509} (\bibinfo {year}
  {2003})}\BibitemShut {NoStop}%
\bibitem [{\citenamefont {Fasel}\ \emph {et~al.}(2004)\citenamefont {Fasel},
  \citenamefont {Gisin}, \citenamefont {Ribordy},\ and\ \citenamefont
  {Zbinden}}]{fasel2004quantum}%
  \BibitemOpen
  \bibfield  {author} {\bibinfo {author} {\bibfnamefont {S.}~\bibnamefont
  {Fasel}}, \bibinfo {author} {\bibfnamefont {N.}~\bibnamefont {Gisin}},
  \bibinfo {author} {\bibfnamefont {G.}~\bibnamefont {Ribordy}}, \ and\
  \bibinfo {author} {\bibfnamefont {H.}~\bibnamefont {Zbinden}},\ }\href@noop
  {} {\bibfield  {journal} {\bibinfo  {journal} {Eur. Phys. J. D - At. Mol.
  Opt. Plasma Phys.}\ }\textbf {\bibinfo {volume} {30}},\ \bibinfo {pages}
  {143} (\bibinfo {year} {2004})}\BibitemShut {NoStop}%
\bibitem [{\citenamefont {Poppe}\ \emph {et~al.}(2004)\citenamefont {Poppe},
  \citenamefont {Fedrizzi}, \citenamefont {Ursin}, \citenamefont {B{\"o}hm},
  \citenamefont {Lor{\"u}nser}, \citenamefont {Maurhardt}, \citenamefont
  {Peev}, \citenamefont {Suda}, \citenamefont {Kurtsiefer}, \citenamefont
  {Weinfurter} \emph {et~al.}}]{poppe2004practical}%
  \BibitemOpen
  \bibfield  {author} {\bibinfo {author} {\bibfnamefont {A.}~\bibnamefont
  {Poppe}}, \bibinfo {author} {\bibfnamefont {A.}~\bibnamefont {Fedrizzi}},
  \bibinfo {author} {\bibfnamefont {R.}~\bibnamefont {Ursin}}, \bibinfo
  {author} {\bibfnamefont {H.}~\bibnamefont {B{\"o}hm}}, \bibinfo {author}
  {\bibfnamefont {T.}~\bibnamefont {Lor{\"u}nser}}, \bibinfo {author}
  {\bibfnamefont {O.}~\bibnamefont {Maurhardt}}, \bibinfo {author}
  {\bibfnamefont {M.}~\bibnamefont {Peev}}, \bibinfo {author} {\bibfnamefont
  {M.}~\bibnamefont {Suda}}, \bibinfo {author} {\bibfnamefont {C.}~\bibnamefont
  {Kurtsiefer}}, \bibinfo {author} {\bibfnamefont {H.}~\bibnamefont
  {Weinfurter}},  \emph {et~al.},\ }\href@noop {} {\bibfield  {journal}
  {\bibinfo  {journal} {Opt. Express}\ }\textbf {\bibinfo {volume} {12}},\
  \bibinfo {pages} {3865} (\bibinfo {year} {2004})}\BibitemShut {NoStop}%
\bibitem [{\citenamefont {Ursin}\ \emph {et~al.}(2004)\citenamefont {Ursin},
  \citenamefont {Jennewein}, \citenamefont {Aspelmeyer}, \citenamefont
  {Kaltenbaek}, \citenamefont {Lindenthal}, \citenamefont {Walther},\ and\
  \citenamefont {Zeilinger}}]{ursin2004quantum}%
  \BibitemOpen
  \bibfield  {author} {\bibinfo {author} {\bibfnamefont {R.}~\bibnamefont
  {Ursin}}, \bibinfo {author} {\bibfnamefont {T.}~\bibnamefont {Jennewein}},
  \bibinfo {author} {\bibfnamefont {M.}~\bibnamefont {Aspelmeyer}}, \bibinfo
  {author} {\bibfnamefont {R.}~\bibnamefont {Kaltenbaek}}, \bibinfo {author}
  {\bibfnamefont {M.}~\bibnamefont {Lindenthal}}, \bibinfo {author}
  {\bibfnamefont {P.}~\bibnamefont {Walther}}, \ and\ \bibinfo {author}
  {\bibfnamefont {A.}~\bibnamefont {Zeilinger}},\ }\href@noop {} {\bibfield
  {journal} {\bibinfo  {journal} {Nature}\ }\textbf {\bibinfo {volume} {430}},\
  \bibinfo {pages} {849} (\bibinfo {year} {2004})}\BibitemShut {NoStop}%
\bibitem [{\citenamefont {Honjo}\ \emph {et~al.}(2007)\citenamefont {Honjo},
  \citenamefont {Takesue},\ and\ \citenamefont
  {Inoue}}]{honjo2007differential}%
  \BibitemOpen
  \bibfield  {author} {\bibinfo {author} {\bibfnamefont {T.}~\bibnamefont
  {Honjo}}, \bibinfo {author} {\bibfnamefont {H.}~\bibnamefont {Takesue}}, \
  and\ \bibinfo {author} {\bibfnamefont {K.}~\bibnamefont {Inoue}},\
  }\href@noop {} {\bibfield  {journal} {\bibinfo  {journal} {Opt. Lett.}\
  }\textbf {\bibinfo {volume} {32}},\ \bibinfo {pages} {1165} (\bibinfo {year}
  {2007})}\BibitemShut {NoStop}%
\bibitem [{\citenamefont {Honjo}\ \emph {et~al.}(2008)\citenamefont {Honjo},
  \citenamefont {Nam}, \citenamefont {Takesue}, \citenamefont {Zhang},
  \citenamefont {Kamada}, \citenamefont {Nishida}, \citenamefont {Tadanaga},
  \citenamefont {Asobe}, \citenamefont {Baek}, \citenamefont {Hadfield} \emph
  {et~al.}}]{honjo2008long}%
  \BibitemOpen
  \bibfield  {author} {\bibinfo {author} {\bibfnamefont {T.}~\bibnamefont
  {Honjo}}, \bibinfo {author} {\bibfnamefont {S.~W.}\ \bibnamefont {Nam}},
  \bibinfo {author} {\bibfnamefont {H.}~\bibnamefont {Takesue}}, \bibinfo
  {author} {\bibfnamefont {Q.}~\bibnamefont {Zhang}}, \bibinfo {author}
  {\bibfnamefont {H.}~\bibnamefont {Kamada}}, \bibinfo {author} {\bibfnamefont
  {Y.}~\bibnamefont {Nishida}}, \bibinfo {author} {\bibfnamefont
  {O.}~\bibnamefont {Tadanaga}}, \bibinfo {author} {\bibfnamefont
  {M.}~\bibnamefont {Asobe}}, \bibinfo {author} {\bibfnamefont
  {B.}~\bibnamefont {Baek}}, \bibinfo {author} {\bibfnamefont {R.}~\bibnamefont
  {Hadfield}},  \emph {et~al.},\ }\href@noop {} {\bibfield  {journal} {\bibinfo
   {journal} {Opt. Express}\ }\textbf {\bibinfo {volume} {16}},\ \bibinfo
  {pages} {19118} (\bibinfo {year} {2008})}\BibitemShut {NoStop}%
\bibitem [{\citenamefont {Zhang}\ \emph
  {et~al.}(2008{\natexlab{b}})\citenamefont {Zhang}, \citenamefont {Takesue},
  \citenamefont {Nam}, \citenamefont {Langrock}, \citenamefont {Xie},
  \citenamefont {Baek}, \citenamefont {Fejer},\ and\ \citenamefont
  {Yamamoto}}]{zhang2008distribution}%
  \BibitemOpen
  \bibfield  {author} {\bibinfo {author} {\bibfnamefont {Q.}~\bibnamefont
  {Zhang}}, \bibinfo {author} {\bibfnamefont {H.}~\bibnamefont {Takesue}},
  \bibinfo {author} {\bibfnamefont {S.~W.}\ \bibnamefont {Nam}}, \bibinfo
  {author} {\bibfnamefont {C.}~\bibnamefont {Langrock}}, \bibinfo {author}
  {\bibfnamefont {X.}~\bibnamefont {Xie}}, \bibinfo {author} {\bibfnamefont
  {B.}~\bibnamefont {Baek}}, \bibinfo {author} {\bibfnamefont {M.~M.}\
  \bibnamefont {Fejer}}, \ and\ \bibinfo {author} {\bibfnamefont
  {Y.}~\bibnamefont {Yamamoto}},\ }\href@noop {} {\bibfield  {journal}
  {\bibinfo  {journal} {Opt. Express}\ }\textbf {\bibinfo {volume} {16}},\
  \bibinfo {pages} {5776} (\bibinfo {year} {2008}{\natexlab{b}})}\BibitemShut
  {NoStop}%
\bibitem [{\citenamefont {Dynes}\ \emph {et~al.}(2009)\citenamefont {Dynes},
  \citenamefont {Takesue}, \citenamefont {Yuan}, \citenamefont {Sharpe},
  \citenamefont {Harada}, \citenamefont {Honjo}, \citenamefont {Kamada},
  \citenamefont {Tadanaga}, \citenamefont {Nishida}, \citenamefont {Asobe}
  \emph {et~al.}}]{dynes2009efficient}%
  \BibitemOpen
  \bibfield  {author} {\bibinfo {author} {\bibfnamefont {J.~F.}\ \bibnamefont
  {Dynes}}, \bibinfo {author} {\bibfnamefont {H.}~\bibnamefont {Takesue}},
  \bibinfo {author} {\bibfnamefont {Z.~L.}\ \bibnamefont {Yuan}}, \bibinfo
  {author} {\bibfnamefont {A.~W.}\ \bibnamefont {Sharpe}}, \bibinfo {author}
  {\bibfnamefont {K.}~\bibnamefont {Harada}}, \bibinfo {author} {\bibfnamefont
  {T.}~\bibnamefont {Honjo}}, \bibinfo {author} {\bibfnamefont
  {H.}~\bibnamefont {Kamada}}, \bibinfo {author} {\bibfnamefont
  {O.}~\bibnamefont {Tadanaga}}, \bibinfo {author} {\bibfnamefont
  {Y.}~\bibnamefont {Nishida}}, \bibinfo {author} {\bibfnamefont
  {M.}~\bibnamefont {Asobe}},  \emph {et~al.},\ }\href@noop {} {\bibfield
  {journal} {\bibinfo  {journal} {Opt. Express}\ }\textbf {\bibinfo {volume}
  {17}},\ \bibinfo {pages} {11440} (\bibinfo {year} {2009})}\BibitemShut
  {NoStop}%
\bibitem [{\citenamefont {Takesue}\ \emph {et~al.}(2010)\citenamefont
  {Takesue}, \citenamefont {Harada}, \citenamefont {Tamaki}, \citenamefont
  {Fukuda}, \citenamefont {Tsuchizawa}, \citenamefont {Watanabe}, \citenamefont
  {Yamada},\ and\ \citenamefont {Itabashi}}]{takesue2010long}%
  \BibitemOpen
  \bibfield  {author} {\bibinfo {author} {\bibfnamefont {H.}~\bibnamefont
  {Takesue}}, \bibinfo {author} {\bibfnamefont {K.-i.}\ \bibnamefont {Harada}},
  \bibinfo {author} {\bibfnamefont {K.}~\bibnamefont {Tamaki}}, \bibinfo
  {author} {\bibfnamefont {H.}~\bibnamefont {Fukuda}}, \bibinfo {author}
  {\bibfnamefont {T.}~\bibnamefont {Tsuchizawa}}, \bibinfo {author}
  {\bibfnamefont {T.}~\bibnamefont {Watanabe}}, \bibinfo {author}
  {\bibfnamefont {K.}~\bibnamefont {Yamada}}, \ and\ \bibinfo {author}
  {\bibfnamefont {S.-i.}\ \bibnamefont {Itabashi}},\ }\href@noop {} {\bibfield
  {journal} {\bibinfo  {journal} {Opt. Express}\ }\textbf {\bibinfo {volume}
  {18}},\ \bibinfo {pages} {16777} (\bibinfo {year} {2010})}\BibitemShut
  {NoStop}%
\bibitem [{\citenamefont {Inagaki}\ \emph {et~al.}(2013)\citenamefont
  {Inagaki}, \citenamefont {Matsuda}, \citenamefont {Tadanaga}, \citenamefont
  {Asobe},\ and\ \citenamefont {Takesue}}]{inagaki2013entanglement}%
  \BibitemOpen
  \bibfield  {author} {\bibinfo {author} {\bibfnamefont {T.}~\bibnamefont
  {Inagaki}}, \bibinfo {author} {\bibfnamefont {N.}~\bibnamefont {Matsuda}},
  \bibinfo {author} {\bibfnamefont {O.}~\bibnamefont {Tadanaga}}, \bibinfo
  {author} {\bibfnamefont {M.}~\bibnamefont {Asobe}}, \ and\ \bibinfo {author}
  {\bibfnamefont {H.}~\bibnamefont {Takesue}},\ }\href@noop {} {\bibfield
  {journal} {\bibinfo  {journal} {Opt. Express}\ }\textbf {\bibinfo {volume}
  {21}},\ \bibinfo {pages} {23241} (\bibinfo {year} {2013})}\BibitemShut
  {NoStop}%
\bibitem [{\citenamefont {Takesue}\ \emph {et~al.}(2015)\citenamefont
  {Takesue}, \citenamefont {Dyer}, \citenamefont {Stevens}, \citenamefont
  {Verma}, \citenamefont {Mirin},\ and\ \citenamefont
  {Nam}}]{takesue2015quantum}%
  \BibitemOpen
  \bibfield  {author} {\bibinfo {author} {\bibfnamefont {H.}~\bibnamefont
  {Takesue}}, \bibinfo {author} {\bibfnamefont {S.~D.}\ \bibnamefont {Dyer}},
  \bibinfo {author} {\bibfnamefont {M.~J.}\ \bibnamefont {Stevens}}, \bibinfo
  {author} {\bibfnamefont {V.}~\bibnamefont {Verma}}, \bibinfo {author}
  {\bibfnamefont {R.~P.}\ \bibnamefont {Mirin}}, \ and\ \bibinfo {author}
  {\bibfnamefont {S.~W.}\ \bibnamefont {Nam}},\ }\href@noop {} {\bibfield
  {journal} {\bibinfo  {journal} {Optica}\ }\textbf {\bibinfo {volume} {2}},\
  \bibinfo {pages} {832} (\bibinfo {year} {2015})}\BibitemShut {NoStop}%
\bibitem [{\citenamefont {Sun}\ \emph {et~al.}(2017)\citenamefont {Sun},
  \citenamefont {Jiang}, \citenamefont {Mao}, \citenamefont {You},
  \citenamefont {Zhang}, \citenamefont {Zhang}, \citenamefont {Jiang},
  \citenamefont {Chen}, \citenamefont {Li}, \citenamefont {Huang} \emph
  {et~al.}}]{sun2017entanglement}%
  \BibitemOpen
  \bibfield  {author} {\bibinfo {author} {\bibfnamefont {Q.-C.}\ \bibnamefont
  {Sun}}, \bibinfo {author} {\bibfnamefont {Y.-F.}\ \bibnamefont {Jiang}},
  \bibinfo {author} {\bibfnamefont {Y.-L.}\ \bibnamefont {Mao}}, \bibinfo
  {author} {\bibfnamefont {L.-X.}\ \bibnamefont {You}}, \bibinfo {author}
  {\bibfnamefont {W.}~\bibnamefont {Zhang}}, \bibinfo {author} {\bibfnamefont
  {W.-J.}\ \bibnamefont {Zhang}}, \bibinfo {author} {\bibfnamefont
  {X.}~\bibnamefont {Jiang}}, \bibinfo {author} {\bibfnamefont {T.-Y.}\
  \bibnamefont {Chen}}, \bibinfo {author} {\bibfnamefont {H.}~\bibnamefont
  {Li}}, \bibinfo {author} {\bibfnamefont {Y.-D.}\ \bibnamefont {Huang}},
  \emph {et~al.},\ }\href@noop {} {\bibfield  {journal} {\bibinfo  {journal}
  {Optica}\ }\textbf {\bibinfo {volume} {4}},\ \bibinfo {pages} {1214}
  (\bibinfo {year} {2017})}\BibitemShut {NoStop}%
\bibitem [{\citenamefont {Huo}\ \emph {et~al.}(2018)\citenamefont {Huo},
  \citenamefont {Qin}, \citenamefont {Cheng}, \citenamefont {Yan},
  \citenamefont {Qin}, \citenamefont {Su}, \citenamefont {Jia}, \citenamefont
  {Xie},\ and\ \citenamefont {Peng}}]{huo2018deterministic}%
  \BibitemOpen
  \bibfield  {author} {\bibinfo {author} {\bibfnamefont {M.}~\bibnamefont
  {Huo}}, \bibinfo {author} {\bibfnamefont {J.}~\bibnamefont {Qin}}, \bibinfo
  {author} {\bibfnamefont {J.}~\bibnamefont {Cheng}}, \bibinfo {author}
  {\bibfnamefont {Z.}~\bibnamefont {Yan}}, \bibinfo {author} {\bibfnamefont
  {Z.}~\bibnamefont {Qin}}, \bibinfo {author} {\bibfnamefont {X.}~\bibnamefont
  {Su}}, \bibinfo {author} {\bibfnamefont {X.}~\bibnamefont {Jia}}, \bibinfo
  {author} {\bibfnamefont {C.}~\bibnamefont {Xie}}, \ and\ \bibinfo {author}
  {\bibfnamefont {K.}~\bibnamefont {Peng}},\ }\href@noop {} {\bibfield
  {journal} {\bibinfo  {journal} {Sci. Adv.}\ }\textbf {\bibinfo {volume}
  {4}},\ \bibinfo {pages} {eaas9401} (\bibinfo {year} {2018})}\BibitemShut
  {NoStop}%
\bibitem [{\citenamefont {Wengerowsky}\ \emph {et~al.}(2019)\citenamefont
  {Wengerowsky}, \citenamefont {Joshi}, \citenamefont {Steinlechner},
  \citenamefont {Zichi}, \citenamefont {Dobrovolskiy}, \citenamefont {van~der
  Molen}, \citenamefont {Los}, \citenamefont {Zwiller}, \citenamefont
  {Versteegh}, \citenamefont {Mura} \emph
  {et~al.}}]{wengerowsky2019entanglement}%
  \BibitemOpen
  \bibfield  {author} {\bibinfo {author} {\bibfnamefont {S.}~\bibnamefont
  {Wengerowsky}}, \bibinfo {author} {\bibfnamefont {S.~K.}\ \bibnamefont
  {Joshi}}, \bibinfo {author} {\bibfnamefont {F.}~\bibnamefont {Steinlechner}},
  \bibinfo {author} {\bibfnamefont {J.~R.}\ \bibnamefont {Zichi}}, \bibinfo
  {author} {\bibfnamefont {S.~M.}\ \bibnamefont {Dobrovolskiy}}, \bibinfo
  {author} {\bibfnamefont {R.}~\bibnamefont {van~der Molen}}, \bibinfo {author}
  {\bibfnamefont {J.~W.~N.}\ \bibnamefont {Los}}, \bibinfo {author}
  {\bibfnamefont {V.}~\bibnamefont {Zwiller}}, \bibinfo {author} {\bibfnamefont
  {M.~A.~M.}\ \bibnamefont {Versteegh}}, \bibinfo {author} {\bibfnamefont
  {A.}~\bibnamefont {Mura}},  \emph {et~al.},\ }\href@noop {} {\bibfield
  {journal} {\bibinfo  {journal} {Proc. Natl. Acad. Sci.}\ }\textbf {\bibinfo
  {volume} {116}},\ \bibinfo {pages} {6684} (\bibinfo {year}
  {2019})}\BibitemShut {NoStop}%
\bibitem [{\citenamefont {Wengerowsky}\ \emph {et~al.}(2020)\citenamefont
  {Wengerowsky}, \citenamefont {Joshi}, \citenamefont {Steinlechner},
  \citenamefont {Zichi}, \citenamefont {Liu}, \citenamefont {Scheidl},
  \citenamefont {Dobrovolskiy}, \citenamefont {van~der Molen}, \citenamefont
  {Los}, \citenamefont {Zwiller} \emph {et~al.}}]{wengerowsky2020passively}%
  \BibitemOpen
  \bibfield  {author} {\bibinfo {author} {\bibfnamefont {S.}~\bibnamefont
  {Wengerowsky}}, \bibinfo {author} {\bibfnamefont {S.~K.}\ \bibnamefont
  {Joshi}}, \bibinfo {author} {\bibfnamefont {F.}~\bibnamefont {Steinlechner}},
  \bibinfo {author} {\bibfnamefont {J.~R.}\ \bibnamefont {Zichi}}, \bibinfo
  {author} {\bibfnamefont {B.}~\bibnamefont {Liu}}, \bibinfo {author}
  {\bibfnamefont {T.}~\bibnamefont {Scheidl}}, \bibinfo {author} {\bibfnamefont
  {S.~M.}\ \bibnamefont {Dobrovolskiy}}, \bibinfo {author} {\bibfnamefont
  {R.}~\bibnamefont {van~der Molen}}, \bibinfo {author} {\bibfnamefont {J.~W.}\
  \bibnamefont {Los}}, \bibinfo {author} {\bibfnamefont {V.}~\bibnamefont
  {Zwiller}},  \emph {et~al.},\ }\href@noop {} {\bibfield  {journal} {\bibinfo
  {journal} {npj Quant. Inf.}\ }\textbf {\bibinfo {volume} {6}},\ \bibinfo
  {pages} {1} (\bibinfo {year} {2020})}\BibitemShut {NoStop}%
\bibitem [{\citenamefont {Ikuta}\ and\ \citenamefont
  {Takesue}(2018)}]{ikuta2018four}%
  \BibitemOpen
  \bibfield  {author} {\bibinfo {author} {\bibfnamefont {T.}~\bibnamefont
  {Ikuta}}\ and\ \bibinfo {author} {\bibfnamefont {H.}~\bibnamefont
  {Takesue}},\ }\href@noop {} {\bibfield  {journal} {\bibinfo  {journal} {Sci.
  Rep.}\ }\textbf {\bibinfo {volume} {8}},\ \bibinfo {pages} {1} (\bibinfo
  {year} {2018})}\BibitemShut {NoStop}%
\bibitem [{\citenamefont {Liu}\ \emph {et~al.}(2020{\natexlab{a}})\citenamefont
  {Liu}, \citenamefont {Nape}, \citenamefont {Wang}, \citenamefont
  {Vall{\'e}s}, \citenamefont {Wang},\ and\ \citenamefont
  {Forbes}}]{liu2020multidimensional}%
  \BibitemOpen
  \bibfield  {author} {\bibinfo {author} {\bibfnamefont {J.}~\bibnamefont
  {Liu}}, \bibinfo {author} {\bibfnamefont {I.}~\bibnamefont {Nape}}, \bibinfo
  {author} {\bibfnamefont {Q.}~\bibnamefont {Wang}}, \bibinfo {author}
  {\bibfnamefont {A.}~\bibnamefont {Vall{\'e}s}}, \bibinfo {author}
  {\bibfnamefont {J.}~\bibnamefont {Wang}}, \ and\ \bibinfo {author}
  {\bibfnamefont {A.}~\bibnamefont {Forbes}},\ }\href@noop {} {\bibfield
  {journal} {\bibinfo  {journal} {Sci. Adv.}\ }\textbf {\bibinfo {volume}
  {6}},\ \bibinfo {pages} {eaay0837} (\bibinfo {year}
  {2020}{\natexlab{a}})}\BibitemShut {NoStop}%
\bibitem [{\citenamefont {Cao}\ \emph {et~al.}(2020)\citenamefont {Cao},
  \citenamefont {Gao}, \citenamefont {Zhang}, \citenamefont {Wang},
  \citenamefont {He}, \citenamefont {Liu}, \citenamefont {Zhou}, \citenamefont
  {Chen}, \citenamefont {Li}, \citenamefont {Yu} \emph
  {et~al.}}]{cao2020distribution}%
  \BibitemOpen
  \bibfield  {author} {\bibinfo {author} {\bibfnamefont {H.}~\bibnamefont
  {Cao}}, \bibinfo {author} {\bibfnamefont {S.-C.}\ \bibnamefont {Gao}},
  \bibinfo {author} {\bibfnamefont {C.}~\bibnamefont {Zhang}}, \bibinfo
  {author} {\bibfnamefont {J.}~\bibnamefont {Wang}}, \bibinfo {author}
  {\bibfnamefont {D.-Y.}\ \bibnamefont {He}}, \bibinfo {author} {\bibfnamefont
  {B.-H.}\ \bibnamefont {Liu}}, \bibinfo {author} {\bibfnamefont {Z.-W.}\
  \bibnamefont {Zhou}}, \bibinfo {author} {\bibfnamefont {Y.-J.}\ \bibnamefont
  {Chen}}, \bibinfo {author} {\bibfnamefont {Z.-H.}\ \bibnamefont {Li}},
  \bibinfo {author} {\bibfnamefont {S.-Y.}\ \bibnamefont {Yu}},  \emph
  {et~al.},\ }\href@noop {} {\bibfield  {journal} {\bibinfo  {journal}
  {Optica}\ }\textbf {\bibinfo {volume} {7}},\ \bibinfo {pages} {232} (\bibinfo
  {year} {2020})}\BibitemShut {NoStop}%
\bibitem [{\citenamefont {Aspelmeyer}\ \emph {et~al.}(2003)\citenamefont
  {Aspelmeyer}, \citenamefont {B{\"o}hm}, \citenamefont {Gyatso}, \citenamefont
  {Jennewein}, \citenamefont {Kaltenbaek}, \citenamefont {Lindenthal},
  \citenamefont {Molina-Terriza}, \citenamefont {Poppe}, \citenamefont {Resch},
  \citenamefont {Taraba} \emph {et~al.}}]{aspelmeyer2003long}%
  \BibitemOpen
  \bibfield  {author} {\bibinfo {author} {\bibfnamefont {M.}~\bibnamefont
  {Aspelmeyer}}, \bibinfo {author} {\bibfnamefont {H.~R.}\ \bibnamefont
  {B{\"o}hm}}, \bibinfo {author} {\bibfnamefont {T.}~\bibnamefont {Gyatso}},
  \bibinfo {author} {\bibfnamefont {T.}~\bibnamefont {Jennewein}}, \bibinfo
  {author} {\bibfnamefont {R.}~\bibnamefont {Kaltenbaek}}, \bibinfo {author}
  {\bibfnamefont {M.}~\bibnamefont {Lindenthal}}, \bibinfo {author}
  {\bibfnamefont {G.}~\bibnamefont {Molina-Terriza}}, \bibinfo {author}
  {\bibfnamefont {A.}~\bibnamefont {Poppe}}, \bibinfo {author} {\bibfnamefont
  {K.}~\bibnamefont {Resch}}, \bibinfo {author} {\bibfnamefont
  {M.}~\bibnamefont {Taraba}},  \emph {et~al.},\ }\href@noop {} {\bibfield
  {journal} {\bibinfo  {journal} {Science}\ }\textbf {\bibinfo {volume}
  {301}},\ \bibinfo {pages} {621} (\bibinfo {year} {2003})}\BibitemShut
  {NoStop}%
\bibitem [{\citenamefont {Resch}\ \emph {et~al.}(2005)\citenamefont {Resch},
  \citenamefont {Lindenthal}, \citenamefont {Blauensteiner}, \citenamefont
  {B{\"o}hm}, \citenamefont {Fedrizzi}, \citenamefont {Kurtsiefer},
  \citenamefont {Poppe}, \citenamefont {Schmitt-Manderbach}, \citenamefont
  {Taraba}, \citenamefont {Ursin} \emph {et~al.}}]{resch2005distributing}%
  \BibitemOpen
  \bibfield  {author} {\bibinfo {author} {\bibfnamefont {K.~J.}\ \bibnamefont
  {Resch}}, \bibinfo {author} {\bibfnamefont {M.}~\bibnamefont {Lindenthal}},
  \bibinfo {author} {\bibfnamefont {B.}~\bibnamefont {Blauensteiner}}, \bibinfo
  {author} {\bibfnamefont {H.}~\bibnamefont {B{\"o}hm}}, \bibinfo {author}
  {\bibfnamefont {A.}~\bibnamefont {Fedrizzi}}, \bibinfo {author}
  {\bibfnamefont {C.}~\bibnamefont {Kurtsiefer}}, \bibinfo {author}
  {\bibfnamefont {A.}~\bibnamefont {Poppe}}, \bibinfo {author} {\bibfnamefont
  {T.}~\bibnamefont {Schmitt-Manderbach}}, \bibinfo {author} {\bibfnamefont
  {M.}~\bibnamefont {Taraba}}, \bibinfo {author} {\bibfnamefont
  {R.}~\bibnamefont {Ursin}},  \emph {et~al.},\ }\href@noop {} {\bibfield
  {journal} {\bibinfo  {journal} {Opt. Express}\ }\textbf {\bibinfo {volume}
  {13}},\ \bibinfo {pages} {202} (\bibinfo {year} {2005})}\BibitemShut
  {NoStop}%
\bibitem [{\citenamefont {Marcikic}\ \emph {et~al.}(2006)\citenamefont
  {Marcikic}, \citenamefont {Lamas-Linares},\ and\ \citenamefont
  {Kurtsiefer}}]{marcikic2006free}%
  \BibitemOpen
  \bibfield  {author} {\bibinfo {author} {\bibfnamefont {I.}~\bibnamefont
  {Marcikic}}, \bibinfo {author} {\bibfnamefont {A.}~\bibnamefont
  {Lamas-Linares}}, \ and\ \bibinfo {author} {\bibfnamefont {C.}~\bibnamefont
  {Kurtsiefer}},\ }\href@noop {} {\bibfield  {journal} {\bibinfo  {journal}
  {Appl. Phys. Lett.}\ }\textbf {\bibinfo {volume} {89}},\ \bibinfo {pages}
  {101122} (\bibinfo {year} {2006})}\BibitemShut {NoStop}%
\bibitem [{\citenamefont {Jin}\ \emph {et~al.}(2010)\citenamefont {Jin},
  \citenamefont {Ren}, \citenamefont {Yang}, \citenamefont {Yi}, \citenamefont
  {Zhou}, \citenamefont {Xu}, \citenamefont {Wang}, \citenamefont {Yang},
  \citenamefont {Hu}, \citenamefont {Jiang} \emph
  {et~al.}}]{jin2010experimental}%
  \BibitemOpen
  \bibfield  {author} {\bibinfo {author} {\bibfnamefont {X.-M.}\ \bibnamefont
  {Jin}}, \bibinfo {author} {\bibfnamefont {J.-G.}\ \bibnamefont {Ren}},
  \bibinfo {author} {\bibfnamefont {B.}~\bibnamefont {Yang}}, \bibinfo {author}
  {\bibfnamefont {Z.-H.}\ \bibnamefont {Yi}}, \bibinfo {author} {\bibfnamefont
  {F.}~\bibnamefont {Zhou}}, \bibinfo {author} {\bibfnamefont {X.-F.}\
  \bibnamefont {Xu}}, \bibinfo {author} {\bibfnamefont {S.-K.}\ \bibnamefont
  {Wang}}, \bibinfo {author} {\bibfnamefont {D.}~\bibnamefont {Yang}}, \bibinfo
  {author} {\bibfnamefont {Y.-F.}\ \bibnamefont {Hu}}, \bibinfo {author}
  {\bibfnamefont {S.}~\bibnamefont {Jiang}},  \emph {et~al.},\ }\href@noop {}
  {\bibfield  {journal} {\bibinfo  {journal} {Nat. Photonics}\ }\textbf
  {\bibinfo {volume} {4}},\ \bibinfo {pages} {376} (\bibinfo {year}
  {2010})}\BibitemShut {NoStop}%
\bibitem [{\citenamefont {Yin}\ \emph {et~al.}(2012)\citenamefont {Yin},
  \citenamefont {Ren}, \citenamefont {Lu}, \citenamefont {Cao}, \citenamefont
  {Yong}, \citenamefont {Wu}, \citenamefont {Liu}, \citenamefont {Liao},
  \citenamefont {Zhou}, \citenamefont {Jiang} \emph {et~al.}}]{yin2012quantum}%
  \BibitemOpen
  \bibfield  {author} {\bibinfo {author} {\bibfnamefont {J.}~\bibnamefont
  {Yin}}, \bibinfo {author} {\bibfnamefont {J.-G.}\ \bibnamefont {Ren}},
  \bibinfo {author} {\bibfnamefont {H.}~\bibnamefont {Lu}}, \bibinfo {author}
  {\bibfnamefont {Y.}~\bibnamefont {Cao}}, \bibinfo {author} {\bibfnamefont
  {H.-L.}\ \bibnamefont {Yong}}, \bibinfo {author} {\bibfnamefont {Y.-P.}\
  \bibnamefont {Wu}}, \bibinfo {author} {\bibfnamefont {C.}~\bibnamefont
  {Liu}}, \bibinfo {author} {\bibfnamefont {S.-K.}\ \bibnamefont {Liao}},
  \bibinfo {author} {\bibfnamefont {F.}~\bibnamefont {Zhou}}, \bibinfo {author}
  {\bibfnamefont {Y.}~\bibnamefont {Jiang}},  \emph {et~al.},\ }\href@noop {}
  {\bibfield  {journal} {\bibinfo  {journal} {Nature}\ }\textbf {\bibinfo
  {volume} {488}},\ \bibinfo {pages} {185} (\bibinfo {year}
  {2012})}\BibitemShut {NoStop}%
\bibitem [{\citenamefont {Ursin}\ \emph {et~al.}(2007)\citenamefont {Ursin},
  \citenamefont {Tiefenbacher}, \citenamefont {Schmitt-Manderbach},
  \citenamefont {Weier}, \citenamefont {Scheidl}, \citenamefont {Lindenthal},
  \citenamefont {Blauensteiner}, \citenamefont {Jennewein}, \citenamefont
  {Perdigues}, \citenamefont {Trojek} \emph {et~al.}}]{ursin2007entanglement}%
  \BibitemOpen
  \bibfield  {author} {\bibinfo {author} {\bibfnamefont {R.}~\bibnamefont
  {Ursin}}, \bibinfo {author} {\bibfnamefont {F.}~\bibnamefont {Tiefenbacher}},
  \bibinfo {author} {\bibfnamefont {T.}~\bibnamefont {Schmitt-Manderbach}},
  \bibinfo {author} {\bibfnamefont {H.}~\bibnamefont {Weier}}, \bibinfo
  {author} {\bibfnamefont {T.}~\bibnamefont {Scheidl}}, \bibinfo {author}
  {\bibfnamefont {M.}~\bibnamefont {Lindenthal}}, \bibinfo {author}
  {\bibfnamefont {B.}~\bibnamefont {Blauensteiner}}, \bibinfo {author}
  {\bibfnamefont {T.}~\bibnamefont {Jennewein}}, \bibinfo {author}
  {\bibfnamefont {J.}~\bibnamefont {Perdigues}}, \bibinfo {author}
  {\bibfnamefont {P.}~\bibnamefont {Trojek}},  \emph {et~al.},\ }\href@noop {}
  {\bibfield  {journal} {\bibinfo  {journal} {Nat. Phys.}\ }\textbf {\bibinfo
  {volume} {3}},\ \bibinfo {pages} {481} (\bibinfo {year} {2007})}\BibitemShut
  {NoStop}%
\bibitem [{\citenamefont {Ma}\ \emph {et~al.}(2012)\citenamefont {Ma},
  \citenamefont {Herbst}, \citenamefont {Scheidl}, \citenamefont {Wang},
  \citenamefont {Kropatschek}, \citenamefont {Naylor}, \citenamefont
  {Wittmann}, \citenamefont {Mech}, \citenamefont {Kofler}, \citenamefont
  {Anisimova} \emph {et~al.}}]{ma2012quantum}%
  \BibitemOpen
  \bibfield  {author} {\bibinfo {author} {\bibfnamefont {X.-S.}\ \bibnamefont
  {Ma}}, \bibinfo {author} {\bibfnamefont {T.}~\bibnamefont {Herbst}}, \bibinfo
  {author} {\bibfnamefont {T.}~\bibnamefont {Scheidl}}, \bibinfo {author}
  {\bibfnamefont {D.}~\bibnamefont {Wang}}, \bibinfo {author} {\bibfnamefont
  {S.}~\bibnamefont {Kropatschek}}, \bibinfo {author} {\bibfnamefont
  {W.}~\bibnamefont {Naylor}}, \bibinfo {author} {\bibfnamefont
  {B.}~\bibnamefont {Wittmann}}, \bibinfo {author} {\bibfnamefont
  {A.}~\bibnamefont {Mech}}, \bibinfo {author} {\bibfnamefont {J.}~\bibnamefont
  {Kofler}}, \bibinfo {author} {\bibfnamefont {E.}~\bibnamefont {Anisimova}},
  \emph {et~al.},\ }\href@noop {} {\bibfield  {journal} {\bibinfo  {journal}
  {Nature}\ }\textbf {\bibinfo {volume} {489}},\ \bibinfo {pages} {269}
  (\bibinfo {year} {2012})}\BibitemShut {NoStop}%
\bibitem [{\citenamefont {Vallone}\ \emph {et~al.}(2014)\citenamefont
  {Vallone}, \citenamefont {D’Ambrosio}, \citenamefont {Sponselli},
  \citenamefont {Slussarenko}, \citenamefont {Marrucci}, \citenamefont
  {Sciarrino},\ and\ \citenamefont {Villoresi}}]{vallone2014free}%
  \BibitemOpen
  \bibfield  {author} {\bibinfo {author} {\bibfnamefont {G.}~\bibnamefont
  {Vallone}}, \bibinfo {author} {\bibfnamefont {V.}~\bibnamefont
  {D’Ambrosio}}, \bibinfo {author} {\bibfnamefont {A.}~\bibnamefont
  {Sponselli}}, \bibinfo {author} {\bibfnamefont {S.}~\bibnamefont
  {Slussarenko}}, \bibinfo {author} {\bibfnamefont {L.}~\bibnamefont
  {Marrucci}}, \bibinfo {author} {\bibfnamefont {F.}~\bibnamefont {Sciarrino}},
  \ and\ \bibinfo {author} {\bibfnamefont {P.}~\bibnamefont {Villoresi}},\
  }\href@noop {} {\bibfield  {journal} {\bibinfo  {journal} {Phys. Rev. Lett.}\
  }\textbf {\bibinfo {volume} {113}},\ \bibinfo {pages} {060503} (\bibinfo
  {year} {2014})}\BibitemShut {NoStop}%
\bibitem [{\citenamefont {Krenn}\ \emph {et~al.}(2015)\citenamefont {Krenn},
  \citenamefont {Handsteiner}, \citenamefont {Fink}, \citenamefont {Fickler},\
  and\ \citenamefont {Zeilinger}}]{krenn2015twisted}%
  \BibitemOpen
  \bibfield  {author} {\bibinfo {author} {\bibfnamefont {M.}~\bibnamefont
  {Krenn}}, \bibinfo {author} {\bibfnamefont {J.}~\bibnamefont {Handsteiner}},
  \bibinfo {author} {\bibfnamefont {M.}~\bibnamefont {Fink}}, \bibinfo {author}
  {\bibfnamefont {R.}~\bibnamefont {Fickler}}, \ and\ \bibinfo {author}
  {\bibfnamefont {A.}~\bibnamefont {Zeilinger}},\ }\href@noop {} {\bibfield
  {journal} {\bibinfo  {journal} {Proc. Natl. Acad. Sci.}\ }\textbf {\bibinfo
  {volume} {112}},\ \bibinfo {pages} {14197} (\bibinfo {year}
  {2015})}\BibitemShut {NoStop}%
\bibitem [{\citenamefont {Ecker}\ \emph {et~al.}(2019)\citenamefont {Ecker},
  \citenamefont {Bouchard}, \citenamefont {Bulla}, \citenamefont {Brandt},
  \citenamefont {Kohout}, \citenamefont {Steinlechner}, \citenamefont
  {Fickler}, \citenamefont {Malik}, \citenamefont {Guryanova}, \citenamefont
  {Ursin} \emph {et~al.}}]{ecker2019overcoming}%
  \BibitemOpen
  \bibfield  {author} {\bibinfo {author} {\bibfnamefont {S.}~\bibnamefont
  {Ecker}}, \bibinfo {author} {\bibfnamefont {F.}~\bibnamefont {Bouchard}},
  \bibinfo {author} {\bibfnamefont {L.}~\bibnamefont {Bulla}}, \bibinfo
  {author} {\bibfnamefont {F.}~\bibnamefont {Brandt}}, \bibinfo {author}
  {\bibfnamefont {O.}~\bibnamefont {Kohout}}, \bibinfo {author} {\bibfnamefont
  {F.}~\bibnamefont {Steinlechner}}, \bibinfo {author} {\bibfnamefont
  {R.}~\bibnamefont {Fickler}}, \bibinfo {author} {\bibfnamefont
  {M.}~\bibnamefont {Malik}}, \bibinfo {author} {\bibfnamefont
  {Y.}~\bibnamefont {Guryanova}}, \bibinfo {author} {\bibfnamefont
  {R.}~\bibnamefont {Ursin}},  \emph {et~al.},\ }\href@noop {} {\bibfield
  {journal} {\bibinfo  {journal} {Phys. Rev. X}\ }\textbf {\bibinfo {volume}
  {9}},\ \bibinfo {pages} {041042} (\bibinfo {year} {2019})}\BibitemShut
  {NoStop}%
\bibitem [{\citenamefont {Liu}\ \emph {et~al.}(2021)\citenamefont {Liu},
  \citenamefont {Tian}, \citenamefont {Gu}, \citenamefont {Fan}, \citenamefont
  {Ni}, \citenamefont {Yang}, \citenamefont {Zhang}, \citenamefont {Hu},
  \citenamefont {Guo}, \citenamefont {Cao} \emph {et~al.}}]{liu2021optical}%
  \BibitemOpen
  \bibfield  {author} {\bibinfo {author} {\bibfnamefont {H.-Y.}\ \bibnamefont
  {Liu}}, \bibinfo {author} {\bibfnamefont {X.-H.}\ \bibnamefont {Tian}},
  \bibinfo {author} {\bibfnamefont {C.}~\bibnamefont {Gu}}, \bibinfo {author}
  {\bibfnamefont {P.}~\bibnamefont {Fan}}, \bibinfo {author} {\bibfnamefont
  {X.}~\bibnamefont {Ni}}, \bibinfo {author} {\bibfnamefont {R.}~\bibnamefont
  {Yang}}, \bibinfo {author} {\bibfnamefont {J.-N.}\ \bibnamefont {Zhang}},
  \bibinfo {author} {\bibfnamefont {M.}~\bibnamefont {Hu}}, \bibinfo {author}
  {\bibfnamefont {J.}~\bibnamefont {Guo}}, \bibinfo {author} {\bibfnamefont
  {X.}~\bibnamefont {Cao}},  \emph {et~al.},\ }\href@noop {} {\bibfield
  {journal} {\bibinfo  {journal} {Phys. Rev. Lett.}\ }\textbf {\bibinfo
  {volume} {126}},\ \bibinfo {pages} {020503} (\bibinfo {year}
  {2021})}\BibitemShut {NoStop}%
\bibitem [{\citenamefont {Bedington}\ \emph {et~al.}(2017)\citenamefont
  {Bedington}, \citenamefont {Arrazola},\ and\ \citenamefont
  {Ling}}]{bedington2017progress}%
  \BibitemOpen
  \bibfield  {author} {\bibinfo {author} {\bibfnamefont {R.}~\bibnamefont
  {Bedington}}, \bibinfo {author} {\bibfnamefont {J.~M.}\ \bibnamefont
  {Arrazola}}, \ and\ \bibinfo {author} {\bibfnamefont {A.}~\bibnamefont
  {Ling}},\ }\href@noop {} {\bibfield  {journal} {\bibinfo  {journal} {npj
  Quant. Inf.}\ }\textbf {\bibinfo {volume} {3}},\ \bibinfo {pages} {1}
  (\bibinfo {year} {2017})}\BibitemShut {NoStop}%
\bibitem [{\citenamefont {Yin}\ \emph {et~al.}(2017{\natexlab{b}})\citenamefont
  {Yin}, \citenamefont {Cao}, \citenamefont {Li}, \citenamefont {Ren},
  \citenamefont {Liao}, \citenamefont {Zhang}, \citenamefont {Cai},
  \citenamefont {Liu}, \citenamefont {Li}, \citenamefont {Dai} \emph
  {et~al.}}]{yin2017satellitetoground}%
  \BibitemOpen
  \bibfield  {author} {\bibinfo {author} {\bibfnamefont {J.}~\bibnamefont
  {Yin}}, \bibinfo {author} {\bibfnamefont {Y.}~\bibnamefont {Cao}}, \bibinfo
  {author} {\bibfnamefont {Y.-H.}\ \bibnamefont {Li}}, \bibinfo {author}
  {\bibfnamefont {J.-G.}\ \bibnamefont {Ren}}, \bibinfo {author} {\bibfnamefont
  {S.-K.}\ \bibnamefont {Liao}}, \bibinfo {author} {\bibfnamefont
  {L.}~\bibnamefont {Zhang}}, \bibinfo {author} {\bibfnamefont {W.-Q.}\
  \bibnamefont {Cai}}, \bibinfo {author} {\bibfnamefont {W.-Y.}\ \bibnamefont
  {Liu}}, \bibinfo {author} {\bibfnamefont {B.}~\bibnamefont {Li}}, \bibinfo
  {author} {\bibfnamefont {H.}~\bibnamefont {Dai}},  \emph {et~al.},\
  }\href@noop {} {\bibfield  {journal} {\bibinfo  {journal} {Phys. Rev. Lett.}\
  }\textbf {\bibinfo {volume} {119}},\ \bibinfo {pages} {200501} (\bibinfo
  {year} {2017}{\natexlab{b}})}\BibitemShut {NoStop}%
\bibitem [{\citenamefont {Ren}\ \emph {et~al.}(2017)\citenamefont {Ren},
  \citenamefont {Xu}, \citenamefont {Yong}, \citenamefont {Zhang},
  \citenamefont {Liao}, \citenamefont {Yin}, \citenamefont {Liu}, \citenamefont
  {Cai}, \citenamefont {Yang}, \citenamefont {Li} \emph
  {et~al.}}]{ren2017ground}%
  \BibitemOpen
  \bibfield  {author} {\bibinfo {author} {\bibfnamefont {J.-G.}\ \bibnamefont
  {Ren}}, \bibinfo {author} {\bibfnamefont {P.}~\bibnamefont {Xu}}, \bibinfo
  {author} {\bibfnamefont {H.-L.}\ \bibnamefont {Yong}}, \bibinfo {author}
  {\bibfnamefont {L.}~\bibnamefont {Zhang}}, \bibinfo {author} {\bibfnamefont
  {S.-K.}\ \bibnamefont {Liao}}, \bibinfo {author} {\bibfnamefont
  {J.}~\bibnamefont {Yin}}, \bibinfo {author} {\bibfnamefont {W.-Y.}\
  \bibnamefont {Liu}}, \bibinfo {author} {\bibfnamefont {W.-Q.}\ \bibnamefont
  {Cai}}, \bibinfo {author} {\bibfnamefont {M.}~\bibnamefont {Yang}}, \bibinfo
  {author} {\bibfnamefont {L.}~\bibnamefont {Li}},  \emph {et~al.},\
  }\href@noop {} {\bibfield  {journal} {\bibinfo  {journal} {Nature}\ }\textbf
  {\bibinfo {volume} {549}},\ \bibinfo {pages} {70} (\bibinfo {year}
  {2017})}\BibitemShut {NoStop}%
\bibitem [{\citenamefont {Liao}\ \emph {et~al.}(2018)\citenamefont {Liao},
  \citenamefont {Cai}, \citenamefont {Handsteiner}, \citenamefont {Liu},
  \citenamefont {Yin}, \citenamefont {Zhang}, \citenamefont {Rauch},
  \citenamefont {Fink}, \citenamefont {Ren}, \citenamefont {Liu} \emph
  {et~al.}}]{liao2018satellite}%
  \BibitemOpen
  \bibfield  {author} {\bibinfo {author} {\bibfnamefont {S.-K.}\ \bibnamefont
  {Liao}}, \bibinfo {author} {\bibfnamefont {W.-Q.}\ \bibnamefont {Cai}},
  \bibinfo {author} {\bibfnamefont {J.}~\bibnamefont {Handsteiner}}, \bibinfo
  {author} {\bibfnamefont {B.}~\bibnamefont {Liu}}, \bibinfo {author}
  {\bibfnamefont {J.}~\bibnamefont {Yin}}, \bibinfo {author} {\bibfnamefont
  {L.}~\bibnamefont {Zhang}}, \bibinfo {author} {\bibfnamefont
  {D.}~\bibnamefont {Rauch}}, \bibinfo {author} {\bibfnamefont
  {M.}~\bibnamefont {Fink}}, \bibinfo {author} {\bibfnamefont {J.-G.}\
  \bibnamefont {Ren}}, \bibinfo {author} {\bibfnamefont {W.-Y.}\ \bibnamefont
  {Liu}},  \emph {et~al.},\ }\href@noop {} {\bibfield  {journal} {\bibinfo
  {journal} {Phys. Rev. Lett.}\ }\textbf {\bibinfo {volume} {120}},\ \bibinfo
  {pages} {030501} (\bibinfo {year} {2018})}\BibitemShut {NoStop}%
\bibitem [{\citenamefont {Holloway}\ \emph {et~al.}(2013)\citenamefont
  {Holloway}, \citenamefont {Doucette}, \citenamefont {Erven}, \citenamefont
  {Bourgoin},\ and\ \citenamefont {Jennewein}}]{Holloway:2013}%
  \BibitemOpen
  \bibfield  {author} {\bibinfo {author} {\bibfnamefont {C.}~\bibnamefont
  {Holloway}}, \bibinfo {author} {\bibfnamefont {J.~A.}\ \bibnamefont
  {Doucette}}, \bibinfo {author} {\bibfnamefont {C.}~\bibnamefont {Erven}},
  \bibinfo {author} {\bibfnamefont {J.-P.}\ \bibnamefont {Bourgoin}}, \ and\
  \bibinfo {author} {\bibfnamefont {T.}~\bibnamefont {Jennewein}},\ }\href
  {\doibase 10.1103/PhysRevA.87.022342} {\bibfield  {journal} {\bibinfo
  {journal} {Phys. Rev. A}\ }\textbf {\bibinfo {volume} {87}},\ \bibinfo
  {pages} {022342} (\bibinfo {year} {2013})}\BibitemShut {NoStop}%
\bibitem [{\citenamefont {Jennewein}\ \emph
  {et~al.}(2014{\natexlab{a}})\citenamefont {Jennewein}, \citenamefont
  {Bourgoin}, \citenamefont {Higgins}, \citenamefont {Holloway}, \citenamefont
  {Meyer-Scott}, \citenamefont {Erven}, \citenamefont {Heim}, \citenamefont
  {Yan}, \citenamefont {Huebel}, \citenamefont {Weihs}, \citenamefont {Choi},
  \citenamefont {d'Souza}, \citenamefont {Hudson},\ and\ \citenamefont
  {Laflamme}}]{jennewein2014b}%
  \BibitemOpen
  \bibfield  {author} {\bibinfo {author} {\bibfnamefont {T.}~\bibnamefont
  {Jennewein}}, \bibinfo {author} {\bibfnamefont {J.~P.}\ \bibnamefont
  {Bourgoin}}, \bibinfo {author} {\bibfnamefont {B.}~\bibnamefont {Higgins}},
  \bibinfo {author} {\bibfnamefont {C.}~\bibnamefont {Holloway}}, \bibinfo
  {author} {\bibfnamefont {E.}~\bibnamefont {Meyer-Scott}}, \bibinfo {author}
  {\bibfnamefont {C.}~\bibnamefont {Erven}}, \bibinfo {author} {\bibfnamefont
  {B.}~\bibnamefont {Heim}}, \bibinfo {author} {\bibfnamefont {Z.}~\bibnamefont
  {Yan}}, \bibinfo {author} {\bibfnamefont {H.}~\bibnamefont {Huebel}},
  \bibinfo {author} {\bibfnamefont {G.}~\bibnamefont {Weihs}}, \bibinfo
  {author} {\bibfnamefont {E.}~\bibnamefont {Choi}}, \bibinfo {author}
  {\bibfnamefont {I.}~\bibnamefont {d'Souza}}, \bibinfo {author} {\bibfnamefont
  {D.}~\bibnamefont {Hudson}}, \ and\ \bibinfo {author} {\bibfnamefont
  {R.}~\bibnamefont {Laflamme}},\ }in\ \href {\doibase {10.1117/12.2041693}}
  {\emph {\bibinfo {booktitle} {{ADVANCES IN PHOTONICS OF QUANTUM COMPUTING,
  MEMORY, AND COMMUNICATION VII}}}},\ \bibinfo {series} {{Proceedings of
  SPIE}}, Vol.\ \bibinfo {volume} {{8997}},\ \bibinfo {editor} {edited by\
  \bibinfo {editor} {\bibnamefont {{Hasan, ZU and Hemmer, PR and Lee, H and
  Santori, CM}}}}\ (\bibinfo {year} {{2014}})\ \bibinfo {note} {{Conference on
  Advances in Photonics of Quantum Computing, Memory, and Communication VII,
  San Francisco, CA, FEB 04-06, 2014}}\BibitemShut {NoStop}%
\bibitem [{\citenamefont {Jennewein}\ \emph
  {et~al.}(2014{\natexlab{b}})\citenamefont {Jennewein}, \citenamefont {Grant},
  \citenamefont {Choi}, \citenamefont {Pugh}, \citenamefont {Holloway},
  \citenamefont {Bourgoin}, \citenamefont {Hakima}, \citenamefont {Higgins},\
  and\ \citenamefont {Zee}}]{jennewein2014nanoqey}%
  \BibitemOpen
  \bibfield  {author} {\bibinfo {author} {\bibfnamefont {T.}~\bibnamefont
  {Jennewein}}, \bibinfo {author} {\bibfnamefont {C.}~\bibnamefont {Grant}},
  \bibinfo {author} {\bibfnamefont {E.}~\bibnamefont {Choi}}, \bibinfo {author}
  {\bibfnamefont {C.}~\bibnamefont {Pugh}}, \bibinfo {author} {\bibfnamefont
  {C.}~\bibnamefont {Holloway}}, \bibinfo {author} {\bibfnamefont
  {J.}~\bibnamefont {Bourgoin}}, \bibinfo {author} {\bibfnamefont
  {H.}~\bibnamefont {Hakima}}, \bibinfo {author} {\bibfnamefont
  {B.}~\bibnamefont {Higgins}}, \ and\ \bibinfo {author} {\bibfnamefont
  {R.}~\bibnamefont {Zee}},\ }in\ \href@noop {} {\emph {\bibinfo {booktitle}
  {Emerging technologies in security and defence II; and quantum-physics-based
  information security III}}},\ Vol.\ \bibinfo {volume} {9254}\ (\bibinfo
  {organization} {International Society for Optics and Photonics},\ \bibinfo
  {year} {2014})\ p.\ \bibinfo {pages} {925402}\BibitemShut {NoStop}%
\bibitem [{\citenamefont {Oi}\ \emph {et~al.}(2017)\citenamefont {Oi},
  \citenamefont {Ling}, \citenamefont {Grieve}, \citenamefont {Jennewein},
  \citenamefont {Dinkelaker},\ and\ \citenamefont
  {Krutzik}}]{oi2017nanosatellites}%
  \BibitemOpen
  \bibfield  {author} {\bibinfo {author} {\bibfnamefont {D.~K.}\ \bibnamefont
  {Oi}}, \bibinfo {author} {\bibfnamefont {A.}~\bibnamefont {Ling}}, \bibinfo
  {author} {\bibfnamefont {J.~A.}\ \bibnamefont {Grieve}}, \bibinfo {author}
  {\bibfnamefont {T.}~\bibnamefont {Jennewein}}, \bibinfo {author}
  {\bibfnamefont {A.~N.}\ \bibnamefont {Dinkelaker}}, \ and\ \bibinfo {author}
  {\bibfnamefont {M.}~\bibnamefont {Krutzik}},\ }\href@noop {} {\bibfield
  {journal} {\bibinfo  {journal} {Contemp. Phys}\ }\textbf {\bibinfo {volume}
  {58}},\ \bibinfo {pages} {25} (\bibinfo {year} {2017})}\BibitemShut {NoStop}%
\bibitem [{\citenamefont {Lee}\ \emph {et~al.}(2019)\citenamefont {Lee},
  \citenamefont {Vergoossen},\ and\ \citenamefont {Ling}}]{lee2019updated}%
  \BibitemOpen
  \bibfield  {author} {\bibinfo {author} {\bibfnamefont {O.}~\bibnamefont
  {Lee}}, \bibinfo {author} {\bibfnamefont {T.}~\bibnamefont {Vergoossen}}, \
  and\ \bibinfo {author} {\bibfnamefont {A.}~\bibnamefont {Ling}},\ }\href@noop
  {} {\bibfield  {journal} {\bibinfo  {journal} {arXiv preprint
  arXiv:1909.13061}\ } (\bibinfo {year} {2019})}\BibitemShut {NoStop}%
\bibitem [{\citenamefont {Bedington}\ \emph {et~al.}(2016)\citenamefont
  {Bedington}, \citenamefont {Bai}, \citenamefont {Truong-Cao}, \citenamefont
  {Tan}, \citenamefont {Durak}, \citenamefont {Zafra}, \citenamefont {Grieve},
  \citenamefont {Oi},\ and\ \citenamefont {Ling}}]{bedington2016nanosatellite}%
  \BibitemOpen
  \bibfield  {author} {\bibinfo {author} {\bibfnamefont {R.}~\bibnamefont
  {Bedington}}, \bibinfo {author} {\bibfnamefont {X.}~\bibnamefont {Bai}},
  \bibinfo {author} {\bibfnamefont {E.}~\bibnamefont {Truong-Cao}}, \bibinfo
  {author} {\bibfnamefont {Y.~C.}\ \bibnamefont {Tan}}, \bibinfo {author}
  {\bibfnamefont {K.}~\bibnamefont {Durak}}, \bibinfo {author} {\bibfnamefont
  {A.~V.}\ \bibnamefont {Zafra}}, \bibinfo {author} {\bibfnamefont {J.~A.}\
  \bibnamefont {Grieve}}, \bibinfo {author} {\bibfnamefont {D.~K.}\
  \bibnamefont {Oi}}, \ and\ \bibinfo {author} {\bibfnamefont {A.}~\bibnamefont
  {Ling}},\ }\href@noop {} {\bibfield  {journal} {\bibinfo  {journal} {Eur.
  Phys. J. Quant. Tech.}\ }\textbf {\bibinfo {volume} {3}},\ \bibinfo {pages}
  {12} (\bibinfo {year} {2016})}\BibitemShut {NoStop}%
\bibitem [{\citenamefont {Tang}\ \emph {et~al.}(2014)\citenamefont {Tang},
  \citenamefont {Chandrasekara}, \citenamefont {Sean}, \citenamefont {Cheng},
  \citenamefont {Wildfeuer},\ and\ \citenamefont {Ling}}]{tang2014near}%
  \BibitemOpen
  \bibfield  {author} {\bibinfo {author} {\bibfnamefont {Z.}~\bibnamefont
  {Tang}}, \bibinfo {author} {\bibfnamefont {R.}~\bibnamefont {Chandrasekara}},
  \bibinfo {author} {\bibfnamefont {Y.~Y.}\ \bibnamefont {Sean}}, \bibinfo
  {author} {\bibfnamefont {C.}~\bibnamefont {Cheng}}, \bibinfo {author}
  {\bibfnamefont {C.}~\bibnamefont {Wildfeuer}}, \ and\ \bibinfo {author}
  {\bibfnamefont {A.}~\bibnamefont {Ling}},\ }\href@noop {} {\bibfield
  {journal} {\bibinfo  {journal} {Sci. Rep.}\ }\textbf {\bibinfo {volume}
  {4}},\ \bibinfo {pages} {6366} (\bibinfo {year} {2014})}\BibitemShut
  {NoStop}%
\bibitem [{\citenamefont {Tang}\ \emph
  {et~al.}(2016{\natexlab{a}})\citenamefont {Tang}, \citenamefont
  {Chandrasekara}, \citenamefont {Tan}, \citenamefont {Cheng}, \citenamefont
  {Sha}, \citenamefont {Hiang}, \citenamefont {Oi},\ and\ \citenamefont
  {Ling}}]{tang2016generation}%
  \BibitemOpen
  \bibfield  {author} {\bibinfo {author} {\bibfnamefont {Z.}~\bibnamefont
  {Tang}}, \bibinfo {author} {\bibfnamefont {R.}~\bibnamefont {Chandrasekara}},
  \bibinfo {author} {\bibfnamefont {Y.~C.}\ \bibnamefont {Tan}}, \bibinfo
  {author} {\bibfnamefont {C.}~\bibnamefont {Cheng}}, \bibinfo {author}
  {\bibfnamefont {L.}~\bibnamefont {Sha}}, \bibinfo {author} {\bibfnamefont
  {G.~C.}\ \bibnamefont {Hiang}}, \bibinfo {author} {\bibfnamefont {D.~K.}\
  \bibnamefont {Oi}}, \ and\ \bibinfo {author} {\bibfnamefont {A.}~\bibnamefont
  {Ling}},\ }\href@noop {} {\bibfield  {journal} {\bibinfo  {journal} {Phys.
  Rev. Applied}\ }\textbf {\bibinfo {volume} {5}},\ \bibinfo {pages} {054022}
  (\bibinfo {year} {2016}{\natexlab{a}})}\BibitemShut {NoStop}%
\bibitem [{\citenamefont {Tang}\ \emph
  {et~al.}(2016{\natexlab{b}})\citenamefont {Tang}, \citenamefont
  {Chandrasekara}, \citenamefont {Tan}, \citenamefont {Cheng}, \citenamefont
  {Durak},\ and\ \citenamefont {Ling}}]{tang2016photon}%
  \BibitemOpen
  \bibfield  {author} {\bibinfo {author} {\bibfnamefont {Z.}~\bibnamefont
  {Tang}}, \bibinfo {author} {\bibfnamefont {R.}~\bibnamefont {Chandrasekara}},
  \bibinfo {author} {\bibfnamefont {Y.~C.}\ \bibnamefont {Tan}}, \bibinfo
  {author} {\bibfnamefont {C.}~\bibnamefont {Cheng}}, \bibinfo {author}
  {\bibfnamefont {K.}~\bibnamefont {Durak}}, \ and\ \bibinfo {author}
  {\bibfnamefont {A.}~\bibnamefont {Ling}},\ }\href@noop {} {\bibfield
  {journal} {\bibinfo  {journal} {Sci. Rep.}\ }\textbf {\bibinfo {volume}
  {6}},\ \bibinfo {pages} {1} (\bibinfo {year}
  {2016}{\natexlab{b}})}\BibitemShut {NoStop}%
\bibitem [{\citenamefont {Villar}\ \emph {et~al.}(2020)\citenamefont {Villar},
  \citenamefont {Lohrmann}, \citenamefont {Bai}, \citenamefont {Vergoossen},
  \citenamefont {Bedington}, \citenamefont {Perumangatt}, \citenamefont {Lim},
  \citenamefont {Islam}, \citenamefont {Reezwana}, \citenamefont {Tang} \emph
  {et~al.}}]{villar2020entanglement}%
  \BibitemOpen
  \bibfield  {author} {\bibinfo {author} {\bibfnamefont {A.}~\bibnamefont
  {Villar}}, \bibinfo {author} {\bibfnamefont {A.}~\bibnamefont {Lohrmann}},
  \bibinfo {author} {\bibfnamefont {X.}~\bibnamefont {Bai}}, \bibinfo {author}
  {\bibfnamefont {T.}~\bibnamefont {Vergoossen}}, \bibinfo {author}
  {\bibfnamefont {R.}~\bibnamefont {Bedington}}, \bibinfo {author}
  {\bibfnamefont {C.}~\bibnamefont {Perumangatt}}, \bibinfo {author}
  {\bibfnamefont {H.~Y.}\ \bibnamefont {Lim}}, \bibinfo {author} {\bibfnamefont
  {T.}~\bibnamefont {Islam}}, \bibinfo {author} {\bibfnamefont
  {A.}~\bibnamefont {Reezwana}}, \bibinfo {author} {\bibfnamefont
  {Z.}~\bibnamefont {Tang}},  \emph {et~al.},\ }\href@noop {} {\bibfield
  {journal} {\bibinfo  {journal} {Optica}\ }\textbf {\bibinfo {volume} {7}},\
  \bibinfo {pages} {734} (\bibinfo {year} {2020})}\BibitemShut {NoStop}%
\bibitem [{\citenamefont {Liu}\ \emph {et~al.}(2020{\natexlab{b}})\citenamefont
  {Liu}, \citenamefont {Tian}, \citenamefont {Gu}, \citenamefont {Fan},
  \citenamefont {Ni}, \citenamefont {Yang}, \citenamefont {Zhang},
  \citenamefont {Hu}, \citenamefont {Guo}, \citenamefont {Cao} \emph
  {et~al.}}]{liu2020drone}%
  \BibitemOpen
  \bibfield  {author} {\bibinfo {author} {\bibfnamefont {H.-Y.}\ \bibnamefont
  {Liu}}, \bibinfo {author} {\bibfnamefont {X.-H.}\ \bibnamefont {Tian}},
  \bibinfo {author} {\bibfnamefont {C.}~\bibnamefont {Gu}}, \bibinfo {author}
  {\bibfnamefont {P.}~\bibnamefont {Fan}}, \bibinfo {author} {\bibfnamefont
  {X.}~\bibnamefont {Ni}}, \bibinfo {author} {\bibfnamefont {R.}~\bibnamefont
  {Yang}}, \bibinfo {author} {\bibfnamefont {J.-N.}\ \bibnamefont {Zhang}},
  \bibinfo {author} {\bibfnamefont {M.}~\bibnamefont {Hu}}, \bibinfo {author}
  {\bibfnamefont {J.}~\bibnamefont {Guo}}, \bibinfo {author} {\bibfnamefont
  {X.}~\bibnamefont {Cao}},  \emph {et~al.},\ }\href@noop {} {\bibfield
  {journal} {\bibinfo  {journal} {Nat. Sci. Rev.}\ }\textbf {\bibinfo {volume}
  {7}},\ \bibinfo {pages} {921} (\bibinfo {year}
  {2020}{\natexlab{b}})}\BibitemShut {NoStop}%
\bibitem [{\citenamefont {Gariano}\ and\ \citenamefont
  {Djordjevic}(2019)}]{gariano2019theoretical}%
  \BibitemOpen
  \bibfield  {author} {\bibinfo {author} {\bibfnamefont {J.}~\bibnamefont
  {Gariano}}\ and\ \bibinfo {author} {\bibfnamefont {I.~B.}\ \bibnamefont
  {Djordjevic}},\ }\href@noop {} {\bibfield  {journal} {\bibinfo  {journal}
  {Opt. Express}\ }\textbf {\bibinfo {volume} {27}},\ \bibinfo {pages} {3055}
  (\bibinfo {year} {2019})}\BibitemShut {NoStop}%
\bibitem [{\citenamefont {Ji}\ \emph {et~al.}(2017)\citenamefont {Ji},
  \citenamefont {Gao}, \citenamefont {Yang}, \citenamefont {Feng},
  \citenamefont {Lin}, \citenamefont {Li},\ and\ \citenamefont
  {Jin}}]{ji2017towards}%
  \BibitemOpen
  \bibfield  {author} {\bibinfo {author} {\bibfnamefont {L.}~\bibnamefont
  {Ji}}, \bibinfo {author} {\bibfnamefont {J.}~\bibnamefont {Gao}}, \bibinfo
  {author} {\bibfnamefont {A.-L.}\ \bibnamefont {Yang}}, \bibinfo {author}
  {\bibfnamefont {Z.}~\bibnamefont {Feng}}, \bibinfo {author} {\bibfnamefont
  {X.-F.}\ \bibnamefont {Lin}}, \bibinfo {author} {\bibfnamefont {Z.-G.}\
  \bibnamefont {Li}}, \ and\ \bibinfo {author} {\bibfnamefont {X.-M.}\
  \bibnamefont {Jin}},\ }\href@noop {} {\bibfield  {journal} {\bibinfo
  {journal} {Opt. Express}\ }\textbf {\bibinfo {volume} {25}},\ \bibinfo
  {pages} {19795} (\bibinfo {year} {2017})}\BibitemShut {NoStop}%
\bibitem [{\citenamefont {Bouchard}\ \emph {et~al.}(2018)\citenamefont
  {Bouchard}, \citenamefont {Sit}, \citenamefont {Hufnagel}, \citenamefont
  {Abbas}, \citenamefont {Zhang}, \citenamefont {Heshami}, \citenamefont
  {Fickler}, \citenamefont {Marquardt}, \citenamefont {Leuchs}, \citenamefont
  {Karimi} \emph {et~al.}}]{bouchard2018quantum}%
  \BibitemOpen
  \bibfield  {author} {\bibinfo {author} {\bibfnamefont {F.}~\bibnamefont
  {Bouchard}}, \bibinfo {author} {\bibfnamefont {A.}~\bibnamefont {Sit}},
  \bibinfo {author} {\bibfnamefont {F.}~\bibnamefont {Hufnagel}}, \bibinfo
  {author} {\bibfnamefont {A.}~\bibnamefont {Abbas}}, \bibinfo {author}
  {\bibfnamefont {Y.}~\bibnamefont {Zhang}}, \bibinfo {author} {\bibfnamefont
  {K.}~\bibnamefont {Heshami}}, \bibinfo {author} {\bibfnamefont
  {R.}~\bibnamefont {Fickler}}, \bibinfo {author} {\bibfnamefont
  {C.}~\bibnamefont {Marquardt}}, \bibinfo {author} {\bibfnamefont
  {G.}~\bibnamefont {Leuchs}}, \bibinfo {author} {\bibfnamefont
  {E.}~\bibnamefont {Karimi}},  \emph {et~al.},\ }\href@noop {} {\bibfield
  {journal} {\bibinfo  {journal} {Opt. Express}\ }\textbf {\bibinfo {volume}
  {26}},\ \bibinfo {pages} {22563} (\bibinfo {year} {2018})}\BibitemShut
  {NoStop}%
\bibitem [{\citenamefont {Lopaeva}\ \emph {et~al.}(2013)\citenamefont
  {Lopaeva}, \citenamefont {Berchera}, \citenamefont {Degiovanni},
  \citenamefont {Olivares}, \citenamefont {Brida},\ and\ \citenamefont
  {Genovese}}]{lopaeva2013experimental}%
  \BibitemOpen
  \bibfield  {author} {\bibinfo {author} {\bibfnamefont {E.~D.}\ \bibnamefont
  {Lopaeva}}, \bibinfo {author} {\bibfnamefont {I.~R.}\ \bibnamefont
  {Berchera}}, \bibinfo {author} {\bibfnamefont {I.~P.}\ \bibnamefont
  {Degiovanni}}, \bibinfo {author} {\bibfnamefont {S.}~\bibnamefont
  {Olivares}}, \bibinfo {author} {\bibfnamefont {G.}~\bibnamefont {Brida}}, \
  and\ \bibinfo {author} {\bibfnamefont {M.}~\bibnamefont {Genovese}},\
  }\href@noop {} {\bibfield  {journal} {\bibinfo  {journal} {Phys. Rev. Lett.}\
  }\textbf {\bibinfo {volume} {110}},\ \bibinfo {pages} {153603} (\bibinfo
  {year} {2013})}\BibitemShut {NoStop}%
\bibitem [{\citenamefont {Zhang}\ \emph
  {et~al.}(2015{\natexlab{b}})\citenamefont {Zhang}, \citenamefont {Mouradian},
  \citenamefont {Wong},\ and\ \citenamefont {Shapiro}}]{zhang2015entanglement}%
  \BibitemOpen
  \bibfield  {author} {\bibinfo {author} {\bibfnamefont {Z.}~\bibnamefont
  {Zhang}}, \bibinfo {author} {\bibfnamefont {S.}~\bibnamefont {Mouradian}},
  \bibinfo {author} {\bibfnamefont {F.~N.~C.}\ \bibnamefont {Wong}}, \ and\
  \bibinfo {author} {\bibfnamefont {J.~H.}\ \bibnamefont {Shapiro}},\
  }\href@noop {} {\bibfield  {journal} {\bibinfo  {journal} {Phys. Rev. Lett.}\
  }\textbf {\bibinfo {volume} {114}},\ \bibinfo {pages} {110506} (\bibinfo
  {year} {2015}{\natexlab{b}})}\BibitemShut {NoStop}%
\bibitem [{\citenamefont {Gregory}\ \emph {et~al.}(2020)\citenamefont
  {Gregory}, \citenamefont {Moreau}, \citenamefont {Toninelli},\ and\
  \citenamefont {Padgett}}]{gregory2020imaging}%
  \BibitemOpen
  \bibfield  {author} {\bibinfo {author} {\bibfnamefont {T.}~\bibnamefont
  {Gregory}}, \bibinfo {author} {\bibfnamefont {P.-A.}\ \bibnamefont {Moreau}},
  \bibinfo {author} {\bibfnamefont {E.}~\bibnamefont {Toninelli}}, \ and\
  \bibinfo {author} {\bibfnamefont {M.~J.}\ \bibnamefont {Padgett}},\
  }\href@noop {} {\bibfield  {journal} {\bibinfo  {journal} {Sci. Adv.}\
  }\textbf {\bibinfo {volume} {6}},\ \bibinfo {pages} {eaay2652} (\bibinfo
  {year} {2020})}\BibitemShut {NoStop}%
\bibitem [{\citenamefont {Tan}\ \emph {et~al.}(2008)\citenamefont {Tan},
  \citenamefont {Erkmen}, \citenamefont {Giovannetti}, \citenamefont {Guha},
  \citenamefont {Lloyd}, \citenamefont {Maccone}, \citenamefont {Pirandola},\
  and\ \citenamefont {Shapiro}}]{tan2008quantum}%
  \BibitemOpen
  \bibfield  {author} {\bibinfo {author} {\bibfnamefont {S.-H.}\ \bibnamefont
  {Tan}}, \bibinfo {author} {\bibfnamefont {B.~I.}\ \bibnamefont {Erkmen}},
  \bibinfo {author} {\bibfnamefont {V.}~\bibnamefont {Giovannetti}}, \bibinfo
  {author} {\bibfnamefont {S.}~\bibnamefont {Guha}}, \bibinfo {author}
  {\bibfnamefont {S.}~\bibnamefont {Lloyd}}, \bibinfo {author} {\bibfnamefont
  {L.}~\bibnamefont {Maccone}}, \bibinfo {author} {\bibfnamefont
  {S.}~\bibnamefont {Pirandola}}, \ and\ \bibinfo {author} {\bibfnamefont
  {J.~H.}\ \bibnamefont {Shapiro}},\ }\href@noop {} {\bibfield  {journal}
  {\bibinfo  {journal} {Phys. Rev. Lett.}\ }\textbf {\bibinfo {volume} {101}},\
  \bibinfo {pages} {253601} (\bibinfo {year} {2008})}\BibitemShut {NoStop}%
\bibitem [{\citenamefont {Lloyd}(2008)}]{lloyd2008enhanced}%
  \BibitemOpen
  \bibfield  {author} {\bibinfo {author} {\bibfnamefont {S.}~\bibnamefont
  {Lloyd}},\ }\href@noop {} {\bibfield  {journal} {\bibinfo  {journal}
  {Science}\ }\textbf {\bibinfo {volume} {321}},\ \bibinfo {pages} {1463}
  (\bibinfo {year} {2008})}\BibitemShut {NoStop}%
\bibitem [{\citenamefont {Slussarenko}\ \emph {et~al.}(2017)\citenamefont
  {Slussarenko}, \citenamefont {Weston}, \citenamefont {Chrzanowski},
  \citenamefont {Shalm}, \citenamefont {Verma}, \citenamefont {Nam},\ and\
  \citenamefont {Pryde}}]{slussarenko2017unconditional}%
  \BibitemOpen
  \bibfield  {author} {\bibinfo {author} {\bibfnamefont {S.}~\bibnamefont
  {Slussarenko}}, \bibinfo {author} {\bibfnamefont {M.~M.}\ \bibnamefont
  {Weston}}, \bibinfo {author} {\bibfnamefont {H.~M.}\ \bibnamefont
  {Chrzanowski}}, \bibinfo {author} {\bibfnamefont {L.~K.}\ \bibnamefont
  {Shalm}}, \bibinfo {author} {\bibfnamefont {V.~B.}\ \bibnamefont {Verma}},
  \bibinfo {author} {\bibfnamefont {S.~W.}\ \bibnamefont {Nam}}, \ and\
  \bibinfo {author} {\bibfnamefont {G.~J.}\ \bibnamefont {Pryde}},\ }\href@noop
  {} {\bibfield  {journal} {\bibinfo  {journal} {Nat. Photonics}\ }\textbf
  {\bibinfo {volume} {11}},\ \bibinfo {pages} {700} (\bibinfo {year}
  {2017})}\BibitemShut {NoStop}%
\bibitem [{\citenamefont {Braunstein}(1992)}]{braunstein1992quantum}%
  \BibitemOpen
  \bibfield  {author} {\bibinfo {author} {\bibfnamefont {S.~L.}\ \bibnamefont
  {Braunstein}},\ }\href@noop {} {\bibfield  {journal} {\bibinfo  {journal}
  {Phys. Rev. Lett.}\ }\textbf {\bibinfo {volume} {69}},\ \bibinfo {pages}
  {3598} (\bibinfo {year} {1992})}\BibitemShut {NoStop}%
\bibitem [{\citenamefont {Braunstein}\ and\ \citenamefont
  {Caves}(1994)}]{braunstein1994statistical}%
  \BibitemOpen
  \bibfield  {author} {\bibinfo {author} {\bibfnamefont {S.~L.}\ \bibnamefont
  {Braunstein}}\ and\ \bibinfo {author} {\bibfnamefont {C.~M.}\ \bibnamefont
  {Caves}},\ }\href@noop {} {\bibfield  {journal} {\bibinfo  {journal} {Phys.
  Rev. Lett.}\ }\textbf {\bibinfo {volume} {72}},\ \bibinfo {pages} {3439}
  (\bibinfo {year} {1994})}\BibitemShut {NoStop}%
\bibitem [{\citenamefont {Giovannetti}\ \emph {et~al.}(2004)\citenamefont
  {Giovannetti}, \citenamefont {Lloyd},\ and\ \citenamefont
  {Maccone}}]{giovannetti2004quantum}%
  \BibitemOpen
  \bibfield  {author} {\bibinfo {author} {\bibfnamefont {V.}~\bibnamefont
  {Giovannetti}}, \bibinfo {author} {\bibfnamefont {S.}~\bibnamefont {Lloyd}},
  \ and\ \bibinfo {author} {\bibfnamefont {L.}~\bibnamefont {Maccone}},\
  }\href@noop {} {\bibfield  {journal} {\bibinfo  {journal} {Science}\ }\textbf
  {\bibinfo {volume} {306}},\ \bibinfo {pages} {1330} (\bibinfo {year}
  {2004})}\BibitemShut {NoStop}%
\bibitem [{\citenamefont {Giovannetti}\ \emph {et~al.}(2006)\citenamefont
  {Giovannetti}, \citenamefont {Lloyd},\ and\ \citenamefont
  {Maccone}}]{giovannetti2006quantum}%
  \BibitemOpen
  \bibfield  {author} {\bibinfo {author} {\bibfnamefont {V.}~\bibnamefont
  {Giovannetti}}, \bibinfo {author} {\bibfnamefont {S.}~\bibnamefont {Lloyd}},
  \ and\ \bibinfo {author} {\bibfnamefont {L.}~\bibnamefont {Maccone}},\
  }\href@noop {} {\bibfield  {journal} {\bibinfo  {journal} {Phys. Rev. Lett.}\
  }\textbf {\bibinfo {volume} {96}},\ \bibinfo {pages} {010401} (\bibinfo
  {year} {2006})}\BibitemShut {NoStop}%
\bibitem [{\citenamefont {Paris}(2009)}]{paris2009quantum}%
  \BibitemOpen
  \bibfield  {author} {\bibinfo {author} {\bibfnamefont {M.~G.~A.}\
  \bibnamefont {Paris}},\ }\href@noop {} {\bibfield  {journal} {\bibinfo
  {journal} {Intl. J. Quant. Inf.}\ }\textbf {\bibinfo {volume} {7}},\ \bibinfo
  {pages} {125} (\bibinfo {year} {2009})}\BibitemShut {NoStop}%
\bibitem [{\citenamefont {Demkowicz-Dobrza{\'n}ski}\ \emph
  {et~al.}(2015)\citenamefont {Demkowicz-Dobrza{\'n}ski}, \citenamefont
  {Jarzyna},\ and\ \citenamefont {Ko{\l}ody{\'n}ski}}]{demkowicz2015quantum}%
  \BibitemOpen
  \bibfield  {author} {\bibinfo {author} {\bibfnamefont {R.}~\bibnamefont
  {Demkowicz-Dobrza{\'n}ski}}, \bibinfo {author} {\bibfnamefont
  {M.}~\bibnamefont {Jarzyna}}, \ and\ \bibinfo {author} {\bibfnamefont
  {J.}~\bibnamefont {Ko{\l}ody{\'n}ski}},\ }in\ \href@noop {} {\emph {\bibinfo
  {booktitle} {Prog. Opt.}}},\ Vol.~\bibinfo {volume} {60}\ (\bibinfo
  {publisher} {Elsevier},\ \bibinfo {year} {2015})\ pp.\ \bibinfo {pages}
  {345--435}\BibitemShut {NoStop}%
\bibitem [{\citenamefont {Wang}\ \emph {et~al.}(2018)\citenamefont {Wang},
  \citenamefont {Wang}, \citenamefont {Zhan}, \citenamefont {Bian},
  \citenamefont {Li}, \citenamefont {Sanders},\ and\ \citenamefont
  {Xue}}]{wang2018entanglement}%
  \BibitemOpen
  \bibfield  {author} {\bibinfo {author} {\bibfnamefont {K.}~\bibnamefont
  {Wang}}, \bibinfo {author} {\bibfnamefont {X.}~\bibnamefont {Wang}}, \bibinfo
  {author} {\bibfnamefont {X.}~\bibnamefont {Zhan}}, \bibinfo {author}
  {\bibfnamefont {Z.}~\bibnamefont {Bian}}, \bibinfo {author} {\bibfnamefont
  {J.}~\bibnamefont {Li}}, \bibinfo {author} {\bibfnamefont {B.~C.}\
  \bibnamefont {Sanders}}, \ and\ \bibinfo {author} {\bibfnamefont
  {P.}~\bibnamefont {Xue}},\ }\href@noop {} {\bibfield  {journal} {\bibinfo
  {journal} {Phys. Rev. A}\ }\textbf {\bibinfo {volume} {97}},\ \bibinfo
  {pages} {042112} (\bibinfo {year} {2018})}\BibitemShut {NoStop}%
\bibitem [{\citenamefont {Courtial}\ \emph {et~al.}(1998)\citenamefont
  {Courtial}, \citenamefont {Dholakia}, \citenamefont {Robertson},
  \citenamefont {Allen},\ and\ \citenamefont
  {Padgett}}]{courtial1998measurement}%
  \BibitemOpen
  \bibfield  {author} {\bibinfo {author} {\bibfnamefont {J.}~\bibnamefont
  {Courtial}}, \bibinfo {author} {\bibfnamefont {K.}~\bibnamefont {Dholakia}},
  \bibinfo {author} {\bibfnamefont {D.~A.}\ \bibnamefont {Robertson}}, \bibinfo
  {author} {\bibfnamefont {L.}~\bibnamefont {Allen}}, \ and\ \bibinfo {author}
  {\bibfnamefont {M.~J.}\ \bibnamefont {Padgett}},\ }\href@noop {} {\bibfield
  {journal} {\bibinfo  {journal} {Phys. Rev. Lett.}\ }\textbf {\bibinfo
  {volume} {80}},\ \bibinfo {pages} {3217} (\bibinfo {year}
  {1998})}\BibitemShut {NoStop}%
\bibitem [{\citenamefont {Lavery}\ \emph {et~al.}(2013)\citenamefont {Lavery},
  \citenamefont {Speirits}, \citenamefont {Barnett},\ and\ \citenamefont
  {Padgett}}]{lavery2013detection}%
  \BibitemOpen
  \bibfield  {author} {\bibinfo {author} {\bibfnamefont {M.~P.~J.}\
  \bibnamefont {Lavery}}, \bibinfo {author} {\bibfnamefont {F.~C.}\
  \bibnamefont {Speirits}}, \bibinfo {author} {\bibfnamefont {S.~M.}\
  \bibnamefont {Barnett}}, \ and\ \bibinfo {author} {\bibfnamefont {M.~J.}\
  \bibnamefont {Padgett}},\ }\href@noop {} {\bibfield  {journal} {\bibinfo
  {journal} {Science}\ }\textbf {\bibinfo {volume} {341}},\ \bibinfo {pages}
  {537} (\bibinfo {year} {2013})}\BibitemShut {NoStop}%
\bibitem [{\citenamefont {Jha}\ \emph {et~al.}(2011)\citenamefont {Jha},
  \citenamefont {Agarwal},\ and\ \citenamefont {Boyd}}]{jha2011supersensitive}%
  \BibitemOpen
  \bibfield  {author} {\bibinfo {author} {\bibfnamefont {A.~K.}\ \bibnamefont
  {Jha}}, \bibinfo {author} {\bibfnamefont {G.~S.}\ \bibnamefont {Agarwal}}, \
  and\ \bibinfo {author} {\bibfnamefont {R.~W.}\ \bibnamefont {Boyd}},\
  }\href@noop {} {\bibfield  {journal} {\bibinfo  {journal} {Phys. Rev. A}\
  }\textbf {\bibinfo {volume} {83}},\ \bibinfo {pages} {053829} (\bibinfo
  {year} {2011})}\BibitemShut {NoStop}%
\bibitem [{\citenamefont {Fickler}\ \emph {et~al.}(2012)\citenamefont
  {Fickler}, \citenamefont {Lapkiewicz}, \citenamefont {Plick}, \citenamefont
  {Krenn}, \citenamefont {Schaeff}, \citenamefont {Ramelow},\ and\
  \citenamefont {Zeilinger}}]{fickler2012quantum}%
  \BibitemOpen
  \bibfield  {author} {\bibinfo {author} {\bibfnamefont {R.}~\bibnamefont
  {Fickler}}, \bibinfo {author} {\bibfnamefont {R.}~\bibnamefont {Lapkiewicz}},
  \bibinfo {author} {\bibfnamefont {W.~N.}\ \bibnamefont {Plick}}, \bibinfo
  {author} {\bibfnamefont {M.}~\bibnamefont {Krenn}}, \bibinfo {author}
  {\bibfnamefont {C.}~\bibnamefont {Schaeff}}, \bibinfo {author} {\bibfnamefont
  {S.}~\bibnamefont {Ramelow}}, \ and\ \bibinfo {author} {\bibfnamefont
  {A.}~\bibnamefont {Zeilinger}},\ }\href@noop {} {\bibfield  {journal}
  {\bibinfo  {journal} {Science}\ }\textbf {\bibinfo {volume} {338}},\ \bibinfo
  {pages} {640} (\bibinfo {year} {2012})}\BibitemShut {NoStop}%
\bibitem [{\citenamefont {D'ambrosio}\ \emph {et~al.}(2013)\citenamefont
  {D'ambrosio}, \citenamefont {Spagnolo}, \citenamefont {Del~Re}, \citenamefont
  {Slussarenko}, \citenamefont {Li}, \citenamefont {Kwek}, \citenamefont
  {Marrucci}, \citenamefont {Walborn}, \citenamefont {Aolita},\ and\
  \citenamefont {Sciarrino}}]{d2013photonic}%
  \BibitemOpen
  \bibfield  {author} {\bibinfo {author} {\bibfnamefont {V.}~\bibnamefont
  {D'ambrosio}}, \bibinfo {author} {\bibfnamefont {N.}~\bibnamefont
  {Spagnolo}}, \bibinfo {author} {\bibfnamefont {L.}~\bibnamefont {Del~Re}},
  \bibinfo {author} {\bibfnamefont {S.}~\bibnamefont {Slussarenko}}, \bibinfo
  {author} {\bibfnamefont {Y.}~\bibnamefont {Li}}, \bibinfo {author}
  {\bibfnamefont {L.~C.}\ \bibnamefont {Kwek}}, \bibinfo {author}
  {\bibfnamefont {L.}~\bibnamefont {Marrucci}}, \bibinfo {author}
  {\bibfnamefont {S.~P.}\ \bibnamefont {Walborn}}, \bibinfo {author}
  {\bibfnamefont {L.}~\bibnamefont {Aolita}}, \ and\ \bibinfo {author}
  {\bibfnamefont {F.}~\bibnamefont {Sciarrino}},\ }\href@noop {} {\bibfield
  {journal} {\bibinfo  {journal} {Nat. Commun.}\ }\textbf {\bibinfo {volume}
  {4}},\ \bibinfo {pages} {1} (\bibinfo {year} {2013})}\BibitemShut {NoStop}%
\bibitem [{\citenamefont {Zhang}\ \emph {et~al.}(2019)\citenamefont {Zhang},
  \citenamefont {Zhang}, \citenamefont {Qiu},\ and\ \citenamefont
  {Chen}}]{zhang2019quantum}%
  \BibitemOpen
  \bibfield  {author} {\bibinfo {author} {\bibfnamefont {W.}~\bibnamefont
  {Zhang}}, \bibinfo {author} {\bibfnamefont {D.}~\bibnamefont {Zhang}},
  \bibinfo {author} {\bibfnamefont {X.}~\bibnamefont {Qiu}}, \ and\ \bibinfo
  {author} {\bibfnamefont {L.}~\bibnamefont {Chen}},\ }\href@noop {} {\bibfield
   {journal} {\bibinfo  {journal} {Phys. Rev. A}\ }\textbf {\bibinfo {volume}
  {100}},\ \bibinfo {pages} {043832} (\bibinfo {year} {2019})}\BibitemShut
  {NoStop}%
\bibitem [{\citenamefont {Gilaberte~Basset}\ \emph {et~al.}(2019)\citenamefont
  {Gilaberte~Basset}, \citenamefont {Setzpfandt}, \citenamefont {Steinlechner},
  \citenamefont {Beckert}, \citenamefont {Pertsch},\ and\ \citenamefont
  {Gr{\"a}fe}}]{gilaberte2019perspectives}%
  \BibitemOpen
  \bibfield  {author} {\bibinfo {author} {\bibfnamefont {M.}~\bibnamefont
  {Gilaberte~Basset}}, \bibinfo {author} {\bibfnamefont {F.}~\bibnamefont
  {Setzpfandt}}, \bibinfo {author} {\bibfnamefont {F.}~\bibnamefont
  {Steinlechner}}, \bibinfo {author} {\bibfnamefont {E.}~\bibnamefont
  {Beckert}}, \bibinfo {author} {\bibfnamefont {T.}~\bibnamefont {Pertsch}}, \
  and\ \bibinfo {author} {\bibfnamefont {M.}~\bibnamefont {Gr{\"a}fe}},\
  }\href@noop {} {\bibfield  {journal} {\bibinfo  {journal} {Laser Photon.
  Rev.}\ }\textbf {\bibinfo {volume} {13}},\ \bibinfo {pages} {1900097}
  (\bibinfo {year} {2019})}\BibitemShut {NoStop}%
\bibitem [{\citenamefont {Pittman}\ \emph {et~al.}(1995)\citenamefont
  {Pittman}, \citenamefont {Shih}, \citenamefont {Strekalov},\ and\
  \citenamefont {Sergienko}}]{pittman1995optical}%
  \BibitemOpen
  \bibfield  {author} {\bibinfo {author} {\bibfnamefont {T.~B.}\ \bibnamefont
  {Pittman}}, \bibinfo {author} {\bibfnamefont {Y.~H.}\ \bibnamefont {Shih}},
  \bibinfo {author} {\bibfnamefont {D.~V.}\ \bibnamefont {Strekalov}}, \ and\
  \bibinfo {author} {\bibfnamefont {A.~V.}\ \bibnamefont {Sergienko}},\
  }\href@noop {} {\bibfield  {journal} {\bibinfo  {journal} {Phys. Rev. A}\
  }\textbf {\bibinfo {volume} {52}},\ \bibinfo {pages} {R3429} (\bibinfo {year}
  {1995})}\BibitemShut {NoStop}%
\bibitem [{\citenamefont {Simon}\ and\ \citenamefont
  {Sergienko}(2012)}]{simon2012two}%
  \BibitemOpen
  \bibfield  {author} {\bibinfo {author} {\bibfnamefont {D.~S.}\ \bibnamefont
  {Simon}}\ and\ \bibinfo {author} {\bibfnamefont {A.~V.}\ \bibnamefont
  {Sergienko}},\ }\href@noop {} {\bibfield  {journal} {\bibinfo  {journal}
  {Phys. Rev. A}\ }\textbf {\bibinfo {volume} {85}},\ \bibinfo {pages} {043825}
  (\bibinfo {year} {2012})}\BibitemShut {NoStop}%
\bibitem [{\citenamefont {Fickler}\ \emph {et~al.}(2013)\citenamefont
  {Fickler}, \citenamefont {Krenn}, \citenamefont {Lapkiewicz}, \citenamefont
  {Ramelow},\ and\ \citenamefont {Zeilinger}}]{fickler2013real}%
  \BibitemOpen
  \bibfield  {author} {\bibinfo {author} {\bibfnamefont {R.}~\bibnamefont
  {Fickler}}, \bibinfo {author} {\bibfnamefont {M.}~\bibnamefont {Krenn}},
  \bibinfo {author} {\bibfnamefont {R.}~\bibnamefont {Lapkiewicz}}, \bibinfo
  {author} {\bibfnamefont {S.}~\bibnamefont {Ramelow}}, \ and\ \bibinfo
  {author} {\bibfnamefont {A.}~\bibnamefont {Zeilinger}},\ }\href@noop {}
  {\bibfield  {journal} {\bibinfo  {journal} {Sci. Rep.}\ }\textbf {\bibinfo
  {volume} {3}},\ \bibinfo {pages} {1914} (\bibinfo {year} {2013})}\BibitemShut
  {NoStop}%
\bibitem [{\citenamefont {Blanchet}\ \emph {et~al.}(2008)\citenamefont
  {Blanchet}, \citenamefont {Devaux}, \citenamefont {Furfaro},\ and\
  \citenamefont {Lantz}}]{blanchet2008measurement}%
  \BibitemOpen
  \bibfield  {author} {\bibinfo {author} {\bibfnamefont {J.-L.}\ \bibnamefont
  {Blanchet}}, \bibinfo {author} {\bibfnamefont {F.}~\bibnamefont {Devaux}},
  \bibinfo {author} {\bibfnamefont {L.}~\bibnamefont {Furfaro}}, \ and\
  \bibinfo {author} {\bibfnamefont {E.}~\bibnamefont {Lantz}},\ }\href@noop {}
  {\bibfield  {journal} {\bibinfo  {journal} {Phys. Rev. Lett.}\ }\textbf
  {\bibinfo {volume} {101}},\ \bibinfo {pages} {233604} (\bibinfo {year}
  {2008})}\BibitemShut {NoStop}%
\bibitem [{\citenamefont {Brida}\ \emph {et~al.}(2010)\citenamefont {Brida},
  \citenamefont {Genovese},\ and\ \citenamefont
  {Berchera}}]{brida2010experimental}%
  \BibitemOpen
  \bibfield  {author} {\bibinfo {author} {\bibfnamefont {G.}~\bibnamefont
  {Brida}}, \bibinfo {author} {\bibfnamefont {M.}~\bibnamefont {Genovese}}, \
  and\ \bibinfo {author} {\bibfnamefont {I.~R.}\ \bibnamefont {Berchera}},\
  }\href@noop {} {\bibfield  {journal} {\bibinfo  {journal} {Nat. Photonics}\
  }\textbf {\bibinfo {volume} {4}},\ \bibinfo {pages} {227} (\bibinfo {year}
  {2010})}\BibitemShut {NoStop}%
\bibitem [{\citenamefont {Toninelli}\ \emph {et~al.}(2017)\citenamefont
  {Toninelli}, \citenamefont {Edgar}, \citenamefont {Moreau}, \citenamefont
  {Gibson}, \citenamefont {Hammond},\ and\ \citenamefont
  {Padgett}}]{toninelli2017sub}%
  \BibitemOpen
  \bibfield  {author} {\bibinfo {author} {\bibfnamefont {E.}~\bibnamefont
  {Toninelli}}, \bibinfo {author} {\bibfnamefont {M.~P.}\ \bibnamefont
  {Edgar}}, \bibinfo {author} {\bibfnamefont {P.-A.}\ \bibnamefont {Moreau}},
  \bibinfo {author} {\bibfnamefont {G.~M.}\ \bibnamefont {Gibson}}, \bibinfo
  {author} {\bibfnamefont {G.~D.}\ \bibnamefont {Hammond}}, \ and\ \bibinfo
  {author} {\bibfnamefont {M.~J.}\ \bibnamefont {Padgett}},\ }\href@noop {}
  {\bibfield  {journal} {\bibinfo  {journal} {Opt. Express}\ }\textbf {\bibinfo
  {volume} {25}},\ \bibinfo {pages} {21826} (\bibinfo {year}
  {2017})}\BibitemShut {NoStop}%
\bibitem [{\citenamefont {Strekalov}\ \emph {et~al.}(1995)\citenamefont
  {Strekalov}, \citenamefont {Sergienko}, \citenamefont {Klyshko},\ and\
  \citenamefont {Shih}}]{strekalov1995observation}%
  \BibitemOpen
  \bibfield  {author} {\bibinfo {author} {\bibfnamefont {D.~V.}\ \bibnamefont
  {Strekalov}}, \bibinfo {author} {\bibfnamefont {A.~V.}\ \bibnamefont
  {Sergienko}}, \bibinfo {author} {\bibfnamefont {D.~N.}\ \bibnamefont
  {Klyshko}}, \ and\ \bibinfo {author} {\bibfnamefont {Y.~H.}\ \bibnamefont
  {Shih}},\ }\href@noop {} {\bibfield  {journal} {\bibinfo  {journal} {Phys.
  Rev. Lett.}\ }\textbf {\bibinfo {volume} {74}},\ \bibinfo {pages} {3600}
  (\bibinfo {year} {1995})}\BibitemShut {NoStop}%
\bibitem [{\citenamefont {Morris}\ \emph {et~al.}(2015)\citenamefont {Morris},
  \citenamefont {Aspden}, \citenamefont {Bell}, \citenamefont {Boyd},\ and\
  \citenamefont {Padgett}}]{morris2015imaging}%
  \BibitemOpen
  \bibfield  {author} {\bibinfo {author} {\bibfnamefont {P.~A.}\ \bibnamefont
  {Morris}}, \bibinfo {author} {\bibfnamefont {R.~S.}\ \bibnamefont {Aspden}},
  \bibinfo {author} {\bibfnamefont {J.~E.~C.}\ \bibnamefont {Bell}}, \bibinfo
  {author} {\bibfnamefont {R.~W.}\ \bibnamefont {Boyd}}, \ and\ \bibinfo
  {author} {\bibfnamefont {M.~J.}\ \bibnamefont {Padgett}},\ }\href@noop {}
  {\bibfield  {journal} {\bibinfo  {journal} {Nat. Commun.}\ }\textbf {\bibinfo
  {volume} {6}},\ \bibinfo {pages} {1} (\bibinfo {year} {2015})}\BibitemShut
  {NoStop}%
\bibitem [{\citenamefont {Padgett}\ and\ \citenamefont
  {Boyd}(2017)}]{padgett2017introduction}%
  \BibitemOpen
  \bibfield  {author} {\bibinfo {author} {\bibfnamefont {M.~J.}\ \bibnamefont
  {Padgett}}\ and\ \bibinfo {author} {\bibfnamefont {R.~W.}\ \bibnamefont
  {Boyd}},\ }\href@noop {} {\bibfield  {journal} {\bibinfo  {journal} {Phil.
  Trans. Roy. Soc. A: Math. Phys. Engg. Sci.}\ }\textbf {\bibinfo {volume}
  {375}},\ \bibinfo {pages} {20160233} (\bibinfo {year} {2017})}\BibitemShut
  {NoStop}%
\bibitem [{\citenamefont {Boto}\ \emph {et~al.}(2000)\citenamefont {Boto},
  \citenamefont {Kok}, \citenamefont {Abrams}, \citenamefont {Braunstein},
  \citenamefont {Williams},\ and\ \citenamefont {Dowling}}]{boto2000quantum}%
  \BibitemOpen
  \bibfield  {author} {\bibinfo {author} {\bibfnamefont {A.~N.}\ \bibnamefont
  {Boto}}, \bibinfo {author} {\bibfnamefont {P.}~\bibnamefont {Kok}}, \bibinfo
  {author} {\bibfnamefont {D.~S.}\ \bibnamefont {Abrams}}, \bibinfo {author}
  {\bibfnamefont {S.~L.}\ \bibnamefont {Braunstein}}, \bibinfo {author}
  {\bibfnamefont {C.~P.}\ \bibnamefont {Williams}}, \ and\ \bibinfo {author}
  {\bibfnamefont {J.~P.}\ \bibnamefont {Dowling}},\ }\href@noop {} {\bibfield
  {journal} {\bibinfo  {journal} {Phys. Rev. Lett.}\ }\textbf {\bibinfo
  {volume} {85}},\ \bibinfo {pages} {2733} (\bibinfo {year}
  {2000})}\BibitemShut {NoStop}%
\bibitem [{\citenamefont {Tsang}(2009)}]{tsang2009quantum}%
  \BibitemOpen
  \bibfield  {author} {\bibinfo {author} {\bibfnamefont {M.}~\bibnamefont
  {Tsang}},\ }\href@noop {} {\bibfield  {journal} {\bibinfo  {journal} {Phys.
  Rev. Lett.}\ }\textbf {\bibinfo {volume} {102}},\ \bibinfo {pages} {253601}
  (\bibinfo {year} {2009})}\BibitemShut {NoStop}%
\bibitem [{\citenamefont {Shin}\ \emph {et~al.}(2011)\citenamefont {Shin},
  \citenamefont {Chan}, \citenamefont {Chang},\ and\ \citenamefont
  {Boyd}}]{shin2011quantum}%
  \BibitemOpen
  \bibfield  {author} {\bibinfo {author} {\bibfnamefont {H.}~\bibnamefont
  {Shin}}, \bibinfo {author} {\bibfnamefont {K.~W.~C.}\ \bibnamefont {Chan}},
  \bibinfo {author} {\bibfnamefont {H.~J.}\ \bibnamefont {Chang}}, \ and\
  \bibinfo {author} {\bibfnamefont {R.~W.}\ \bibnamefont {Boyd}},\ }\href@noop
  {} {\bibfield  {journal} {\bibinfo  {journal} {Phys. Rev. Lett.}\ }\textbf
  {\bibinfo {volume} {107}},\ \bibinfo {pages} {083603} (\bibinfo {year}
  {2011})}\BibitemShut {NoStop}%
\bibitem [{\citenamefont {Rozema}\ \emph {et~al.}(2014)\citenamefont {Rozema},
  \citenamefont {Bateman}, \citenamefont {Mahler}, \citenamefont {Okamoto},
  \citenamefont {Feizpour}, \citenamefont {Hayat},\ and\ \citenamefont
  {Steinberg}}]{rozema2014scalable}%
  \BibitemOpen
  \bibfield  {author} {\bibinfo {author} {\bibfnamefont {L.~A.}\ \bibnamefont
  {Rozema}}, \bibinfo {author} {\bibfnamefont {J.~D.}\ \bibnamefont {Bateman}},
  \bibinfo {author} {\bibfnamefont {D.~H.}\ \bibnamefont {Mahler}}, \bibinfo
  {author} {\bibfnamefont {R.}~\bibnamefont {Okamoto}}, \bibinfo {author}
  {\bibfnamefont {A.}~\bibnamefont {Feizpour}}, \bibinfo {author}
  {\bibfnamefont {A.}~\bibnamefont {Hayat}}, \ and\ \bibinfo {author}
  {\bibfnamefont {A.~M.}\ \bibnamefont {Steinberg}},\ }\href@noop {} {\bibfield
   {journal} {\bibinfo  {journal} {Phys. Rev. Lett.}\ }\textbf {\bibinfo
  {volume} {112}},\ \bibinfo {pages} {223602} (\bibinfo {year}
  {2014})}\BibitemShut {NoStop}%
\bibitem [{\citenamefont {Xu}\ \emph {et~al.}(2015)\citenamefont {Xu},
  \citenamefont {Song}, \citenamefont {Li}, \citenamefont {Zhang},
  \citenamefont {Wang}, \citenamefont {Xiong},\ and\ \citenamefont
  {Wang}}]{xu2015experimental}%
  \BibitemOpen
  \bibfield  {author} {\bibinfo {author} {\bibfnamefont {D.-Q.}\ \bibnamefont
  {Xu}}, \bibinfo {author} {\bibfnamefont {X.-B.}\ \bibnamefont {Song}},
  \bibinfo {author} {\bibfnamefont {H.-G.}\ \bibnamefont {Li}}, \bibinfo
  {author} {\bibfnamefont {D.-J.}\ \bibnamefont {Zhang}}, \bibinfo {author}
  {\bibfnamefont {H.-B.}\ \bibnamefont {Wang}}, \bibinfo {author}
  {\bibfnamefont {J.}~\bibnamefont {Xiong}}, \ and\ \bibinfo {author}
  {\bibfnamefont {K.}~\bibnamefont {Wang}},\ }\href@noop {} {\bibfield
  {journal} {\bibinfo  {journal} {Appl. Phys. Lett.}\ }\textbf {\bibinfo
  {volume} {106}},\ \bibinfo {pages} {171104} (\bibinfo {year}
  {2015})}\BibitemShut {NoStop}%
\bibitem [{\citenamefont {Hong}\ and\ \citenamefont
  {Zhang}(2017)}]{hong2017heisenberg}%
  \BibitemOpen
  \bibfield  {author} {\bibinfo {author} {\bibfnamefont {P.}~\bibnamefont
  {Hong}}\ and\ \bibinfo {author} {\bibfnamefont {G.}~\bibnamefont {Zhang}},\
  }\href@noop {} {\bibfield  {journal} {\bibinfo  {journal} {Opt. Express}\
  }\textbf {\bibinfo {volume} {25}},\ \bibinfo {pages} {22789} (\bibinfo {year}
  {2017})}\BibitemShut {NoStop}%
\bibitem [{\citenamefont {Bj{\"o}rk}\ \emph {et~al.}(2001)\citenamefont
  {Bj{\"o}rk}, \citenamefont {S{\'a}nchez-Soto},\ and\ \citenamefont
  {S{\"o}derholm}}]{bjork2001entangled}%
  \BibitemOpen
  \bibfield  {author} {\bibinfo {author} {\bibfnamefont {G.}~\bibnamefont
  {Bj{\"o}rk}}, \bibinfo {author} {\bibfnamefont {L.~L.}\ \bibnamefont
  {S{\'a}nchez-Soto}}, \ and\ \bibinfo {author} {\bibfnamefont
  {J.}~\bibnamefont {S{\"o}derholm}},\ }\href@noop {} {\bibfield  {journal}
  {\bibinfo  {journal} {Phys. Rev. Lett.}\ }\textbf {\bibinfo {volume} {86}},\
  \bibinfo {pages} {4516} (\bibinfo {year} {2001})}\BibitemShut {NoStop}%
\bibitem [{\citenamefont {D'Angelo}\ \emph {et~al.}(2001)\citenamefont
  {D'Angelo}, \citenamefont {Chekhova},\ and\ \citenamefont {Shih}}]{d2001two}%
  \BibitemOpen
  \bibfield  {author} {\bibinfo {author} {\bibfnamefont {M.}~\bibnamefont
  {D'Angelo}}, \bibinfo {author} {\bibfnamefont {M.~V.}\ \bibnamefont
  {Chekhova}}, \ and\ \bibinfo {author} {\bibfnamefont {Y.}~\bibnamefont
  {Shih}},\ }\href@noop {} {\bibfield  {journal} {\bibinfo  {journal} {Phys.
  Rev. Lett.}\ }\textbf {\bibinfo {volume} {87}},\ \bibinfo {pages} {013602}
  (\bibinfo {year} {2001})}\BibitemShut {NoStop}%
\bibitem [{\citenamefont {Kothe}\ \emph {et~al.}(2011)\citenamefont {Kothe},
  \citenamefont {Bj{\"o}rk}, \citenamefont {Inoue},\ and\ \citenamefont
  {Bourennane}}]{kothe2011efficiency}%
  \BibitemOpen
  \bibfield  {author} {\bibinfo {author} {\bibfnamefont {C.}~\bibnamefont
  {Kothe}}, \bibinfo {author} {\bibfnamefont {G.}~\bibnamefont {Bj{\"o}rk}},
  \bibinfo {author} {\bibfnamefont {S.}~\bibnamefont {Inoue}}, \ and\ \bibinfo
  {author} {\bibfnamefont {M.}~\bibnamefont {Bourennane}},\ }\href@noop {}
  {\bibfield  {journal} {\bibinfo  {journal} {New J. Phys.}\ }\textbf {\bibinfo
  {volume} {13}},\ \bibinfo {pages} {043028} (\bibinfo {year}
  {2011})}\BibitemShut {NoStop}%
\bibitem [{\citenamefont {Abouraddy}\ \emph {et~al.}(2002)\citenamefont
  {Abouraddy}, \citenamefont {Toussaint}, \citenamefont {Sergienko},
  \citenamefont {Saleh},\ and\ \citenamefont {Teich}}]{abouraddy2002entangled}%
  \BibitemOpen
  \bibfield  {author} {\bibinfo {author} {\bibfnamefont {A.~F.}\ \bibnamefont
  {Abouraddy}}, \bibinfo {author} {\bibfnamefont {K.~C.}\ \bibnamefont
  {Toussaint}}, \bibinfo {author} {\bibfnamefont {A.~V.}\ \bibnamefont
  {Sergienko}}, \bibinfo {author} {\bibfnamefont {B.~E.~A.}\ \bibnamefont
  {Saleh}}, \ and\ \bibinfo {author} {\bibfnamefont {M.~C.}\ \bibnamefont
  {Teich}},\ }\href@noop {} {\bibfield  {journal} {\bibinfo  {journal} {J. Opt.
  Soc. Am. B}\ }\textbf {\bibinfo {volume} {19}},\ \bibinfo {pages} {656}
  (\bibinfo {year} {2002})}\BibitemShut {NoStop}%
\bibitem [{\citenamefont {Graham}\ \emph {et~al.}(2006)\citenamefont {Graham},
  \citenamefont {Parkins},\ and\ \citenamefont
  {Watkins}}]{graham2006ellipsometry}%
  \BibitemOpen
  \bibfield  {author} {\bibinfo {author} {\bibfnamefont {D.~J.~L.}\
  \bibnamefont {Graham}}, \bibinfo {author} {\bibfnamefont {A.~S.}\
  \bibnamefont {Parkins}}, \ and\ \bibinfo {author} {\bibfnamefont {L.~R.}\
  \bibnamefont {Watkins}},\ }\href@noop {} {\bibfield  {journal} {\bibinfo
  {journal} {Opt. Express}\ }\textbf {\bibinfo {volume} {14}},\ \bibinfo
  {pages} {7037} (\bibinfo {year} {2006})}\BibitemShut {NoStop}%
\bibitem [{\citenamefont {Nasr}\ \emph {et~al.}(2003)\citenamefont {Nasr},
  \citenamefont {Saleh}, \citenamefont {Sergienko},\ and\ \citenamefont
  {Teich}}]{nasr2003demonstration}%
  \BibitemOpen
  \bibfield  {author} {\bibinfo {author} {\bibfnamefont {M.~B.}\ \bibnamefont
  {Nasr}}, \bibinfo {author} {\bibfnamefont {B.~E.~A.}\ \bibnamefont {Saleh}},
  \bibinfo {author} {\bibfnamefont {A.~V.}\ \bibnamefont {Sergienko}}, \ and\
  \bibinfo {author} {\bibfnamefont {M.~C.}\ \bibnamefont {Teich}},\ }\href@noop
  {} {\bibfield  {journal} {\bibinfo  {journal} {Phys. Rev. Lett.}\ }\textbf
  {\bibinfo {volume} {91}},\ \bibinfo {pages} {083601} (\bibinfo {year}
  {2003})}\BibitemShut {NoStop}%
\bibitem [{\citenamefont {Booth}\ \emph {et~al.}(2011)\citenamefont {Booth},
  \citenamefont {Saleh},\ and\ \citenamefont {Teich}}]{booth2011polarization}%
  \BibitemOpen
  \bibfield  {author} {\bibinfo {author} {\bibfnamefont {M.~C.}\ \bibnamefont
  {Booth}}, \bibinfo {author} {\bibfnamefont {B.~E.~A.}\ \bibnamefont {Saleh}},
  \ and\ \bibinfo {author} {\bibfnamefont {M.~C.}\ \bibnamefont {Teich}},\
  }\href@noop {} {\bibfield  {journal} {\bibinfo  {journal} {Opt. Commun.}\
  }\textbf {\bibinfo {volume} {284}},\ \bibinfo {pages} {2542} (\bibinfo {year}
  {2011})}\BibitemShut {NoStop}%
\bibitem [{\citenamefont {Mazurek}\ \emph {et~al.}(2013)\citenamefont
  {Mazurek}, \citenamefont {Schreiter}, \citenamefont {Prevedel}, \citenamefont
  {Kaltenbaek},\ and\ \citenamefont {Resch}}]{mazurek2013dispersion}%
  \BibitemOpen
  \bibfield  {author} {\bibinfo {author} {\bibfnamefont {M.~D.}\ \bibnamefont
  {Mazurek}}, \bibinfo {author} {\bibfnamefont {K.~M.}\ \bibnamefont
  {Schreiter}}, \bibinfo {author} {\bibfnamefont {R.}~\bibnamefont {Prevedel}},
  \bibinfo {author} {\bibfnamefont {R.}~\bibnamefont {Kaltenbaek}}, \ and\
  \bibinfo {author} {\bibfnamefont {K.~J.}\ \bibnamefont {Resch}},\ }\href@noop
  {} {\bibfield  {journal} {\bibinfo  {journal} {Sci. Rep.}\ }\textbf {\bibinfo
  {volume} {3}},\ \bibinfo {pages} {1} (\bibinfo {year} {2013})}\BibitemShut
  {NoStop}%
\bibitem [{\citenamefont {Graciano}\ \emph {et~al.}(2019)\citenamefont
  {Graciano}, \citenamefont {Mart{\'\i}nez}, \citenamefont {Lopez-Mago},
  \citenamefont {Castro-Olvera}, \citenamefont {Rosete-Aguilar}, \citenamefont
  {Gardu{\~n}o-Mej{\'\i}a}, \citenamefont {Alarc{\'o}n}, \citenamefont
  {Ram{\'\i}rez},\ and\ \citenamefont {U’Ren}}]{graciano2019interference}%
  \BibitemOpen
  \bibfield  {author} {\bibinfo {author} {\bibfnamefont {P.~Y.}\ \bibnamefont
  {Graciano}}, \bibinfo {author} {\bibfnamefont {A.~M.~A.}\ \bibnamefont
  {Mart{\'\i}nez}}, \bibinfo {author} {\bibfnamefont {D.}~\bibnamefont
  {Lopez-Mago}}, \bibinfo {author} {\bibfnamefont {G.}~\bibnamefont
  {Castro-Olvera}}, \bibinfo {author} {\bibfnamefont {M.}~\bibnamefont
  {Rosete-Aguilar}}, \bibinfo {author} {\bibfnamefont {J.}~\bibnamefont
  {Gardu{\~n}o-Mej{\'\i}a}}, \bibinfo {author} {\bibfnamefont {R.~R.}\
  \bibnamefont {Alarc{\'o}n}}, \bibinfo {author} {\bibfnamefont {H.~C.}\
  \bibnamefont {Ram{\'\i}rez}}, \ and\ \bibinfo {author} {\bibfnamefont
  {A.~B.}\ \bibnamefont {U’Ren}},\ }\href@noop {} {\bibfield  {journal}
  {\bibinfo  {journal} {Sci. Rep.}\ }\textbf {\bibinfo {volume} {9}},\ \bibinfo
  {pages} {1} (\bibinfo {year} {2019})}\BibitemShut {NoStop}%
\bibitem [{\citenamefont {Valencia}\ \emph {et~al.}(2004)\citenamefont
  {Valencia}, \citenamefont {Scarcelli},\ and\ \citenamefont
  {Shih}}]{valencia2004distant}%
  \BibitemOpen
  \bibfield  {author} {\bibinfo {author} {\bibfnamefont {A.}~\bibnamefont
  {Valencia}}, \bibinfo {author} {\bibfnamefont {G.}~\bibnamefont {Scarcelli}},
  \ and\ \bibinfo {author} {\bibfnamefont {Y.}~\bibnamefont {Shih}},\
  }\href@noop {} {\bibfield  {journal} {\bibinfo  {journal} {Appl. Phys.
  Lett.}\ }\textbf {\bibinfo {volume} {85}},\ \bibinfo {pages} {2655} (\bibinfo
  {year} {2004})}\BibitemShut {NoStop}%
\bibitem [{\citenamefont {Fink}\ \emph {et~al.}(2019)\citenamefont {Fink},
  \citenamefont {Steinlechner}, \citenamefont {Handsteiner}, \citenamefont
  {Dowling}, \citenamefont {Scheidl},\ and\ \citenamefont
  {Ursin}}]{fink2019entanglement}%
  \BibitemOpen
  \bibfield  {author} {\bibinfo {author} {\bibfnamefont {M.}~\bibnamefont
  {Fink}}, \bibinfo {author} {\bibfnamefont {F.}~\bibnamefont {Steinlechner}},
  \bibinfo {author} {\bibfnamefont {J.}~\bibnamefont {Handsteiner}}, \bibinfo
  {author} {\bibfnamefont {J.~P.}\ \bibnamefont {Dowling}}, \bibinfo {author}
  {\bibfnamefont {T.}~\bibnamefont {Scheidl}}, \ and\ \bibinfo {author}
  {\bibfnamefont {R.}~\bibnamefont {Ursin}},\ }\href@noop {} {\bibfield
  {journal} {\bibinfo  {journal} {New J. Phys.}\ }\textbf {\bibinfo {volume}
  {21}},\ \bibinfo {pages} {053010} (\bibinfo {year} {2019})}\BibitemShut
  {NoStop}%
\bibitem [{\citenamefont {Jeong}\ \emph {et~al.}(2016)\citenamefont {Jeong},
  \citenamefont {Hong},\ and\ \citenamefont {Kim}}]{jeong2016bright}%
  \BibitemOpen
  \bibfield  {author} {\bibinfo {author} {\bibfnamefont {Y.-C.}\ \bibnamefont
  {Jeong}}, \bibinfo {author} {\bibfnamefont {K.-H.}\ \bibnamefont {Hong}}, \
  and\ \bibinfo {author} {\bibfnamefont {Y.-H.}\ \bibnamefont {Kim}},\
  }\href@noop {} {\bibfield  {journal} {\bibinfo  {journal} {Opt. Express}\
  }\textbf {\bibinfo {volume} {24}},\ \bibinfo {pages} {1165} (\bibinfo {year}
  {2016})}\BibitemShut {NoStop}%
\bibitem [{\citenamefont {Altepeter}\ \emph
  {et~al.}(2005{\natexlab{b}})\citenamefont {Altepeter}, \citenamefont
  {Jeffrey},\ and\ \citenamefont {Kwiat}}]{altepeter2005phase}%
  \BibitemOpen
  \bibfield  {author} {\bibinfo {author} {\bibfnamefont {J.~B.}\ \bibnamefont
  {Altepeter}}, \bibinfo {author} {\bibfnamefont {E.~R.}\ \bibnamefont
  {Jeffrey}}, \ and\ \bibinfo {author} {\bibfnamefont {P.~G.}\ \bibnamefont
  {Kwiat}},\ }\href@noop {} {\bibfield  {journal} {\bibinfo  {journal} {Opt.
  Express}\ }\textbf {\bibinfo {volume} {13}},\ \bibinfo {pages} {8951}
  (\bibinfo {year} {2005}{\natexlab{b}})}\BibitemShut {NoStop}%
\bibitem [{\citenamefont {Lohrmann}\ \emph {et~al.}(2018)\citenamefont
  {Lohrmann}, \citenamefont {Villar}, \citenamefont {Stolk},\ and\
  \citenamefont {Ling}}]{lohrmann2018high}%
  \BibitemOpen
  \bibfield  {author} {\bibinfo {author} {\bibfnamefont {A.}~\bibnamefont
  {Lohrmann}}, \bibinfo {author} {\bibfnamefont {A.}~\bibnamefont {Villar}},
  \bibinfo {author} {\bibfnamefont {A.}~\bibnamefont {Stolk}}, \ and\ \bibinfo
  {author} {\bibfnamefont {A.}~\bibnamefont {Ling}},\ }\href@noop {} {\bibfield
   {journal} {\bibinfo  {journal} {Appl. Phys. Lett.}\ }\textbf {\bibinfo
  {volume} {113}},\ \bibinfo {pages} {171109} (\bibinfo {year}
  {2018})}\BibitemShut {NoStop}%
\bibitem [{\citenamefont {Smith}\ \emph {et~al.}(2012)\citenamefont {Smith},
  \citenamefont {Gillett}, \citenamefont {De~Almeida}, \citenamefont
  {Branciard}, \citenamefont {Fedrizzi}, \citenamefont {Weinhold},
  \citenamefont {Lita}, \citenamefont {Calkins}, \citenamefont {Gerrits},
  \citenamefont {Wiseman} \emph {et~al.}}]{smith2012conclusive}%
  \BibitemOpen
  \bibfield  {author} {\bibinfo {author} {\bibfnamefont {D.~H.}\ \bibnamefont
  {Smith}}, \bibinfo {author} {\bibfnamefont {G.}~\bibnamefont {Gillett}},
  \bibinfo {author} {\bibfnamefont {M.~P.}\ \bibnamefont {De~Almeida}},
  \bibinfo {author} {\bibfnamefont {C.}~\bibnamefont {Branciard}}, \bibinfo
  {author} {\bibfnamefont {A.}~\bibnamefont {Fedrizzi}}, \bibinfo {author}
  {\bibfnamefont {T.~J.}\ \bibnamefont {Weinhold}}, \bibinfo {author}
  {\bibfnamefont {A.}~\bibnamefont {Lita}}, \bibinfo {author} {\bibfnamefont
  {B.}~\bibnamefont {Calkins}}, \bibinfo {author} {\bibfnamefont
  {T.}~\bibnamefont {Gerrits}}, \bibinfo {author} {\bibfnamefont {H.~M.}\
  \bibnamefont {Wiseman}},  \emph {et~al.},\ }\href@noop {} {\bibfield
  {journal} {\bibinfo  {journal} {Nat. Commun.}\ }\textbf {\bibinfo {volume}
  {3}},\ \bibinfo {pages} {625} (\bibinfo {year} {2012})}\BibitemShut {NoStop}%
\bibitem [{\citenamefont {Ramelow}\ \emph {et~al.}(2013)\citenamefont
  {Ramelow}, \citenamefont {Mech}, \citenamefont {Giustina}, \citenamefont
  {Gr{\"o}blacher}, \citenamefont {Wieczorek}, \citenamefont {Beyer},
  \citenamefont {Lita}, \citenamefont {Calkins}, \citenamefont {Gerrits},
  \citenamefont {Nam} \emph {et~al.}}]{ramelow2013highly}%
  \BibitemOpen
  \bibfield  {author} {\bibinfo {author} {\bibfnamefont {S.}~\bibnamefont
  {Ramelow}}, \bibinfo {author} {\bibfnamefont {A.}~\bibnamefont {Mech}},
  \bibinfo {author} {\bibfnamefont {M.}~\bibnamefont {Giustina}}, \bibinfo
  {author} {\bibfnamefont {S.}~\bibnamefont {Gr{\"o}blacher}}, \bibinfo
  {author} {\bibfnamefont {W.}~\bibnamefont {Wieczorek}}, \bibinfo {author}
  {\bibfnamefont {J.}~\bibnamefont {Beyer}}, \bibinfo {author} {\bibfnamefont
  {A.}~\bibnamefont {Lita}}, \bibinfo {author} {\bibfnamefont {B.}~\bibnamefont
  {Calkins}}, \bibinfo {author} {\bibfnamefont {T.}~\bibnamefont {Gerrits}},
  \bibinfo {author} {\bibfnamefont {S.~W.}\ \bibnamefont {Nam}},  \emph
  {et~al.},\ }\href@noop {} {\bibfield  {journal} {\bibinfo  {journal} {Opt.
  Express}\ }\textbf {\bibinfo {volume} {21}},\ \bibinfo {pages} {6707}
  (\bibinfo {year} {2013})}\BibitemShut {NoStop}%
\bibitem [{\citenamefont {Joshi}(2014)}]{joshi2014entangled}%
  \BibitemOpen
  \bibfield  {author} {\bibinfo {author} {\bibfnamefont {S.~K.}\ \bibnamefont
  {Joshi}},\ }\emph {\bibinfo {title} {Entangled photon pairs: Efficient
  generation and detection, and bit commitment}},\ \href@noop {} {Ph.D. thesis}
  (\bibinfo {year} {2014})\BibitemShut {NoStop}%
\bibitem [{\citenamefont {Chapman}\ \emph {et~al.}(2020)\citenamefont
  {Chapman}, \citenamefont {Au{\ss}erlechner}, \citenamefont {Frick},
  \citenamefont {Prilm{\"u}ller},\ and\ \citenamefont
  {Weihs}}]{ausserlechner2020approaching}%
  \BibitemOpen
  \bibfield  {author} {\bibinfo {author} {\bibfnamefont {R.~J.}\ \bibnamefont
  {Chapman}}, \bibinfo {author} {\bibfnamefont {S.}~\bibnamefont
  {Au{\ss}erlechner}}, \bibinfo {author} {\bibfnamefont {S.}~\bibnamefont
  {Frick}}, \bibinfo {author} {\bibfnamefont {M.}~\bibnamefont
  {Prilm{\"u}ller}}, \ and\ \bibinfo {author} {\bibfnamefont {G.}~\bibnamefont
  {Weihs}},\ }in\ \href@noop {} {\emph {\bibinfo {booktitle} {Frontiers in
  Optics}}}\ (\bibinfo {organization} {Optical Society of America},\ \bibinfo
  {year} {2020})\ pp.\ \bibinfo {pages} {FTu6D--3}\BibitemShut {NoStop}%
\bibitem [{\citenamefont {Noh}\ \emph {et~al.}(2007)\citenamefont {Noh},
  \citenamefont {Kim}, \citenamefont {Zyung},\ and\ \citenamefont
  {Kim}}]{noh2007efficient}%
  \BibitemOpen
  \bibfield  {author} {\bibinfo {author} {\bibfnamefont {T.-G.}\ \bibnamefont
  {Noh}}, \bibinfo {author} {\bibfnamefont {H.}~\bibnamefont {Kim}}, \bibinfo
  {author} {\bibfnamefont {T.}~\bibnamefont {Zyung}}, \ and\ \bibinfo {author}
  {\bibfnamefont {J.}~\bibnamefont {Kim}},\ }\href@noop {} {\bibfield
  {journal} {\bibinfo  {journal} {Appl. Phys. Lett.}\ }\textbf {\bibinfo
  {volume} {90}},\ \bibinfo {pages} {011116} (\bibinfo {year}
  {2007})}\BibitemShut {NoStop}%
\bibitem [{\citenamefont {Sauge}\ \emph {et~al.}(2008)\citenamefont {Sauge},
  \citenamefont {Swillo}, \citenamefont {Tengner},\ and\ \citenamefont
  {Karlsson}}]{sauge2008single}%
  \BibitemOpen
  \bibfield  {author} {\bibinfo {author} {\bibfnamefont {S.}~\bibnamefont
  {Sauge}}, \bibinfo {author} {\bibfnamefont {M.}~\bibnamefont {Swillo}},
  \bibinfo {author} {\bibfnamefont {M.}~\bibnamefont {Tengner}}, \ and\
  \bibinfo {author} {\bibfnamefont {A.}~\bibnamefont {Karlsson}},\ }\href@noop
  {} {\bibfield  {journal} {\bibinfo  {journal} {Opt. Express}\ }\textbf
  {\bibinfo {volume} {16}},\ \bibinfo {pages} {9701} (\bibinfo {year}
  {2008})}\BibitemShut {NoStop}%
\bibitem [{\citenamefont {Jin}\ \emph {et~al.}(2014)\citenamefont {Jin},
  \citenamefont {Shimizu}, \citenamefont {Wakui}, \citenamefont {Fujiwara},
  \citenamefont {Yamashita}, \citenamefont {Miki}, \citenamefont {Terai},
  \citenamefont {Wang},\ and\ \citenamefont {Sasaki}}]{jin2014pulsed}%
  \BibitemOpen
  \bibfield  {author} {\bibinfo {author} {\bibfnamefont {R.-B.}\ \bibnamefont
  {Jin}}, \bibinfo {author} {\bibfnamefont {R.}~\bibnamefont {Shimizu}},
  \bibinfo {author} {\bibfnamefont {K.}~\bibnamefont {Wakui}}, \bibinfo
  {author} {\bibfnamefont {M.}~\bibnamefont {Fujiwara}}, \bibinfo {author}
  {\bibfnamefont {T.}~\bibnamefont {Yamashita}}, \bibinfo {author}
  {\bibfnamefont {S.}~\bibnamefont {Miki}}, \bibinfo {author} {\bibfnamefont
  {H.}~\bibnamefont {Terai}}, \bibinfo {author} {\bibfnamefont
  {Z.}~\bibnamefont {Wang}}, \ and\ \bibinfo {author} {\bibfnamefont
  {M.}~\bibnamefont {Sasaki}},\ }\href@noop {} {\bibfield  {journal} {\bibinfo
  {journal} {Opt. Express}\ }\textbf {\bibinfo {volume} {22}},\ \bibinfo
  {pages} {11498} (\bibinfo {year} {2014})}\BibitemShut {NoStop}%
\bibitem [{\citenamefont {Li}\ \emph {et~al.}(2015)\citenamefont {Li},
  \citenamefont {Zhou}, \citenamefont {Ding},\ and\ \citenamefont
  {Shi}}]{li2015cw}%
  \BibitemOpen
  \bibfield  {author} {\bibinfo {author} {\bibfnamefont {Y.}~\bibnamefont
  {Li}}, \bibinfo {author} {\bibfnamefont {Z.-Y.}\ \bibnamefont {Zhou}},
  \bibinfo {author} {\bibfnamefont {D.-S.}\ \bibnamefont {Ding}}, \ and\
  \bibinfo {author} {\bibfnamefont {B.-S.}\ \bibnamefont {Shi}},\ }\href@noop
  {} {\bibfield  {journal} {\bibinfo  {journal} {Opt. Express}\ }\textbf
  {\bibinfo {volume} {23}},\ \bibinfo {pages} {28792} (\bibinfo {year}
  {2015})}\BibitemShut {NoStop}%
\bibitem [{\citenamefont {Weston}\ \emph {et~al.}(2016)\citenamefont {Weston},
  \citenamefont {Chrzanowski}, \citenamefont {Wollmann}, \citenamefont
  {Boston}, \citenamefont {Ho}, \citenamefont {Shalm}, \citenamefont {Verma},
  \citenamefont {Allman}, \citenamefont {Nam}, \citenamefont {Patel} \emph
  {et~al.}}]{weston2016efficient}%
  \BibitemOpen
  \bibfield  {author} {\bibinfo {author} {\bibfnamefont {M.~M.}\ \bibnamefont
  {Weston}}, \bibinfo {author} {\bibfnamefont {H.~M.}\ \bibnamefont
  {Chrzanowski}}, \bibinfo {author} {\bibfnamefont {S.}~\bibnamefont
  {Wollmann}}, \bibinfo {author} {\bibfnamefont {A.}~\bibnamefont {Boston}},
  \bibinfo {author} {\bibfnamefont {J.}~\bibnamefont {Ho}}, \bibinfo {author}
  {\bibfnamefont {L.~K.}\ \bibnamefont {Shalm}}, \bibinfo {author}
  {\bibfnamefont {V.~B.}\ \bibnamefont {Verma}}, \bibinfo {author}
  {\bibfnamefont {M.~S.}\ \bibnamefont {Allman}}, \bibinfo {author}
  {\bibfnamefont {S.~W.}\ \bibnamefont {Nam}}, \bibinfo {author} {\bibfnamefont
  {R.~B.}\ \bibnamefont {Patel}},  \emph {et~al.},\ }\href@noop {} {\bibfield
  {journal} {\bibinfo  {journal} {Opt. Express}\ }\textbf {\bibinfo {volume}
  {24}},\ \bibinfo {pages} {10869} (\bibinfo {year} {2016})}\BibitemShut
  {NoStop}%
\bibitem [{\citenamefont {Wang}\ \emph {et~al.}(2004)\citenamefont {Wang},
  \citenamefont {Horikiri},\ and\ \citenamefont
  {Kobayashi}}]{wang2004polarization}%
  \BibitemOpen
  \bibfield  {author} {\bibinfo {author} {\bibfnamefont {H.}~\bibnamefont
  {Wang}}, \bibinfo {author} {\bibfnamefont {T.}~\bibnamefont {Horikiri}}, \
  and\ \bibinfo {author} {\bibfnamefont {T.}~\bibnamefont {Kobayashi}},\
  }\href@noop {} {\bibfield  {journal} {\bibinfo  {journal} {Phys. Rev. A}\
  }\textbf {\bibinfo {volume} {70}},\ \bibinfo {pages} {043804} (\bibinfo
  {year} {2004})}\BibitemShut {NoStop}%
\bibitem [{\citenamefont {Martin}\ \emph {et~al.}(2010)\citenamefont {Martin},
  \citenamefont {Issautier}, \citenamefont {Herrmann}, \citenamefont {Sohler},
  \citenamefont {Ostrowsky}, \citenamefont {Alibart},\ and\ \citenamefont
  {Tanzilli}}]{martin2010polarization}%
  \BibitemOpen
  \bibfield  {author} {\bibinfo {author} {\bibfnamefont {A.}~\bibnamefont
  {Martin}}, \bibinfo {author} {\bibfnamefont {A.}~\bibnamefont {Issautier}},
  \bibinfo {author} {\bibfnamefont {H.}~\bibnamefont {Herrmann}}, \bibinfo
  {author} {\bibfnamefont {W.}~\bibnamefont {Sohler}}, \bibinfo {author}
  {\bibfnamefont {D.~B.}\ \bibnamefont {Ostrowsky}}, \bibinfo {author}
  {\bibfnamefont {O.}~\bibnamefont {Alibart}}, \ and\ \bibinfo {author}
  {\bibfnamefont {S.}~\bibnamefont {Tanzilli}},\ }\href@noop {} {\bibfield
  {journal} {\bibinfo  {journal} {New J. Phys.}\ }\textbf {\bibinfo {volume}
  {12}},\ \bibinfo {pages} {103005} (\bibinfo {year} {2010})}\BibitemShut
  {NoStop}%
\bibitem [{\citenamefont {Zhong}\ \emph {et~al.}(2010)\citenamefont {Zhong},
  \citenamefont {Hu}, \citenamefont {Wong}, \citenamefont {Berggren},
  \citenamefont {Roberts},\ and\ \citenamefont {Battle}}]{zhong2010high}%
  \BibitemOpen
  \bibfield  {author} {\bibinfo {author} {\bibfnamefont {T.}~\bibnamefont
  {Zhong}}, \bibinfo {author} {\bibfnamefont {X.}~\bibnamefont {Hu}}, \bibinfo
  {author} {\bibfnamefont {F.~N.~C.}\ \bibnamefont {Wong}}, \bibinfo {author}
  {\bibfnamefont {K.~K.}\ \bibnamefont {Berggren}}, \bibinfo {author}
  {\bibfnamefont {T.~D.}\ \bibnamefont {Roberts}}, \ and\ \bibinfo {author}
  {\bibfnamefont {P.}~\bibnamefont {Battle}},\ }\href@noop {} {\bibfield
  {journal} {\bibinfo  {journal} {Opt. Lett.}\ }\textbf {\bibinfo {volume}
  {35}},\ \bibinfo {pages} {1392} (\bibinfo {year} {2010})}\BibitemShut
  {NoStop}%
\bibitem [{\citenamefont {Clausen}\ \emph {et~al.}(2014)\citenamefont
  {Clausen}, \citenamefont {Bussieres}, \citenamefont {Tiranov}, \citenamefont
  {Herrmann}, \citenamefont {Silberhorn}, \citenamefont {Sohler}, \citenamefont
  {Afzelius},\ and\ \citenamefont {Gisin}}]{clausen2014source}%
  \BibitemOpen
  \bibfield  {author} {\bibinfo {author} {\bibfnamefont {C.}~\bibnamefont
  {Clausen}}, \bibinfo {author} {\bibfnamefont {F.}~\bibnamefont {Bussieres}},
  \bibinfo {author} {\bibfnamefont {A.}~\bibnamefont {Tiranov}}, \bibinfo
  {author} {\bibfnamefont {H.}~\bibnamefont {Herrmann}}, \bibinfo {author}
  {\bibfnamefont {C.}~\bibnamefont {Silberhorn}}, \bibinfo {author}
  {\bibfnamefont {W.}~\bibnamefont {Sohler}}, \bibinfo {author} {\bibfnamefont
  {M.}~\bibnamefont {Afzelius}}, \ and\ \bibinfo {author} {\bibfnamefont
  {N.}~\bibnamefont {Gisin}},\ }\href@noop {} {\bibfield  {journal} {\bibinfo
  {journal} {New J. Phys.}\ }\textbf {\bibinfo {volume} {16}},\ \bibinfo
  {pages} {093058} (\bibinfo {year} {2014})}\BibitemShut {NoStop}%
\bibitem [{\citenamefont {Autebert}\ \emph {et~al.}(2016)\citenamefont
  {Autebert}, \citenamefont {Trapateau}, \citenamefont {Orieux}, \citenamefont
  {Lema{\^\i}tre}, \citenamefont {Gomez-Carbonell}, \citenamefont {Diamanti},
  \citenamefont {Zaquine},\ and\ \citenamefont {Ducci}}]{autebert2016multi}%
  \BibitemOpen
  \bibfield  {author} {\bibinfo {author} {\bibfnamefont {C.}~\bibnamefont
  {Autebert}}, \bibinfo {author} {\bibfnamefont {J.}~\bibnamefont {Trapateau}},
  \bibinfo {author} {\bibfnamefont {A.}~\bibnamefont {Orieux}}, \bibinfo
  {author} {\bibfnamefont {A.}~\bibnamefont {Lema{\^\i}tre}}, \bibinfo {author}
  {\bibfnamefont {C.}~\bibnamefont {Gomez-Carbonell}}, \bibinfo {author}
  {\bibfnamefont {E.}~\bibnamefont {Diamanti}}, \bibinfo {author}
  {\bibfnamefont {I.}~\bibnamefont {Zaquine}}, \ and\ \bibinfo {author}
  {\bibfnamefont {S.}~\bibnamefont {Ducci}},\ }\href@noop {} {\bibfield
  {journal} {\bibinfo  {journal} {Quantum Science and Technology}\ }\textbf
  {\bibinfo {volume} {1}},\ \bibinfo {pages} {01LT02} (\bibinfo {year}
  {2016})}\BibitemShut {NoStop}%
\bibitem [{\citenamefont {Atzeni}\ \emph {et~al.}(2018)\citenamefont {Atzeni},
  \citenamefont {Rab}, \citenamefont {Corrielli}, \citenamefont {Polino},
  \citenamefont {Valeri}, \citenamefont {Mataloni}, \citenamefont {Spagnolo},
  \citenamefont {Crespi}, \citenamefont {Sciarrino},\ and\ \citenamefont
  {Osellame}}]{atzeni2018integrated}%
  \BibitemOpen
  \bibfield  {author} {\bibinfo {author} {\bibfnamefont {S.}~\bibnamefont
  {Atzeni}}, \bibinfo {author} {\bibfnamefont {A.~S.}\ \bibnamefont {Rab}},
  \bibinfo {author} {\bibfnamefont {G.}~\bibnamefont {Corrielli}}, \bibinfo
  {author} {\bibfnamefont {E.}~\bibnamefont {Polino}}, \bibinfo {author}
  {\bibfnamefont {M.}~\bibnamefont {Valeri}}, \bibinfo {author} {\bibfnamefont
  {P.}~\bibnamefont {Mataloni}}, \bibinfo {author} {\bibfnamefont
  {N.}~\bibnamefont {Spagnolo}}, \bibinfo {author} {\bibfnamefont
  {A.}~\bibnamefont {Crespi}}, \bibinfo {author} {\bibfnamefont
  {F.}~\bibnamefont {Sciarrino}}, \ and\ \bibinfo {author} {\bibfnamefont
  {R.}~\bibnamefont {Osellame}},\ }\href@noop {} {\bibfield  {journal}
  {\bibinfo  {journal} {Optica}\ }\textbf {\bibinfo {volume} {5}},\ \bibinfo
  {pages} {311} (\bibinfo {year} {2018})}\BibitemShut {NoStop}%
\bibitem [{\citenamefont {Jiang}\ and\ \citenamefont
  {Tomita}(2007)}]{jiang2007generation}%
  \BibitemOpen
  \bibfield  {author} {\bibinfo {author} {\bibfnamefont {Y.-K.}\ \bibnamefont
  {Jiang}}\ and\ \bibinfo {author} {\bibfnamefont {A.}~\bibnamefont {Tomita}},\
  }\href@noop {} {\bibfield  {journal} {\bibinfo  {journal} {J. Phys. B: At.
  Mol. Opt. Phys.}\ }\textbf {\bibinfo {volume} {40}},\ \bibinfo {pages} {437}
  (\bibinfo {year} {2007})}\BibitemShut {NoStop}%
\bibitem [{\citenamefont {Lim}\ \emph {et~al.}(2008)\citenamefont {Lim},
  \citenamefont {Yoshizawa}, \citenamefont {Tsuchida},\ and\ \citenamefont
  {Kikuchi}}]{lim2008stable}%
  \BibitemOpen
  \bibfield  {author} {\bibinfo {author} {\bibfnamefont {H.~C.}\ \bibnamefont
  {Lim}}, \bibinfo {author} {\bibfnamefont {A.}~\bibnamefont {Yoshizawa}},
  \bibinfo {author} {\bibfnamefont {H.}~\bibnamefont {Tsuchida}}, \ and\
  \bibinfo {author} {\bibfnamefont {K.}~\bibnamefont {Kikuchi}},\ }\href@noop
  {} {\bibfield  {journal} {\bibinfo  {journal} {Opt. Express}\ }\textbf
  {\bibinfo {volume} {16}},\ \bibinfo {pages} {12460} (\bibinfo {year}
  {2008})}\BibitemShut {NoStop}%
\bibitem [{\citenamefont {Arahira}\ \emph {et~al.}(2011)\citenamefont
  {Arahira}, \citenamefont {Namekata}, \citenamefont {Kishimoto}, \citenamefont
  {Yaegashi},\ and\ \citenamefont {Inoue}}]{arahira2011generation}%
  \BibitemOpen
  \bibfield  {author} {\bibinfo {author} {\bibfnamefont {S.}~\bibnamefont
  {Arahira}}, \bibinfo {author} {\bibfnamefont {N.}~\bibnamefont {Namekata}},
  \bibinfo {author} {\bibfnamefont {T.}~\bibnamefont {Kishimoto}}, \bibinfo
  {author} {\bibfnamefont {H.}~\bibnamefont {Yaegashi}}, \ and\ \bibinfo
  {author} {\bibfnamefont {S.}~\bibnamefont {Inoue}},\ }\href@noop {}
  {\bibfield  {journal} {\bibinfo  {journal} {Opt. Express}\ }\textbf {\bibinfo
  {volume} {19}},\ \bibinfo {pages} {16032} (\bibinfo {year}
  {2011})}\BibitemShut {NoStop}%
\bibitem [{\citenamefont {Sansoni}\ \emph {et~al.}(2017)\citenamefont
  {Sansoni}, \citenamefont {Luo}, \citenamefont {Eigner}, \citenamefont
  {Ricken}, \citenamefont {Quiring}, \citenamefont {Herrmann},\ and\
  \citenamefont {Silberhorn}}]{sansoni2017two}%
  \BibitemOpen
  \bibfield  {author} {\bibinfo {author} {\bibfnamefont {L.}~\bibnamefont
  {Sansoni}}, \bibinfo {author} {\bibfnamefont {K.~H.}\ \bibnamefont {Luo}},
  \bibinfo {author} {\bibfnamefont {C.}~\bibnamefont {Eigner}}, \bibinfo
  {author} {\bibfnamefont {R.}~\bibnamefont {Ricken}}, \bibinfo {author}
  {\bibfnamefont {V.}~\bibnamefont {Quiring}}, \bibinfo {author} {\bibfnamefont
  {H.}~\bibnamefont {Herrmann}}, \ and\ \bibinfo {author} {\bibfnamefont
  {C.}~\bibnamefont {Silberhorn}},\ }\href@noop {} {\bibfield  {journal}
  {\bibinfo  {journal} {npj Quant. Inf.}\ }\textbf {\bibinfo {volume} {3}},\
  \bibinfo {pages} {1} (\bibinfo {year} {2017})}\BibitemShut {NoStop}%
\bibitem [{\citenamefont {Schlager}\ \emph {et~al.}(2017)\citenamefont
  {Schlager}, \citenamefont {Pressl}, \citenamefont {Laiho}, \citenamefont
  {Suchomel}, \citenamefont {Kamp}, \citenamefont {H{\"o}fling}, \citenamefont
  {Schneider},\ and\ \citenamefont {Weihs}}]{schlager2017temporally}%
  \BibitemOpen
  \bibfield  {author} {\bibinfo {author} {\bibfnamefont {A.}~\bibnamefont
  {Schlager}}, \bibinfo {author} {\bibfnamefont {B.}~\bibnamefont {Pressl}},
  \bibinfo {author} {\bibfnamefont {K.}~\bibnamefont {Laiho}}, \bibinfo
  {author} {\bibfnamefont {H.}~\bibnamefont {Suchomel}}, \bibinfo {author}
  {\bibfnamefont {M.}~\bibnamefont {Kamp}}, \bibinfo {author} {\bibfnamefont
  {S.}~\bibnamefont {H{\"o}fling}}, \bibinfo {author} {\bibfnamefont
  {C.}~\bibnamefont {Schneider}}, \ and\ \bibinfo {author} {\bibfnamefont
  {G.}~\bibnamefont {Weihs}},\ }\href@noop {} {\bibfield  {journal} {\bibinfo
  {journal} {Opt. Lett.}\ }\textbf {\bibinfo {volume} {42}},\ \bibinfo {pages}
  {2102} (\bibinfo {year} {2017})}\BibitemShut {NoStop}%
\bibitem [{\citenamefont {Chen}\ \emph
  {et~al.}(2018{\natexlab{b}})\citenamefont {Chen}, \citenamefont {Riazi},
  \citenamefont {Zhu}, \citenamefont {Ng}, \citenamefont {Gladyshev},
  \citenamefont {Kazansky},\ and\ \citenamefont {Qian}}]{chen2018turn}%
  \BibitemOpen
  \bibfield  {author} {\bibinfo {author} {\bibfnamefont {C.}~\bibnamefont
  {Chen}}, \bibinfo {author} {\bibfnamefont {A.}~\bibnamefont {Riazi}},
  \bibinfo {author} {\bibfnamefont {E.~Y.}\ \bibnamefont {Zhu}}, \bibinfo
  {author} {\bibfnamefont {M.}~\bibnamefont {Ng}}, \bibinfo {author}
  {\bibfnamefont {A.~V.}\ \bibnamefont {Gladyshev}}, \bibinfo {author}
  {\bibfnamefont {P.~G.}\ \bibnamefont {Kazansky}}, \ and\ \bibinfo {author}
  {\bibfnamefont {L.}~\bibnamefont {Qian}},\ }\href@noop {} {\bibfield
  {journal} {\bibinfo  {journal} {OSA Cont.}\ }\textbf {\bibinfo {volume}
  {1}},\ \bibinfo {pages} {981} (\bibinfo {year}
  {2018}{\natexlab{b}})}\BibitemShut {NoStop}%
\bibitem [{\citenamefont {Ono}\ \emph {et~al.}(2013)\citenamefont {Ono},
  \citenamefont {Okamoto},\ and\ \citenamefont
  {Takeuchi}}]{ono2013entanglement}%
  \BibitemOpen
  \bibfield  {author} {\bibinfo {author} {\bibfnamefont {T.}~\bibnamefont
  {Ono}}, \bibinfo {author} {\bibfnamefont {R.}~\bibnamefont {Okamoto}}, \ and\
  \bibinfo {author} {\bibfnamefont {S.}~\bibnamefont {Takeuchi}},\ }\href@noop
  {} {\bibfield  {journal} {\bibinfo  {journal} {Nat. Commun.}\ }\textbf
  {\bibinfo {volume} {4}},\ \bibinfo {pages} {1} (\bibinfo {year}
  {2013})}\BibitemShut {NoStop}%
\bibitem [{\citenamefont {Liu}\ \emph {et~al.}(2016)\citenamefont {Liu},
  \citenamefont {Cao}, \citenamefont {Wu}, \citenamefont {Fukuda},
  \citenamefont {You}, \citenamefont {Zhong}, \citenamefont {Numata},
  \citenamefont {Chen}, \citenamefont {Zhang}, \citenamefont {Shi} \emph
  {et~al.}}]{liu2016experimental}%
  \BibitemOpen
  \bibfield  {author} {\bibinfo {author} {\bibfnamefont {Y.}~\bibnamefont
  {Liu}}, \bibinfo {author} {\bibfnamefont {Z.}~\bibnamefont {Cao}}, \bibinfo
  {author} {\bibfnamefont {C.}~\bibnamefont {Wu}}, \bibinfo {author}
  {\bibfnamefont {D.}~\bibnamefont {Fukuda}}, \bibinfo {author} {\bibfnamefont
  {L.}~\bibnamefont {You}}, \bibinfo {author} {\bibfnamefont {J.}~\bibnamefont
  {Zhong}}, \bibinfo {author} {\bibfnamefont {T.}~\bibnamefont {Numata}},
  \bibinfo {author} {\bibfnamefont {S.}~\bibnamefont {Chen}}, \bibinfo {author}
  {\bibfnamefont {W.}~\bibnamefont {Zhang}}, \bibinfo {author} {\bibfnamefont
  {S.-C.}\ \bibnamefont {Shi}},  \emph {et~al.},\ }\href@noop {} {\bibfield
  {journal} {\bibinfo  {journal} {Phys. Rev. A}\ }\textbf {\bibinfo {volume}
  {94}},\ \bibinfo {pages} {020301} (\bibinfo {year} {2016})}\BibitemShut
  {NoStop}%
\bibitem [{\citenamefont {Shi}\ \emph {et~al.}(2020)\citenamefont {Shi},
  \citenamefont {Thar}, \citenamefont {Poh}, \citenamefont {Grieve},
  \citenamefont {Kurtsiefer},\ and\ \citenamefont {Ling}}]{shi2020stable}%
  \BibitemOpen
  \bibfield  {author} {\bibinfo {author} {\bibfnamefont {Y.}~\bibnamefont
  {Shi}}, \bibinfo {author} {\bibfnamefont {S.~M.}\ \bibnamefont {Thar}},
  \bibinfo {author} {\bibfnamefont {H.~S.}\ \bibnamefont {Poh}}, \bibinfo
  {author} {\bibfnamefont {J.~A.}\ \bibnamefont {Grieve}}, \bibinfo {author}
  {\bibfnamefont {C.}~\bibnamefont {Kurtsiefer}}, \ and\ \bibinfo {author}
  {\bibfnamefont {A.}~\bibnamefont {Ling}},\ }\href@noop {} {\bibfield
  {journal} {\bibinfo  {journal} {arXiv preprint arXiv:2007.01989}\ } (\bibinfo
  {year} {2020})}\BibitemShut {NoStop}%
\bibitem [{\citenamefont {Ibarra-Borja}\ \emph {et~al.}(2020)\citenamefont
  {Ibarra-Borja}, \citenamefont {Sevilla-Guti{\'e}rrez}, \citenamefont
  {Ram{\'\i}rez-Alarc{\'o}n}, \citenamefont {Cruz-Ram{\'\i}rez},\ and\
  \citenamefont {U’Ren}}]{ibarra2020experimental}%
  \BibitemOpen
  \bibfield  {author} {\bibinfo {author} {\bibfnamefont {Z.}~\bibnamefont
  {Ibarra-Borja}}, \bibinfo {author} {\bibfnamefont {C.}~\bibnamefont
  {Sevilla-Guti{\'e}rrez}}, \bibinfo {author} {\bibfnamefont {R.}~\bibnamefont
  {Ram{\'\i}rez-Alarc{\'o}n}}, \bibinfo {author} {\bibfnamefont
  {H.}~\bibnamefont {Cruz-Ram{\'\i}rez}}, \ and\ \bibinfo {author}
  {\bibfnamefont {A.~B.}\ \bibnamefont {U’Ren}},\ }\href@noop {} {\bibfield
  {journal} {\bibinfo  {journal} {Photon. Res.}\ }\textbf {\bibinfo {volume}
  {8}},\ \bibinfo {pages} {51} (\bibinfo {year} {2020})}\BibitemShut {NoStop}%
\bibitem [{\citenamefont {Kim}\ and\ \citenamefont {Grice}(2002)}]{Kim:2002aa}%
  \BibitemOpen
  \bibfield  {author} {\bibinfo {author} {\bibfnamefont {Y.-H.}\ \bibnamefont
  {Kim}}\ and\ \bibinfo {author} {\bibfnamefont {W.~P.}\ \bibnamefont
  {Grice}},\ }\href@noop {} {\bibfield  {journal} {\bibinfo  {journal} {J. Mod.
  Opt.}\ }\textbf {\bibinfo {volume} {49}},\ \bibinfo {pages} {2309} (\bibinfo
  {year} {2002})}\BibitemShut {NoStop}%
\bibitem [{\citenamefont {Yepiz-Graciano}\ \emph {et~al.}(2020)\citenamefont
  {Yepiz-Graciano}, \citenamefont {Mart{\'\i}nez}, \citenamefont {Lopez-Mago},
  \citenamefont {Cruz-Ramirez},\ and\ \citenamefont
  {U’Ren}}]{yepiz2020spectrally}%
  \BibitemOpen
  \bibfield  {author} {\bibinfo {author} {\bibfnamefont {P.}~\bibnamefont
  {Yepiz-Graciano}}, \bibinfo {author} {\bibfnamefont {A.~M.~A.}\ \bibnamefont
  {Mart{\'\i}nez}}, \bibinfo {author} {\bibfnamefont {D.}~\bibnamefont
  {Lopez-Mago}}, \bibinfo {author} {\bibfnamefont {H.}~\bibnamefont
  {Cruz-Ramirez}}, \ and\ \bibinfo {author} {\bibfnamefont {A.~B.}\
  \bibnamefont {U’Ren}},\ }\href@noop {} {\bibfield  {journal} {\bibinfo
  {journal} {Photon. Res.}\ }\textbf {\bibinfo {volume} {8}},\ \bibinfo {pages}
  {1023} (\bibinfo {year} {2020})}\BibitemShut {NoStop}%
\bibitem [{\citenamefont {Jin}\ \emph {et~al.}(2015)\citenamefont {Jin},
  \citenamefont {Takeoka}, \citenamefont {Takagi}, \citenamefont {Shimizu},\
  and\ \citenamefont {Sasaki}}]{Jin:2015aa}%
  \BibitemOpen
  \bibfield  {author} {\bibinfo {author} {\bibfnamefont {R.-B.}\ \bibnamefont
  {Jin}}, \bibinfo {author} {\bibfnamefont {M.}~\bibnamefont {Takeoka}},
  \bibinfo {author} {\bibfnamefont {U.}~\bibnamefont {Takagi}}, \bibinfo
  {author} {\bibfnamefont {R.}~\bibnamefont {Shimizu}}, \ and\ \bibinfo
  {author} {\bibfnamefont {M.}~\bibnamefont {Sasaki}},\ }\href {\doibase
  10.1038/srep09333} {\bibfield  {journal} {\bibinfo  {journal} {Sci. Rep.}\
  }\textbf {\bibinfo {volume} {5}},\ \bibinfo {pages} {9333} (\bibinfo {year}
  {2015})}\BibitemShut {NoStop}%
\bibitem [{\citenamefont {Pirandola}\ \emph {et~al.}(2015)\citenamefont
  {Pirandola}, \citenamefont {Eisert}, \citenamefont {Weedbrook}, \citenamefont
  {Furusawa},\ and\ \citenamefont {Braunstein}}]{pirandola2015advances}%
  \BibitemOpen
  \bibfield  {author} {\bibinfo {author} {\bibfnamefont {S.}~\bibnamefont
  {Pirandola}}, \bibinfo {author} {\bibfnamefont {J.}~\bibnamefont {Eisert}},
  \bibinfo {author} {\bibfnamefont {C.}~\bibnamefont {Weedbrook}}, \bibinfo
  {author} {\bibfnamefont {A.}~\bibnamefont {Furusawa}}, \ and\ \bibinfo
  {author} {\bibfnamefont {S.~L.}\ \bibnamefont {Braunstein}},\ }\href@noop {}
  {\bibfield  {journal} {\bibinfo  {journal} {Nat. Photonics}\ }\textbf
  {\bibinfo {volume} {9}},\ \bibinfo {pages} {641} (\bibinfo {year}
  {2015})}\BibitemShut {NoStop}%
\bibitem [{\citenamefont {Hamel}\ \emph {et~al.}(2014)\citenamefont {Hamel},
  \citenamefont {Shalm}, \citenamefont {H{\"u}bel}, \citenamefont {Miller},
  \citenamefont {Marsili}, \citenamefont {Verma}, \citenamefont {Mirin},
  \citenamefont {Nam}, \citenamefont {Resch},\ and\ \citenamefont
  {Jennewein}}]{Hamel:2014qf}%
  \BibitemOpen
  \bibfield  {author} {\bibinfo {author} {\bibfnamefont {D.~R.}\ \bibnamefont
  {Hamel}}, \bibinfo {author} {\bibfnamefont {L.~K.}\ \bibnamefont {Shalm}},
  \bibinfo {author} {\bibfnamefont {H.}~\bibnamefont {H{\"u}bel}}, \bibinfo
  {author} {\bibfnamefont {A.~J.}\ \bibnamefont {Miller}}, \bibinfo {author}
  {\bibfnamefont {F.}~\bibnamefont {Marsili}}, \bibinfo {author} {\bibfnamefont
  {V.~B.}\ \bibnamefont {Verma}}, \bibinfo {author} {\bibfnamefont {R.~P.}\
  \bibnamefont {Mirin}}, \bibinfo {author} {\bibfnamefont {S.~W.}\ \bibnamefont
  {Nam}}, \bibinfo {author} {\bibfnamefont {K.~J.}\ \bibnamefont {Resch}}, \
  and\ \bibinfo {author} {\bibfnamefont {T.}~\bibnamefont {Jennewein}},\ }\href
  {http://dx.doi.org/10.1038/nphoton.2014.218} {\bibfield  {journal} {\bibinfo
  {journal} {Nat. Photon}\ }\textbf {\bibinfo {volume} {8}},\ \bibinfo {pages}
  {801} (\bibinfo {year} {2014})}\BibitemShut {NoStop}%
\bibitem [{\citenamefont {Lopez-Mago}\ and\ \citenamefont
  {Novotny}(2012)}]{lopez2012quantum}%
  \BibitemOpen
  \bibfield  {author} {\bibinfo {author} {\bibfnamefont {D.}~\bibnamefont
  {Lopez-Mago}}\ and\ \bibinfo {author} {\bibfnamefont {L.}~\bibnamefont
  {Novotny}},\ }\href@noop {} {\bibfield  {journal} {\bibinfo  {journal} {Opt.
  Lett.}\ }\textbf {\bibinfo {volume} {37}},\ \bibinfo {pages} {4077} (\bibinfo
  {year} {2012})}\BibitemShut {NoStop}%
\bibitem [{\citenamefont {Gerrits}\ \emph {et~al.}(2015)\citenamefont
  {Gerrits}, \citenamefont {Marsili}, \citenamefont {Verma}, \citenamefont
  {Shalm}, \citenamefont {Shaw}, \citenamefont {Mirin},\ and\ \citenamefont
  {Nam}}]{Gerrits:2015aa}%
  \BibitemOpen
  \bibfield  {author} {\bibinfo {author} {\bibfnamefont {T.}~\bibnamefont
  {Gerrits}}, \bibinfo {author} {\bibfnamefont {F.}~\bibnamefont {Marsili}},
  \bibinfo {author} {\bibfnamefont {V.~B.}\ \bibnamefont {Verma}}, \bibinfo
  {author} {\bibfnamefont {L.~K.}\ \bibnamefont {Shalm}}, \bibinfo {author}
  {\bibfnamefont {M.}~\bibnamefont {Shaw}}, \bibinfo {author} {\bibfnamefont
  {R.~P.}\ \bibnamefont {Mirin}}, \ and\ \bibinfo {author} {\bibfnamefont
  {S.~W.}\ \bibnamefont {Nam}},\ }\href {\doibase 10.1103/PhysRevA.91.013830}
  {\bibfield  {journal} {\bibinfo  {journal} {Phys. Rev. A}\ }\textbf {\bibinfo
  {volume} {91}},\ \bibinfo {pages} {013830} (\bibinfo {year}
  {2015})}\BibitemShut {NoStop}%
\end{thebibliography}%


\providecommand{\noopsort}[1]{}\providecommand{\singleletter}[1]{#1}%
%

\end{document}